%% file: B2M1.tex
\documentclass[11pt]{article}
\usepackage{verbatim,amsmath,amssymb}
\usepackage{epsfig,float,color}
\usepackage[font=small,labelfont=bf]{caption}
\usepackage{geometry}
\usepackage{setspace}
\usepackage{natbib}
\usepackage{amsthm,mathrsfs,amsfonts,dsfont} 
\usepackage{bm}
\usepackage{hyperref}
\usepackage[labelfont=bf]{caption}

\definecolor{darkblue}{rgb}{0,0,1}
\definecolor{col1}{rgb}{1,0,1}
\definecolor{col2}{rgb}{0,0.5,0}
\definecolor{col3}{rgb}{0.5,0,1}
\definecolor{col4}{rgb}{0.1,.75,0}

\hypersetup{pdftex=true, colorlinks=true, breaklinks=true, linkcolor=darkblue, menucolor=darkblue, pagecolor=darkblue, citecolor=darkblue, urlcolor=darkblue}

\newtheoremstyle{rem}
{6pt}
{6pt}
{\small}
{}
{\bf}
{:}
{.5em}
{}

\theoremstyle{rem}
\newtheorem{remark}{Remark}[section]

\input{neco.tex}

\begin{document}

\begin{center}
\Large{\bf{
A simple and efficient hybrid discretization approach to alleviate membrane locking in isogeometric thin shells
}}\\

\end{center}

\renewcommand{\thefootnote}{\fnsymbol{footnote}}

\begin{center}
\large{Roger A.~Sauer$^{\mra,\mrb,\mrc,}$\footnote[1]{corresponding author, email: roger.sauer@pg.edu.pl, sauer@aices.rwth-aachen.de},
Zhihui Zou$^\mrd$ 
and Thomas J.R.~Hughes$^\mrd$
}\\
\vspace{4mm}

\small{\textit{
$^\mra$Faculty of Civil and Environmental Engineering, Gda\'{n}sk University of Technology, Poland \\[1.1mm]
$^\mrb$Aachen Institute for Advanced Study in Computational Engineering Science (AICES), \\ 
RWTH Aachen University, Germany \\[1.1mm]
$^\mrc$Dept.~of Mechanical Engineering, Indian Institute of Technology Guwahati, India \\[1.1mm]
$^\mrd$Oden Institute for Computational Engineering and Sciences, UT Austin, USA
}}

\vspace{3mm}

\small{Published\footnote[2]{This pdf is the personal version of an article whose journal version is available at \href{https://doi.org/10.1016/j.cma.2024.116869}{www.sciencedirect.com}} 
in \textit{Comput.~Methods Appl.~Mech.~Eng.}, \href{https://doi.org/10.1016/j.cma.2024.116869}{DOI: 10.1016/j.cma.2024.116869} \\
Submitted on 28 December 2023; Revised on 7 February 2024; Accepted on 19 February 2024} 

\end{center}

\renewcommand{\thefootnote}{\arabic{footnote}}

\vspace{-4mm}


\rule{\linewidth}{.15mm}
{\bf Abstract:}
This work presents a new hybrid discretization approach to alleviate membrane locking in isogeometric finite element formulations for Kirchhoff-Love shells.
The approach is simple, and requires no additional dofs and no static condensation. 
It does not increase the bandwidth of the tangent matrix and is effective for both linear and nonlinear problems.
It combines isogeometric surface discretizations with classical Lagrange-based surface discretizations, and can thus be run with existing isogeometric finite element codes.
Also, the stresses can be recovered straightforwardly.
The effectiveness of the proposed approach in alleviating, if not eliminating, membrane locking is demonstrated through 
the rigorous study of the convergence behavior of 
several classical benchmark problems.
Accuracy gains are particularly large in the membrane stresses.
The approach is formulated here for quadratic NURBS, but an extension to other discretization types can be anticipated.
The same applies to other constraints and associated locking phenomena.


\vspace{-5mm}
\rule{\linewidth}{.15mm}

\section{Introduction}\label{s:intro}

Locking is an artifact in finite element methods that can be associated with constraints -- or rather the numerical inability to properly capture their (near) enforcement.
The artifact in thin shells is \textit{membrane locking} \citep{stolarski82}, and the associated constraint is surface inextensibility.
This constraint is approached with decreasing shell thickness, when the membrane stiffness dominates more and more over the bending stiffness.
In the discretization of curved shells, the large membrane stiffness can then pollute the bending terms and spuriously increase bending stiffness.
As a result, the deformations are far too small -- hence the structure \textit{locks}.
Additionally, membrane stresses become oscillatory, which is a typical behavior of the stresses associated with kinematic constraints.

Shells have wide-ranging applications in science and engineering, and hence the development of locking-free shell finite element methods has been an important research topic for many decades. 
Prominent early works are those by \citet{belytschko85,bathe86,park86} and \citet{andelfinger93}.
The advent of isogeometric analysis \citep{hughes05} and its incorporation into classical finite element (FE) code structures \citep{borden11} has reinvigorated the study of shells, starting with the works of \citet{kiendl09} and \citet{benson10}.
While isogeometric shell formulations offer many advantages over classical finite element shell formulations -- such as improved accuracy and efficiency, and straightforward rotation-free formulations for Kirchhoff-Love shells -- they still lock.
Therefore the alleviation and elimination of membrane locking in isogeometric shell formulations has received much attention in recent years.

One approach is to use selective reduced integration rules.
This has been first considered for linear Timoshenko beams \citep{bouclier12,adam14} and Reissner-Mindlin (RM) shells \citep{adam15,adam15a}, and was later extended to nonlinear Kirchhoff-Love (KL) shells \citep{leonetti19}. 
Recently, a new selective reduced integration technique based on Greville-abscissae has been developed for KL and RM shells by \citet{zou21,zou22}.

Another approach for alleviating locking is to use mixed (displacement/strain) methods based on the Hellinger-Reissner (HR) variational principle. 
A question there is whether and how to apply static condensation of the extra strain variables.
Early works considered continuous strain interpolations, which lead to fully populated tangent matrices after static condensation (for each isogeometric patch), and thus are expensive to solve \citep{echter13,bouclier15,oesterle16,bieber18}.
Discontinuous strain approximations \citep{bouclier13,bombarde22a}, on the other hand, retain matrix sparsity, but can lose accuracy and still lock \citep{greco17}.
This led to the development of more advanced condensation approaches that are based on global reconstruction \citep{greco17,greco18} or dual shape functions \citep{zou20}, although at the loss of the tangent matrix symmetry \citep{kikis22}.
Another approach considers restricting the discontinuity to the center of isogeometric patches \citep{Hu16,hu20}.

Statically condensed mixed methods can also be interpreted as B-bar methods.
Such methods have been developed for Timoshenko beams \citep{bouclier12,Hu16}, solid-shells \citep{bouclier13}, Kirchhoff rods \citep{greco16,greco17}, Kirchhoff-Love shells  \citep{greco18} and Reissner-Mindlin shells \citep{hu20}.

B-bar methods can also be obtained through Assumed Natural Strain (ANS) formulations.
Such formulations have been developed for solid shells \citep{caseiro14,antolin20} and Reissner-Mindlin shells \citep{Kim22} using discontinuous assumed strains between elements, and for Kirchhoff rods \citep{casquero22} and Kirchhoff-Love shells \citep{casquero23} using continuous assumed strains between elements.
The latter approach has also been extended to Timoshenko rods \citep{golestanian23} and near incompressibility \citep{casquero24}.
ANS is also the basis of the isogeometric MITC shell formulation for Reissner-Mindlin shells \citep{mi21}.

A final approach to note is the use of variable reparameterization, as has been considered for Euler-Bernoulli beams by \citet{bieber22}.
Recently, also a spectral analysis of membrane locking for isogeometric Euler-Bernoulli beams has appeared \citep{nguyen22,hiemstra23}.

Despite the rich literature on membrane locking, there is still a demand for simpler approaches, that are more efficient than mixed methods and, contrary to the HR variational principle, apply to nonlinear material models without restrictions. 
Nonlinear shell formulations in particular have received much less attention in the literature.
Little attention has also been devoted to the examination of convergences rates for both displacements and stresses. 
Examining the latter is essential for assessing locking, since locking can exhibit very large stress errors, even when displacement errors are low.

This motivates the development of the proposed new approach, which is simple and efficient:
It requires no extra implementation effort if one has an existing finite element code for quadratic NURBS-based KL shell elements and linear Lagrange-based membrane elements.
The two element types are simply combined in the new approach, which only requires specifying their connectivity and setting the membrane stiffness of the shell elements to zero.
The approach also works for large deformations without any modifications.

The focus here is placed on basic isogeometric KL shells discretized with quadratic NURBS due to their large popularity, although extensions to other discretizations can be anticipated.  
Further, the approach is presented for the simple and widely used Koiter constitutive shell model, but it can be easily applied to other shell material models, as long as membrane and bending stiffness can be specified separately, as is the case for the bioshell models of \cite{bioshell}, the nonlinear bending models presented in \citet{savitha22} and the viscoelastic shell model of \citet{viscshell}.
It is also expected that it can be extended to isogeometric shell formulations for cellular membranes \citep{liquidshell} and textile membranes \citep{textshell2}.

To summarize, the highlights of the proposed new formulation can be stated as follows: \\[-8mm]
\begin{itemize}
\item It is a simple new hybrid discretization approach. \\[-7mm]
\item It requires no additional dofs and no static condensation. \\[-7mm]
\item It does not increase the bandwidth of the stiffness matrix, nor does it destroy its symmetry. \\[-7mm]
\item It can be run with existing isogeometric FE codes. \\[-7mm]
\item It alleviates, if not eliminates, membrane locking, both in the linear and nonlinear regimes. \\[-7mm]
\item It applies to all material models that allow for a split of membrane and bending parts. \\[-7mm]
\end{itemize}

The remainder of this paper is organized as follows:
Secs.~\ref{s:theo} \& \ref{s:FE} provide a brief overview of the governing equations of thin shells and their finite element formulation.
The proposed new discretization approach is then presented and discussed in detail in Sec.~\ref{s:B2M1}.
This is followed by the detailed investigation of three numerical examples in Sec.~\ref{s:Nex}.
The presentation concludes with Sec.~\ref{s:concl}.

\section{Thin shell theory}\label{s:theo}

This section briefly describes the governing equations of Kirchhoff-Love shells, both in the geometrically linear and nonlinear case.

\subsection{Kinematics}\label{s:kine}

The kinematics of thin shells can be fully derived from the surface description
\eqb{l}
\bx = \bx(\xi^1,\xi^2)\,,
\eqe
that corresponds to a mapping of a 2D domain, described by the two parameters $\xi^\alpha$, $\alpha=1,2$, to a 3D surface.
Such a mapping is natural to the finite element formulation following in Sec.~\ref{s:FE}. 
The set of all surface points is denoted $\sS$. 
The tangent vectors at $\bx\in\sS$ are $\ba_\alpha := \bx_{\!,\alpha}$, where $..._{,\alpha} := \partial .../\partial\xi^\alpha$,
while the surface normal at $\bx\in\sS$ is $\bn := \ba_1\times\ba_2/\norm{\ba_1\times\ba_2}$.
From these follow the surface metric and curvature tensor components
\eqb{l}
a_{\alpha\beta} := \ba_\alpha\cdot\ba_\beta\,,\quad
b_{\alpha\beta} := \ba_{\alpha,\beta}\cdot\bn\,,
\eqe
which are essential in describing in-plane stretching and out-of-plane bending.
Given a surface variation, denoted $\delta\bx$, the variations of $a_{\alpha\beta}$ and $b_{\alpha\beta}$ follow as \citep{shelltheo}
\eqb{lll}
\delta a_{\alpha\beta} \is \delta\ba_\alpha\cdot\ba_\beta + \ba_\alpha\cdot\delta\ba_\beta\,,\\[1.5mm]
\delta b_{\alpha\beta} \is \big(\delta\ba_{\alpha,\beta}-\Gamma^\gamma_{\!\alpha\beta}\,\delta\ba_\gamma\big)\cdot\bn\,,
\label{e:dab}\eqe
where $\Gamma^\gamma_{\!\alpha\beta} := \ba_{\alpha,\beta}\cdot\ba^\gamma$ is the Christoffel symbol and $\delta\ba_\alpha = \delta\bx_{,\alpha}$.
Here, the dual tangent vectors $\ba^\alpha:=a^{\alpha\beta}\ba_\beta$ follow from the inverse surface metric $[a^{\alpha\beta}] := [a_{\alpha\beta}]^{-1}$.
As usual in index notation, summation (from 1 to 2) is implied on all repeated indices within a term.

Surface $\sS$ changes due to mechanical loading.
The description above is used to denote its current (deformed) configuration.
Analogous to this, the undeformed reference configuration $\bX=\bX(\xi^\alpha)$ is introduced.
The mapping $\bX=\bX(\xi^\alpha)$ then leads to a likewise definition of $\bA_\alpha$, $\bN$, $A_{\alpha\beta}$, $B_{\alpha\beta}$, $\bA^\alpha$, and $\mathring\Gamma^\gamma_{\!\alpha\beta} := \bA_{\alpha,\beta}\cdot\bA^\gamma$.
Further, the displacement field $\bu := \bx - \bX$, and the Green-Lagrange strain and relative curvature components
\eqb{lll}
\eps_{\alpha\beta} \dis \big(a_{\alpha\beta} - A_{\alpha\beta}\big)/2\,, \\[1mm]
\kappa_{\alpha\beta} \dis  b_{\alpha\beta} - B_{\alpha\beta} 
\label{e:epsGL}\eqe
can be introduced.
Their linearization leads to the infinitesimal strain and curvature components
\eqb{lll}
\bar\eps_{\alpha\beta} \dis \big(\bu_{,\alpha}\cdot\bA_\beta + \bu_{,\beta}\cdot\bA_\alpha\big)/2\,, \\[1mm]
\bar\kappa_{\alpha\beta} \dis  \big(\bu_{,\alpha\beta}-\mathring\Gamma^\gamma_{\!\alpha\beta}\,\bu_{,\gamma}\big)\cdot\bN\,.
\label{e:epslin}\eqe
This follows from replacing $\delta\bx$ by $\bu$ and $\ba_\alpha$ by $\bA_\alpha$ in the variations $\delta\eps_{\alpha\beta} = \delta a_{\alpha\beta}/2$ and $\delta\kappa_{\alpha\beta} = \delta b_{\alpha\beta}$ defined by Eq.~\eqref{e:dab}, since the displacement $\bu$ of the initial configuration $\bX$ is analogous to the variation $\delta\bx$ of the current configuration $\bx$.

\subsection{Constitution}\label{s:const}

For hyperelastic shells, the membrane stresses $\sig^{\alpha\beta}$ and the bending stress couples $M^{\alpha\beta}$ are fully specified by the strain variables $\eps_{\alpha\beta}$ and $\kappa_{\alpha\beta}$.\footnote{The units of $\sig^{\alpha\beta}$ and $M^{\alpha\beta}$ are [force/length] and [moment/length], respectively. Thus they denote 2D stress states. In contrast to 3D stress states (which are not considered here), they are often referred to \textit{resultant stresses} in the literature. Sometimes they are also denoted as membrane forces and bending moments in the literature.}
This work considers the widely used Koiter shell model \citep{koiter66,ciarlet05}.
It can be obtained by integrating the 3D St.Venant-Kirchhoff model through the thickness, e.g.~see \citet{solidshell}, 
leading to a linear relation between the strain variables $\eps_{\alpha\beta}$ and $\kappa_{\alpha\beta}$, and the Kirchhoff stresses $\tau^{\alpha\beta} := J\sigma^{\alpha\beta}$ and $M_0^{\alpha\beta} := J M^{\alpha\beta}$, i.e. \\[-4mm]
\eqb{l}
\tau^{\alpha\beta} = c^{\alpha\beta\gamma\delta}\,\eps_{\gamma\delta}\,,\quad
M_0^{\alpha\beta} = f^{\alpha\beta\gamma\delta}\,\kappa_{\gamma\delta}\,,
\label{e:Koiter}\eqe
where, for material isotropy,
\eqb{l}
c^{\alpha\beta\gamma\delta} = \Lambda\,A^{\alpha\beta}A^{\gamma\delta}  + \mu\big(A^{\alpha\gamma}A^{\beta\delta} + A^{\alpha\delta}A^{\beta\gamma}\big)\,,\quad
f^{\alpha\beta\gamma\delta} = \ds\frac{T^2}{12}c^{\alpha\beta\gamma\delta}\,.
\eqe
Here, the parameters $\Lambda$, $\mu$ and $T$ denote the surface bulk modulus, surface shear stiffness and shell thickness, respectively, while $J := \sqrt{\det[a_{\alpha\beta}]/\det[A_{\alpha\beta}]}$ is the local surface area change. 
Given the 3D Young's modulus $E$ and Poisson's ratio $\nu$, the former two become
\eqb{l}
\Lambda = \ds\frac{2\mu\nu}{1-\nu}\,,\quad
\mu = \ds\frac{ET}{2(1+\nu)}\,.
\eqe
In the infinitesimal setting, $\eps_{\gamma\delta}$ and $\kappa_{\gamma\delta}$ are replaced by their linearized counterparts, while  $J=1$, leading to
\eqb{l}
\sig^{\alpha\beta} = c^{\alpha\beta\gamma\delta}\,\bar\eps_{\gamma\delta}\,,\quad
M^{\alpha\beta} = f^{\alpha\beta\gamma\delta}\,\bar\kappa_{\gamma\delta}\,.
\label{e:linel}\eqe
The stress $\sigma^{\alpha\beta}$ appearing in \eqref{e:Koiter} and \eqref{e:linel}, and in the weak form below, is termed \textit{effective stress} in \citet{simo89}. 
It is generally not equal to the physical Cauchy stress given by
\eqb{l}
N^{\alpha\beta} := \sigma^{\alpha\beta} + M^{\alpha\gamma}\,b^\beta_\gamma\,.
\label{e:NsbM}\eqe
The latter stress is unsymmetric and appears in the equilibrium equation and Cauchy's formula, while the former is symmetric.
Eq.~\eqref{e:NsbM} arises from the balance of angular momentum, e.g.~see \citet{shelltheo}.  
In Eq.~\eqref{e:NsbM}, $b^\beta_\gamma = a^{\alpha\beta}\,b_{\alpha\gamma}$.
In the infinitesimal setting this is replaced by $B^\beta_\gamma = A^{\alpha\beta} B_{\alpha\gamma}$.

\begin{remark}
The stress notation above follows the notation of \citet{naghdi72} and \citet{steigmann99b}. 
This translates to the notation of \citet{simo89} as $N^{\alpha\beta} = n^{\beta\alpha}$, $\sigma^{\alpha\beta} = \tilde n^{\alpha\beta}$ and $M^{\alpha\beta} = \tilde m^{\alpha\beta}$.
\end{remark}

\begin{remark}
It is important to note that the continuity in $\eps_{\alpha\beta}$ and $\sigma^{\alpha\beta}$ is one order lower than that of $\bu$, e.g.~$C^0$ for quadratic NURBS.
The continuity in $\kappa_{\alpha\beta}$ and $M^{\alpha\beta}$, and hence also $N^{\alpha\beta}$, is two orders lower than that of $\bu$, e.g.~$C^{-1}$ for quadratic NURBS.
\end{remark}

\subsection{Weak form}\label{s:WF}

The quasi-static weak form of the thin shell is given by
\eqb{l}
G_\mathrm{int} - G_\mathrm{ext}= 0 \quad \forall\,\delta\bx\in\sV\,,
\label{e:WF}\eqe
with
\eqb{lll}
G_\mathrm{int} \is \ds\int_{\sS}\delta\eps_{\alpha\beta}\,\sigma^{\alpha\beta}\,\dif a + \int_{\sS}\delta\kappa_{\alpha\beta}\,M^{\alpha\beta}\,\dif a\,, \\[4mm]
G_\mathrm{ext} \is \ds\int_{\sS}\delta\bx\cdot\bff\,\dif a + \int_{\partial_t\sS}\delta\bx\cdot\bT\,\dif s + \int_{\partial_m\sS}\delta\bn\cdot\bM\,\dif s\,,
\label{e:GG}\eqe
e.g.~see \citet{shelltheo}.
The first term denotes the internal virtual work generated by the membrane stresses $\sigma^{\alpha\beta}$ and bending stress couples $M^{\alpha\beta}$. The second term denotes the external virtual work generated by the surface load $\bff$, the boundary traction $\bT$ and the boundary moment $\bM$.
The latter two are applied to the Neumann boundaries $\partial_t\sS$ and $\partial_m\sS$ of the surface.
In the following examples $\bM = \mathbf{0}$. 
$\sV$ denotes the space of kinematically admissible variations.
Those need to satisfy the differentiability requirements of Eq.~\eqref{e:dab}, as well as conform to the Dirichlet boundary conditions $\bu = \bar\bu$ on $\partial_u\sS$ and $\bn = \bar\bn$ on $\partial_n\sS$ for displacements and rotations.
Additionally, their first and second derivatives need to be square integrable, due to Eqs.~\eqref{e:GG}, \eqref{e:linel} and \eqref{e:epslin} (or likewise  \eqref{e:Koiter} and \eqref{e:epsGL}). 
Such functions are denoted $\sH^2$, leading to
\eqb{l}
\sV = \big\{\delta\bx\in\sH^2\,\big|\,\delta\bx = \mathbf{0}$ on $\partial_u\sS,\,\delta\bn = \mathbf{0}$ on $\partial_n\sS \big\}\,.
\eqe

\begin{remark} 
Actually, only the $\delta\bx$ of the second term in $G_\mathrm{int}$ needs to be in $\sH^2$. 
The first term only contains first derivatives, and hence its $\delta\bx$ can be in $\sH^1$.
This observation motivates the proposed B2M1 discretization.
\end{remark}

\section{Finite element formulation}\label{s:FE}

This section proceeds with a summary of the standard isogeometric finite element formulation of the preceding shell model.
In isogeometric as well as classical finite elements, the surface is described by 
a set of $n_\mathrm{no}$ control points (or nodes) and corresponding basis functions, in the form
\eqb{l}
\bX^h = \ds\sum_{I=1}^{n_\mathrm{no}} N_I\,\bX_I\,,
\label{e:Xh}\eqe
where $N_I = N_I(\xi^\alpha)$ denotes the basis function and $\bX_I$ the position of control point $I$.
The basis, or shape functions are subject of Sec.~\ref{s:B2M1}.
In isogeometric FE, $\bX^h$ describes the surface exactly, whereas in classical FE, $\bX^h$ is only an approximation of $\bX$. 
In both cases, the deformed surface and the displacement field are an approximation to the real surface and displacement, i.e.
\eqb{l}
\bx \approx \bx^h = \ds\sum_{I=1}^{n_\mathrm{no}} N_I\,\bx_I\,,\quad
\bu \approx \bu^h = \ds\sum_{I=1}^{n_\mathrm{no}} N_I\,\bu_I\,.
\label{e:xh}\eqe
In classical FE the nodal values are interpolated by \eqref{e:Xh}-\eqref{e:xh}, implying $N_I(\xi^\alpha_J) = \delta_{IJ}$, where $\delta_{IJ}$ is the Konecker delta, while this is generally not true for isogeometric FE.
Expressions \eqref{e:Xh}-\eqref{e:xh} can be evaluated element-wise -- in the form
\eqb{l}
\bX^h = \mN_e\,\mX_e\,,\quad
\bx^h = \mN_e\,\mx_e\,,\quad
\bu^h = \mN_e\,\muu_e\,,
\label{e:xe}\eqe
where $\mX_e$, $\mx_e$ and $\muu_e$ are vectors containing the $n_e$ nodal values $\bX_I$, $\bx_I$ and $\bu_I$ of element $e$, and $\mN_e := [N_{e_1}\bone,\,N_{e_2}\bone,\,...,\,N_{e_{n_e}}\!\bone]$ 
is a corresponding $3\times3n_e$ matrix containing the $n_e$ shape functions of the element.
The set
\eqb{l}
\sC_e := \{e_1,\,e_2,\,...,\,e_{n_e}\}
\label{e:con}\eqe
contains the global node numbers of element $e$ and thus specifies the element connectivity.
Tensor $\bone$ denotes the identity in $\bbR^3$.

Eq.~\eqref{e:xe} is used to evaluate all the kinematical and constitutive quantities of Secs.~\ref{s:kine} and \ref{s:const}, and then discretize weak form \eqref{e:WF}.
This leads to the equation
\eqb{l}
\mf := \mf_\mathrm{int} - \mf_\mathrm{ext} = \mathbf{0}\,, 
\label{e:f}\eqe
describing the force balance at the free finite element nodes (where no displacement BCs are specified).
The force vectors $\mf_\mathrm{int}=\mf_\mrm + \mf_\mrb$ and $\mf_\mathrm{ext}$ are assembled from the elemental 
membrane, bending and external contributions \citep{solidshell}
\eqb{lll}
\mf^e_\mrm \dis \ds\int_{\Omega^e_0}\mN^\mrT_{e,\alpha}\,\tau^{\alpha\beta}\,\ba_\beta\,\dif A\,,\\[4mm]
\mf^e_\mrb \dis \ds\int_{\Omega^e_0}\mN^\mrT_{e;\alpha\beta}\,M_0^{\alpha\beta}\,\bn\,\dif A\,,\\[4mm]
\mf^e_\mathrm{ext} \dis \ds\int_{\Omega^e_0}\mN^\mrT_{e}\bff_0\,\dif A + \int_{\Gamma^e_0}\mN^\mrT_{e}\,\bT_0\,\dif S\,,
\label{e:mf}\eqe
that follow from the virtual work contributions in Eq.~\eqref{e:GG} using $\dif a = J\,\dif A$.
By choice, they are written here as integrals over the reference configuration of the element and its traction boundary, denoted $\Omega^e_0$ and $\Gamma^e_0$.
Thus, loads $\bff_0$ and $\bT_0$ are the counterparts of $\bff$ and $\bT$ with respect to the reference configuration.
They are treated constant here.
The last term from Eq.~(\ref{e:GG}.2) has been omitted as it is not needed in the following examples.
It is anyway not altered by the proposed new hybrid interpolation.
The array
\eqb{l}
\mN_{e;\alpha\beta} := \mN_{e,\alpha\beta} - \Gamma^\gamma_{\alpha\beta}\,\mN_{e,\gamma}
\label{e:N;}\eqe
follows from Eq.~(\ref{e:dab}.2).
The two terms within the internal force vector $\mf^e_\mathrm{int}$ lead to the
material stiffness matrices
\eqb{lll}
\mk^e_\mathrm{mmat} \dis 
\ds\int_{\Omega_0^e}c^{\alpha\beta\gamma\delta}\,\mN^\mrT_{e,\alpha}\,(\ba_\beta\otimes\ba_\gamma)\,\mN_{e,\delta}\,\dif A\,,\\[4mm]
\mk^e_\mathrm{bmat} \dis 
\ds\int_{\Omega_0^e}f^{\alpha\beta\gamma\delta}\,\mN^\mrT_{e;\alpha\beta}\,(\bn\otimes\bn)\,\mN_{e;\gamma\delta}\,\dif A\,,
\label{e:km}\eqe
and geometrical stiffness matrices
\eqb{lll}
\mk^e_\mathrm{mgeo} \dis 
\ds\int_{\Omega^e_0}\mN^\mrT_{e,\alpha}\,\tau^{\alpha\beta}\,\mN_{e,\beta}\,\dif A\,,\\[4mm]
\mk^e_\mathrm{bgeo} \dis \mk^e_\mathrm{M1} + \mk^e_\mathrm{M2} + \mk^{e\,\mrT}_\mathrm{M2}\,,
\label{e:kg}\eqe
that are associated with the membrane and bending deformations, respectively.
Here, the geometrical bending stiffness matrix is composed of
\eqb{lll}
\mk^e_\mathrm{M1} \dis 
-\ds\int_{\Omega_0^e}M_0^{\alpha\beta}\,b_{\alpha\beta}\,a^{\gamma\delta}\,\mN^\mrT_{e,\gamma}\,(\bn\otimes\bn)\,\mN_{e,\delta}\,\dif A\,,\\[4mm]
\mk^e_\mathrm{M2} \dis 
-\ds\int_{\Omega_0^e}M_0^{\alpha\beta} \,\mN^\mrT_{e,\gamma}\,(\bn\otimes\ba^\gamma)\,\mN_{e;\alpha\beta}\,\dif A\,.
\label{e:kM}\eqe
The global force vector $\mf$ and stiffness matrix $\mK = \mK_\mrm + \mK_\mrb = \mK_\mathrm{mmat} + \mK_\mathrm{mgeo} + \mK_\mathrm{bmat} + \mK_\mathrm{bgeo}$ 
are assembled from their elemental contribution in \eqref{e:mf}--\eqref{e:kg} based on connectivity \eqref{e:con}, and then used in the Newton-Raphson iteration to solve Eq.~\eqref{e:f} for the unknown nodal displacement vector $\muu$.

\begin{remark} 
In the small deformation setting, all kinematical objects in Eqs.~\eqref{e:mf}-\eqref{e:km} are taken from the reference configuration, and the stresses are obtained through Eqs.~\eqref{e:epslin} and \eqref{e:linel}.
The geometrical stiffness matrices and internal force vectors are zero then, and the Newton-Raphson iteration converges in a single step, i.e.~$\muu$ follows from $\mK\muu = \mf_\mathrm{ext}$.
\end{remark}

\begin{remark} 
As can be seen, only the second term in \eqref{e:mf} and its stiffness contributions in \eqref{e:km}-\eqref{e:kM} contain second derivatives.
All other terms contain at most first derivatives.
At least $C^1$-continuous functions are required to capture second derivatives, while $C^0$-continuity is sufficient for first derivatives.
\end{remark}

The accuracy of the finite element solution can be assessed by examining the relative $L^2$-error norm of the displacement field, membrane stress and bending moment, defined by
\eqb{llllll}
e_u \dis \ds\frac{\norm{\bu^h-\bu_\mathrm{ref}}_{L^2}}{\norm{\bu_\mathrm{ref}}_{L^2}}\,, 
	& \norm{\bu}_{L^2} \dis \ds\sqrt{\int_{\sS_0} \bu\cdot\bu\,\dif A}\,, \\[5mm]
e_\sig \dis \ds\frac{\norm{\bsig^h-\bsig_\mathrm{ref}}_{L^2}}{\norm{\bsig_\mathrm{ref}}_{L^2}}\,, 
	& \norm{\bsig}_{L^2} \dis \ds\sqrt{\int_{\sS_0} \sig^\alpha_\beta\, \sig^\beta_\alpha\,\dif A} \,, \\[5mm]
e_M \dis \ds\frac{\norm{\bM^h-\bM_\mathrm{ref}}_{L^2}}{\norm{\bM_\mathrm{ref}}_{L^2}}\,, 
	& \norm{\bM}_{L^2} \dis \ds\sqrt{\int_{\sS_0} M^\alpha_\beta\, M^\beta_\alpha\,\dif A} \,,
\label{e:L2}\eqe
where $\sigma^\alpha_\beta = \sigma^{\alpha\gamma}a_{\beta\gamma}$ and $M^\alpha_\beta = M^{\alpha\gamma}a_{\beta\gamma}$.
The $L^2$-error of stress $N^\alpha_\beta$ is defined analogously.
Here ``ref" denotes an accurate (ideally exact) reference solution.
Since the $L^2$-error is not always easy to evaluate, the relative error
\eqb{l}
e_\bullet := \bigg|\ds\frac{\bullet^h}{\bullet_\mathrm{ref}}-1\bigg|
\label{e:maxe}\eqe
between local (maximum) displacements, stresses and moments will be also investigated.

\section{Hybrid membrane-bending interpolation}\label{s:B2M1}

This section presents a new approach to choosing suitable shape functions for the preceding FE formulation.
Using classical Lagrange shape functions is insufficient for describing bending.
Isogeometric shape functions, while being sufficient, still suffer from severe membrane locking of thin shells.
This motivates the following hybrid approach.

\subsection{Basic idea}\label{s:idea}

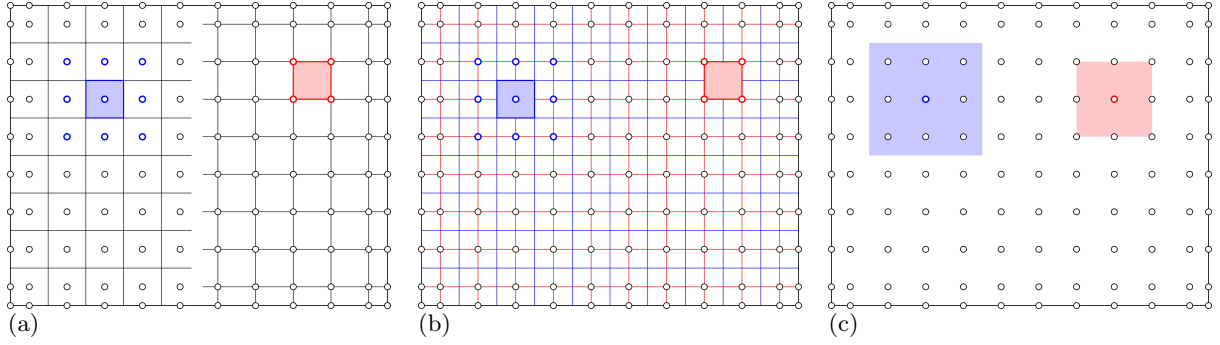
\begin{figure}[h]
\begin{center} \unitlength1cm
\begin{picture}(0,4.3)
\put(-7.95,.25){\includegraphics[height=40.5mm]{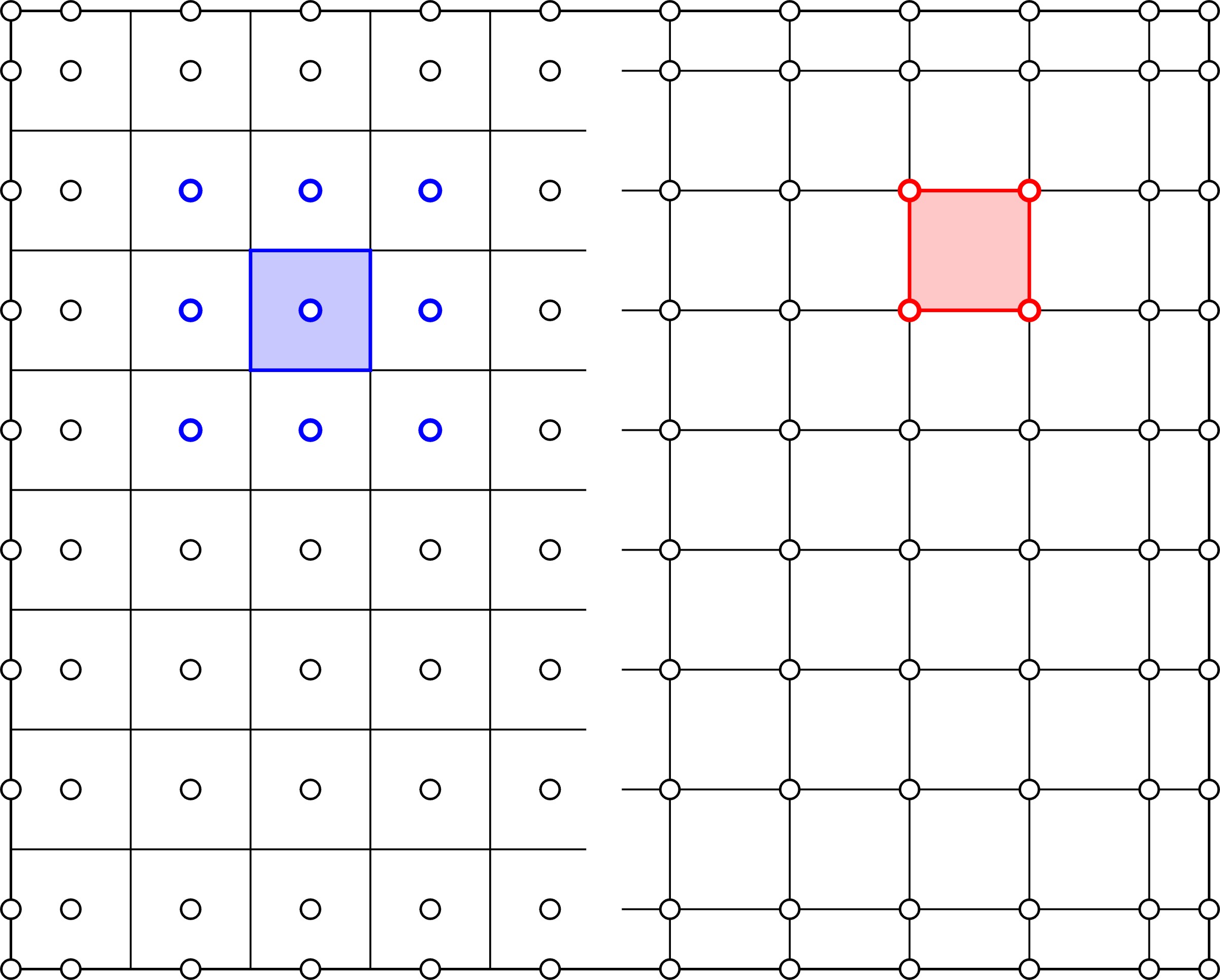}}
\put(-2.55,.25){\includegraphics[height=40.5mm]{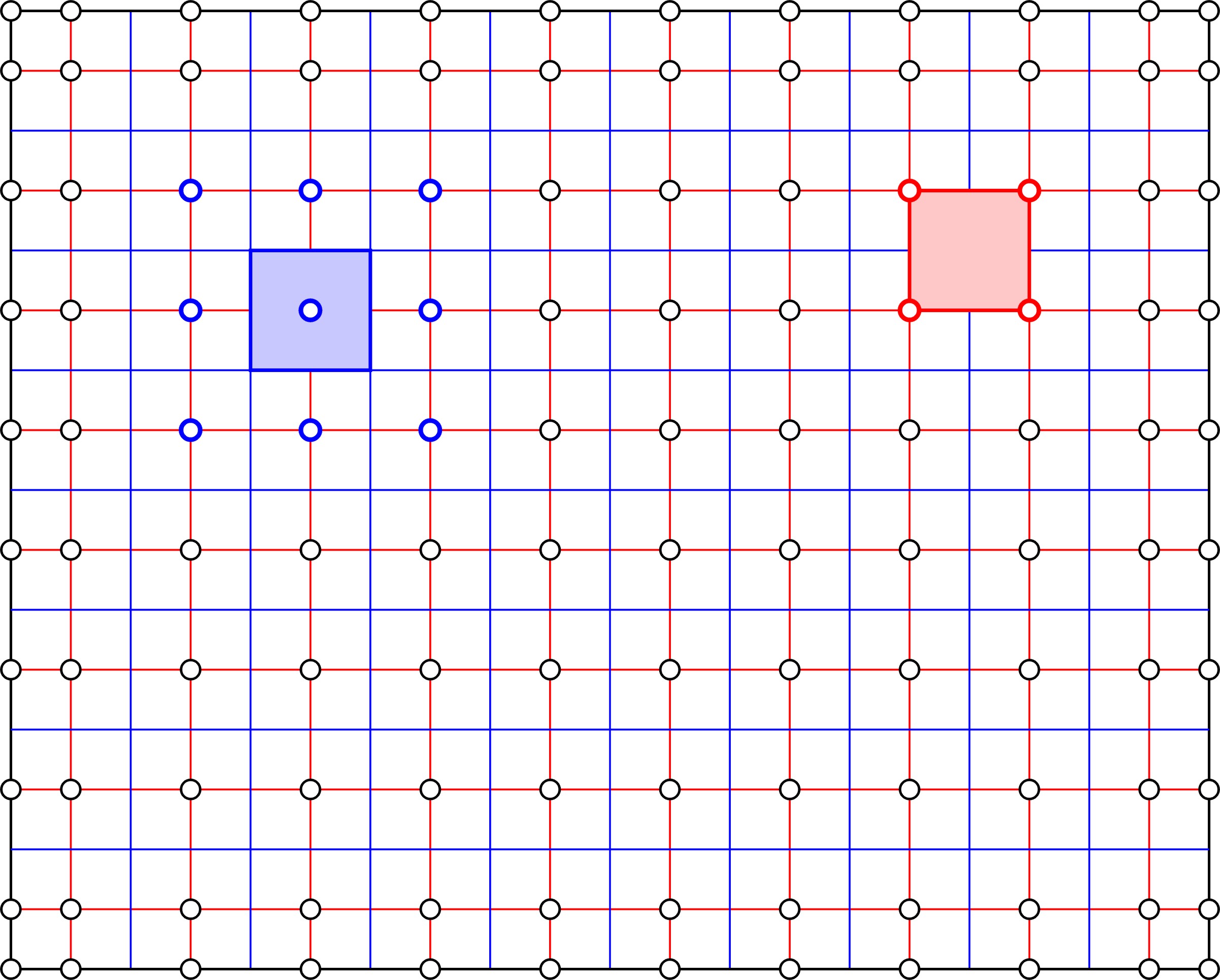}}
\put(2.85,.25){\includegraphics[height=40.5mm]{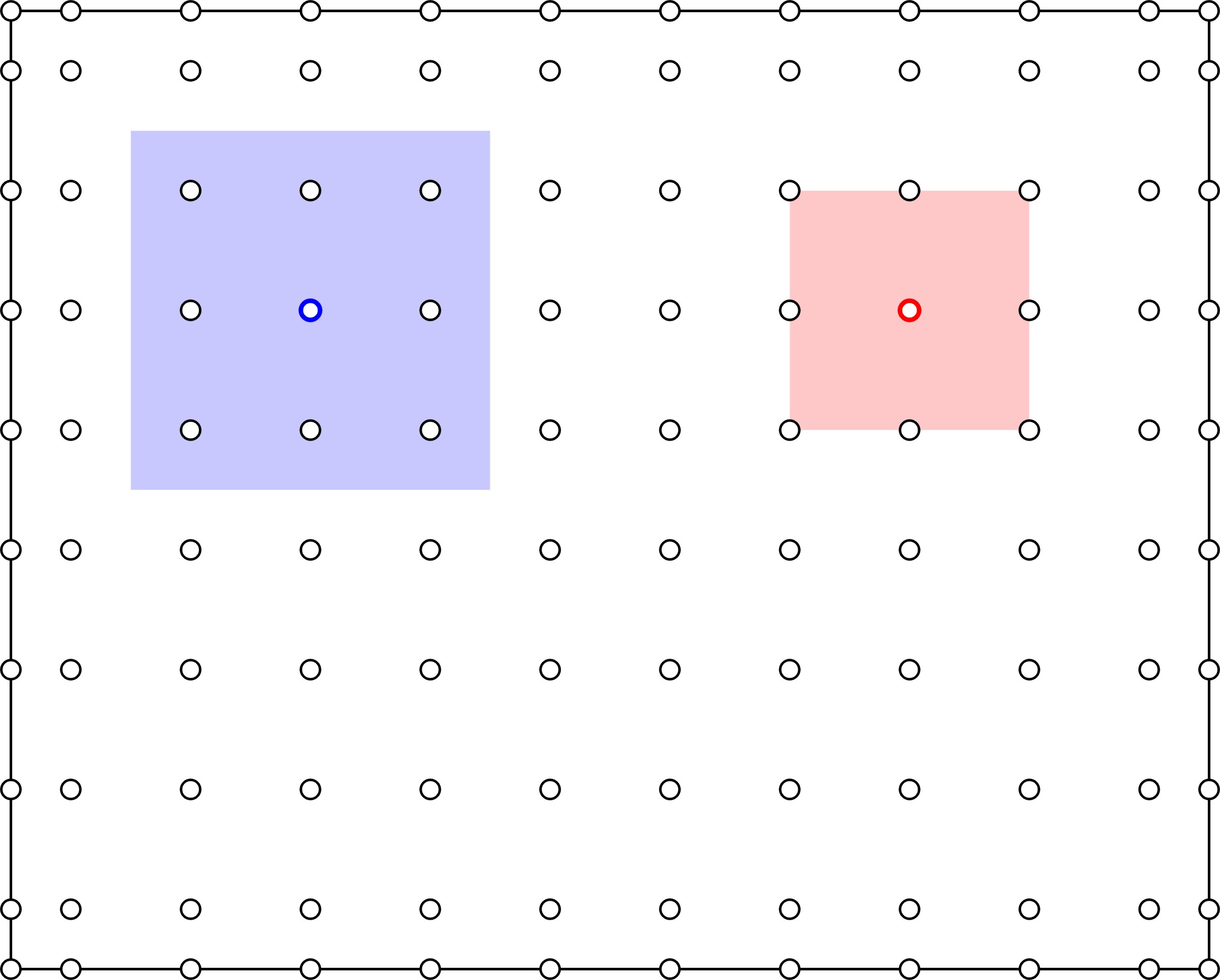}}
\put(-7.95,-.05){\footnotesize (a)}
\put(-2.55,-.05){\footnotesize (b)}
\put(2.85,-.05){\footnotesize (c)}
\end{picture}
\caption{Basic idea of the B2M1 discretization: 
(a) A domain discretized by quadratic NURBS (blue) OR linear Lagrange elements (red). 
(b) The same domain discretized by quadratic NURBS (blue) AND linear Lagrange elements (red) = elemental view of the B2M1 discretization.
(c) Shape function support for quadratic NURBS (blue) and linear Lagrange elements (red) = nodal view of the B2M1 discretization.}
\label{f:B2M1}
\end{center}
\end{figure}

As was noted already, the bending terms in (\ref{e:mf}.2), (\ref{e:km}.2) and (\ref{e:kg}.2) require $C^1$-continuity, while the membrane terms in (\ref{e:mf}.1), (\ref{e:km}.1) and (\ref{e:kg}.1) only require $C^0$-continuity.
The basic elements providing this are quadratic NURBS and linear Lagrange elements, as shown in Fig.~\ref{f:B2M1}a.

\begin{figure}[h]
\begin{center} \unitlength1cm
\begin{picture}(0,5.8)
\put(-7.85,-.2){\includegraphics[height=58mm]{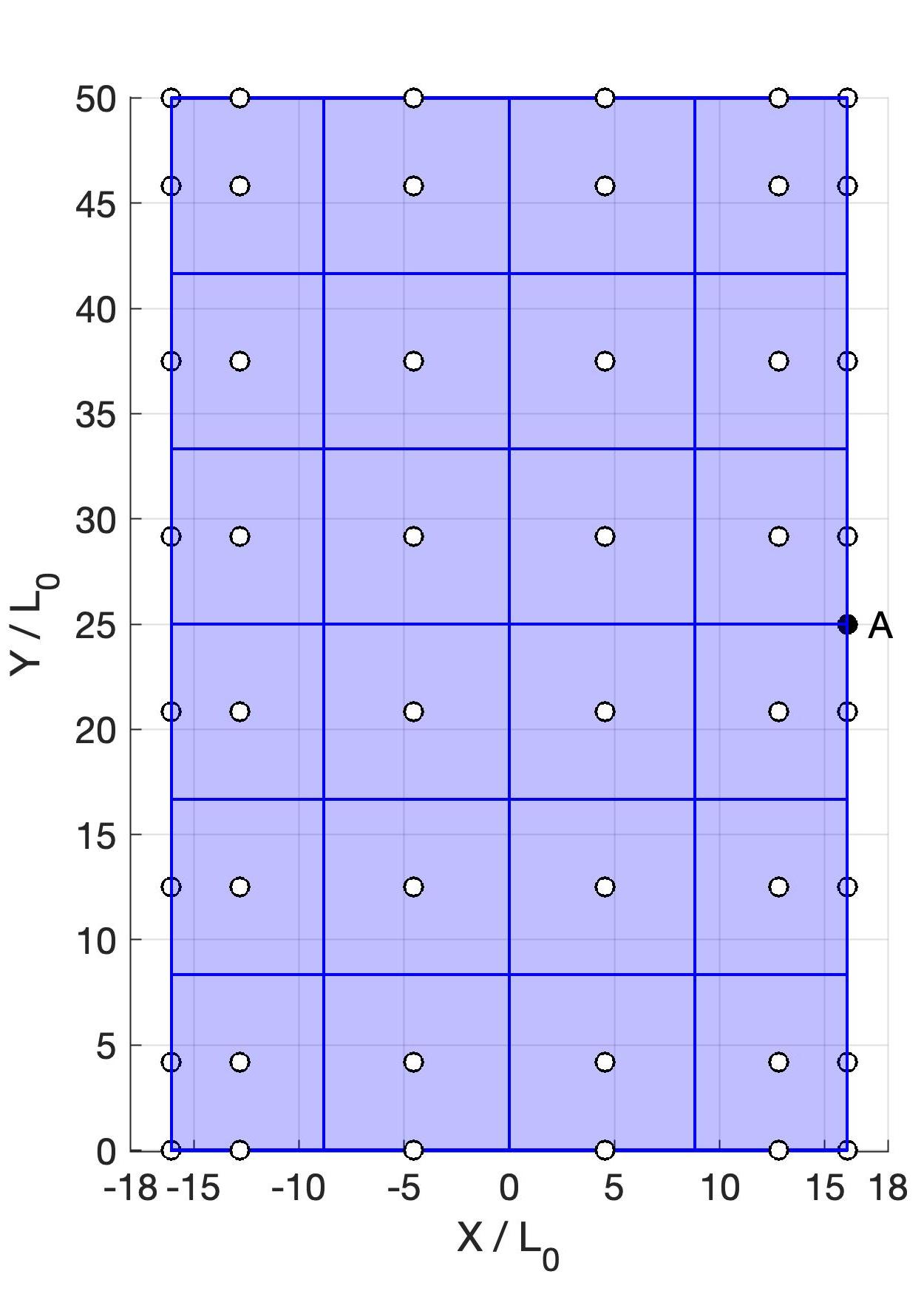}}
\put(-3.05,-.2){\includegraphics[height=58mm]{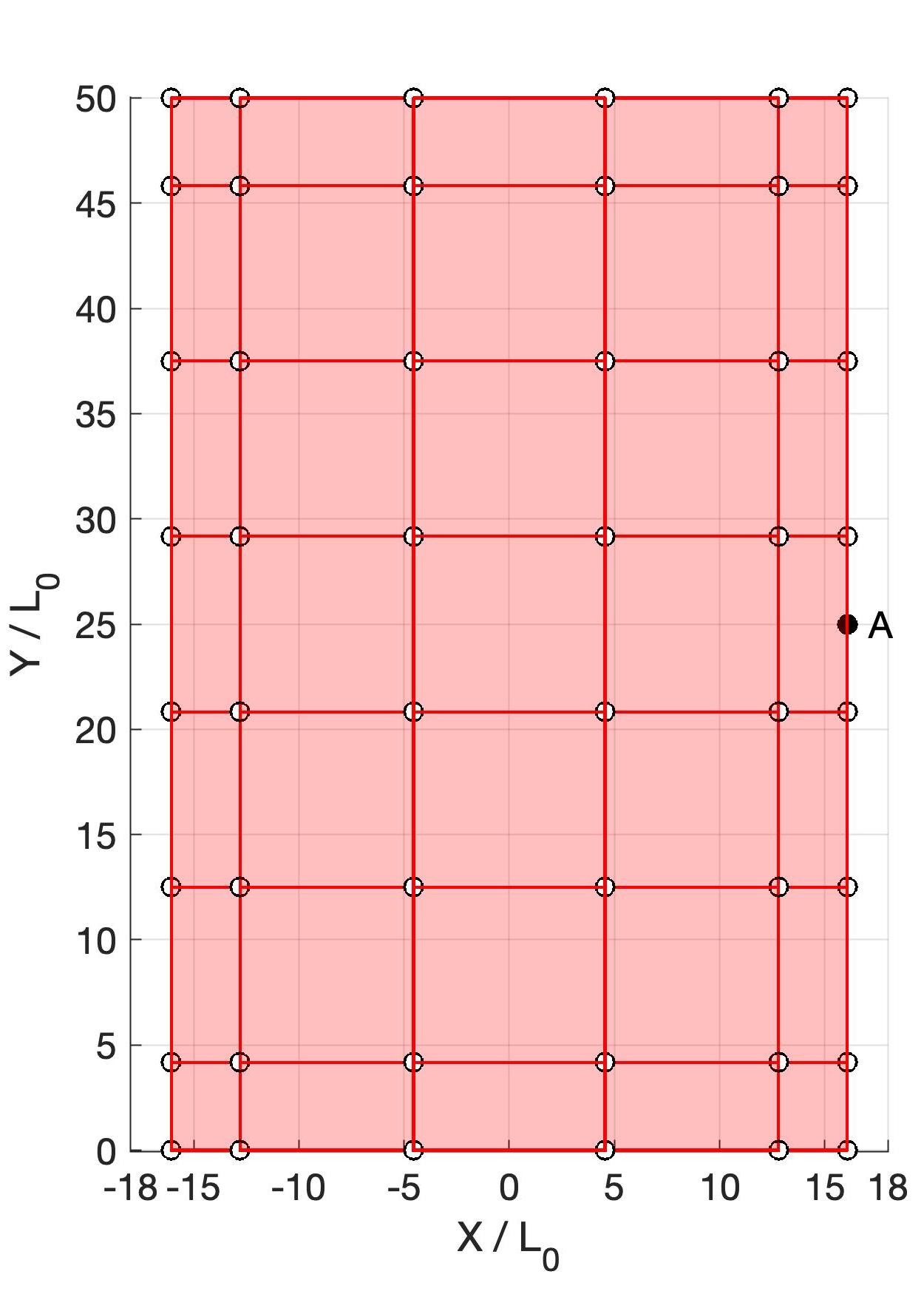}}
\put(1.6,1.85){\includegraphics[height=40mm]{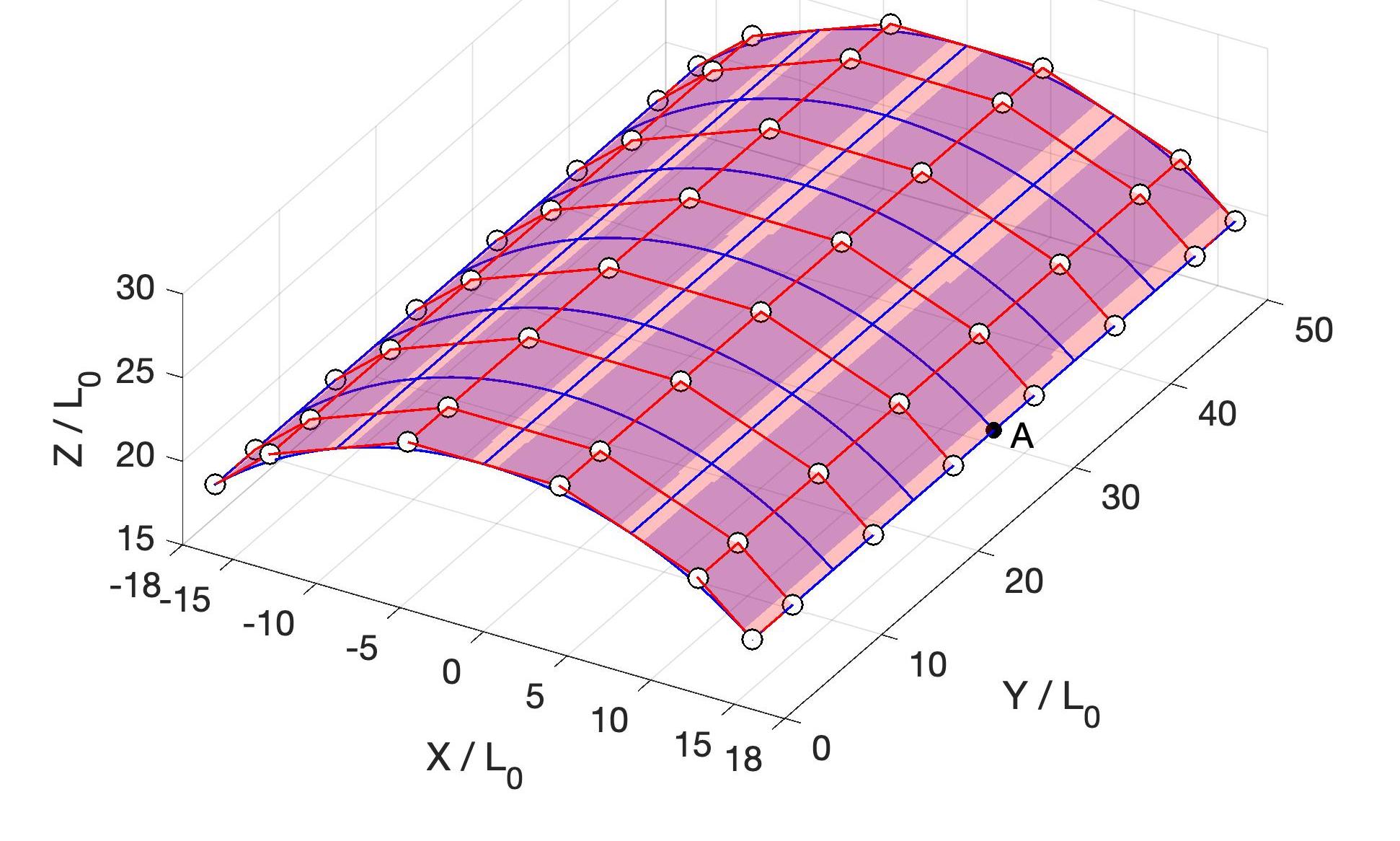}}
\put(1.925,-.1){\includegraphics[height=21.5mm]{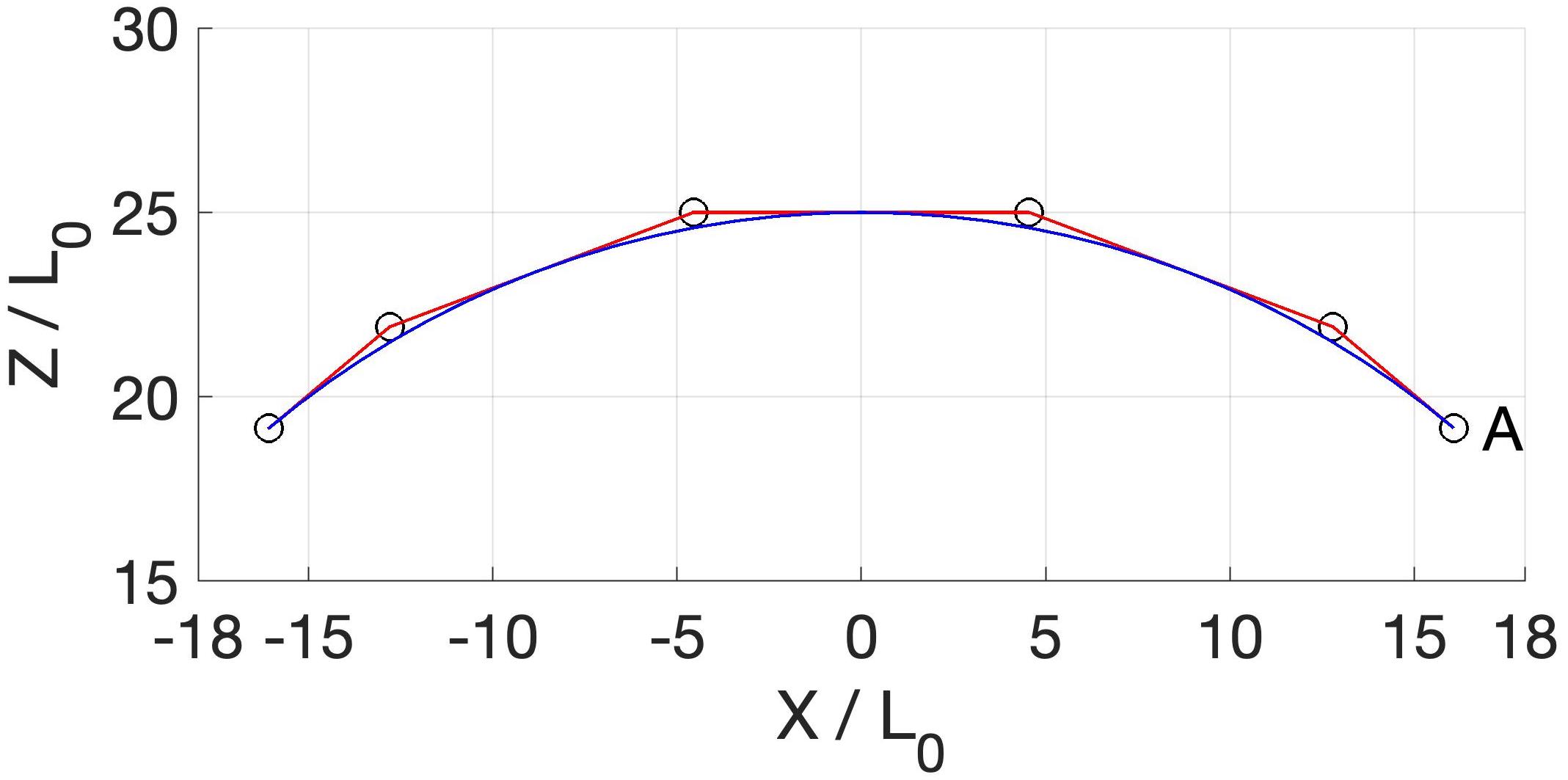}}
\put(-7.95,-.05){\footnotesize (a)}
\put(-3.15,-.05){\footnotesize (b)}
\put(1.8,2.3){\footnotesize (c)}
\put(1.8,-.05){\footnotesize (d)}
\end{picture}
\caption{B2M1 discretization of the Scordelis-Lo roof - here for $6\times 8$ CPs:
(a)~B2 mesh; (b)~M1 mesh; (c-d)~B2M1 mesh in 3D view and front view.
The surface position and tangent plane of the B2 and M1 meshes agree along the vertical B2 element boundaries in (a), but differ elsewhere.
This difference vanishes upon mesh refinement. 
The B2 surface is assumed to be the true one, while the M1 surface is an auxiliary one only required for the membrane strain determination.}
\label{f:B2M1s}
\end{center}
\end{figure}
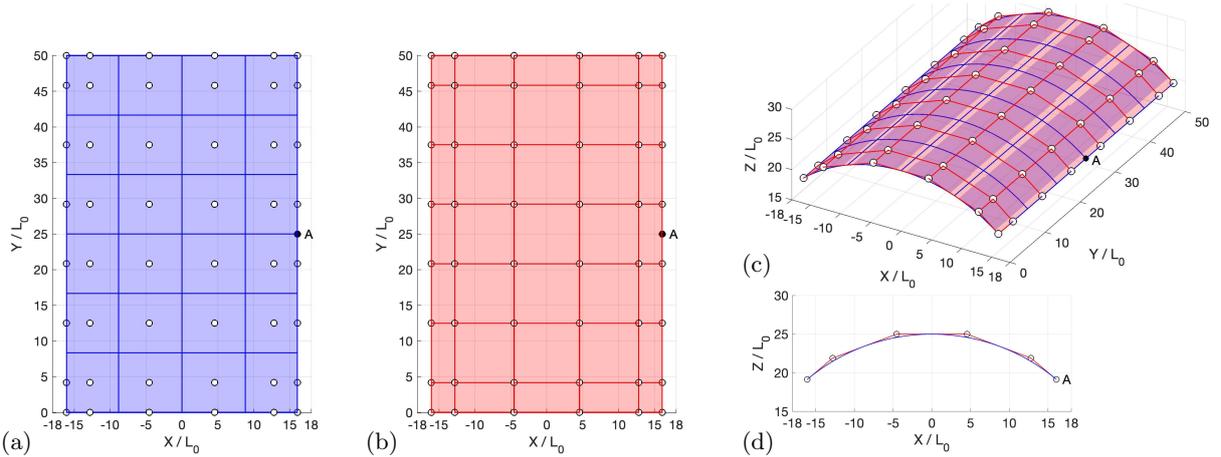

\begin{figure}[h]
\begin{center} \unitlength1cm
\begin{picture}(0,3.6)
\put(-7.95,-.1){\includegraphics[height=40mm]{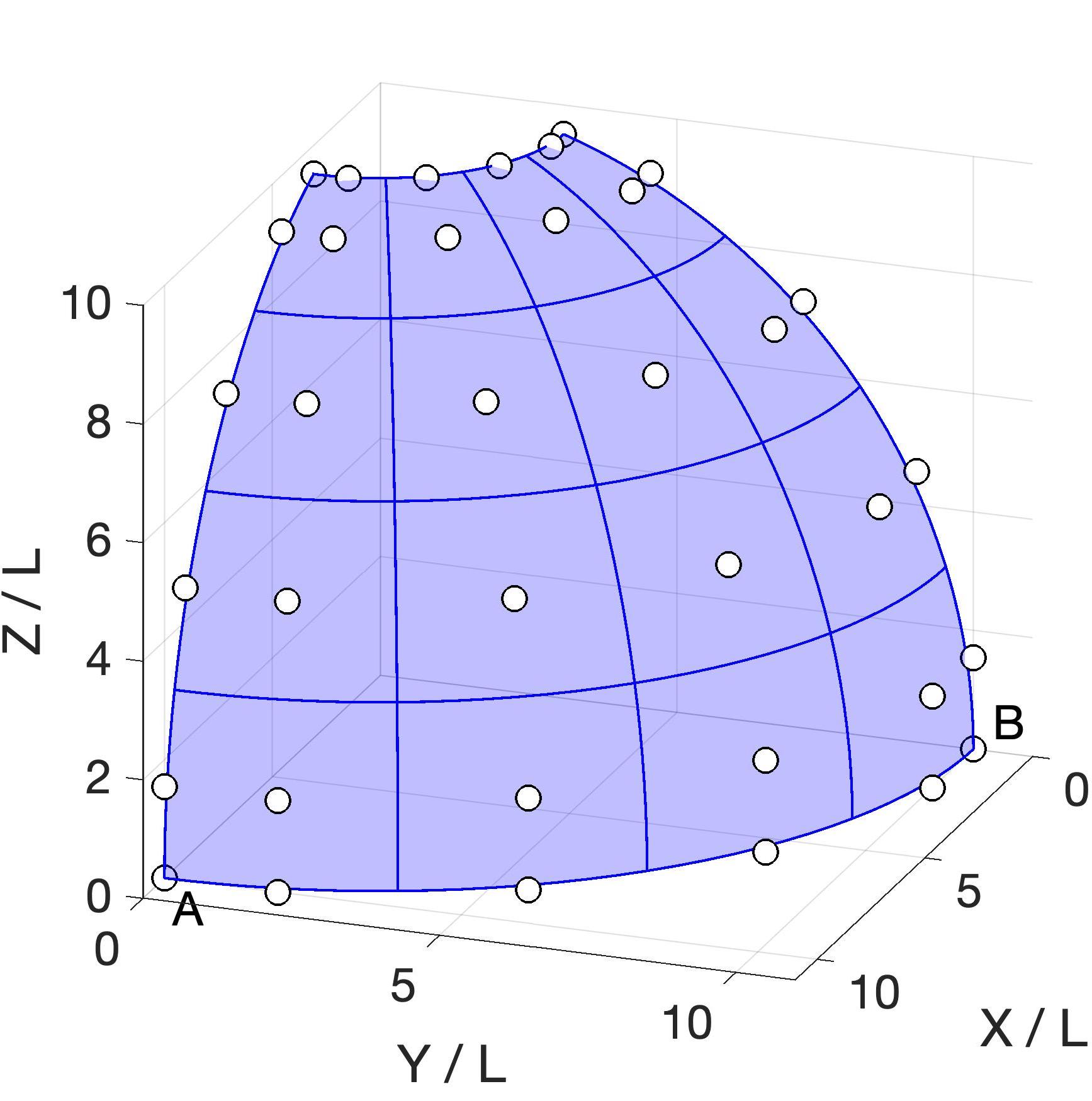}}
\put(-3.75,-.1){\includegraphics[height=40mm]{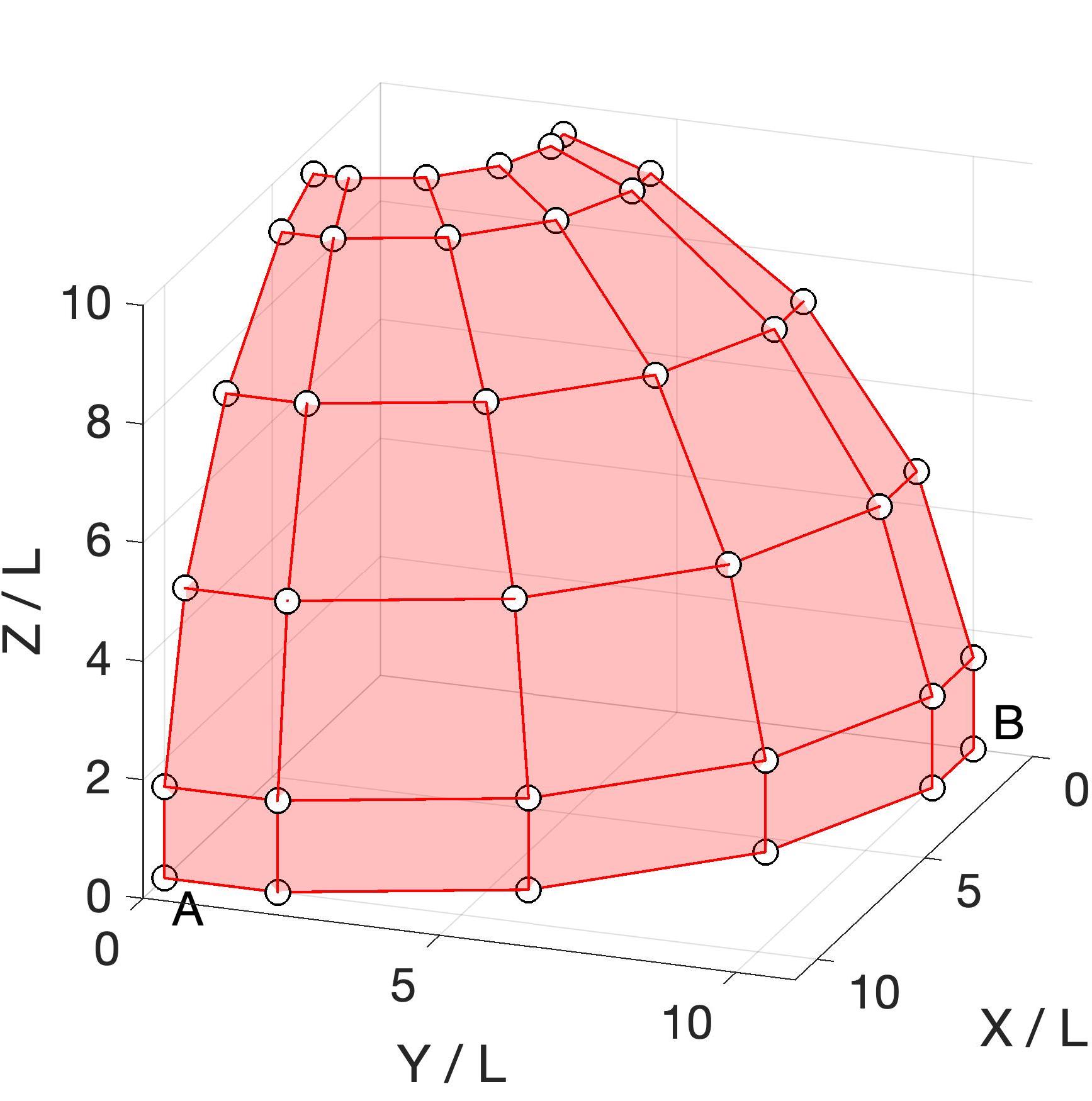}}
\put(0.45,-.1){\includegraphics[height=40mm]{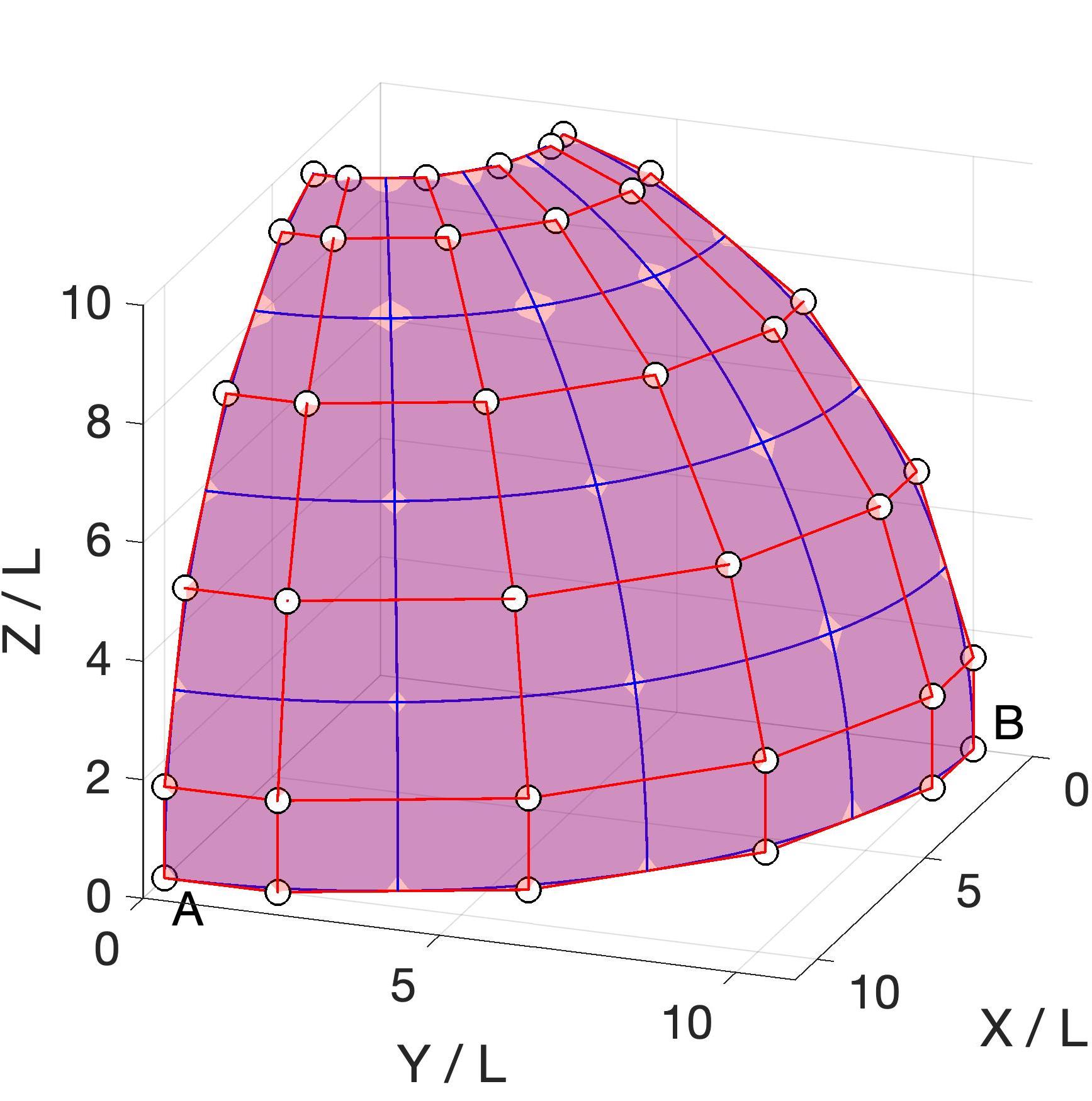}}
\put(4.6,0.1){\includegraphics[height=34mm]{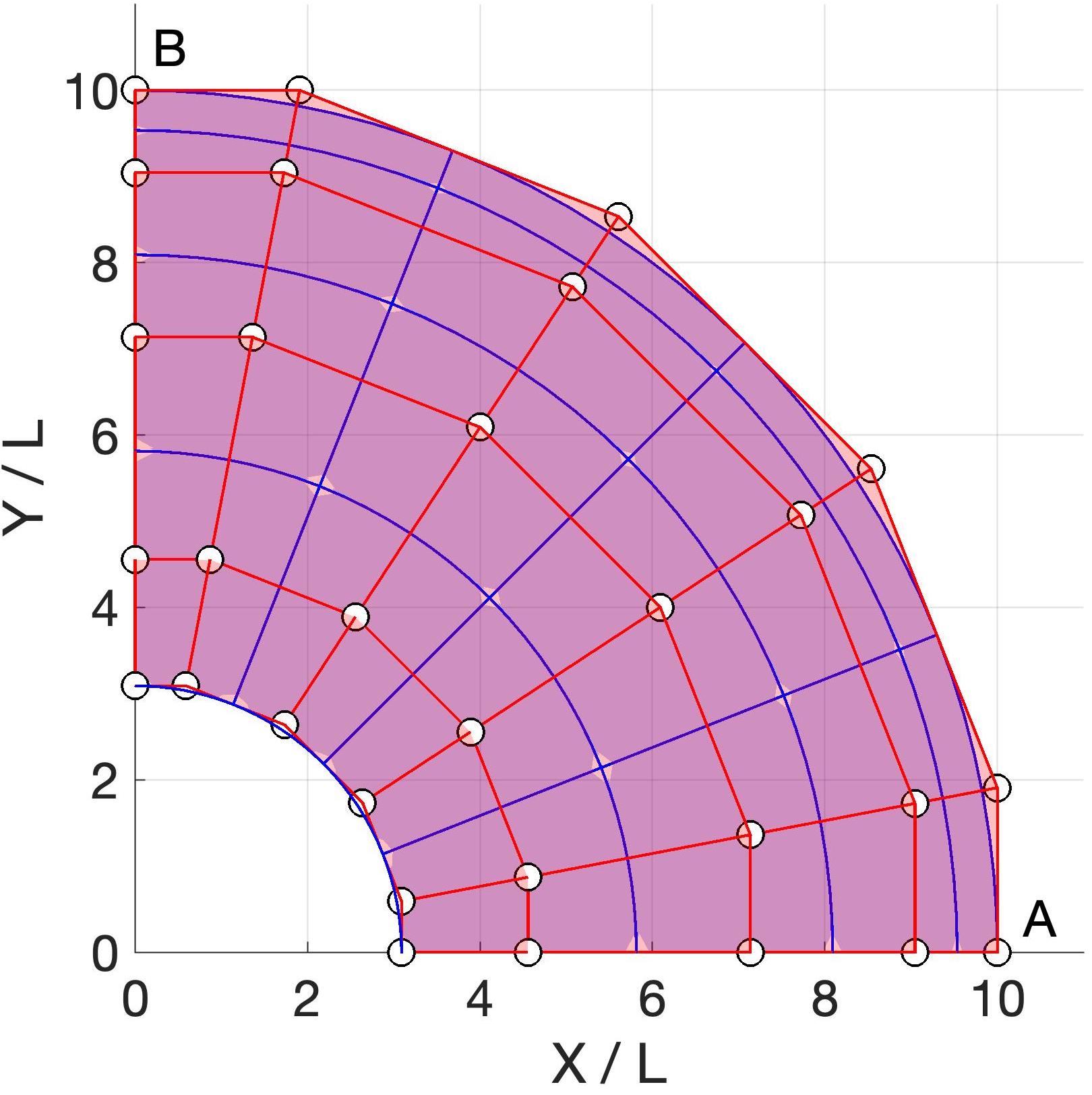}}
\put(-7.95,-.05){\footnotesize (a)}
\put(-3.75,-.05){\footnotesize (b)}
\put(0.45,-.05){\footnotesize (c)}
\put(4.65,-.05){\footnotesize (d)}
\end{picture}
\caption{B2M1 discretization of the hemisphere with hole  - here for $6\times 6$ CPs:
(a)~B2 mesh; (b)~M1 mesh; (c-d)~B2M1 mesh in 3D view and top view.
The surface position and tangent plane of the B2 and M1 meshes agree at the B2 element corners, but differ elsewhere.
This difference vanishes upon mesh refinement.}
\label{f:B2M1h}
\end{center}
\end{figure}
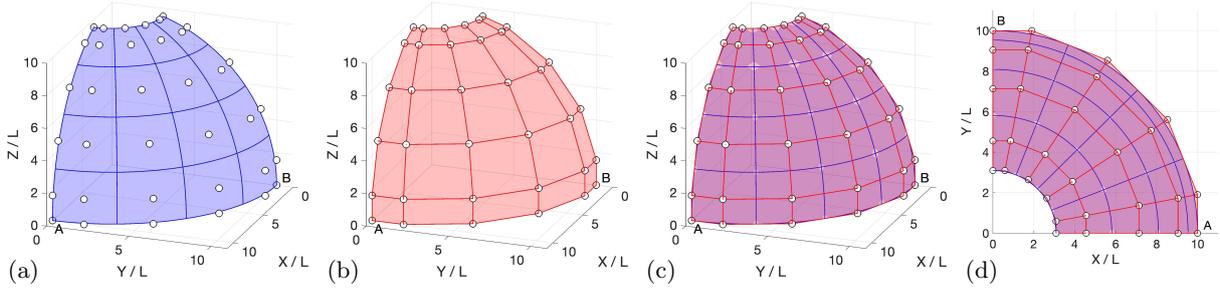

The basic idea of the new approach is now very simple: 
Use quadratic NURBS elements for the bending part and linear Lagrange elements for the membrane part, see Fig.~\ref{f:B2M1}b.
Thus, the original shell surface is discretized with two sets of elements, leading to two separate surface discretizations.
But there is only one set of nodes/control points. 
The new hybrid discretization is denoted \textit{B2M1 discretization} -- quadratic bending / linear membrane discretization.
The resulting surface discretizations for the Scordelis-Lo roof and the hemisphere with hole problem are shown in Figs.~\ref{f:B2M1s} and \ref{f:B2M1h}.
%

The surface difference may seem awkward, but it not problematic. 
On the contrary, it alleviates if not eliminates membrane locking as is demonstrated in the examples of Sec.~\ref{s:Nex}.
In fact, the surface shape only appears in integrals \eqref{e:mf}--\eqref{e:kM} indirectly (as a high order effect in d$A$); 
the evaluation of the integrands themselves only makes use of the shape function derivatives, not the shape functions.
The M1 surface can thus be understood as an auxiliary surface used to construct a lower dimensional membrane strain and stress approximation, and thus is only used for the membrane integrals.
The only physical surface, that is exact and used for all other integrals, in particular the external loads in (\ref{e:mf}.3), is the B2 surface.
In fact, one could work with the B2 surface exclusively, and simply replace the nodal shape functions in the membrane integrals by linear Lagrange shape functions. 
In this nodal view, only additional shape functions are introduced but no additional elements.
The surface thus appears unchanged, and only the shape functions change between bending and membrane contribution, see Fig.~\ref{f:B2M1}c.

It is emphasized that the proposed new discretization approach does not introduce any additional dofs.

\subsection{Implementation}\label{s:impl}

The simplest implementation -- the one used here -- is to define two sets of surface elements: 
Quadratic NURBS elements for the bending terms (B2 elements) and linear Lagrange elements for the membrane terms (M1 elements), see Fig.~\ref{f:B2M1}b.
Since these are well established element types, one can use existing FE codes. 
Only the element connectivity of both element sets needs to be defined.
It is important to ensure that the B2 elements only have bending stiffness but no membrane stiffness, while the M1 elements only have membrane stiffness but no bending stiffness.
This can be easily done with the Koiter model of Eq.~\eqref{e:Koiter}.
In this simple implementation, the number of elements roughly doubles, which increases the computational effort.
But the system matrix and its bandwidth do not increase, as no new dofs and no wider shape functions are introduced.
Solving can thus be expected to be similarly efficient.
This is confirmed by the condition number of the reduced system matrix, which is similarly large for the examples studied in Sec.~\ref{s:Nex}. 
It is natural to use $3\times3$ quadrature points for the B2 elements and $2\times2$ points for the M1 elements, which is done here.

An alternative implementation approach, that can be used to increase efficiency,
is to only use the original B2 surface elements.
In this case the membrane contributions need to be determined by projection onto the M1 surface.  
If the quadrature points for the membrane terms are chosen on the B2 surface such that they coincide with the quadrature points on the M1 surface after projection, and if all required surface quantities ($\bA_\alpha$, $\ba_\alpha$ and $\dif A$) of integrals \eqref{e:mf}--\eqref{e:kM} are then evaluated at these M1 points, the formulation is fully equivalent to the simple two-element-set implementation approach mentioned above.
If the projection is performed at different locations, or if $\dif A$ is taken from the B2 surface, only minor differences can be expected. 
It is important to note that in this projection approach, the tangent vectors $\bA_\alpha$ and $\ba_\alpha$ need to be taken from the M1 surface.
Otherwise it becomes the classical quadratic IGA formulation that locks.

\subsection{Analysis}\label{s:ana}

The proposed new formulation shares similarities with several existing approaches.

The first are strain projection methods that have been developed in the context of mixed, B-bar and ANS methods.
Existing approaches usually project the membrane strains to a basis that is one order lower than the primary field\footnote{The order is lowered in gradient direction, while in perpendicular direction, the original order is maintained.}, e.g.~linear for quadratic IGA, and thus is still continuous between elements.
In ANS this continuity is lost if the classical tying points are used \citep{caseiro14,Kim22}, but it is ensured if tying is done at element corners \citep{casquero22,casquero23}.
This is different to the proposed approach. 
Here the strains become piecewise constant within each B2 element:
Constant in each of the four quadrants of the B2 element, where four different M1 elements overlap, see Figs.~\ref{f:B2M1}-\ref{f:B2M1h}. 
This makes the strains (and corresponding stresses) discontinuous across the M1 element boundaries but continuous across the B2 element boundaries.
The proposed approach can thus be interpreted as an ANS approach with piecewise constant assumed strains that are continuous across element boundaries. 
While these strains are constructed here through the definition of an auxiliary surface, they can be also obtained through a corresponding strain projection.
Strain projection onto discontinuous bases have been (partially) considered for shells in \citet{bouclier13,greco18,hu20,kikis22}, 
but by introducing and condensing-out additional strain variables, which is not done in the approach proposed here.

Alternatively to strain projection, one can also interpret the proposed formulation as a geometric projection of the displacement field onto the M1 surface for the purpose of generating suitable membrane strains.
Thus there are two displacement fields -- the original, $C^1$-continuous one associated with the original B2 surface, and a coarsened, $C^0$-continuous one associated with the M1 surface. 
But the latter is the projection of the former and not a new variable.

The second approach the new formulation shares similarities with, is reduced integration.
As Figs.~\ref{f:B2M1s}-\ref{f:B2M1h} show, the B2 and M1 surfaces as well as their tangent planes agree at the four corners of the B2 elements, because the B2 and M1 shape functions $N_I$ and $N_{I,\alpha}$ agree at these locations.
Therefore, using the corner points for the quadrature of the membrane integrals on the B2 mesh (= trapezoidal rule) is equivalent to using reduced, $1\times1$ integration of the M1 elements.
This, however, is unstable with respect to hourglass modes, such that stabilization schemes are then required.
Quadrature of the M1 elements with $2\times2$ points, used here, naturally provides this.
Reduced stabilized integration similar to the trapezoidal rule\footnote{In \citet{adam15a}, the trapezoidal rule is stabilized by modifying the quadrature points at the boundary.} has been considered by \citet{adam15a}.

\subsection{FE membrane force redistribution}\label{s:redisb}

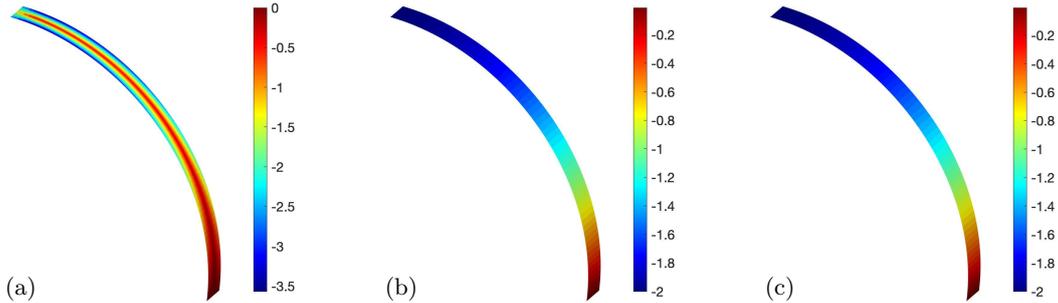
\begin{figure}[h]
\begin{center} \unitlength1cm
\begin{picture}(0,4)
\put(-7,-.5){\includegraphics[height=46mm]{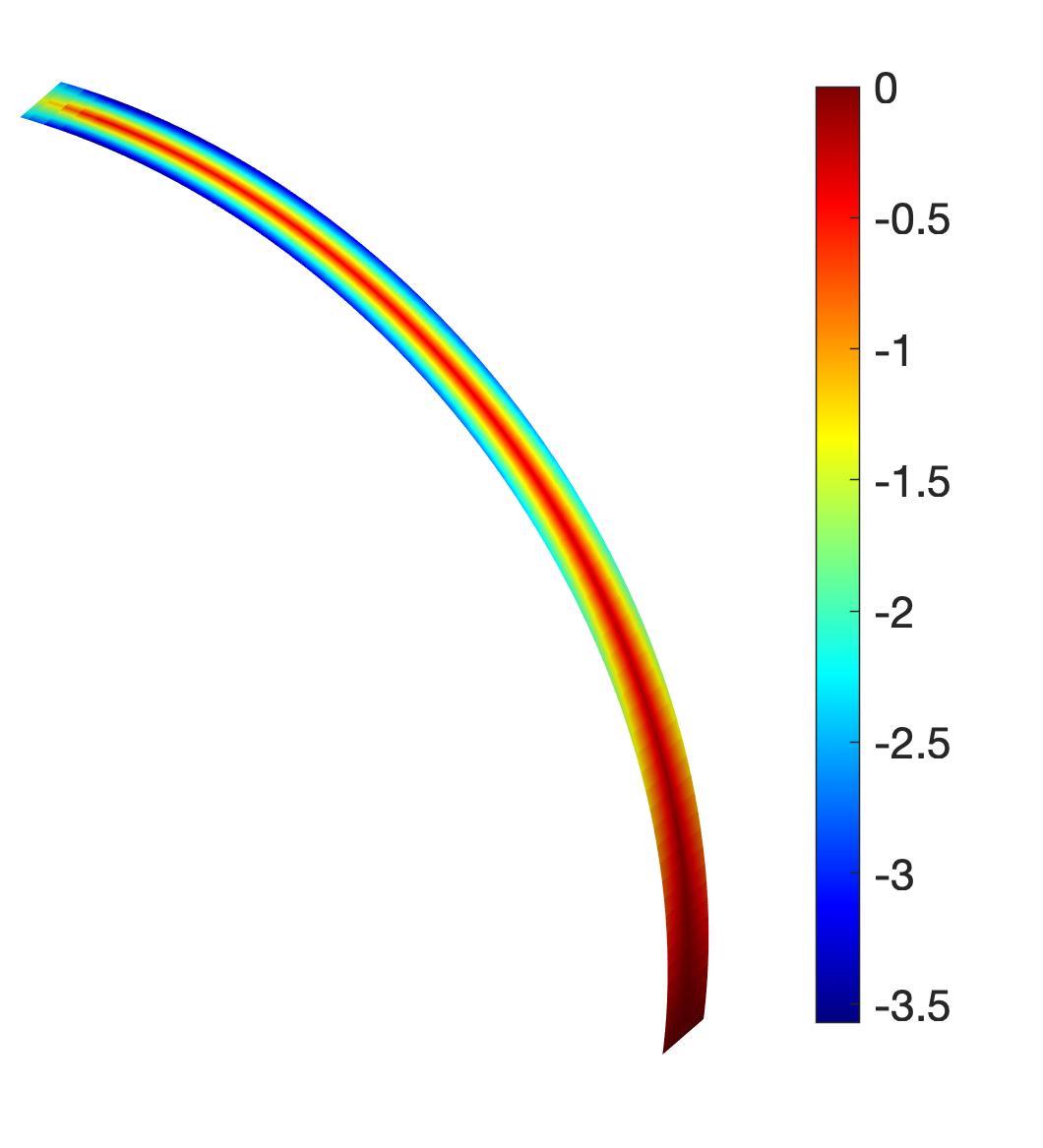}}
\put(-2,-.5){\includegraphics[height=46mm]{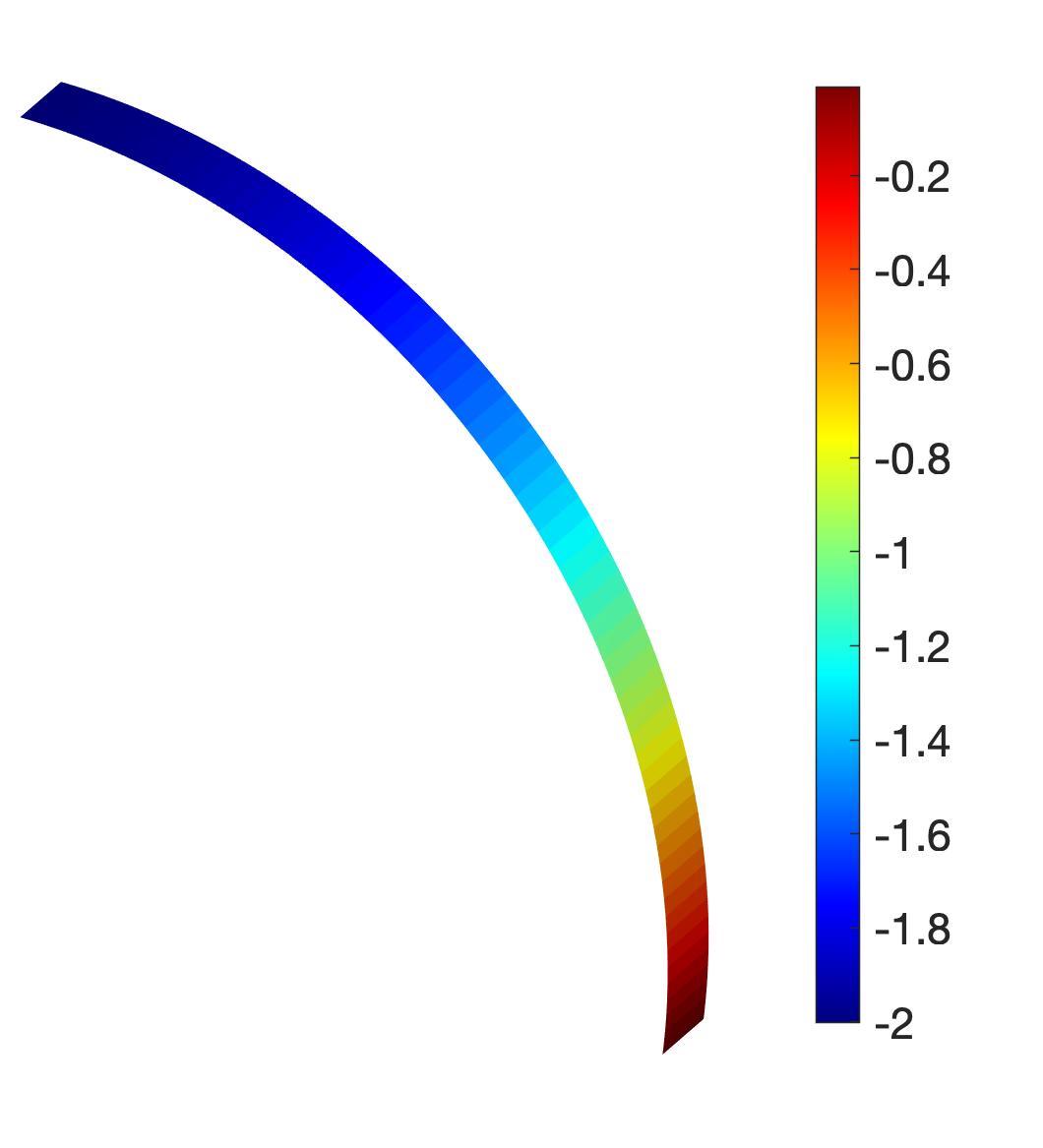}}
\put(3,-.5){\includegraphics[height=46mm]{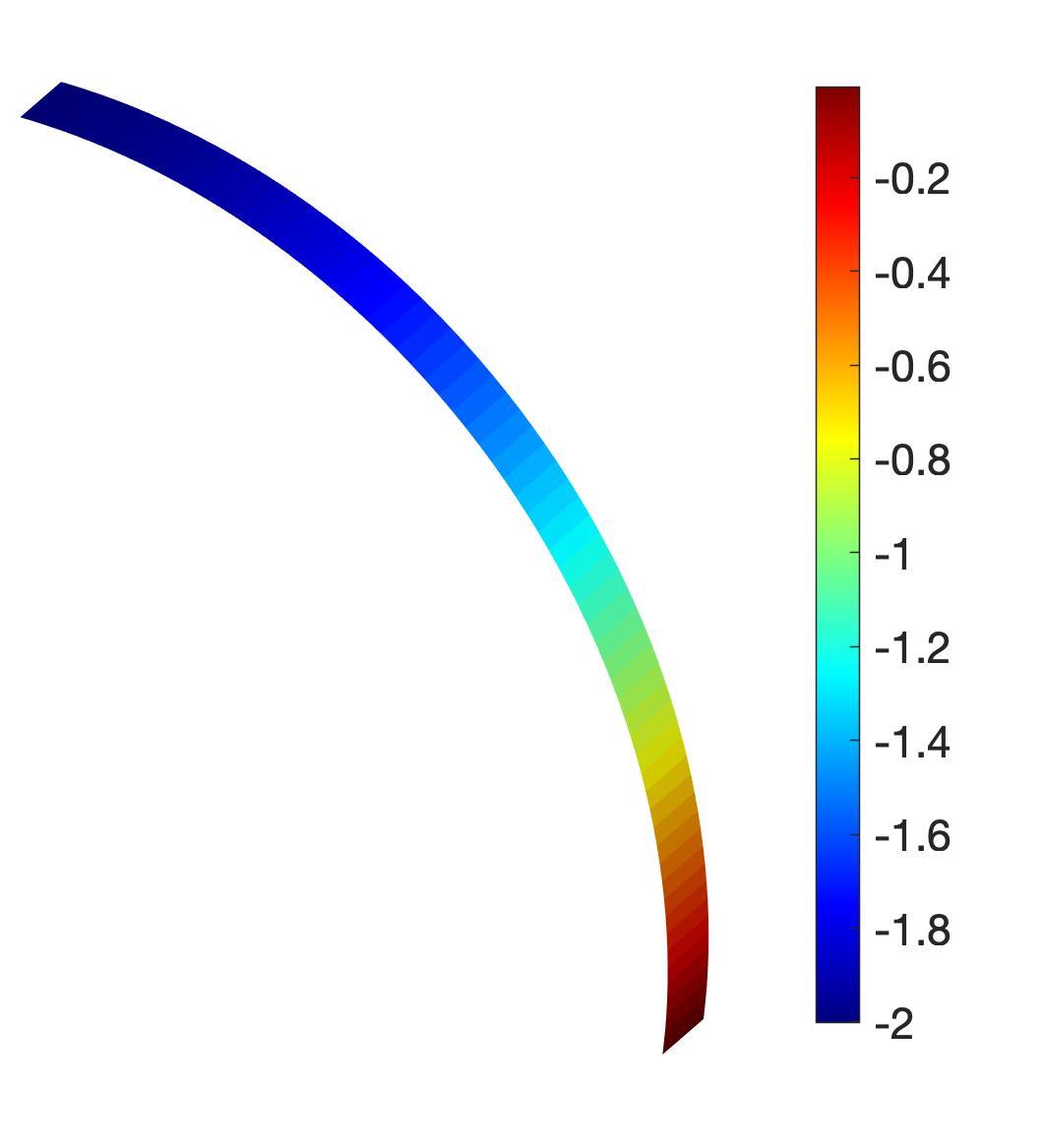}}
\put(-7,-.05){\footnotesize (a)}
\put(-2,-.05){\footnotesize (b)}
\put(3,-.05){\footnotesize (c)}
\end{picture}
\caption{Influence of force redistribution on membrane stress $\sigma$ for the B2M1 discretization: 
(a) No force redistribution; 
(b) Lateral force redistribution; 
(c) Full force redistribution (lateral and axial).
This is the curved cantilever strip problem of Sec.~\ref{s:canti} with $64\times1$ B2 and $65\times 2$ M1 elements.
Shown are the raw stresses.}
\label{f:redisp}
\end{center}
\end{figure}

There is a force mismatch between the B2 and M1 discretizations that is negligible for fine meshes, but relevant for coarse meshes
-- in particular meshes with only a few elements, like the curved cantilever strip mesh from Sec.~\ref{s:canti} that uses only one NURBS element along the lateral strip direction.
The problem arises from the mismatch of nodal forces:
Integrating a constant over a 1D domain leads to the nodal force distribution 1/3, 1/3, 1/3 for one NURBS element, but 1/4, 1/2, 1/4 for two Lagrange elements, which is the lateral discretization used for the curved cantilever strip, see Sec.~\ref{s:canti}.
This mismatch leads to an incorrect stress distribution, even for mesh refinement along the other direction, as is shown in Fig.~\ref{f:redisp}a.

\begin{figure}[h]
\begin{center} \unitlength1cm
\begin{picture}(0,5.8)
\put(-8,-.1){\includegraphics[height=58mm]{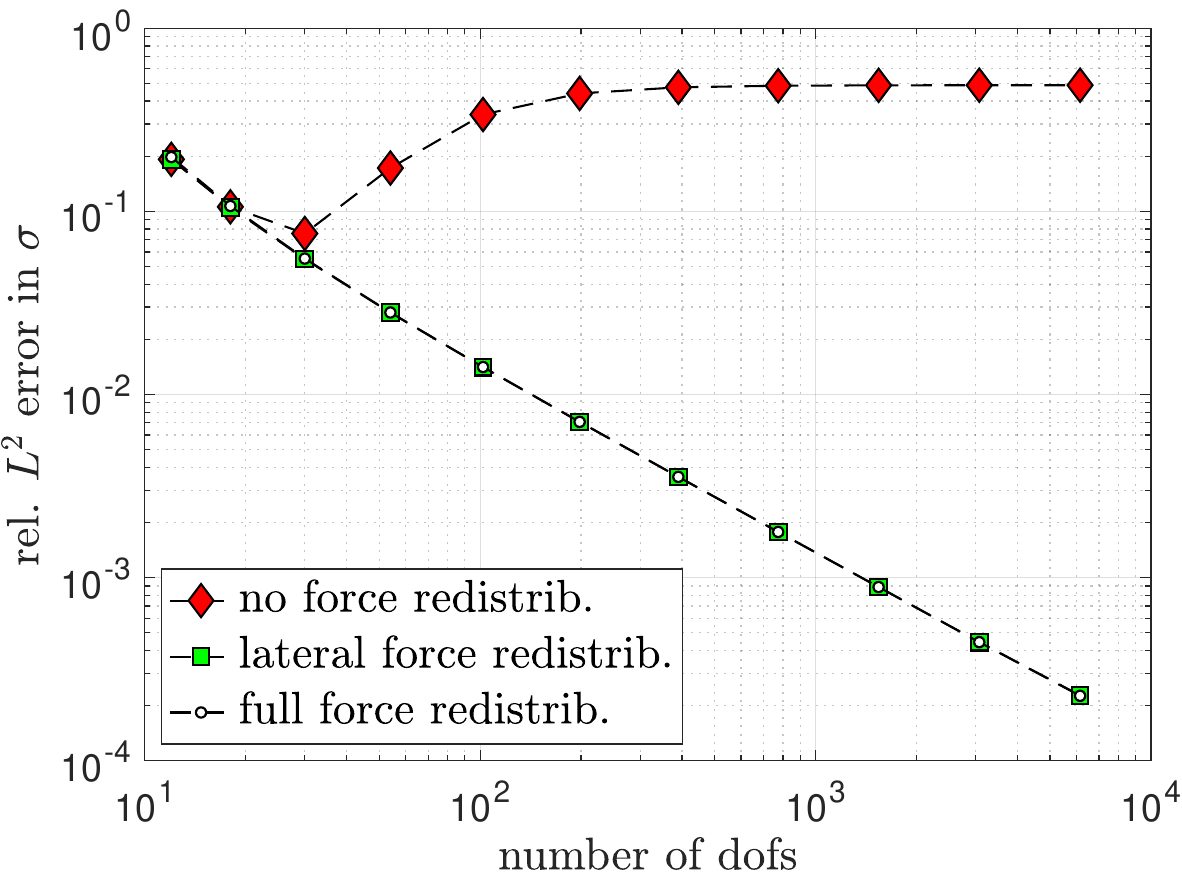}}
\put(0.2,-.1){\includegraphics[height=58mm]{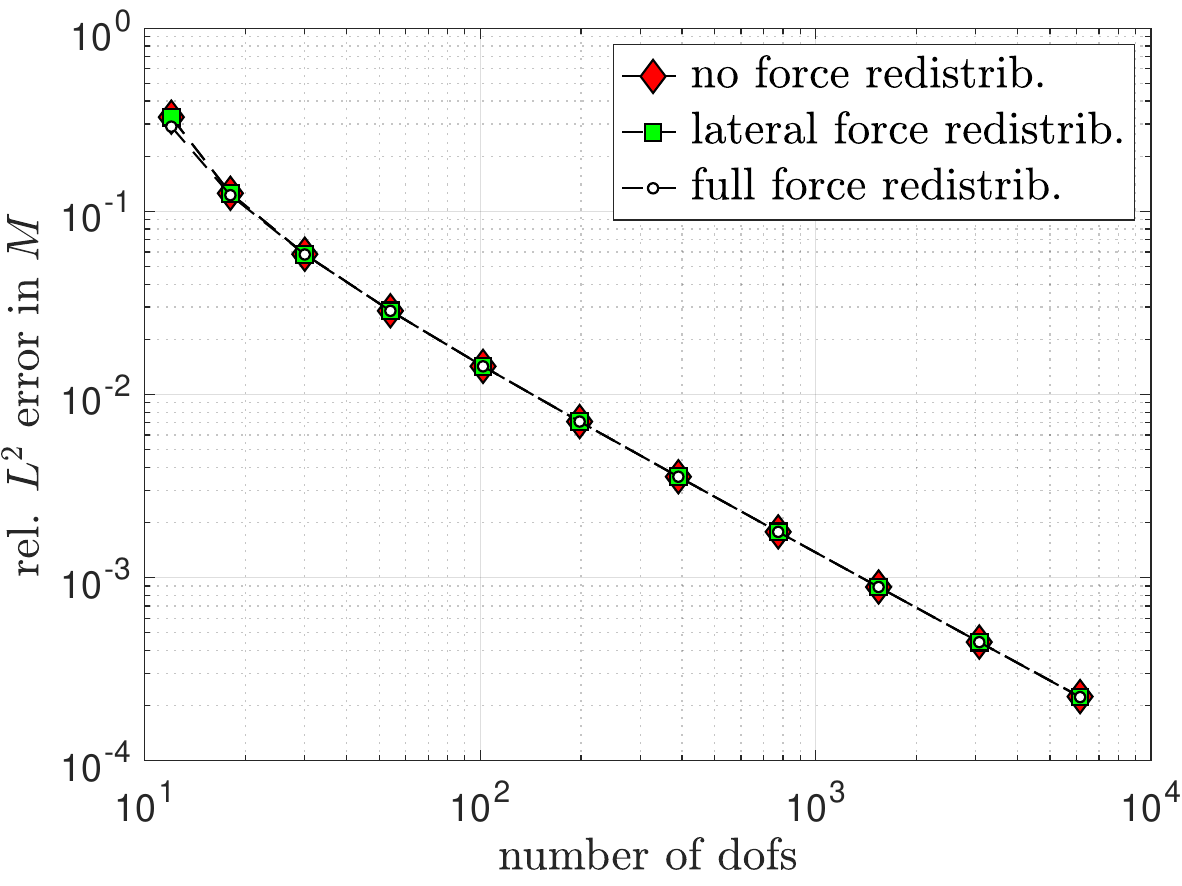}}
\put(-7.95,-.05){\footnotesize (a)}
\put(0.25,-.05){\footnotesize (b)}
\end{picture}
\caption{FE membrane force redistribution in B2M1 for the curved cantilever problem: 
$L^2$-error norm convergence of (a) effective membrane stress $\sigma$, and (b) bending stress $M$;
all for $R/T=1000$. 
For 1D discretizations, the FE membrane forces need to be redistributed at least laterally for $\sigma$ to converge.}
\label{f:CC_FR}
\end{center}
\end{figure}

The problem is easily fixed if 1/6 of the Lagrange center force is redistributed to the outer nodes, such that one gets the new distribution $1/4 + 1/6 \cdot 1/2 = 1/3$ there, and $2/3 \cdot 1/2 = 1/3$ at the center.
This redistribution is illustrated in App.~\ref{s:W} (see Fig.~\ref{f:B2M1FR}).
Applying this redistribution to the FE membrane forces from (\ref{e:mf}.1) leads to the correct stresses shown in Fig.~\ref{f:redisp}b \& \ref{f:redisp}c.
Formally the redistribution procedure can be written as
\eqb{l}
\mf_{\mrr\mrm} = \mW\,\mf_\mrm\,,
\label{e:mfred}\eqe
where $\mW$ is the global redistribution matrix, which can be assembled from the five elemental cases shown in App.~\ref{s:W} (see Fig.~\ref{f:B2M1FR}).
The redistribution leads to the redistributed displacement vector
\eqb{l}
\muu_\mrr = \mW^\mrT\muu 
\label{e:mur}\eqe
used to evaluate $\mf_{\mrr\mrm}$, such that the redistributed membrane stiffness $\mK_{\mrr\mrm} := \partial\mf_{\mrr\mrm}/\partial\muu$ becomes 
\eqb{l}
\mK_{\mrr\mrm} = \mW\,\mK_\mrm\mW^\mrT.
\label{e:mkred}\eqe
Once $\mW$ is assembled from the elemental $\mw^e$ in App.~\ref{s:W}, this redistribution can be easily applied to modify the FE membrane force vector and stiffness according to \eqref{e:mfred} and \eqref{e:mkred}. 
(No redistribution is applied to the bending parts).
As mentioned, force redistribution becomes negligible under mesh refinement (along both directions).
Fig.~\ref{f:CC_FR} shows that for coarse meshes, FE membrane force redistribution is important to obtain correct membrane stresses.
%
The result is for the curved cantilever strip which uses only one B2 and two M1 elements in the lateral direction.
The FE membrane forces therefore need to be redistributed at least in the lateral direction.
Without this, the membrane stresses do not converge, as Fig.~\ref{f:CC_FR}a shows.
The redistribution is only crucial for these stresses. 
Other quantities, such as the bending stresses (see Fig.~\ref{f:CC_FR}b) or the displacements (which look similar to Fig.~\ref{f:CC_FR}b) are not affected significantly -- only for coarse axial discretizations a noticeable difference is seen.

It is concluded that redistribution is essential if very coarse meshes are used in at least one direction. 
If both directions are refined, redistribution is not essential. 
Of course, in this 1D bending example, constant interpolation could also be used in lateral direction, avoiding the need for redistribution altogether.

It is finally noted that the displacement redistribution in \eqref{e:mur} leads to redistributed nodal positions 
\eqb{l}
\mX_\mrr = \mW^\mrT\mX\,,\quad
\mx_\mrr = \mW^\mrT\mx\,.
\label{e:mxr}\eqe
This does not affect the surface shape in the present example, as surface points only move tangentially (and corners do not move at all).
For infinitesimal computations, $\mx_\mrr$ is not needed, and $\mX_\mrr=\mX$ can be used without significant difference in the results.
For large deformations, update (\ref{e:mxr}.2) is required for quadratic convergence during Newton-Raphson iteration.

It is emphasized again that the redistribution becomes insignificant for mesh refinement.

\subsection{Stress recovery}\label{s:sigpp}

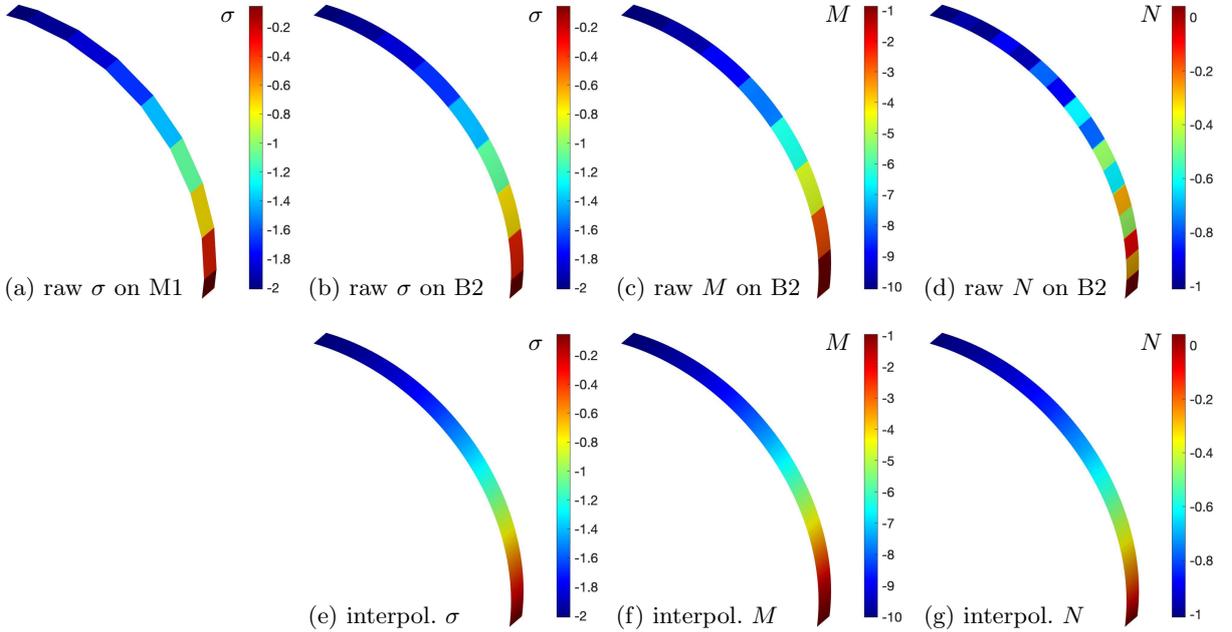
\begin{figure}[h]
\begin{center} \unitlength1cm
\begin{picture}(0,8.4)
\put(-8.05,3.9){\includegraphics[height=46mm]{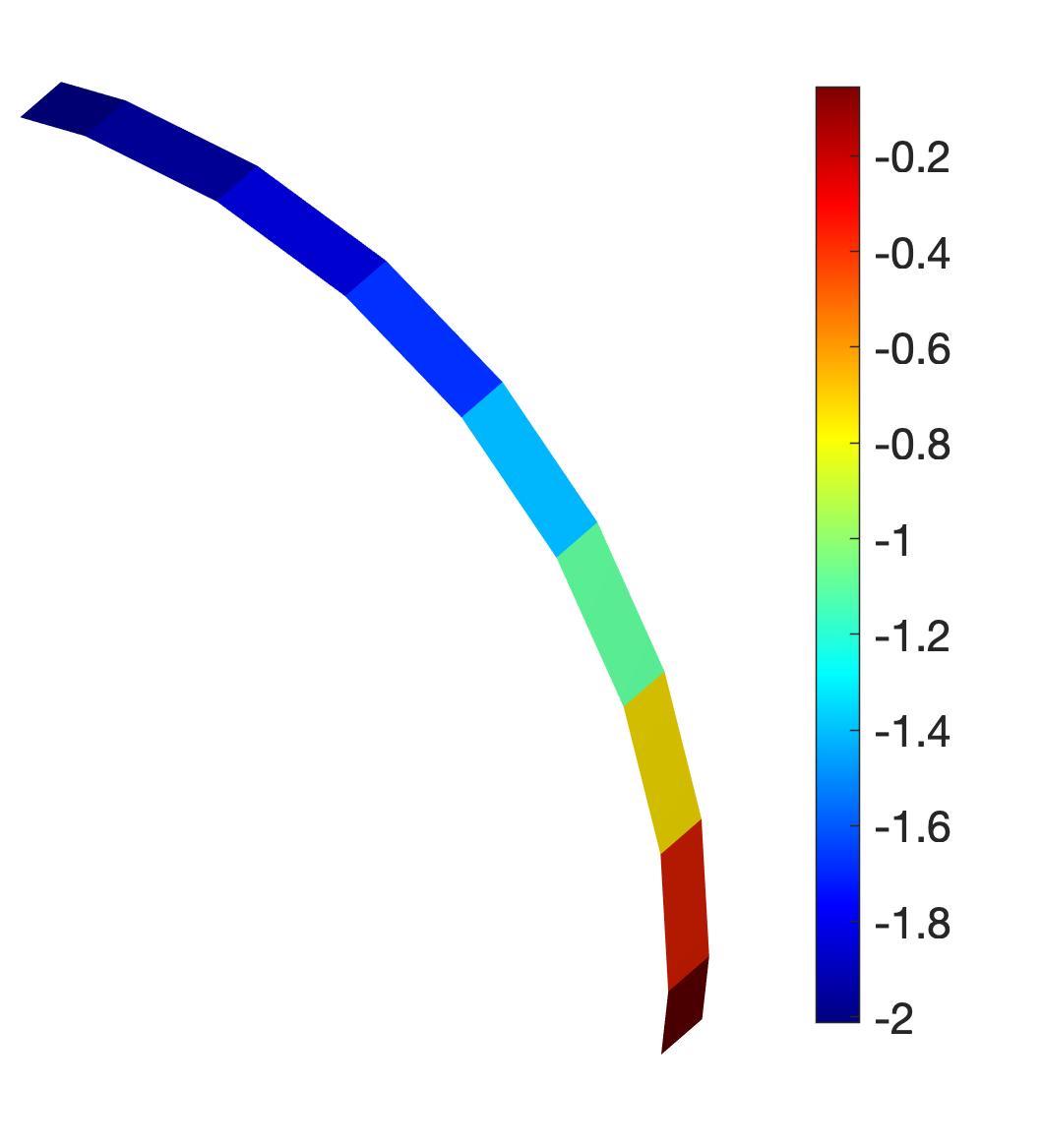}}
\put(-4,3.9){\includegraphics[height=46mm]{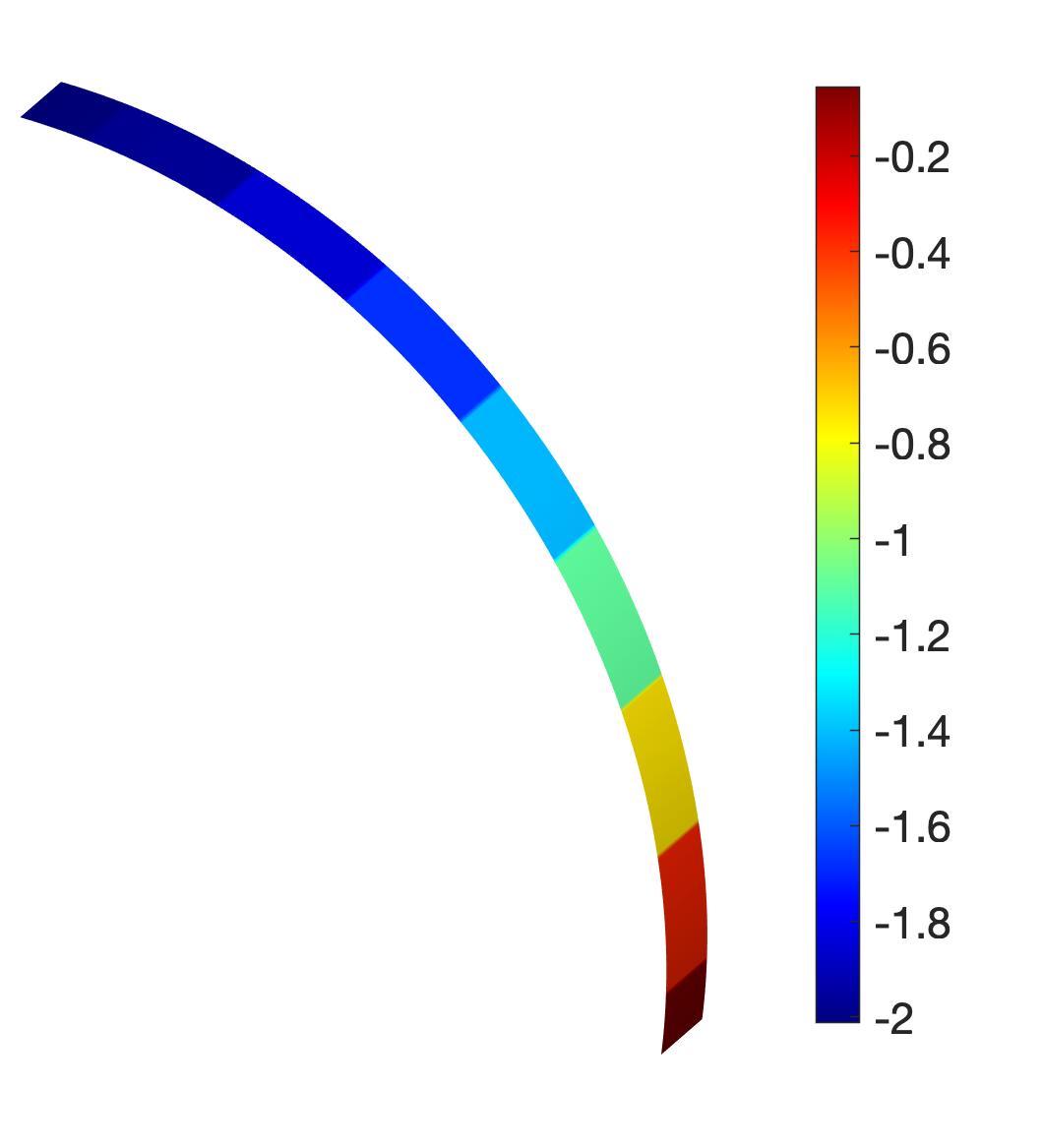}}
\put(0.05,3.9){\includegraphics[height=46mm]{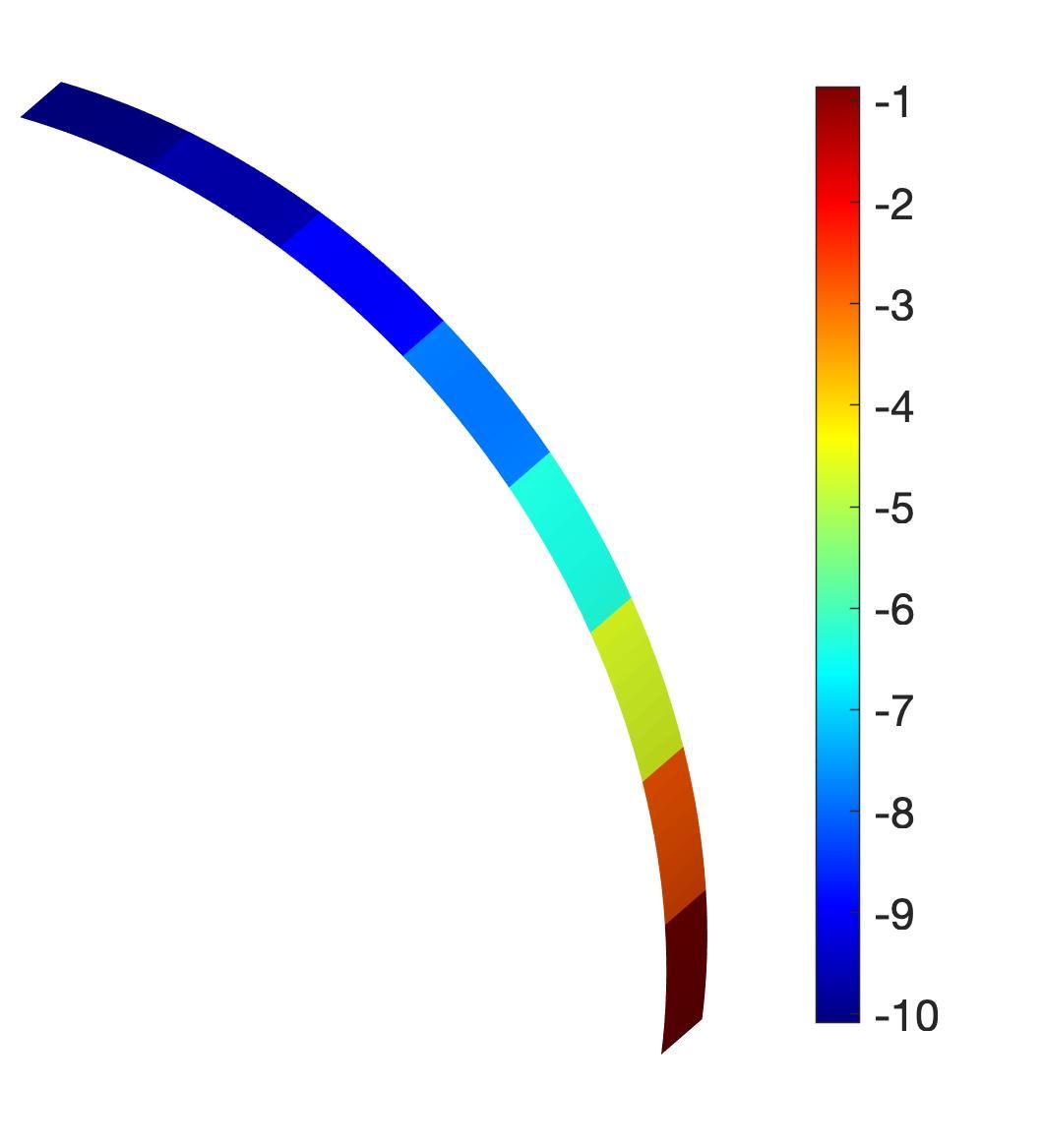}}
\put(4.1,3.9){\includegraphics[height=46mm]{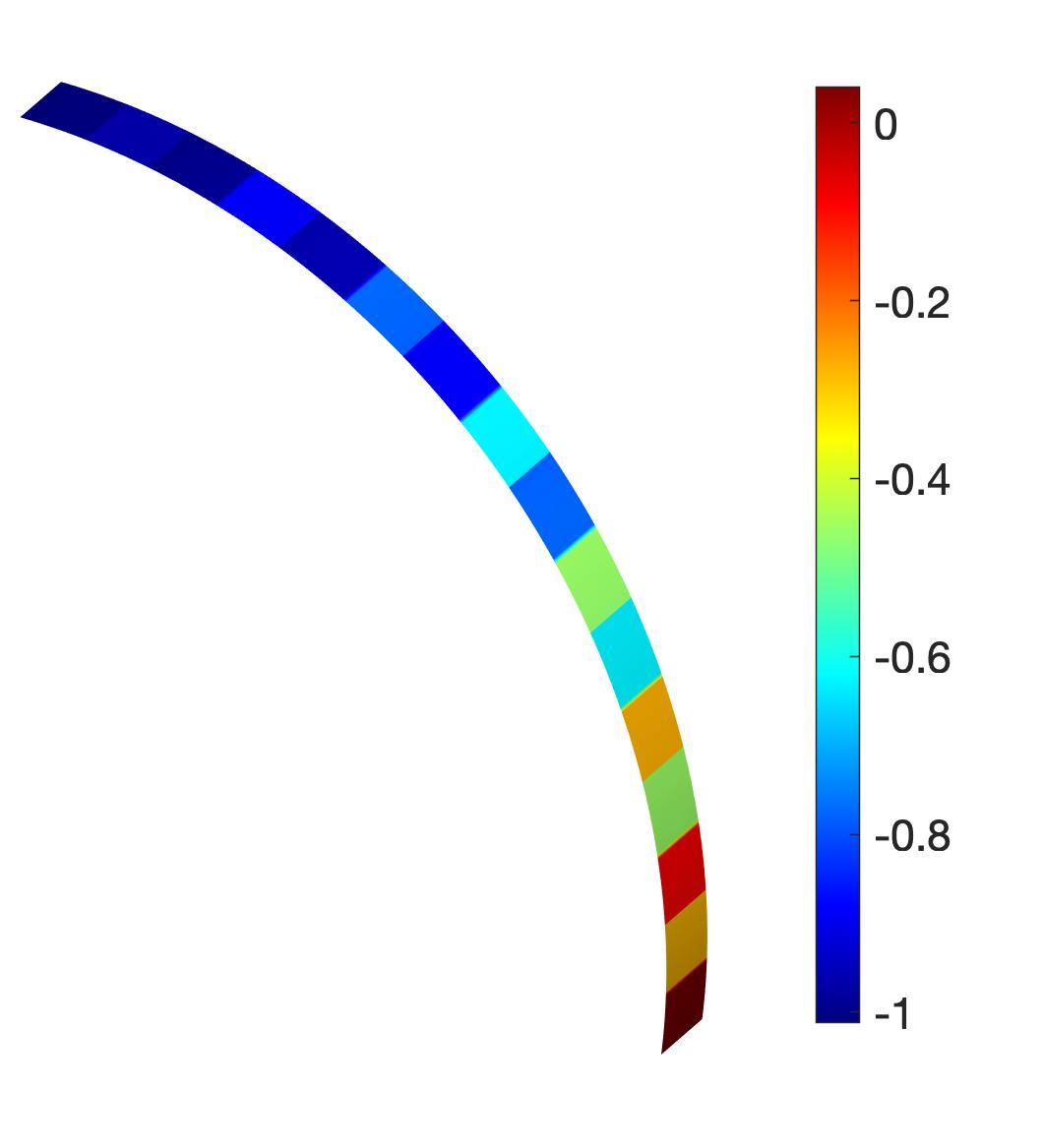}}
\put(-4,-.45){\includegraphics[height=46mm]{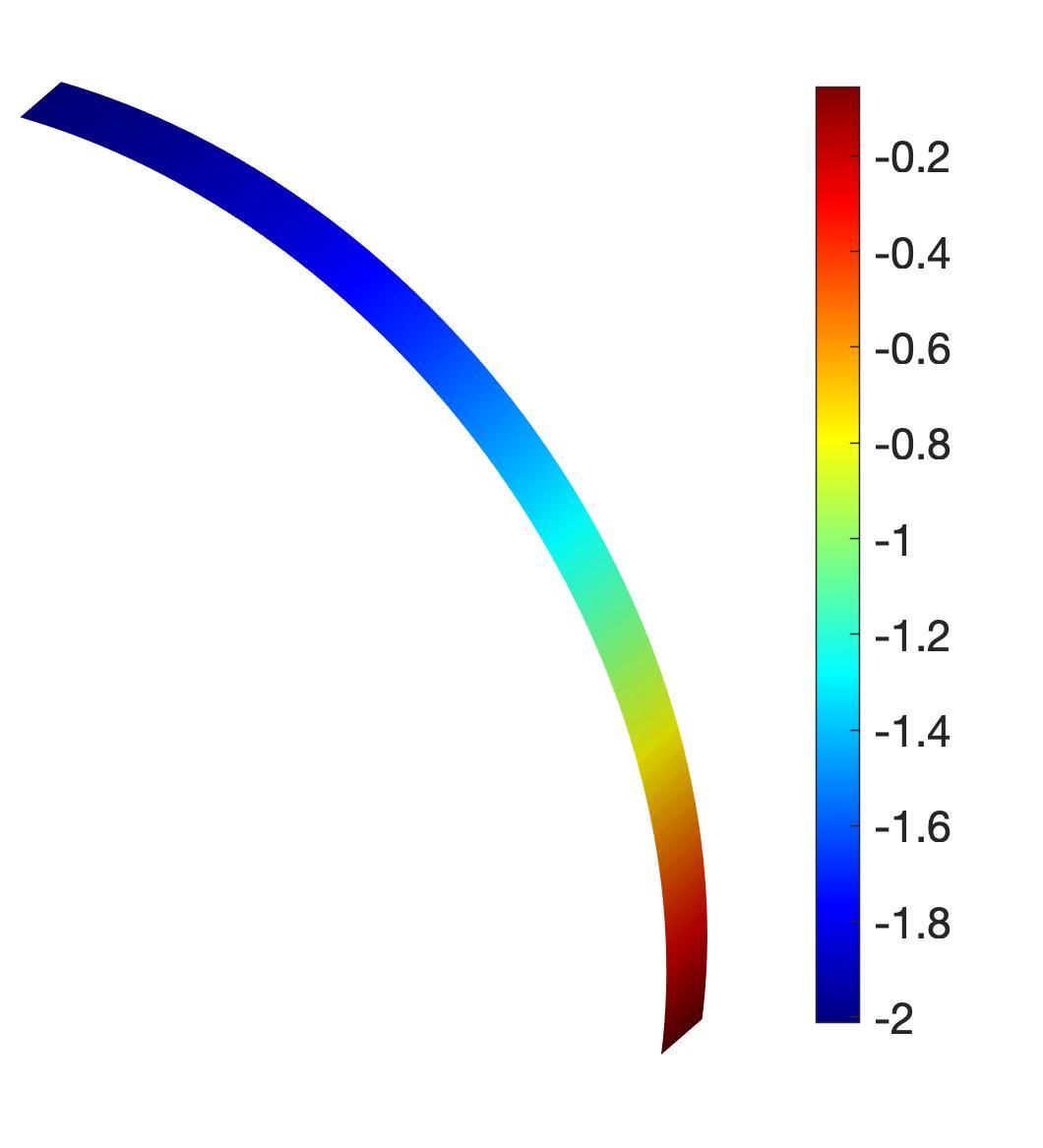}}
\put(0.05,-.45){\includegraphics[height=46mm]{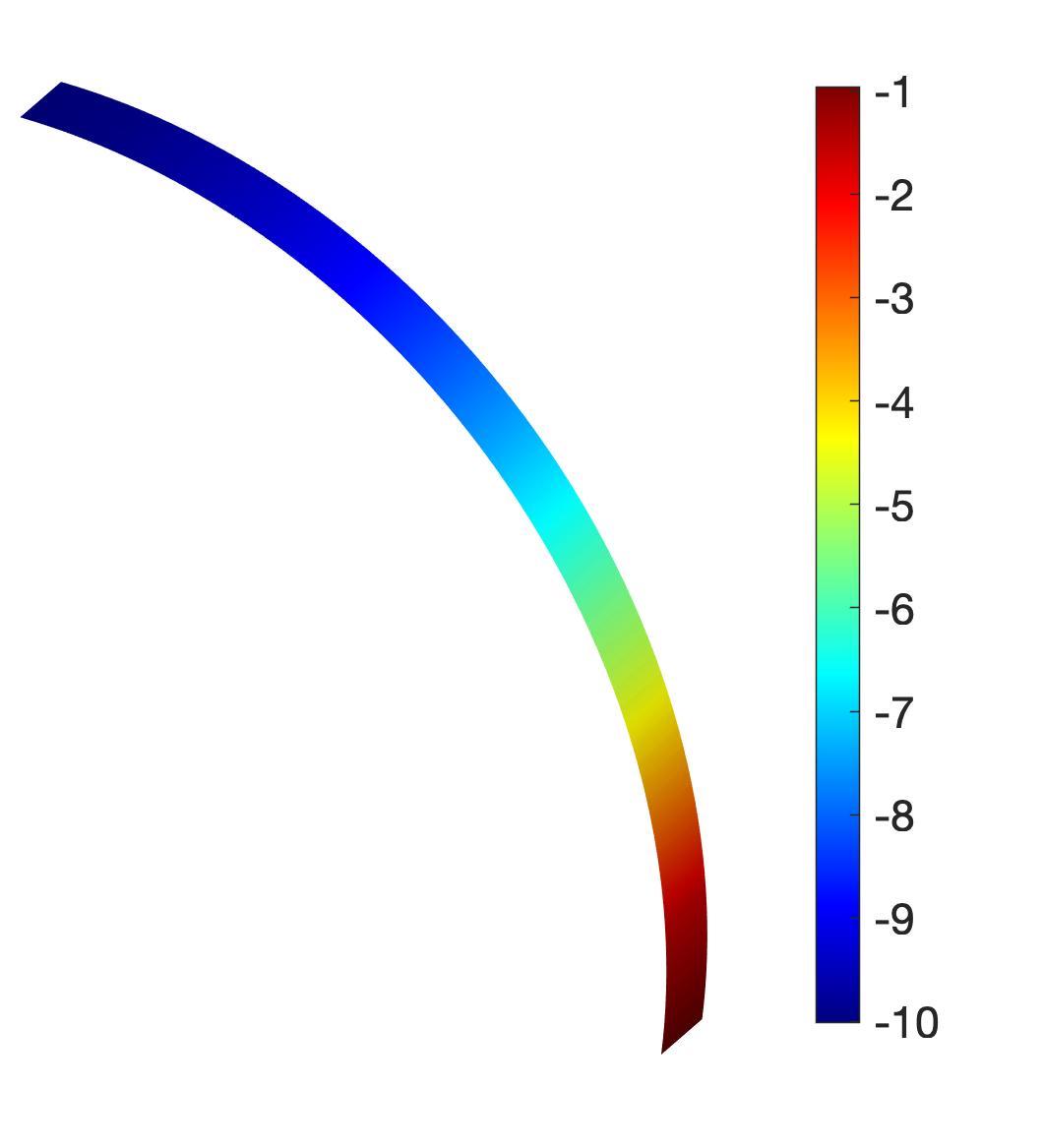}}
\put(4.1,-.45){\includegraphics[height=46mm]{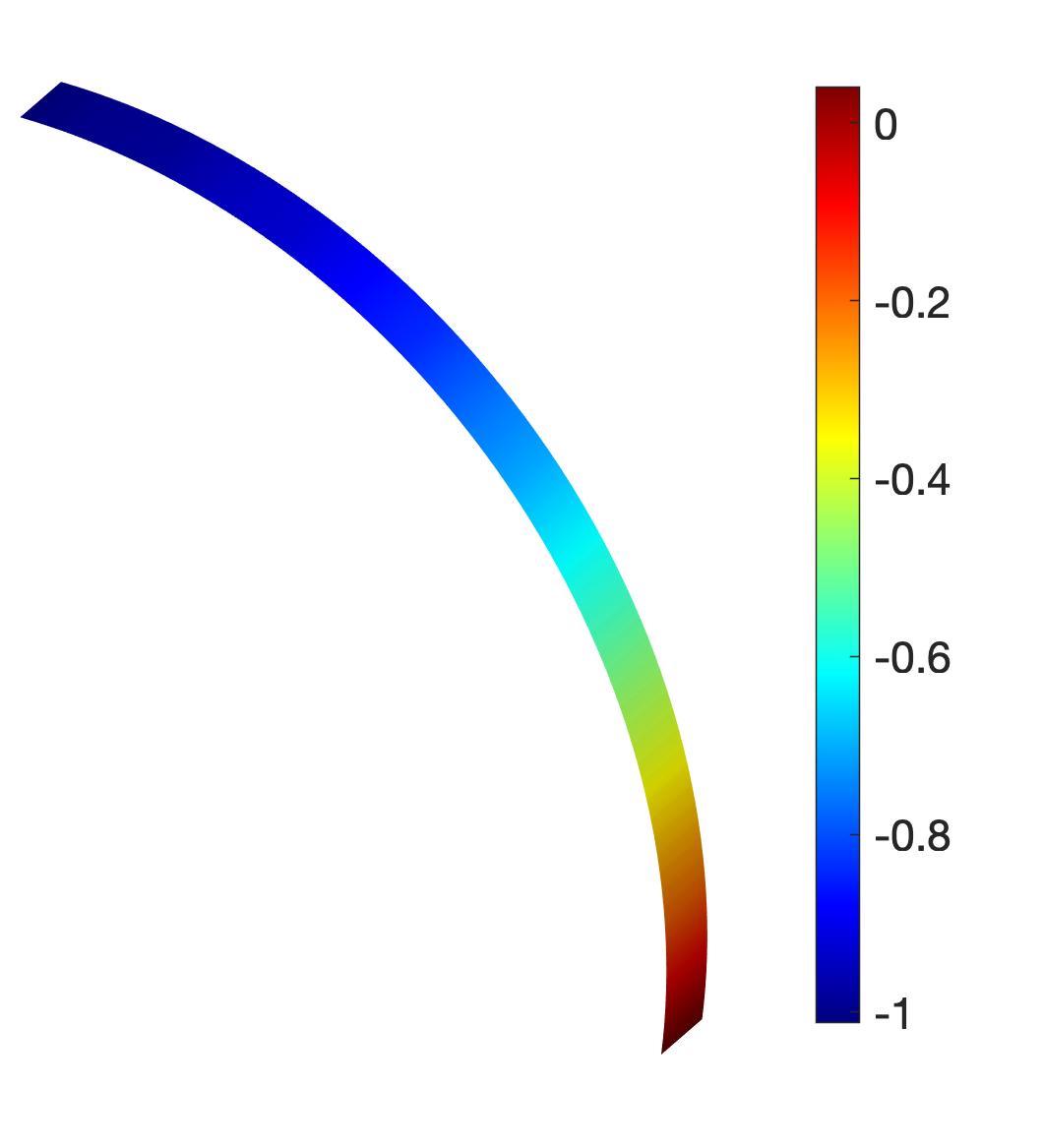}}
\put(-8,4.325){\footnotesize (a) raw $\sigma$ on M1}
\put(-4,4.325){\footnotesize (b) raw $\sigma$ on B2}
\put(0.05,4.325){\footnotesize (c) raw $M$ on B2}
\put(4.1,4.325){\footnotesize (d) raw $N$ on B2}
\put(-4,-.025){\footnotesize (e) interpol.~$\sigma$}
\put(0.05,-.025){\footnotesize (f) interpol.~$M$}
\put(4.1,-.025){\footnotesize (g) interpol.~$N$}
\put(-5.15,7.95){\footnotesize $\sigma$}
\put(-1.1,7.95){\footnotesize $\sigma$}
\put(2.8,7.95){\footnotesize $M$} 
\put(6.95,7.95){\footnotesize $N$} 
\put(-1.1,3.6){\footnotesize $\sigma$}
\put(2.8,3.6){\footnotesize $M$} 
\put(6.95,3.6){\footnotesize $N$} 
\end{picture}
\caption{Stress recovery for B2M1 in case of the curved cantilever strip: 
(a) raw effective membrane stress $\sig$ on M1 mesh; 
(b) raw effective membrane stress $\sig$ projected onto B2 mesh; 
(c) raw bending moment $M$ on B2 mesh; 
(d) raw Cauchy membrane stress $N$ on B2 mesh; 
(e) linearly interpolated effective membrane stress $\sig$ on B2 mesh; 
(f)  linearly interpolated bending moment $M$ on B2 mesh; 
(g) linearly interpolated Cauchy membrane stress $N$ on B2 mesh; 
all for $8 \times 1$ B2 and $9 \times 2$ M1 elements.}
\label{f:pp}
\end{center}
\end{figure}

In order to assess stress accuracy, the raw stresses are examined here, without employing stress smoothing. 
This is important for revealing possible stress oscillations.
The stresses $\sigma$ and $M$ can be evaluated directly and straightforwardly on their corresponding surfaces (M1 and B2), as shown in Figs.~\ref{f:pp}a and \ref{f:pp}c.
This is done here with an $8 \times 8$ visualization grid per element.

Evaluating $N$ from \eqref{e:NsbM} is more tricky, since it involves $\sigma$ on M1 and $M$ on B2.
Two approaches for $N$ can be used: 
Project $M$ onto the M1 mesh, or project $\sigma$ onto the B2 mesh.
The second approach is preferred as it uses the more accurate surface.
The following procedure is used for this: \\[-8mm]
\begin{enumerate}
\item Loop over the B2 elements. \\[-7mm]
\item Obtain $M$ the usual way at chosen visualization locations $\xi^\alpha_\mathrm{B}$. \\[-7mm]
\item Project locations along B2 surface normal onto M1 mesh: Gives corresponding $\xi^\alpha_\mathrm{M}$ on M1. \\[-7mm]
\item Evaluate $\sig$ at these $\xi^\alpha_\mathrm{M}$. \\[-7mm]
\item Use $\sig$ and $M$ to obtain $N$ according to \eqref{e:NsbM}. \\[-7mm]
\item Display both $\sig$ and $N$ at $\xi^\alpha_\mathrm{B}$ on B2 surface. \\[-7mm]
\end{enumerate}
The result is shown in Figs.~\ref{f:pp}b and \ref{f:pp}d.
It works well, but $N$ does not look so nice, as it results from the subtraction of two step functions.
Even though this looks much better than what comes out of IGA (see Fig.~\ref{f:CC_us}), there is a simple way to get even better stresses. 
This is based on the observation that $\sigma$ seems to be most accurate at the B2 element corners (= M1 element center).
Following steps 1-4 from above with $\xi^\alpha_\mathrm{B}$ taken now as the B2 corner points, the alternative procedure is to \\[-8mm]
\begin{itemize}
\item[5.] Linearly interpolate $\sigma$ from the B2 corner values (Fig.~\ref{f:pp}e). \\[-7mm]
\item[6.] Average $M$ at the B2 corners (as it jumps there) and then interpolate it linearly (Fig.~\ref{f:pp}f). \\[-7mm]
\item[7.] Using these $\sigma$ and $M$, obtain $N$ according to \eqref{e:NsbM} (Fig.~\ref{f:pp}g). \\[-7mm]
\end{itemize}
This gives very good stress fields as Figs.~\ref{f:pp}e-g show.
The first procedure still shows the raw stresses, while the second post-processes them based on linear interpolation.
In all subsequent stress plots, $N$ is shown according to procedure 2, while procedure 1 is used for $\sigma$.
In all subsequent convergence plots, procedure 1 is used.
Some of the convergence plots show the $L^2$-error according to \eqref{e:L2}.
For this, the $L^2$-error of $\bu$, $M$ and $N$ is integrated over the B2 surface, while the M1 surface is used for $\sigma$. 
Full Gaussian quadrature is used for this ($3\times3$ on B2 elements, $2\times2$ on M1 elements), even though this may cause quadrature errors for coarse meshes, where elemental stresses tend to vary more strongly.

\section{Numerical examples}\label{s:Nex}

The behavior of the proposed new discretization is illustrated by three numerical examples with increasing complexity:
A 1D cantilever strip with uniaxial curvature and bending, the Scordelis-Lo roof with uniaxial curvature but biaxial bending, and a hemispherical shell with biaxial curvature and bending.

\subsection{Curved cantilever strip}\label{s:canti}

\subsubsection{Problem setup}

\begin{figure}[h]
\begin{center} \unitlength1cm
\begin{picture}(0,4.6)
\put(-7.75,0){\includegraphics[height=44mm]{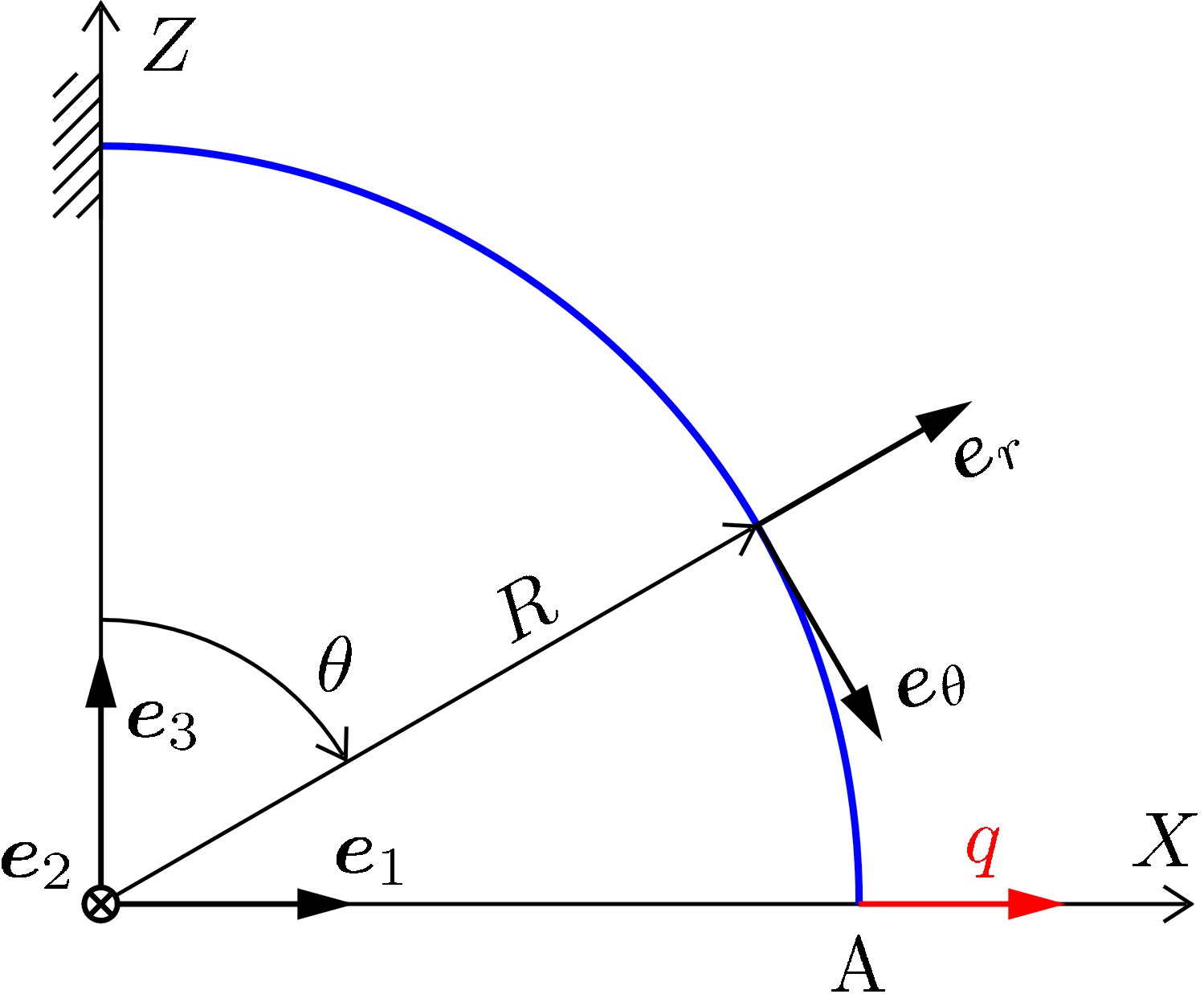}}
\put(-1.75,-.25){\includegraphics[height=50mm]{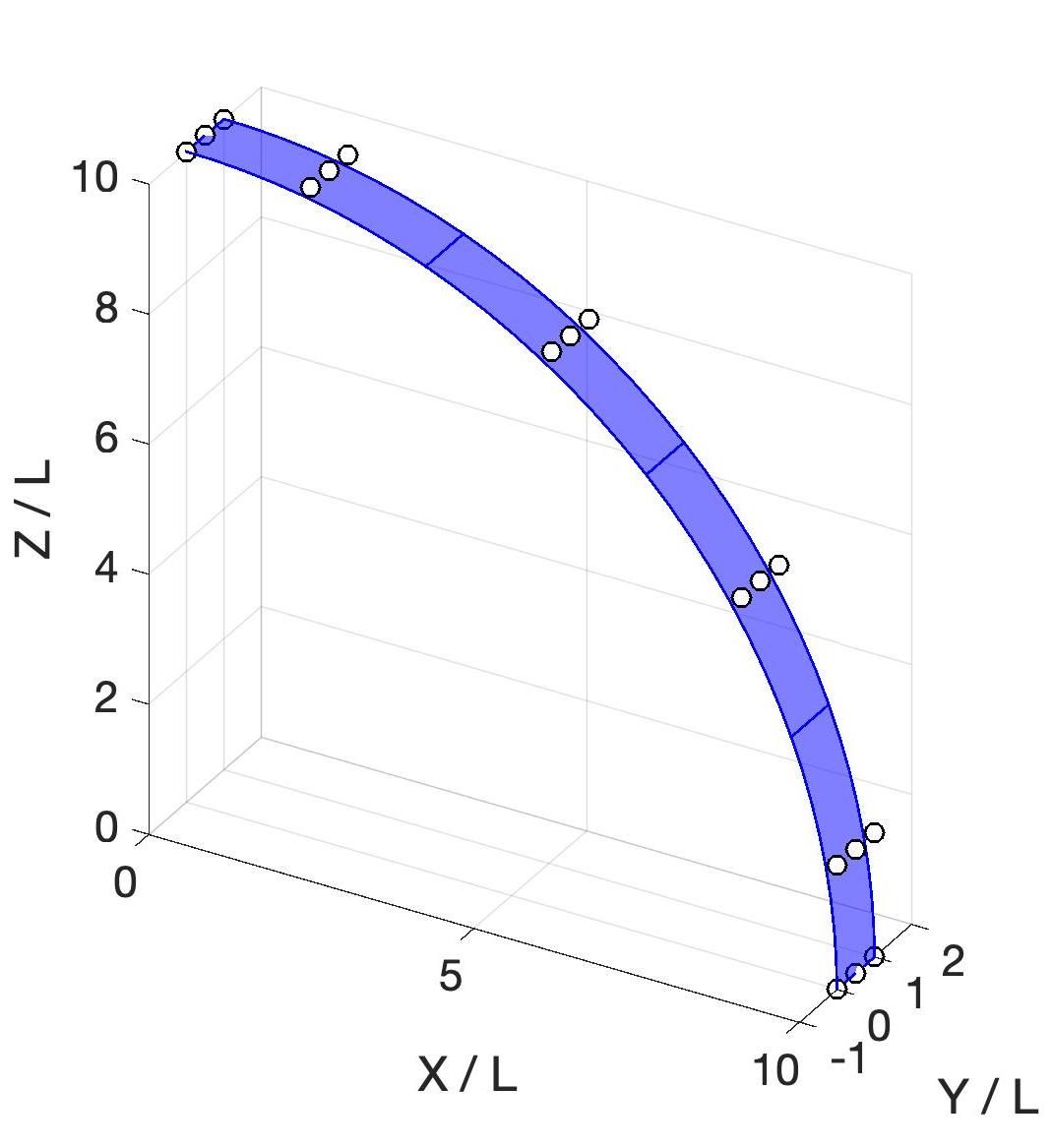}}
\put(3.4,-.25){\includegraphics[height=50mm]{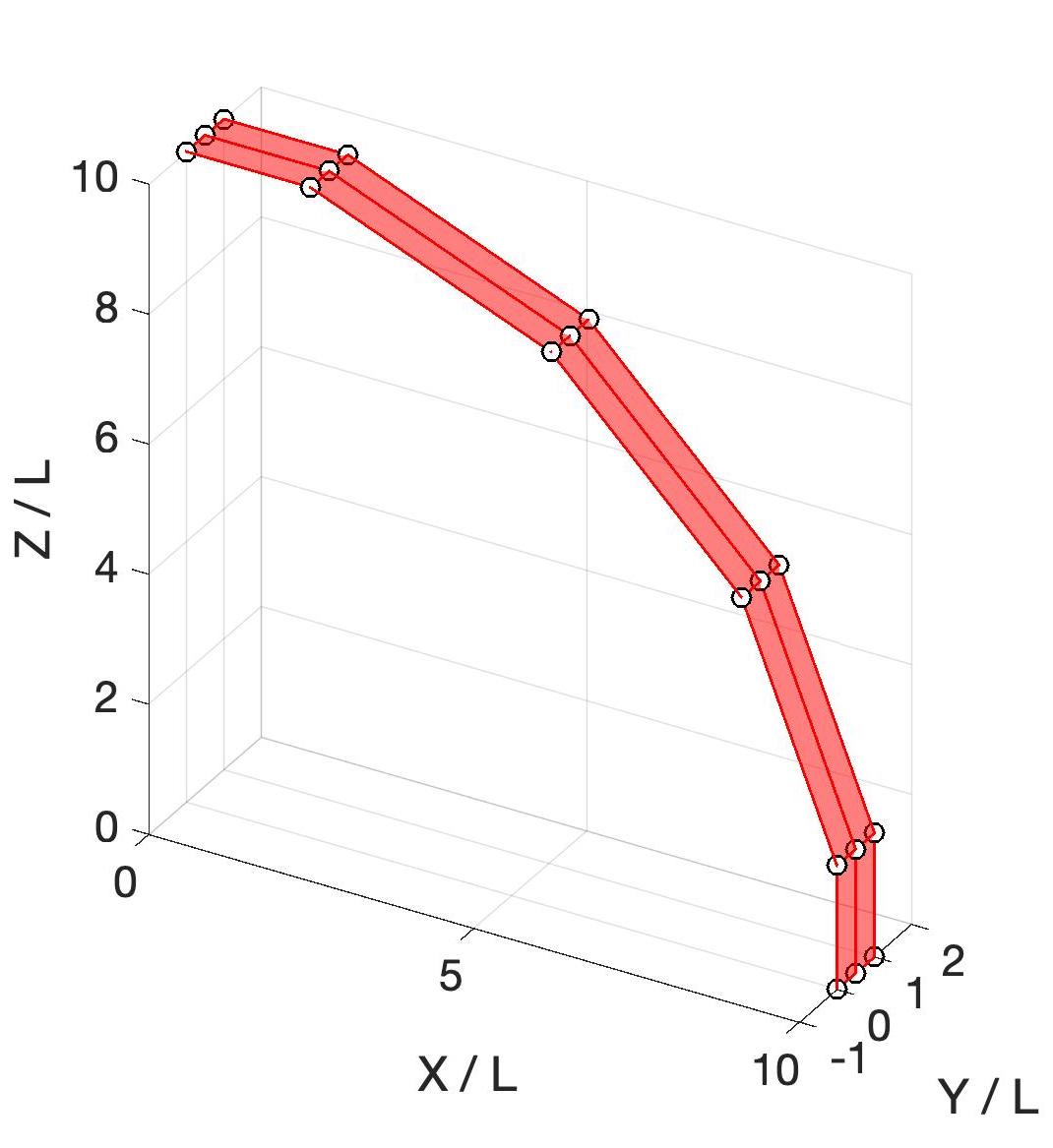}}
\put(-7.95,-.05){\footnotesize (a)}
\put(-1.75,-.05){\footnotesize (b)}
\put(3.4,-.05){\footnotesize (c)}
\end{picture}
\caption{Curved cantilever strip: Geometry, boundary conditions, surface parameterization (a) and discretization into B2 (b) and M1 (c) elements. 
The B2 discretization represents the actual surface, while the M1 surface is just an auxiliary one for the membrane strain determination.}
\label{f:CC0}
\end{center}
\end{figure}
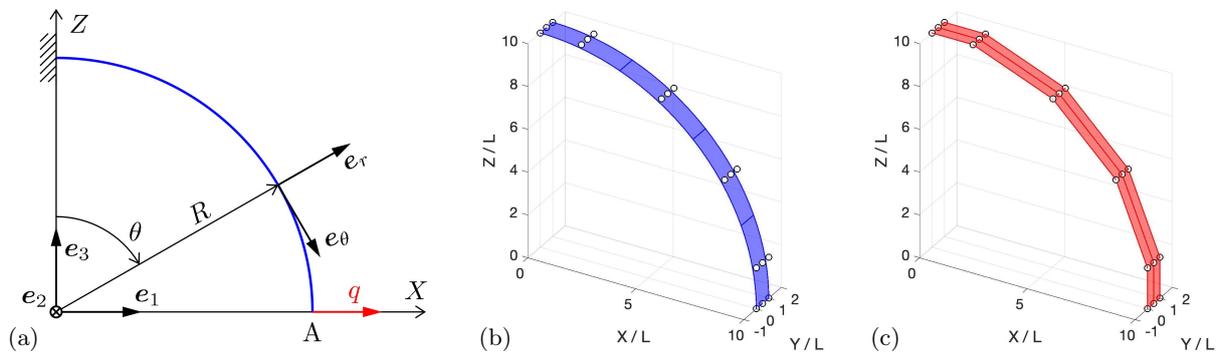

The problem consists of a $90^\circ$ cantilever arc with width $L$, radius $R=10L$, thickness $T$, 3D Young's modulus $E$ and Poisson's ratio $\nu = 0$.
The cantilever is clamped at one end and loaded by the shear traction $q$ along $\be_2$ on the other end, as shown in Fig.~\ref{f:CC0}.
The Cauchy stress and bending moment follow from statics.
For small deformations, one finds
\eqb{l}
N^1_1 = q\cos\theta =: N(\theta)\,,\quad
M^1_1 = qR\cos\theta =: M(\theta)\,.
\label{e:NMana}\eqe
With the beam curvature $-1/R$, the effective stress then becomes
\eqb{l}
\sigma^1_1 = 2q\cos\theta =: \sigma(\theta) \,,\quad
\label{e:sigana}\eqe
according to \eqref{e:NsbM}.
From these stresses one can then determine the exact solution for the displacement field $\bu$, as is shown in App.~\ref{s:anasol}.
One finds that the maximum radial displacement occurs at point A and is $u_\mathrm{A} \approx -0.3\pi L$ for the common choice $qR^3/(ELT^3)=-0.1$ and $T/R\ll1$. 
The extreme stresses occur at the support where $\cos\theta = 1$.
For $q=-1$ and $L=1$ they become, $\sigma_\mathrm{min} = -2$, $M_\mathrm{min} = -10$ and $N_\mathrm{min} = -1$.

For classical IGA, the strip is meshed with $m\times1$ NURBS elements (= B$p$M$p$ discretization).
The B2M1 discretization uses the same NURBS mesh together with $(m+1)\times2$ linear Lagrange elements for the membrane part.
The two meshes are shown in Fig.~\ref{f:CC0}b-c.
Strictly, the lateral discretization is not necessary in this example, as there is no variation along $\be_2$.
Accordingly, the number of degrees-of-freedom is taken from a single line of nodes, i.e.~$n_\mathrm{dof} = 3(m+p)$.
The lateral discretization is still useful for verifying the implementation.

\subsubsection{Infinitesimal case}

\begin{figure}[h]
\begin{center} \unitlength1cm
\begin{picture}(0,8.5)
\put(-8,3.9){\includegraphics[height=46mm]{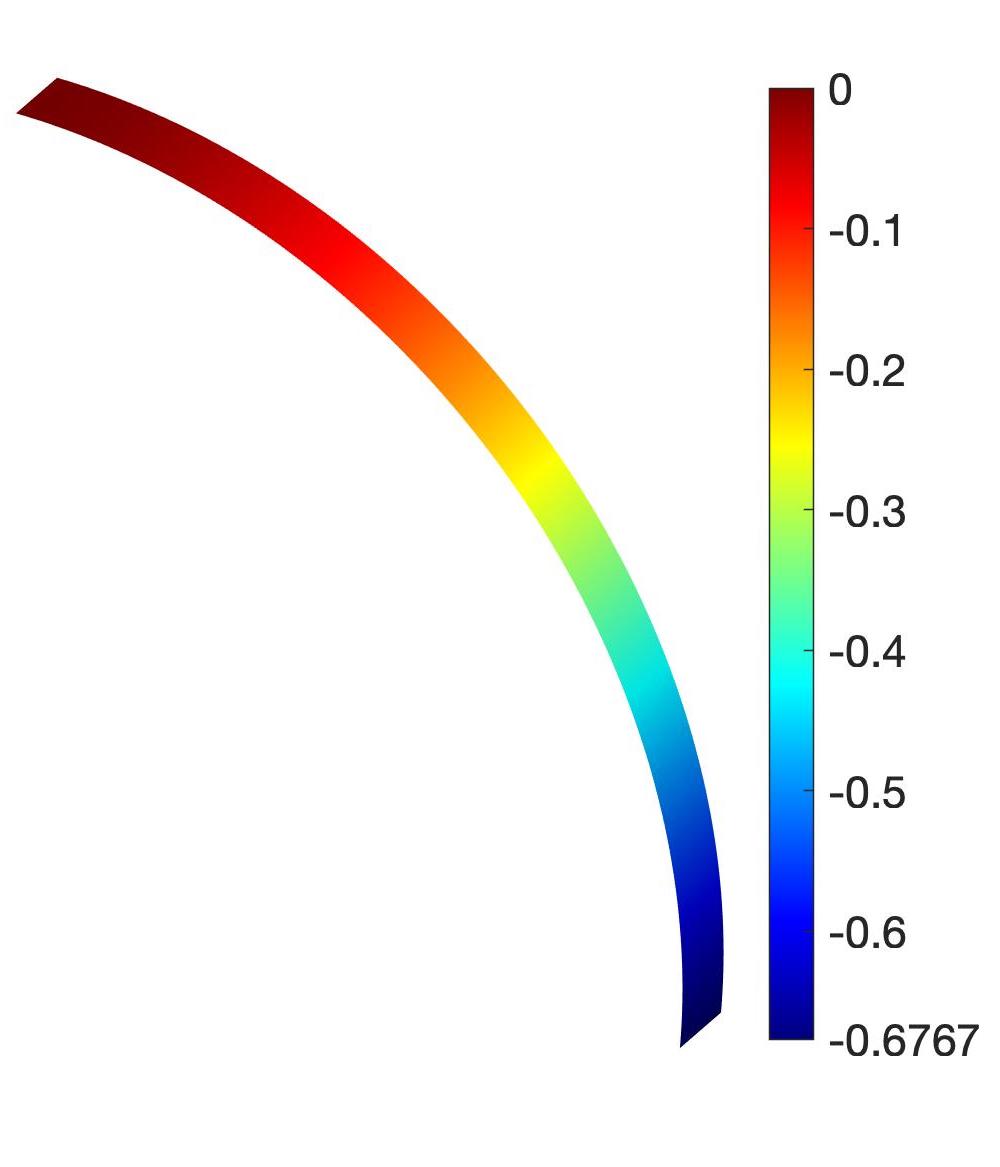}}
\put(-4,3.9){\includegraphics[height=46mm]{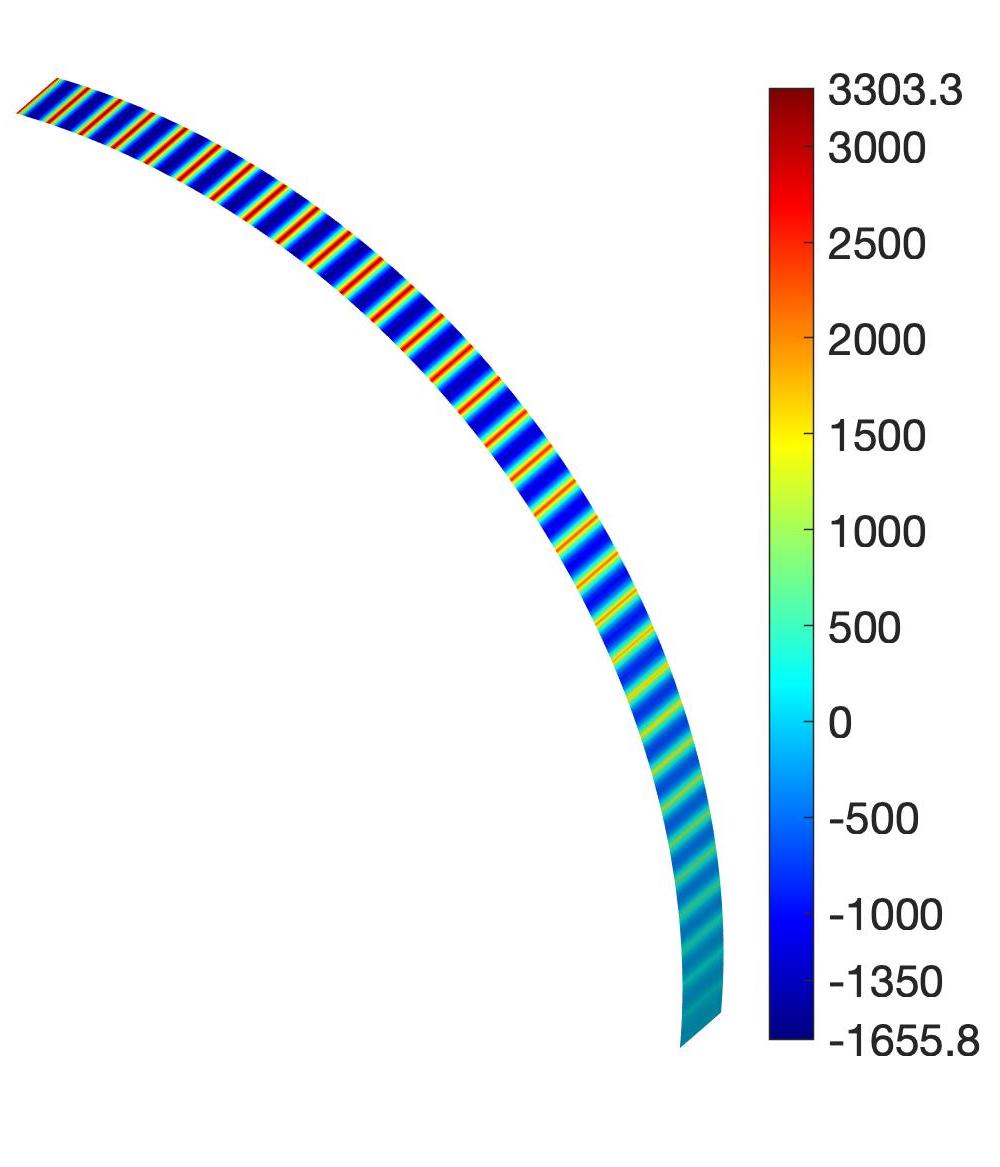}}
\put(0.0,3.9){\includegraphics[height=46mm]{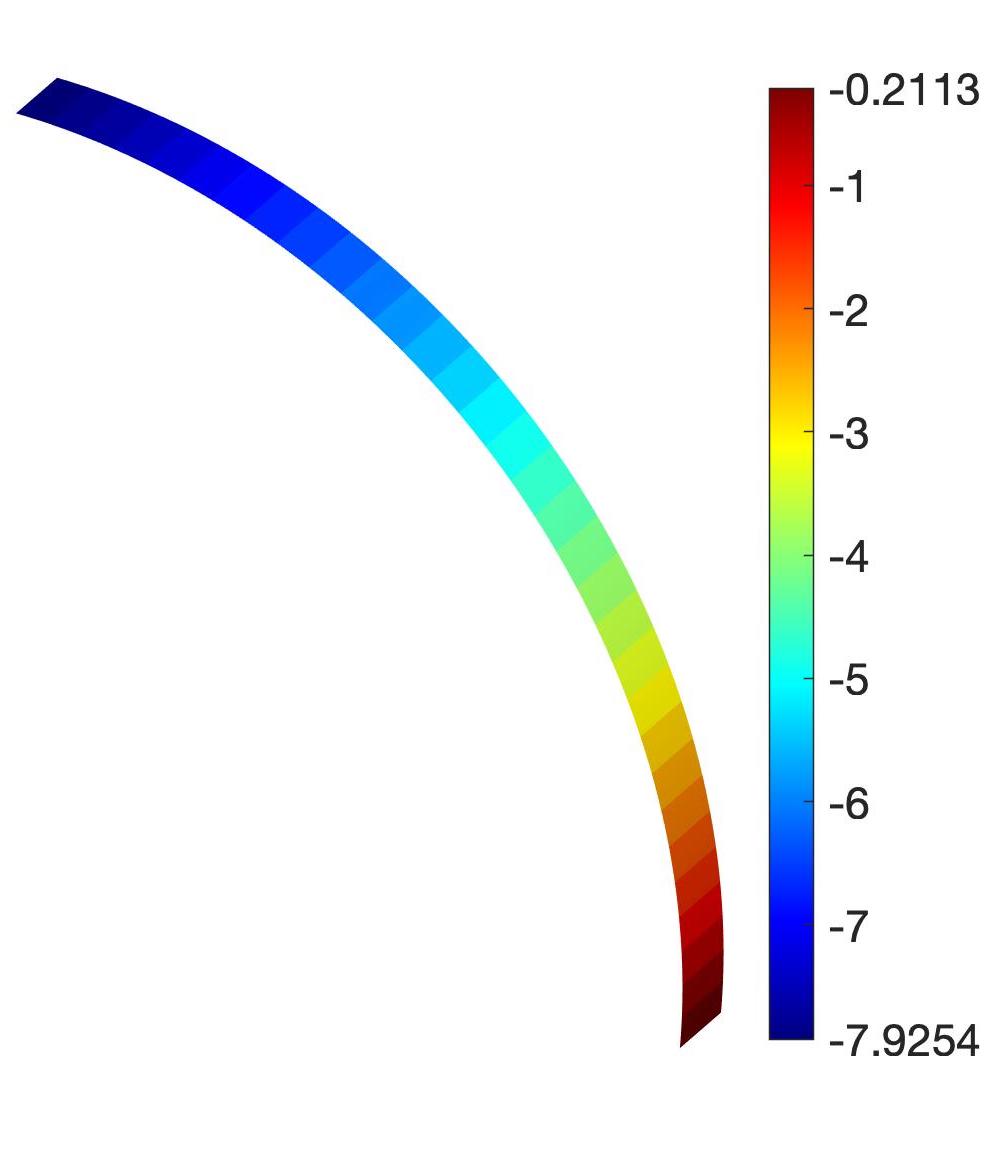}}
\put(4.0,3.9){\includegraphics[height=46mm]{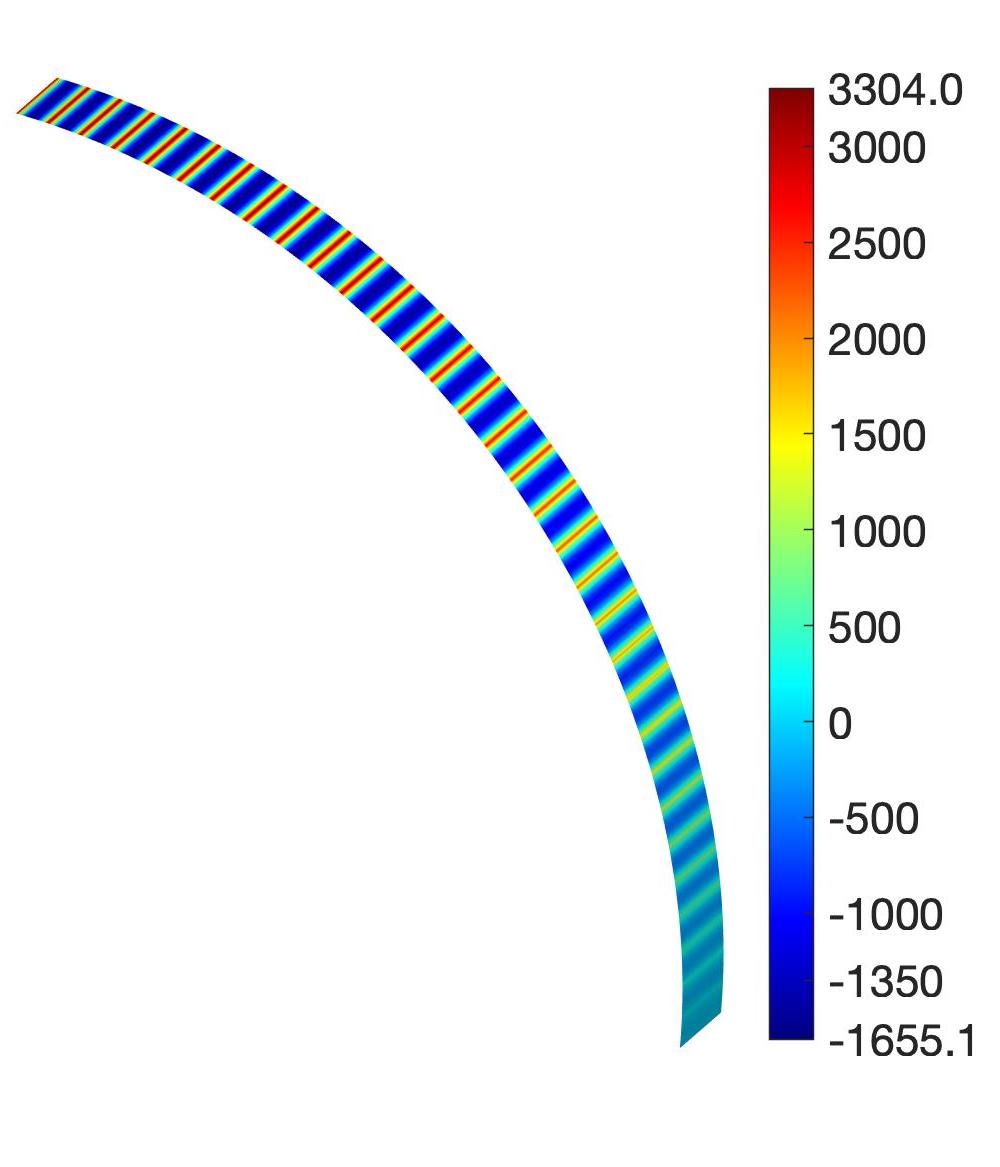}}
\put(-8,-.45){\includegraphics[height=46mm]{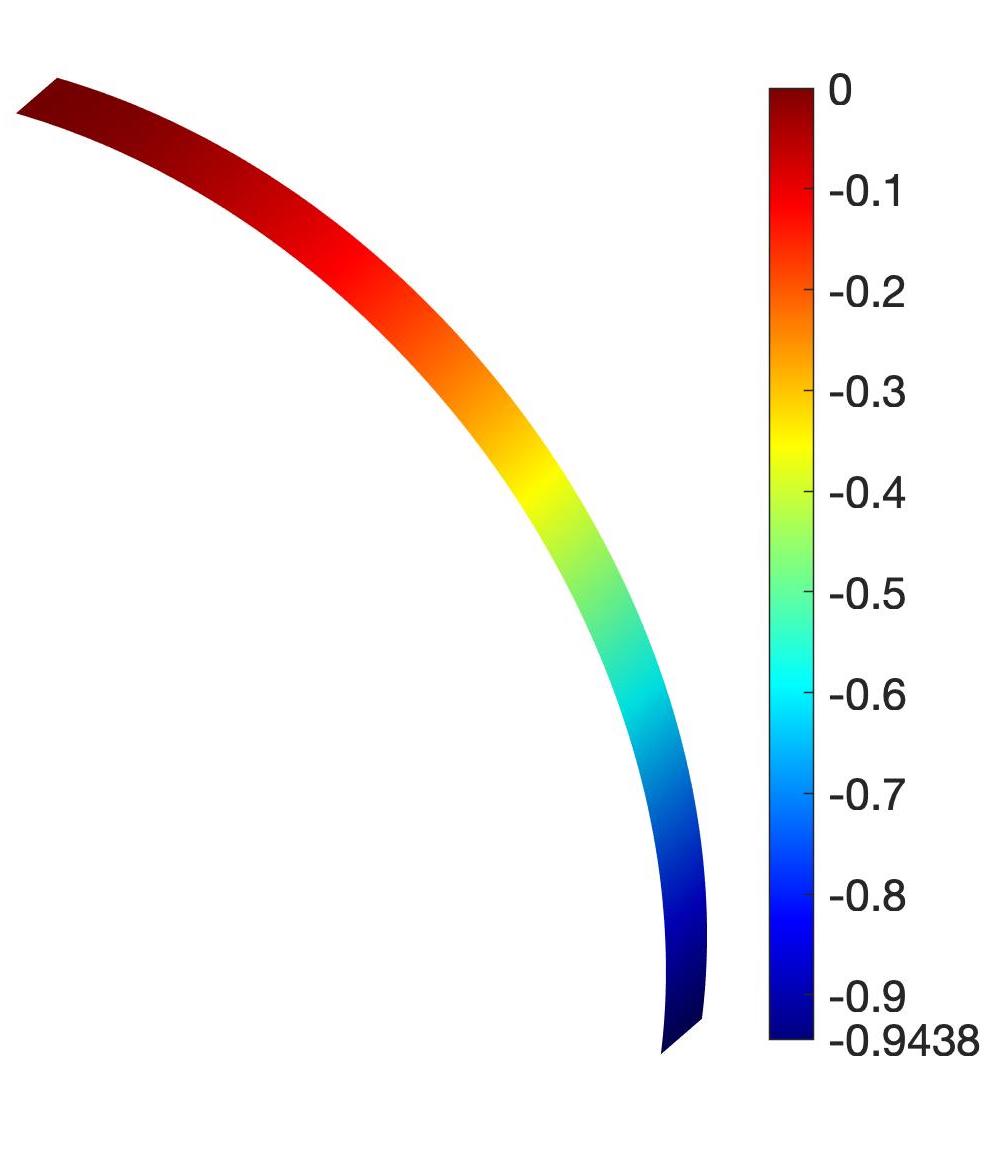}}
\put(-4,-.45){\includegraphics[height=46mm]{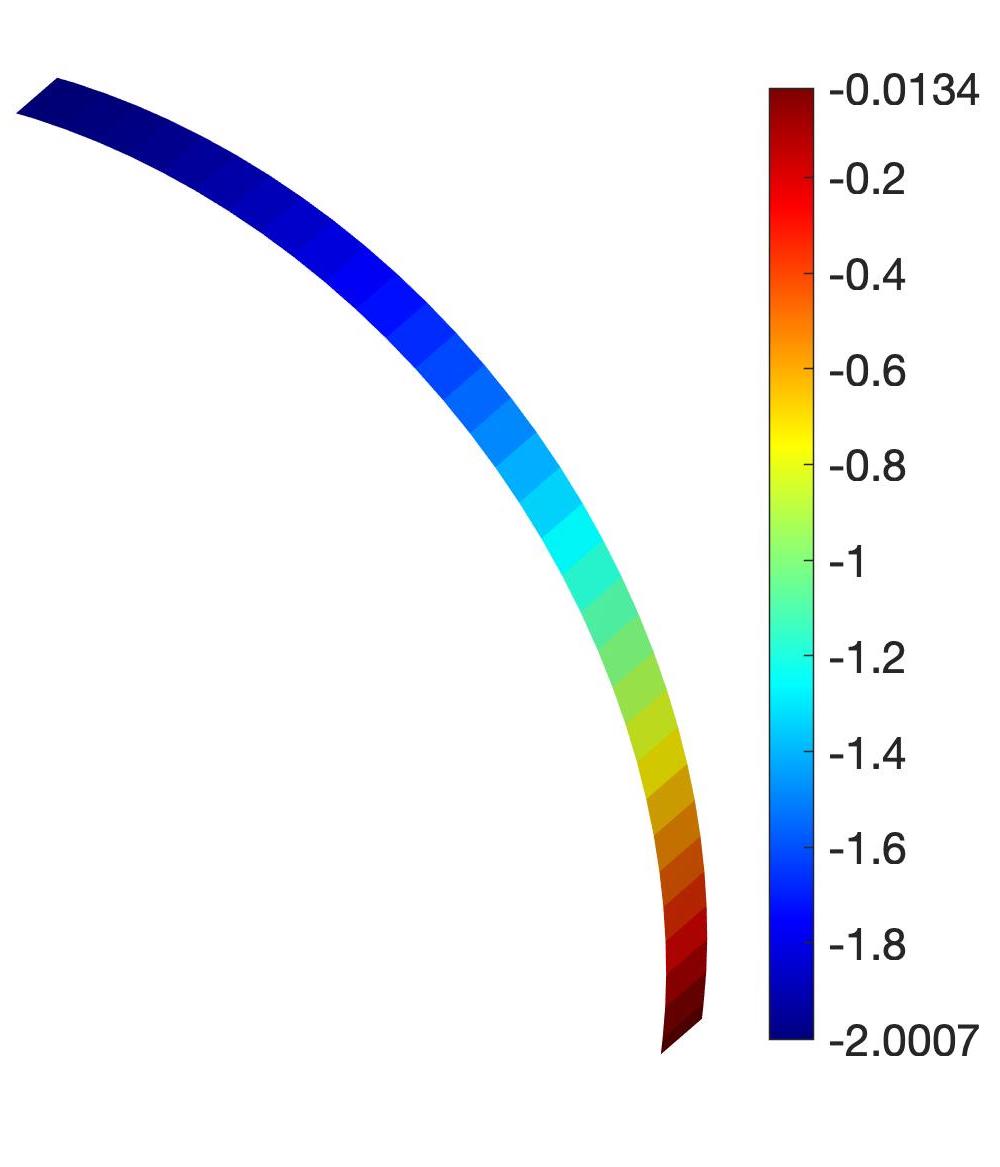}}
\put(0,-.45){\includegraphics[height=46mm]{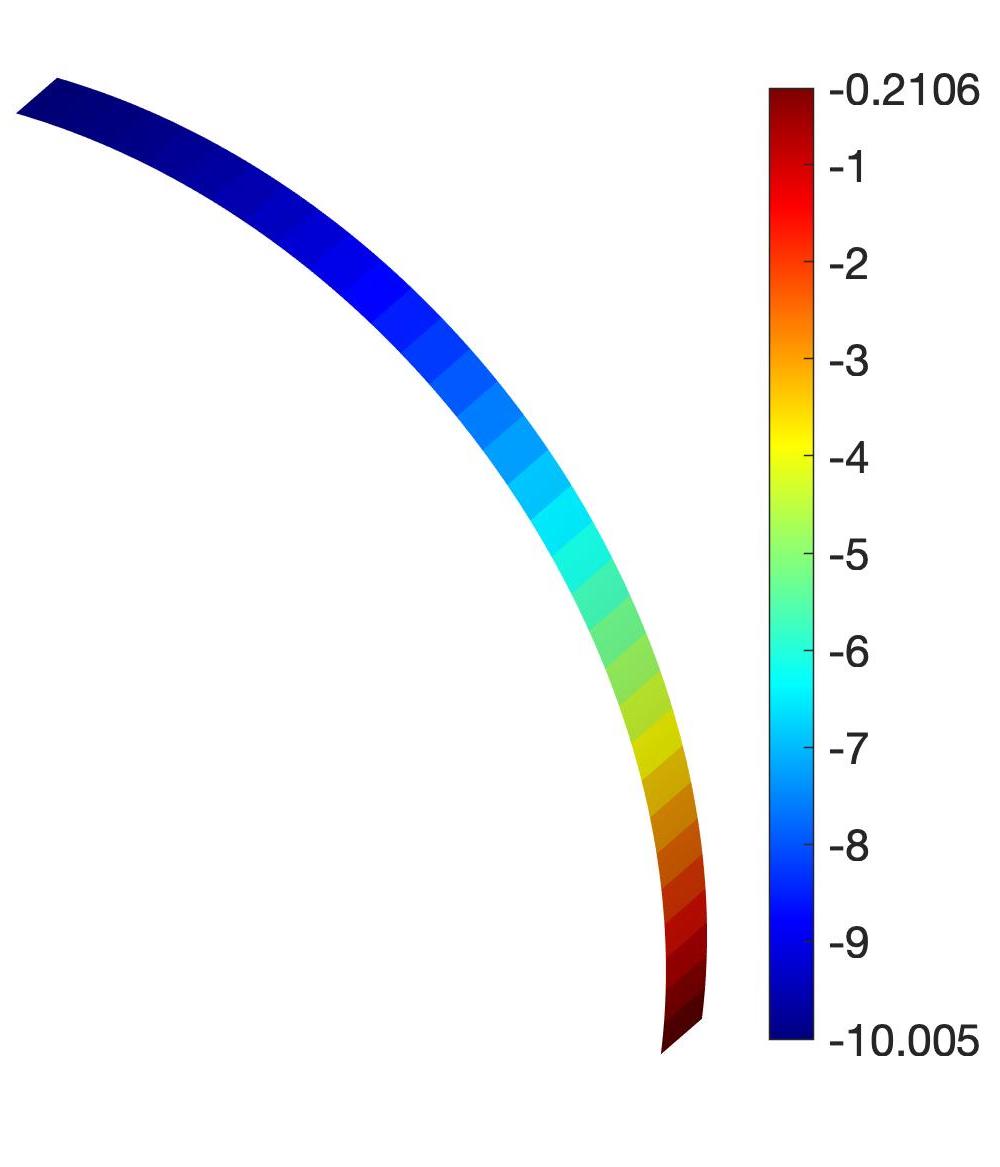}}
\put(4,-.45){\includegraphics[height=46mm]{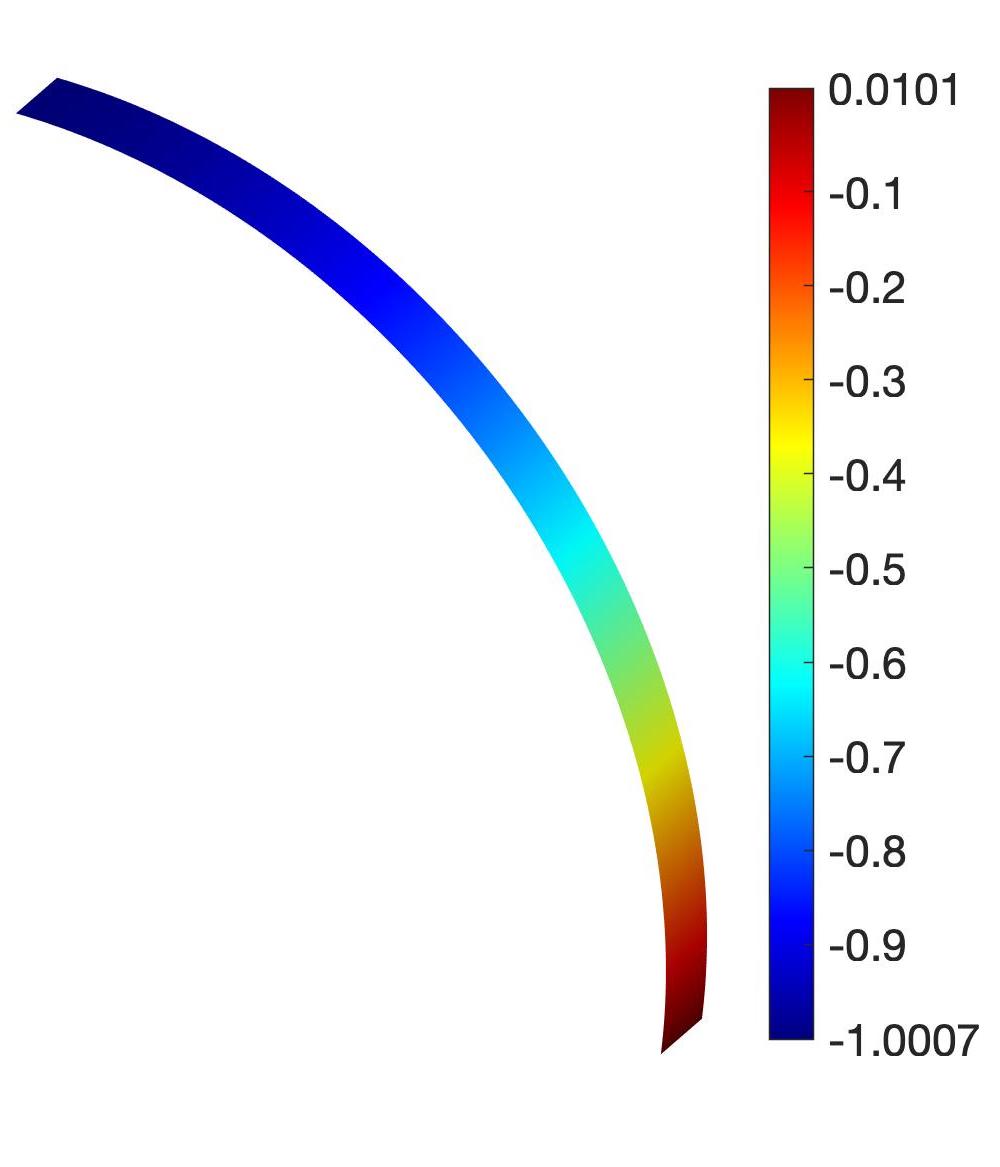}}
\put(-5.45,7.95){\footnotesize $u_r$} 
\put(-1.35,7.95){\footnotesize $\sigma$} 
\put(2.55,7.95){\footnotesize $M$} 
\put(6.6,7.95){\footnotesize $N$} 
\put(-5.45,3.6){\footnotesize $u_r$} 
\put(-1.35,3.6){\footnotesize $\sigma$} 
\put(2.55,3.6){\footnotesize $M$} 
\put(6.6,3.6){\footnotesize $N$} 
\end{picture}
\caption{Curved cantilever strip: 
Radial displacement $u_r$, raw effective membrane stress $\sig$, bending moment $M$ and Cauchy membrane stress $N$ (left to right) for the B2M2 discretization (top row) and B2M1 discretization (bottom row) using slenderness $R/T=1000$ and mesh $m = 32$. 
The B2M1 results are highly accurate, while the B2M2 results show very large errors, especially in $\sigma$ and $N$.
The B2M1 result for $N$ are obtained from post-processing scheme 2.}
\label{f:CC_us}
\end{center}
\end{figure}
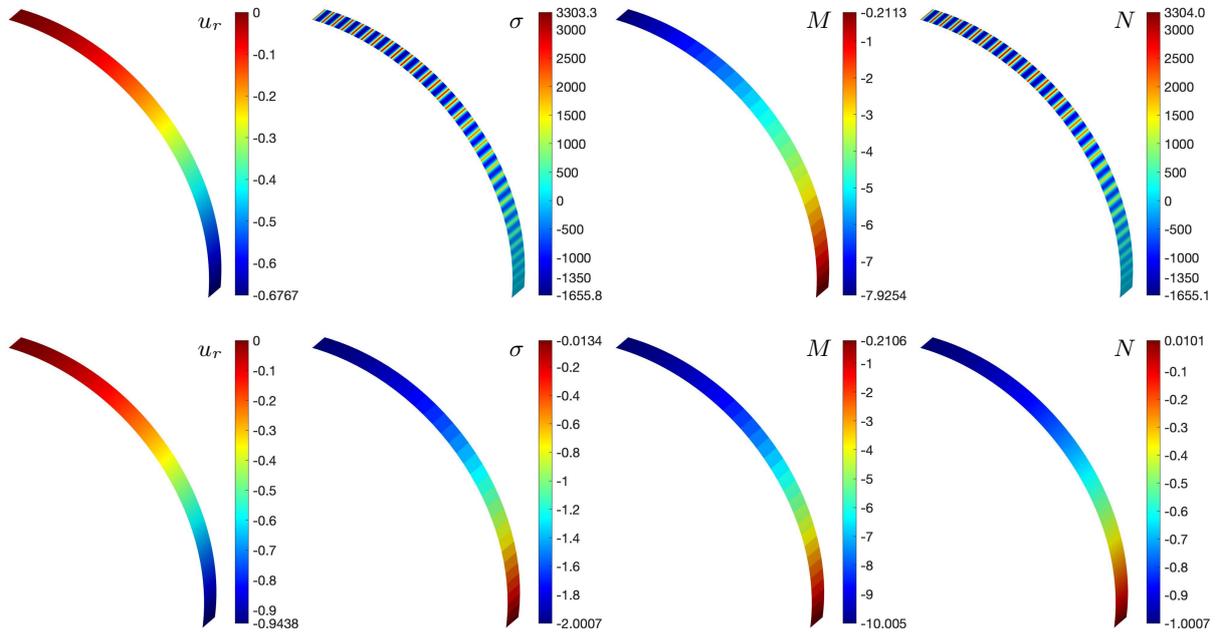

Fig.~\ref{f:CC_us} shows a comparison between the B2M2 and B2M1 results for the slenderness ratio $R/T=1000$.
As seen, the B2M1 results are highly accurate, while the B2M2 results show very large errors, especially in the membrane stresses  $\sigma$ and $N$.
The results here are for mesh $m=32$, although B2M1 already shows very good accuracy for $m=8$, as can be seen from Fig.~\ref{f:pp}.

\begin{figure}[h]
\begin{center} \unitlength1cm
\begin{picture}(0,11.7)
\put(-8,5.9){\includegraphics[height=58mm]{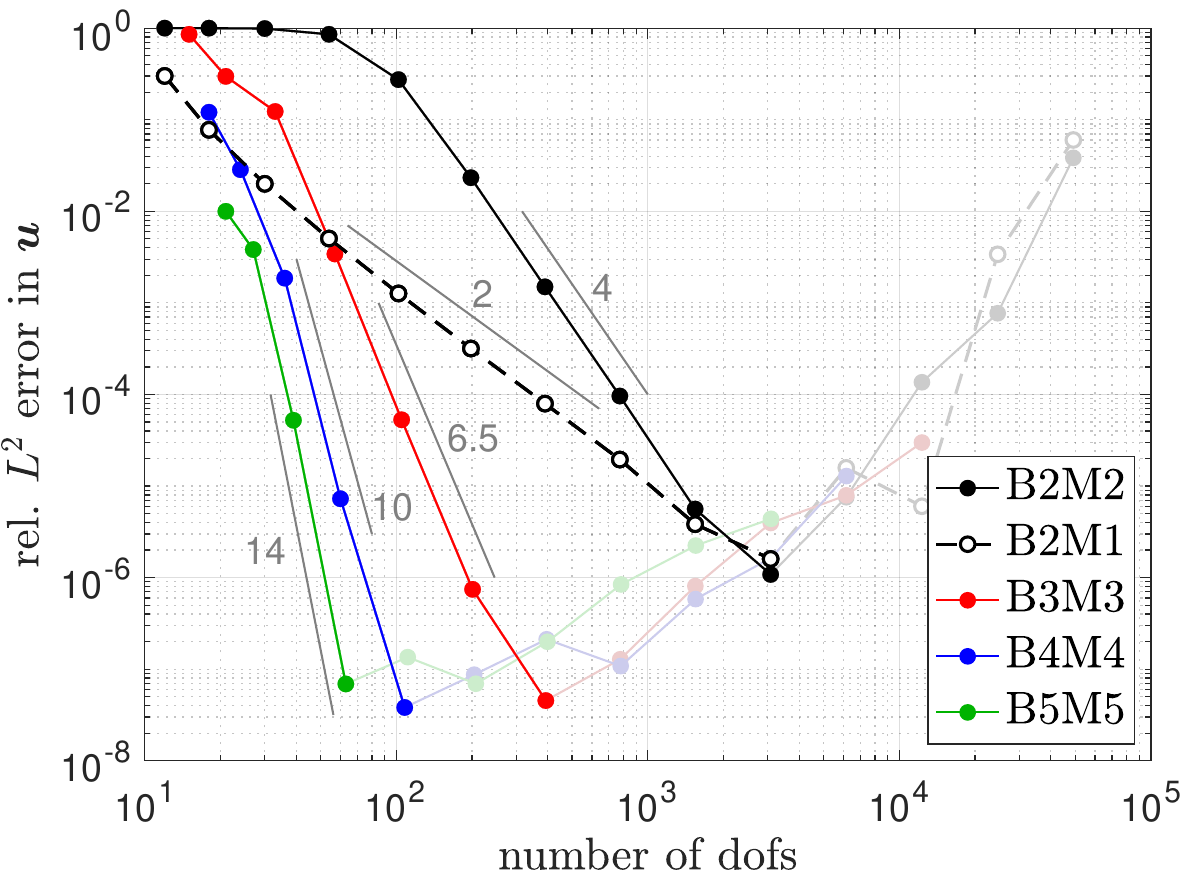}}
\put(0.2,5.9){\includegraphics[height=58mm]{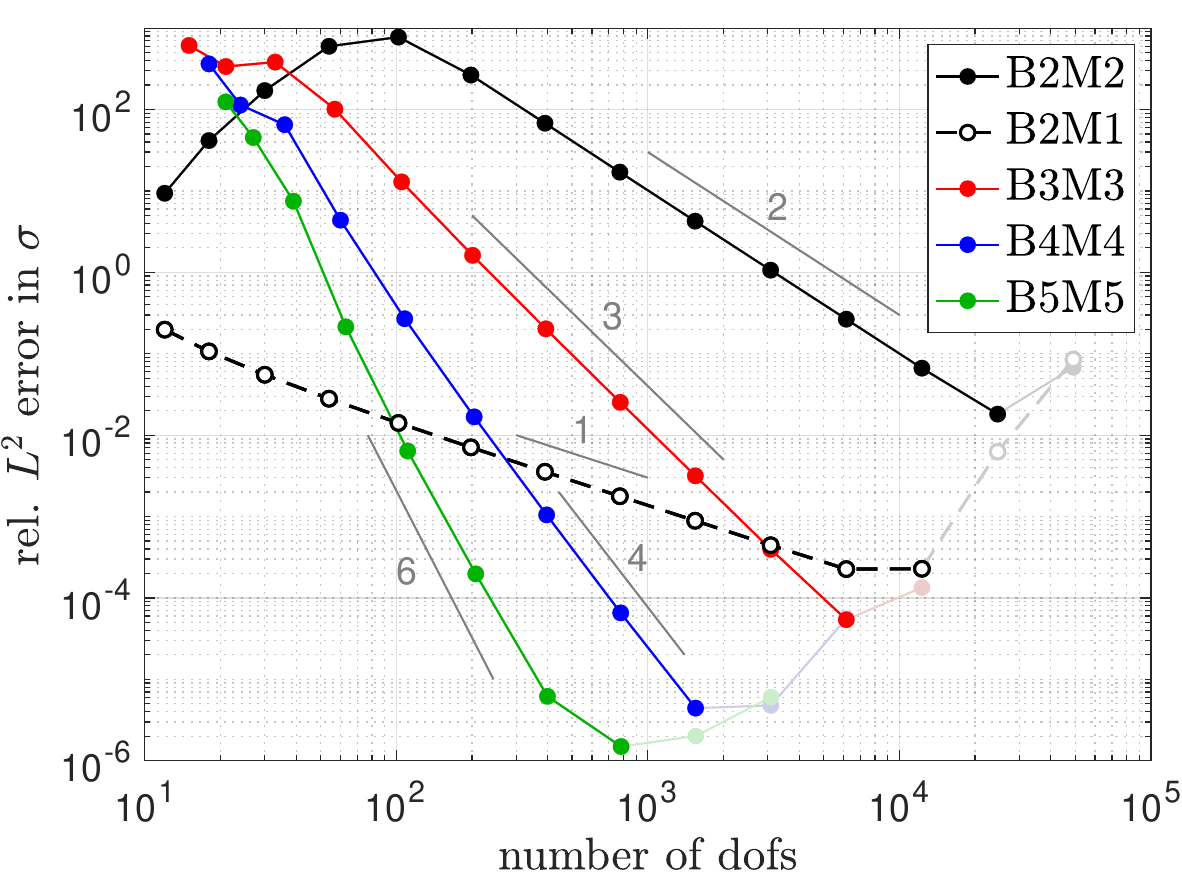}}
\put(-8,-.1){\includegraphics[height=58mm]{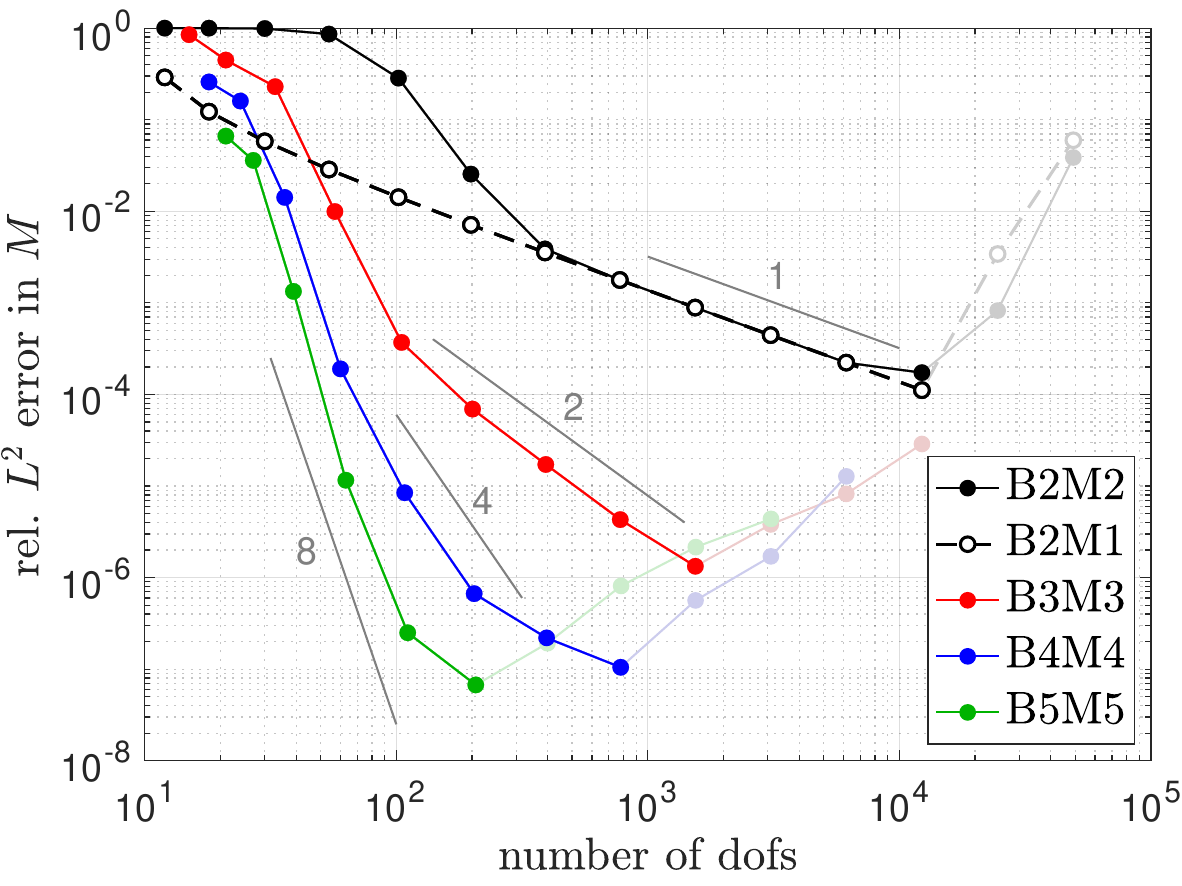}}
\put(0.2,-.1){\includegraphics[height=58mm]{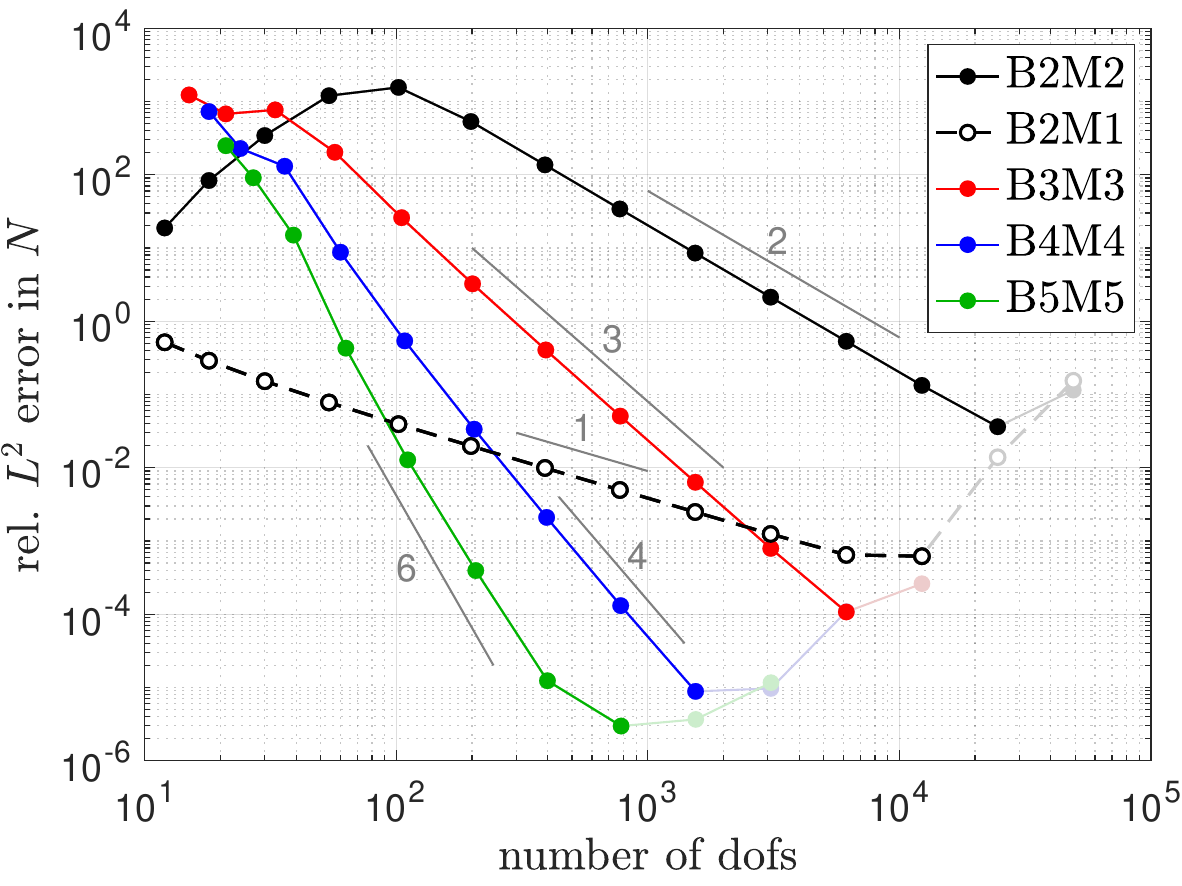}}
\put(-7.95,6.0){\footnotesize (a)}
\put(0.15,6.0){\footnotesize (b)}
\put(-7.95,0){\footnotesize (c)}
\put(0.15,0){\footnotesize (d)}
\end{picture}
\caption{Curved cantilever strip: $L^2$-error norm convergence of (a) displacement $\bu$, 
(b) raw effective membrane stress $\sigma$, (c) raw bending moment $M$ and (d) post-processed Cauchy membrane stress $N$;
all for $R/T=1000$ and with $m=2, 4, 8, 16, ..., 16384$ NURBS elements. 
B2M1 is always more accurate than B2M2 -- in case of $\sigma$ and $N$ up to five orders of magnitude.
For $m<32$ (i.e.~$n_\mathrm{dof}<102$), B2M1 even beats B5M5 in the stresses $\sigma$ and $N$.}
\label{f:CC_L2}
\end{center}
\end{figure}
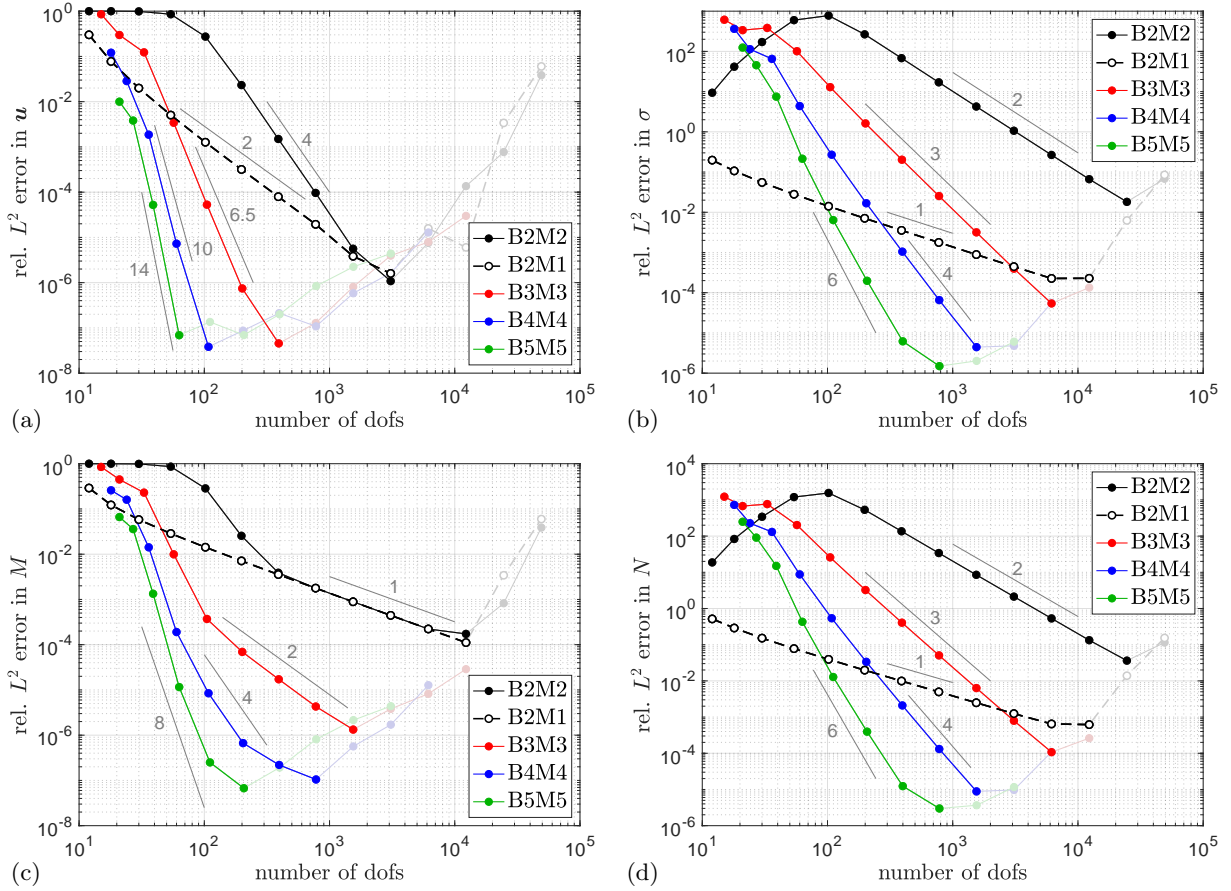

The superior accuracy of B2M1 can also be seen from the $L^2$-error norms of Eq.~\eqref{e:L2} that are shown in Fig.~\ref{f:CC_L2}.
Here B2M1 is compared to quadratic, cubic, quartic and quintic IGA (B2M2, B3M3, B4M4 \& B5M5, respectively).
As seen B2M1 is always more accurate than B2M2 -- up to five orders of magnitude in terms of membrane stresses $\sigma$ and $N$.
For $m<32$, B2M1 even beats B5M5 in the stresses $\sigma$ and $N$.
For B2M2, the stress error increases up to $m=32$, while the displacement error stagnates in this range, which is characteristic for membrane locking.
For B2M1, all errors decrease from the beginning, and the membrane stresses are of the same accuracy as the bending moment $M$.
The figure therefore demonstrates that the piecewise constant membrane stress representation within the B2M1 discretization completely eliminates membrane locking in this example.
The figure also shows that it is important to examine stress errors when assessing locking. 
Those can be very large even when displacement errors are small, which can be seen in all classical IGA cases.
Fig.~\ref{f:CC_L2} further shows that ill-conditioning sets in for large $n_\mathrm{dof}$ -- leading to an increase of the errors.\footnote{The condition number of the reduced tangent matrix increases exponentially with $n_\mathrm{dof}$ and is similarly large for all five discretization cases; also for the nonlinear regime examined in Sec.~\ref{s:CCnl}.}
These results are shown in lighter line colors.

B2M1 achieves the same convergence rate in $M$ as the ANS scheme of \citet{casquero23}, while the rate in $N$ is slightly higher in \citet{casquero23} (1.5 vs.~1).
The convergence rates of the standard IGA formulations B$p$M$p$ are consistent with the rates reported in \citet{greco16,greco17} and \citet{zou21}.

\begin{figure}[h]
\begin{center} \unitlength1cm
\begin{picture}(0,11.7)
\put(-8,5.9){\includegraphics[height=58mm]{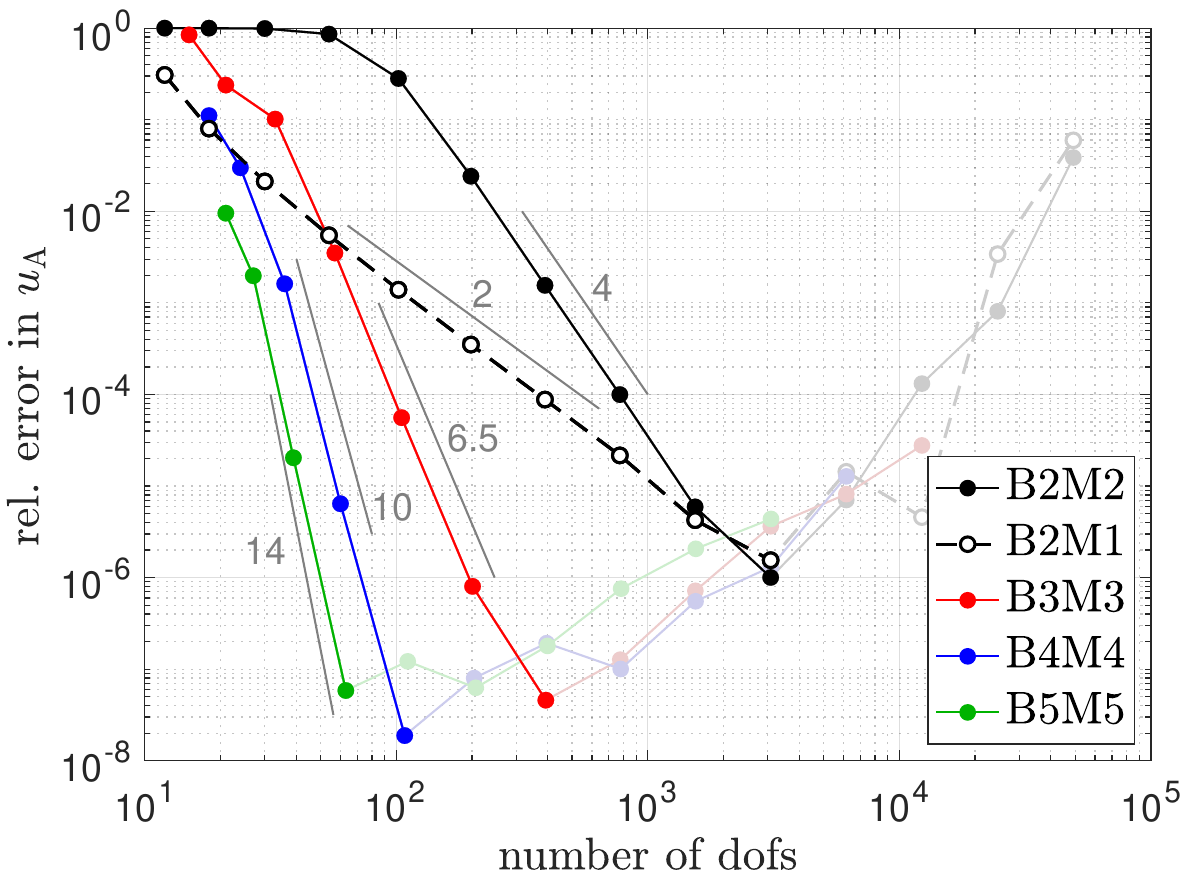}}
\put(0.2,5.9){\includegraphics[height=58mm]{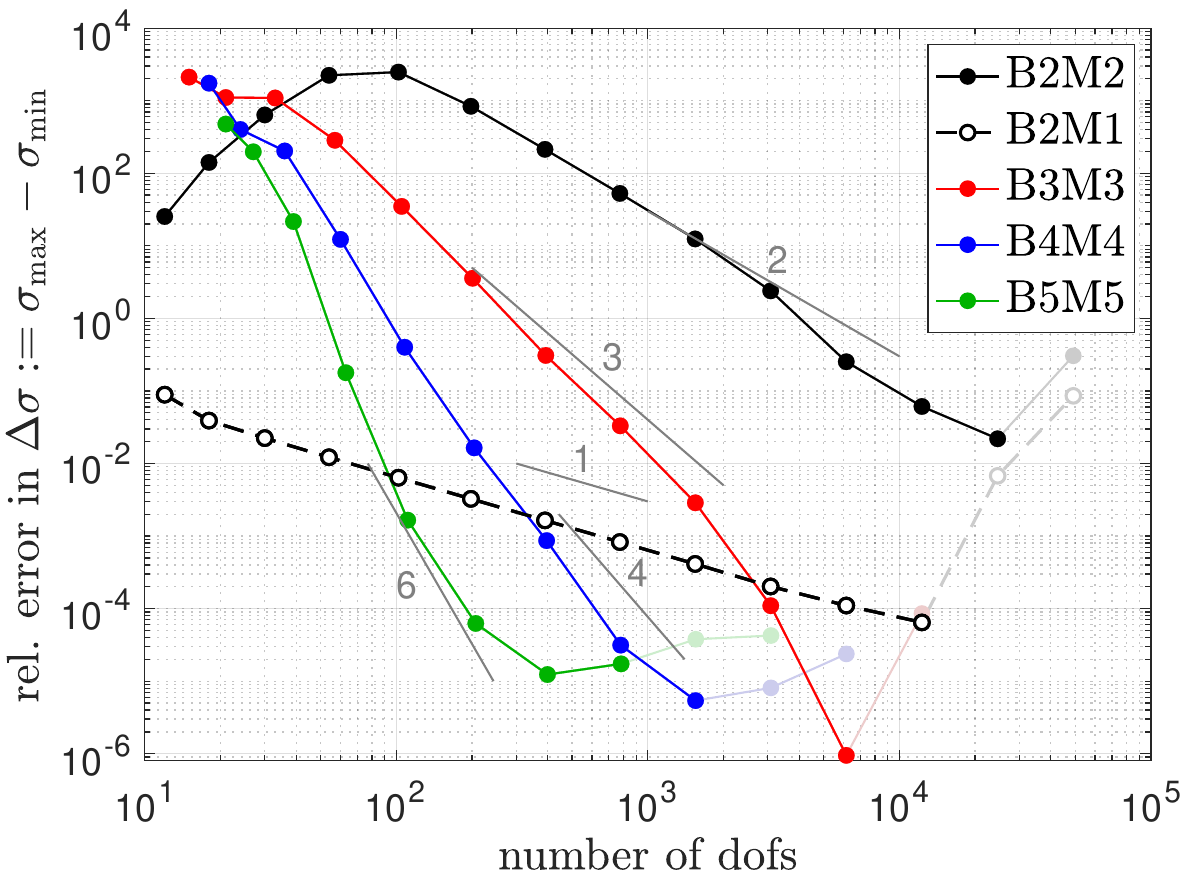}}
\put(-8,-.1){\includegraphics[height=58mm]{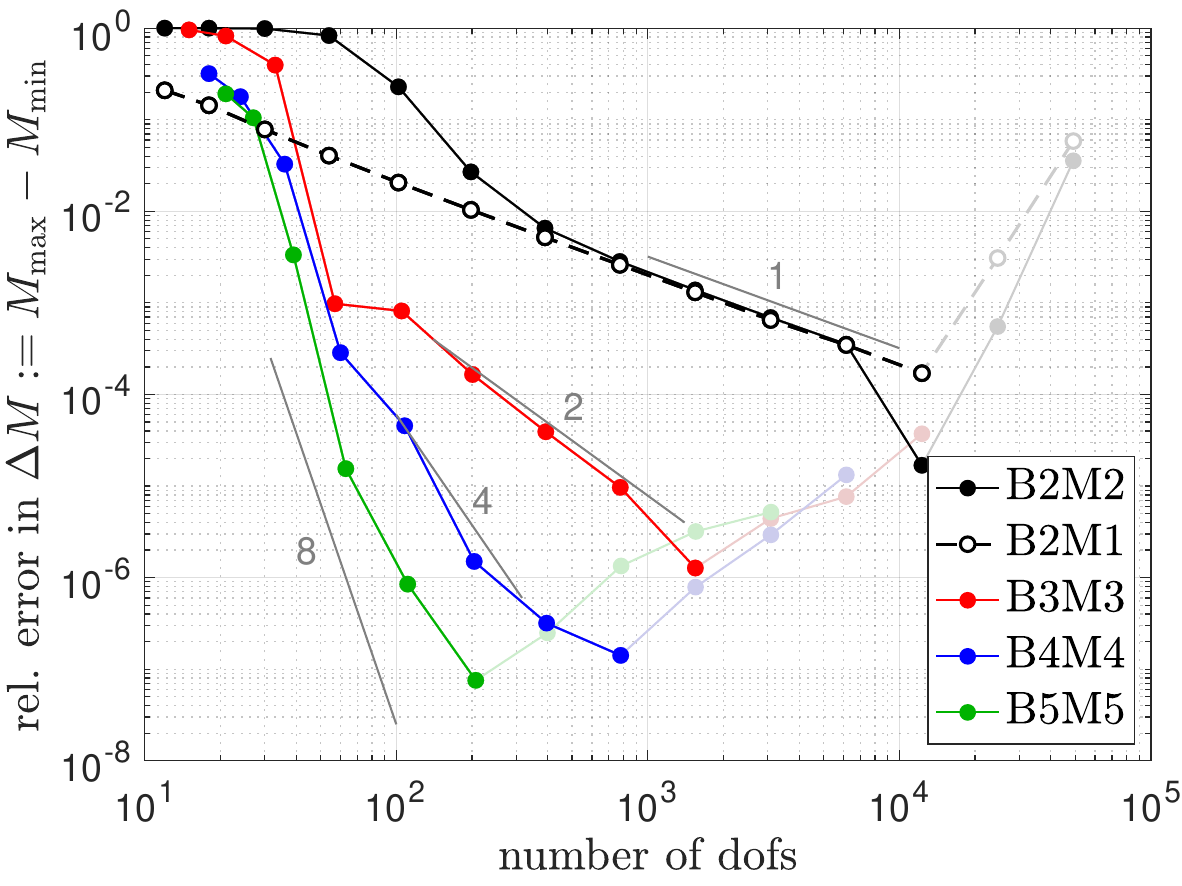}}
\put(0.2,-.1){\includegraphics[height=58mm]{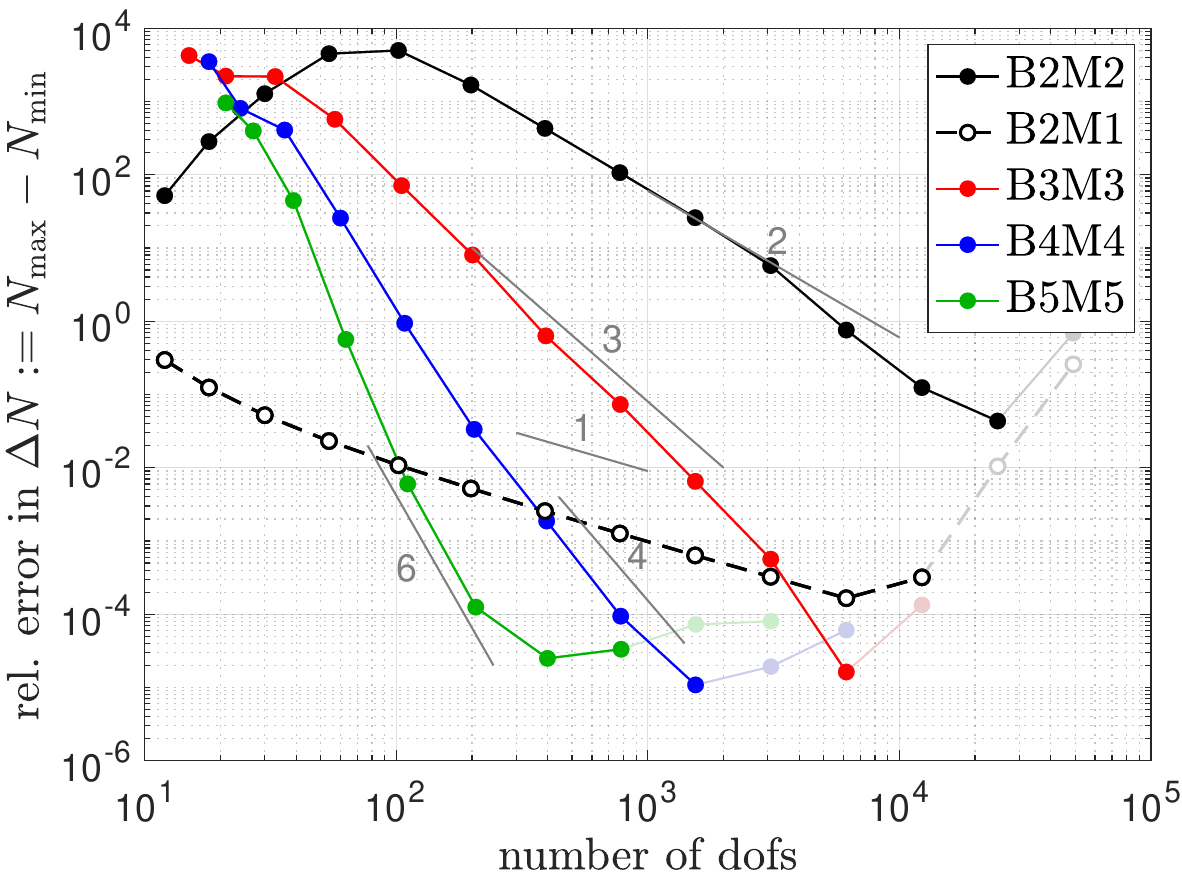}}
\put(-7.95,6.0){\footnotesize (a)}
\put(0.15,6.0){\footnotesize (b)}
\put(-7.95,0.0){\footnotesize (c)}
\put(0.15,0.0){\footnotesize (d)}
\end{picture}
\caption{Curved cantilever strip: Maximum error convergence of (a) displacement $u_\mathrm{A}$, 
(b) raw effective membrane stress $\sigma$, (c) raw bending moment $M$ and (d) post-processed Cauchy membrane stress $N$; all for $R/T=1000$ and with $m=2, 4, 8, 16, ..., 16384$ NURBS elements. 
The overall behavior is equivalent to that of the $L^2$-error norm in Fig.~\ref{f:CC_L2}.
This can also be seen in comparison to the gray convergence slopes, which are copied over from Fig.~\ref{f:CC_L2}.}
\label{f:CC_max}
\end{center}
\end{figure}
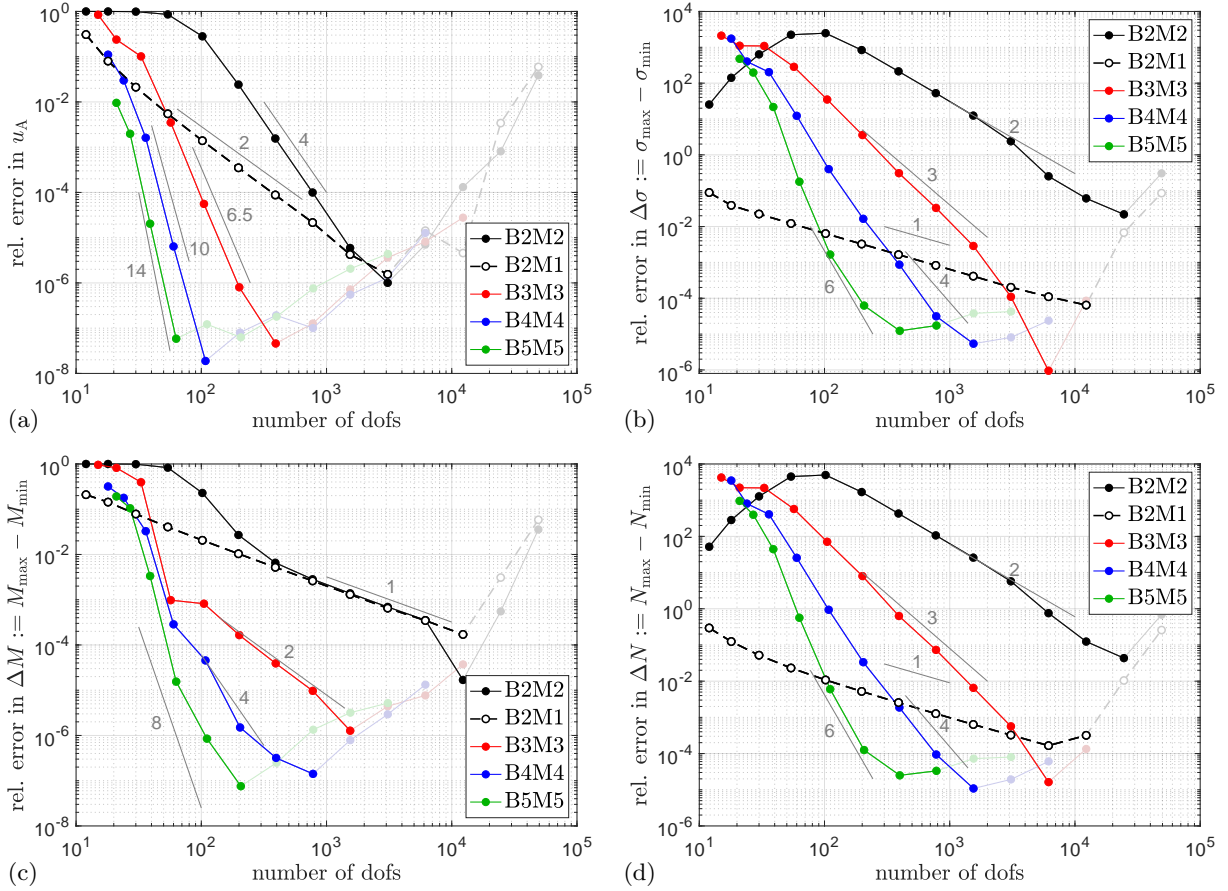

Alternatively to the $L^2$-error norm, one can also examine the maximum error norms of Eq.~\eqref{e:maxe} that are shown in Fig.~\ref{f:CC_max}.
The comparison between Figs.~\ref{f:CC_L2} and \ref{f:CC_max} shows that the two norms give more or less the same picture -- only the slopes are not matched so nicely for the maximum error norm than for the $L^2$-error norm.
It is therefore sufficient to only examine the simpler maximum norm in the following cases.

Figs.~\ref{f:CC_us}-\ref{f:CC_max} also show that $N$ behaves like $\sig$.
(The difference converges to the factor $\sig/N=2$ that follows from Eqs.~\eqref{e:NMana} and \eqref{e:sigana}.)
The reason for this equivalency is that the error in $N$ is coming predominantly from $\sig$.
It is therefore sufficient to only examine $\sig$ in the following examples.

\subsubsection{Efficiency gains of B2M1}\label{s:CCeff}

\begin{figure}[h]
\begin{center} \unitlength1cm
\begin{picture}(0,5.7)
\put(-8,-.1){\includegraphics[height=58mm]{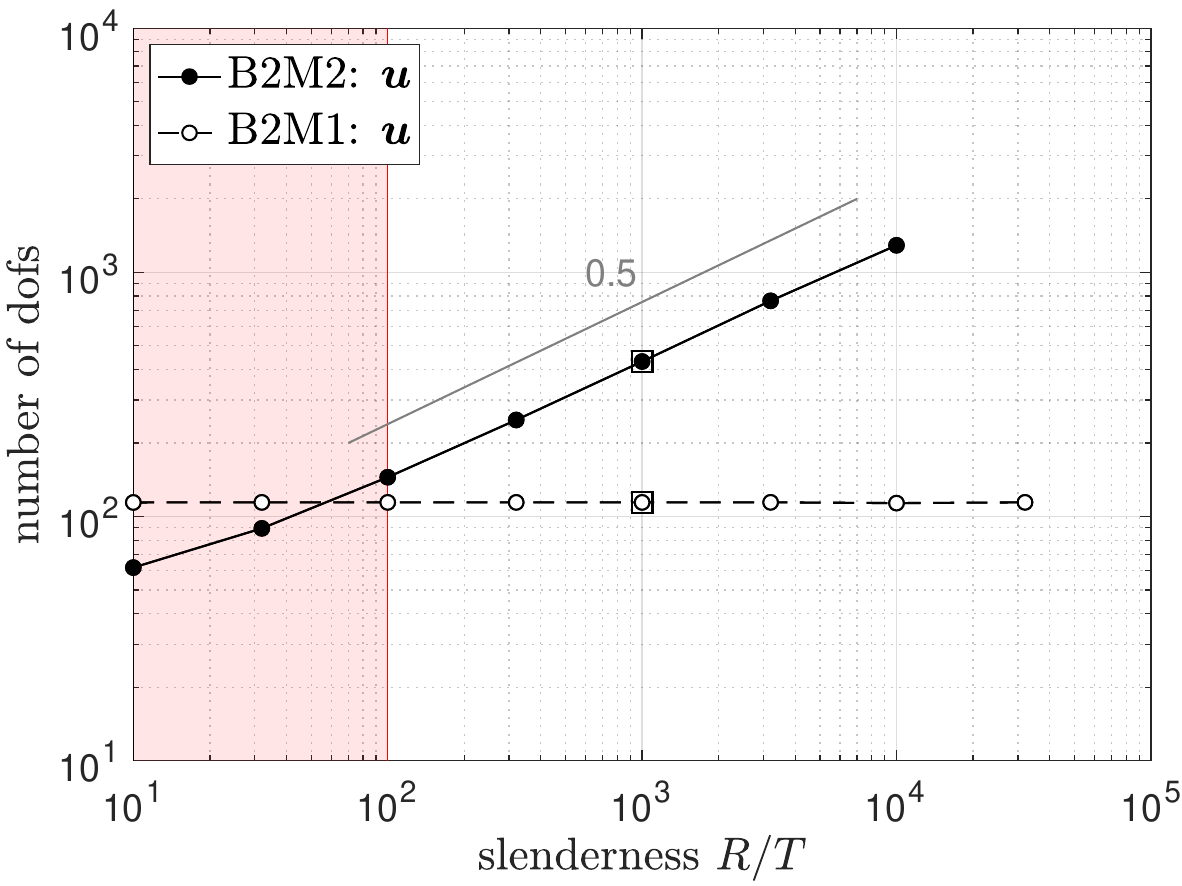}}
\put(0.2,-.1){\includegraphics[height=58mm]{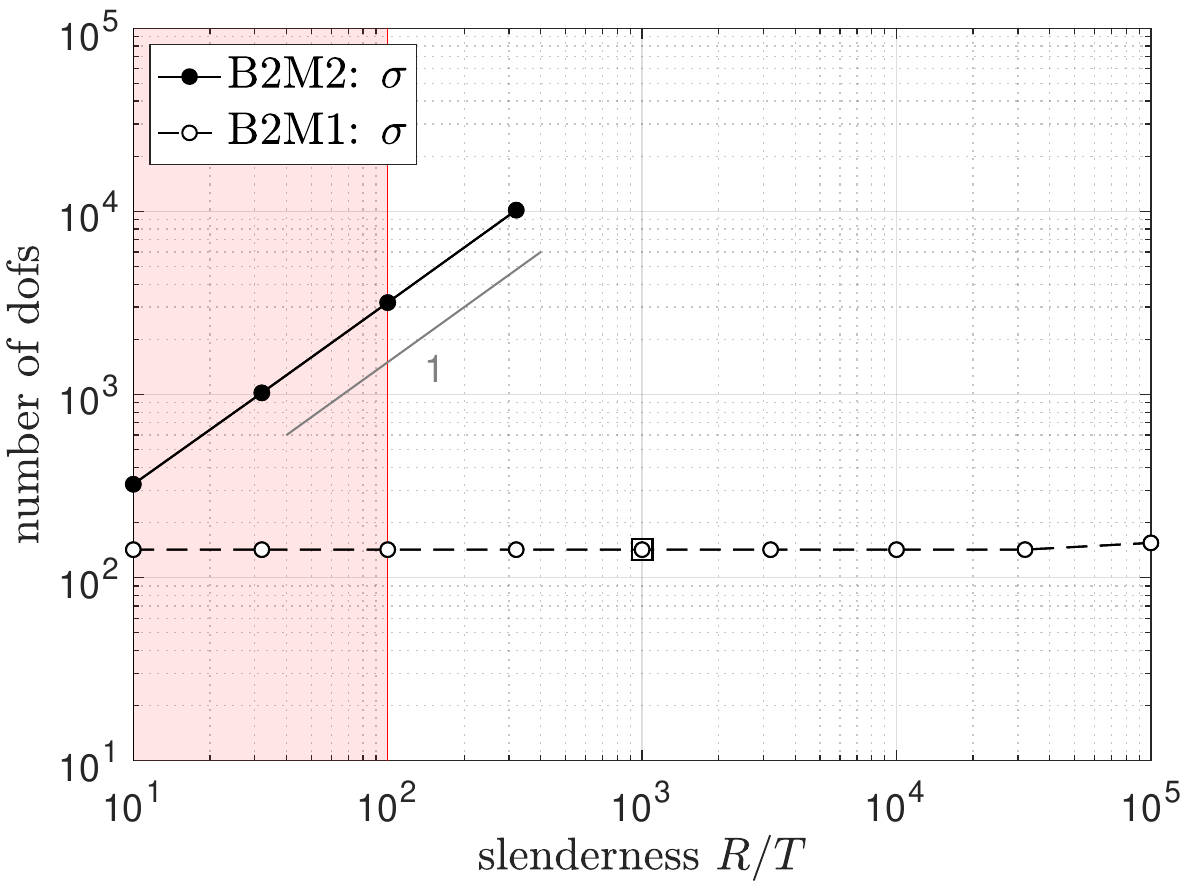}}
\put(-7.95,0.0){\footnotesize (a)}
\put(0.15,0.0){\footnotesize (b)}
\end{picture}
\caption{Efficiency gain of B2M1 vs.~B2M2 for the curved cantilever strip. 
The diagrams show the minimum number of dofs, as a function of slenderness $R/T$, required to achieve a given $L^2$ accuracy in (a) $\bu$ and (b) $\sigma$ -- here for the relative error falling below (a) $10^{-3}$ and (b) $10^{-2}$.
The curves terminate when the accuracy cannot be reached anymore due to ill-conditioning caused by large $R/T$.
For the B2M2 discretization the required dofs increase with $R/T$, while for B2M1 the required dofs are independent of $R/T$. 
In the regime where the Kirchhoff-Love assumptions hold ($R/T>100$, beyond the pink region), B2M1 is always more efficient than B2M2.
The squares at $R/T = 10^3$ show the results from Fig.~\ref{f:CC_L2}. 
In this case, B2M2 requires about 3.7 times more dofs than B2M1 to achieve comparable accuracy in $\bu$, while it is unable to achieve the required stress accuracy altogether.}
\label{f:CC_acc}
\end{center}
\end{figure}

In order to assess the performance of B2M1 vs.~B2M2 across a large range of slenderness ratios, the efficiency of both formulations is examined.
This is done by looking at the required dofs to reach a given accuracy.
This is shown for $\bu$ and $\sigma$ in the accuracy diagrams of Fig.~\ref{f:CC_acc}.
Here, the data points are obtained from the intersection of the B2M2 and B2M1 convergence curves -- like those in Fig.~\ref{f:CC_L2}a \& b -- 
with the $10^{-3}$ \& $10^{-2}$ error level, respectively.
The figure shows that for B2M2 the required dofs increase with $R/T$, while for B2M1 the required dofs are independent of $R/T$. 
This demonstrate that there is no membrane locking in the B2M1 discretization.
In the regime where the Kirchhoff-Love assumptions hold ($R/T>100$, beyond the pink region), B2M1 is always more efficient than B2M2.
Above a certain $R/T$ value, the given accuracy may not be reached anymore due to ill-conditioning\footnote{Here, double precision has been used. Increasing to quadruple precision would allow for higher $R/T$ values.}.
This is the case for B2M2 above $R/T=10^4$ and $R/T=3.2\cdot10^2$ in Fig.~\ref{f:CC_acc}a \& b, respectively.
The B2M1 discretization, on the other hand, is able to meet the desired accuracy with the same number of dofs up to $R/T=3.2\cdot10^4$.

\subsubsection{Large deformation case}\label{s:CCnl}

\begin{figure}[h]
\begin{center} \unitlength1cm
\begin{picture}(0,4)
\put(-7.35,-.45){\includegraphics[height=46mm]{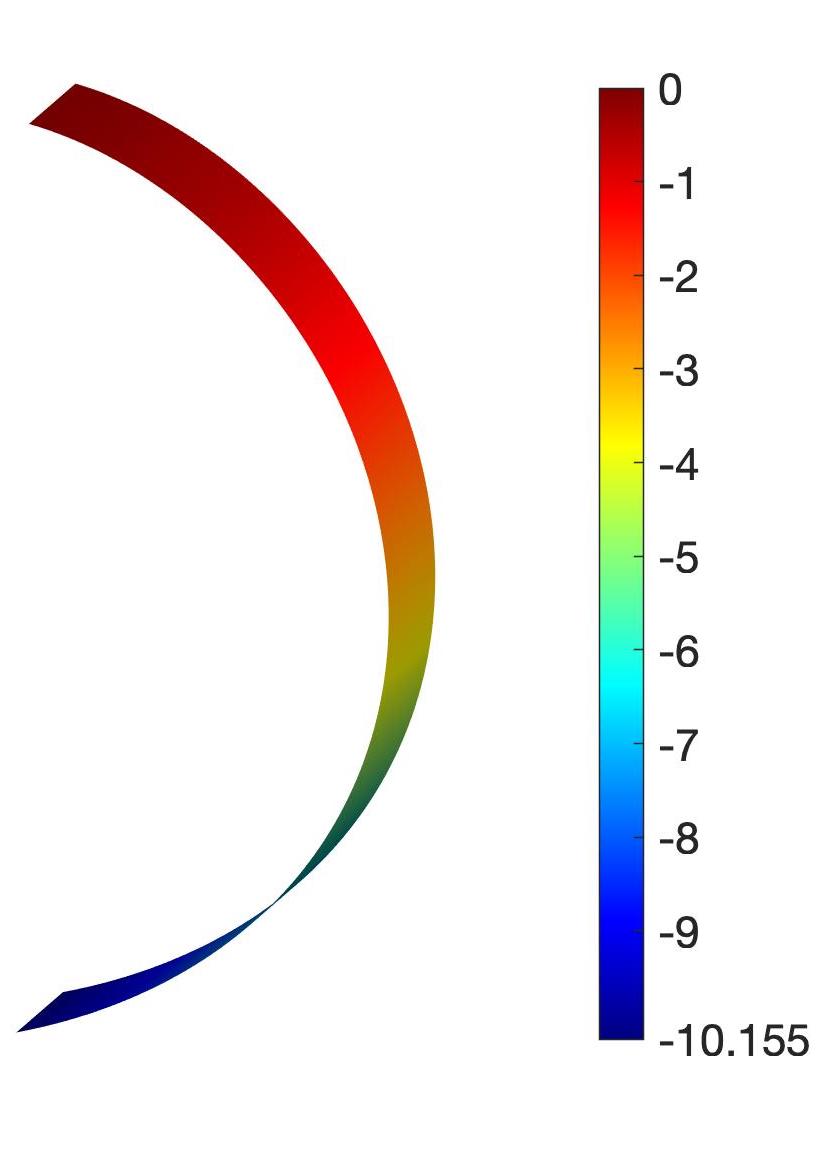}}
\put(-3.35,-.45){\includegraphics[height=46mm]{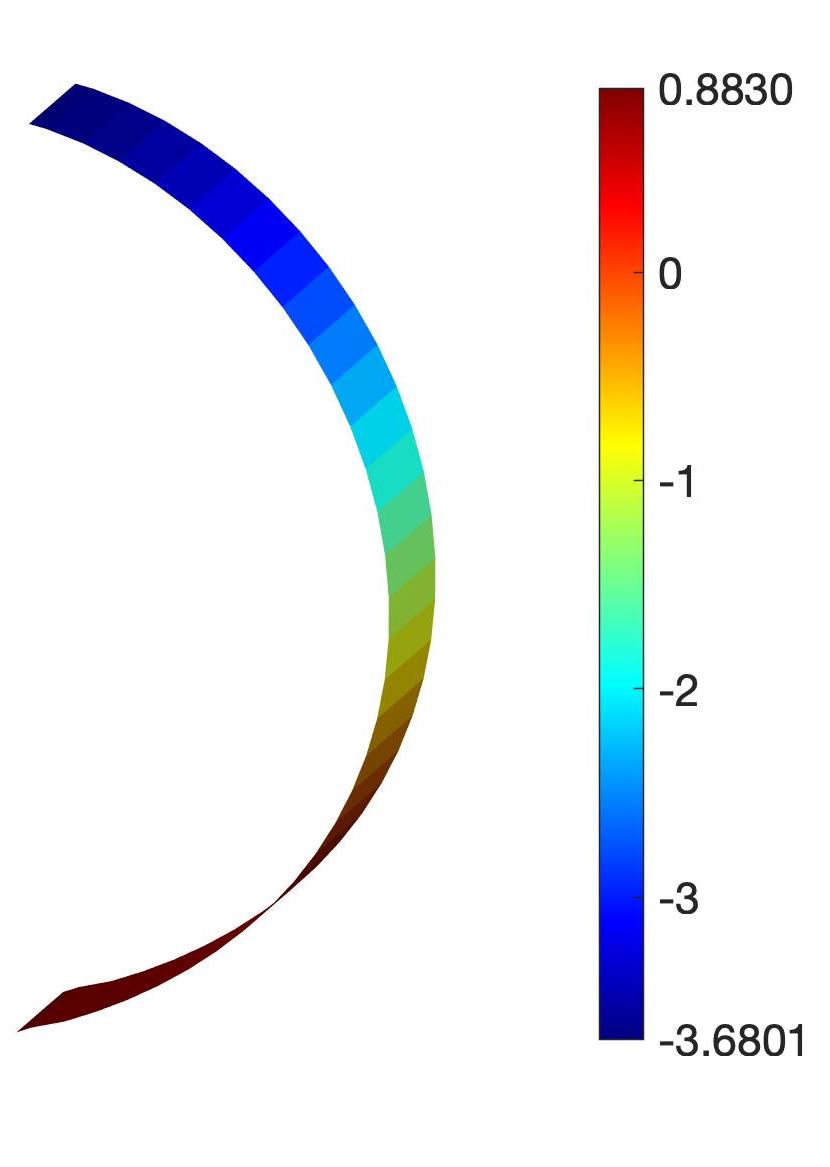}}
\put(.65,-.45){\includegraphics[height=46mm]{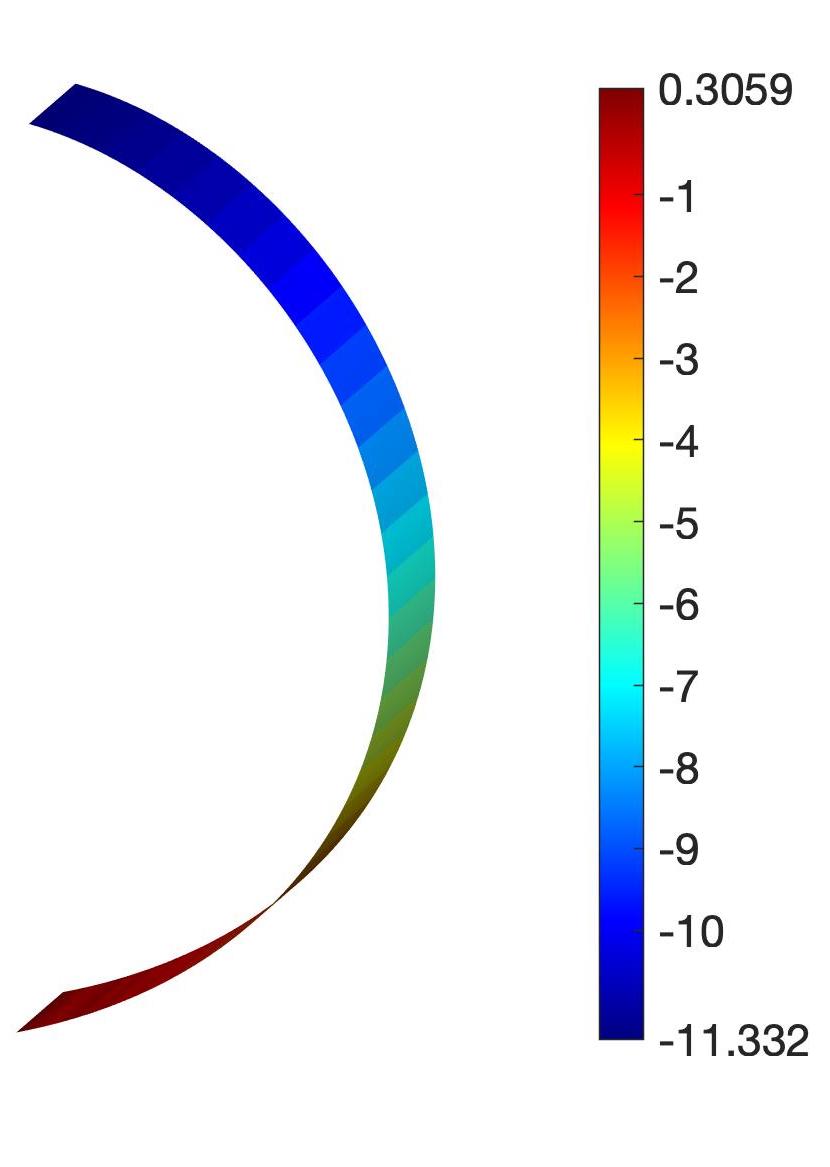}}
\put(4.65,-.45){\includegraphics[height=46mm]{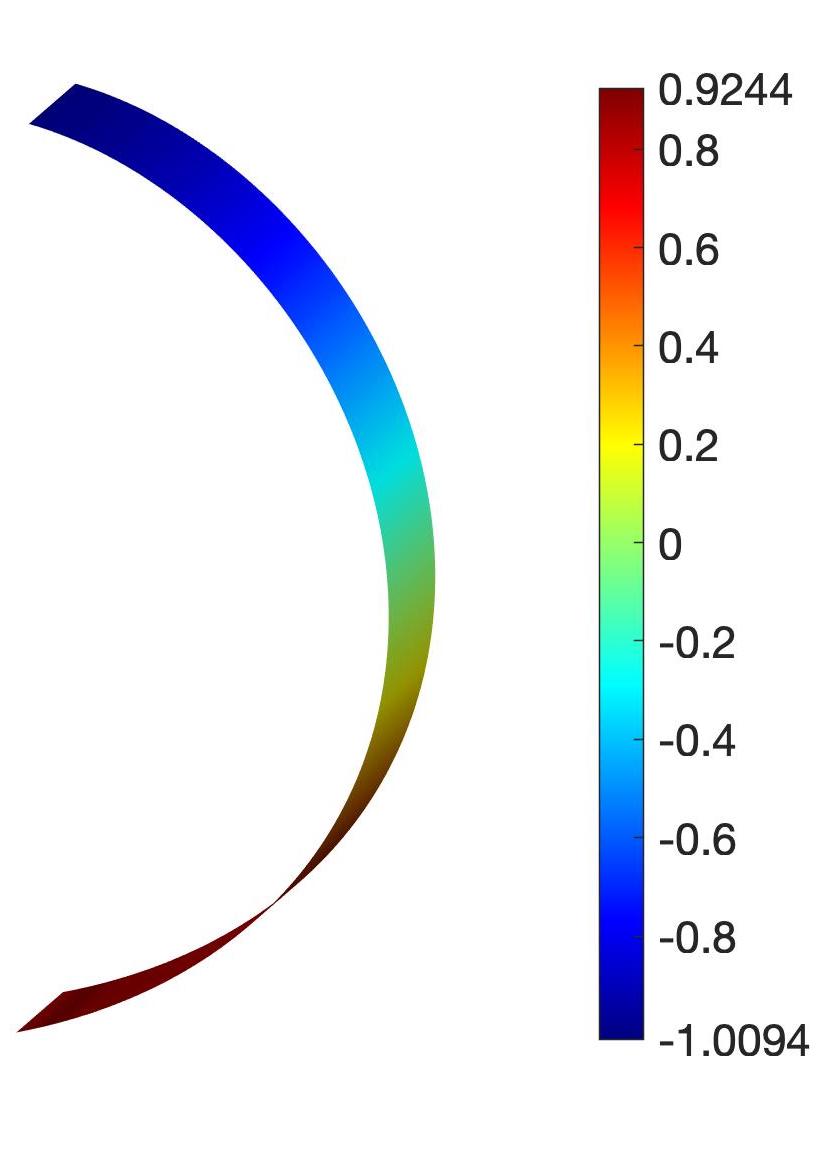}}    	
\put(-5.45,3.6){\footnotesize $u_r$} 
\put(-1.35,3.6){\footnotesize $\sigma$} 
\put(2.55,3.6){\footnotesize $M$} 
\put(6.6,3.6){\footnotesize $N$} 
\end{picture}
\caption{Curved cantilever strip -- large deformation case: 
Radial displacement $u_r$, raw effective membrane stress $\sig$, bending moment $M$ and Cauchy membrane stress $N$ (left to right) for the B2M1 discretization using slenderness $R/T=1000$ and mesh $m = 32$.}
\label{f:CC_usNL}
\end{center}
\end{figure}
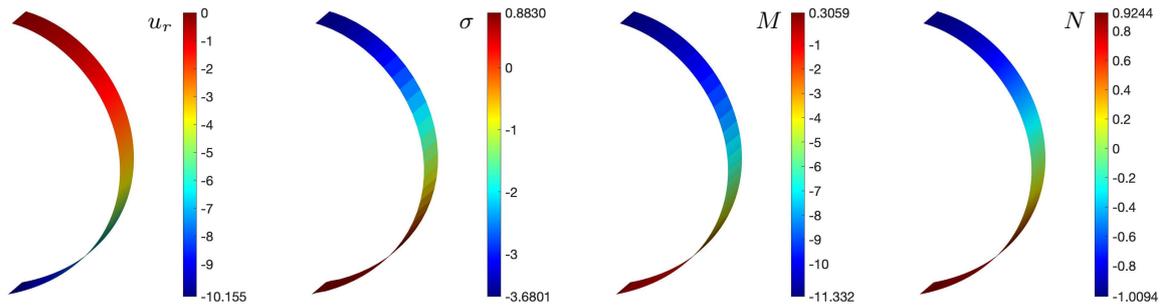

\begin{figure}[h]
\begin{center} \unitlength1cm
\begin{picture}(0,11.7)
\put(-8,5.9){\includegraphics[height=58mm]{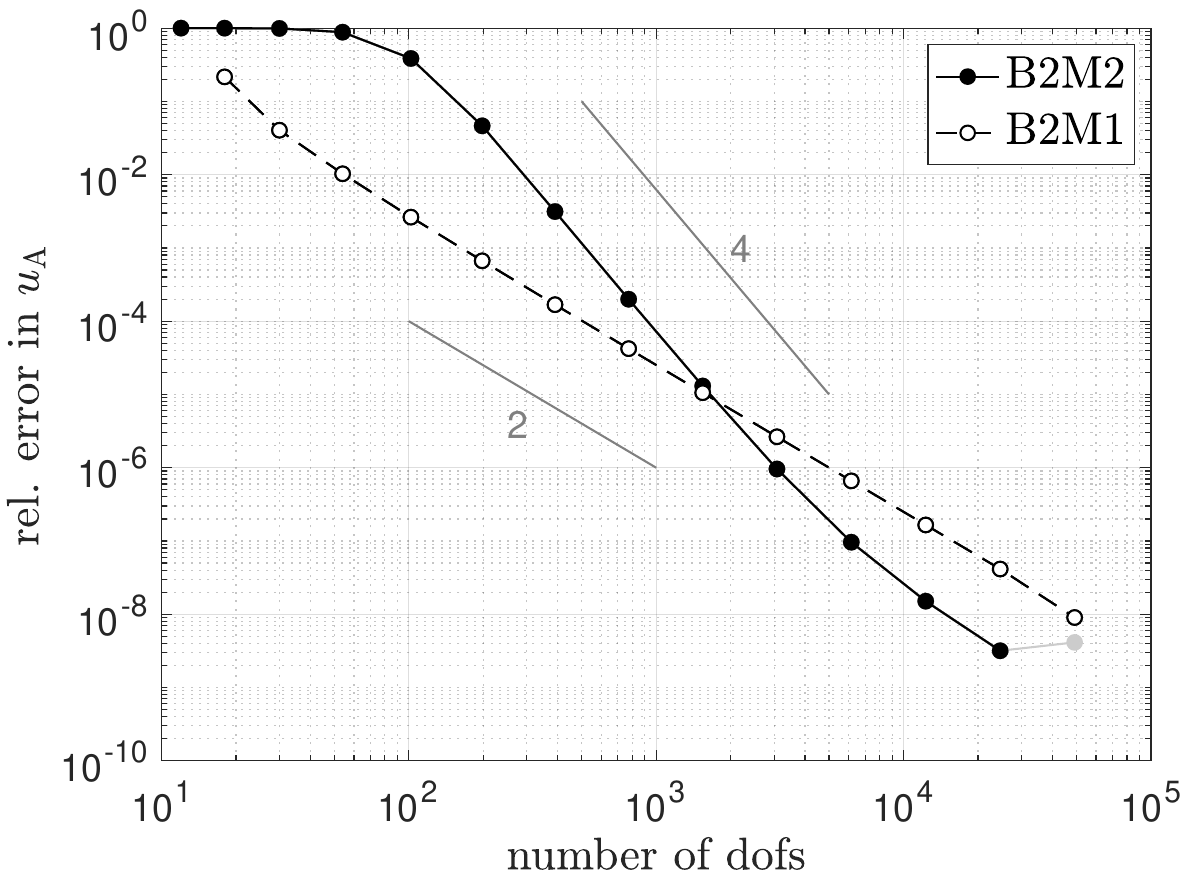}}
\put(0.2,5.9){\includegraphics[height=58mm]{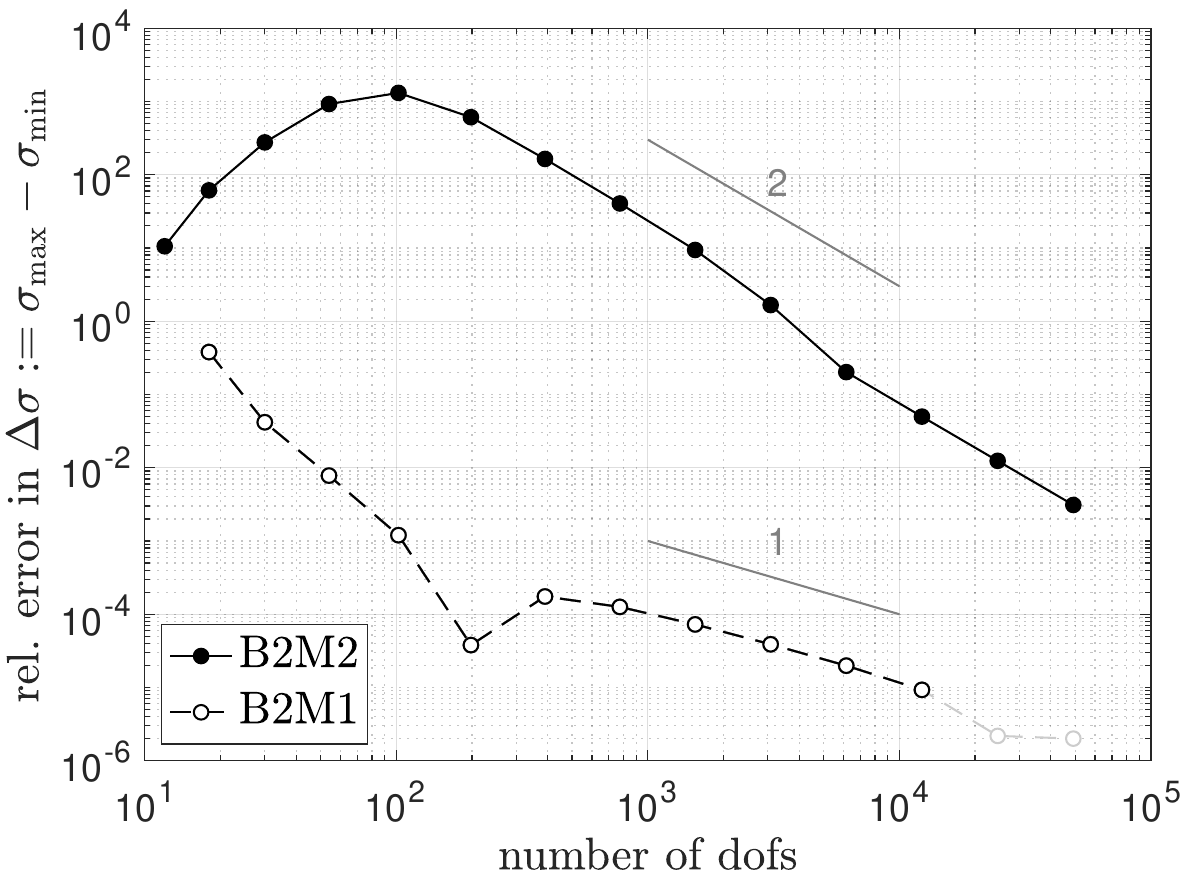}}
\put(-8,-.1){\includegraphics[height=58mm]{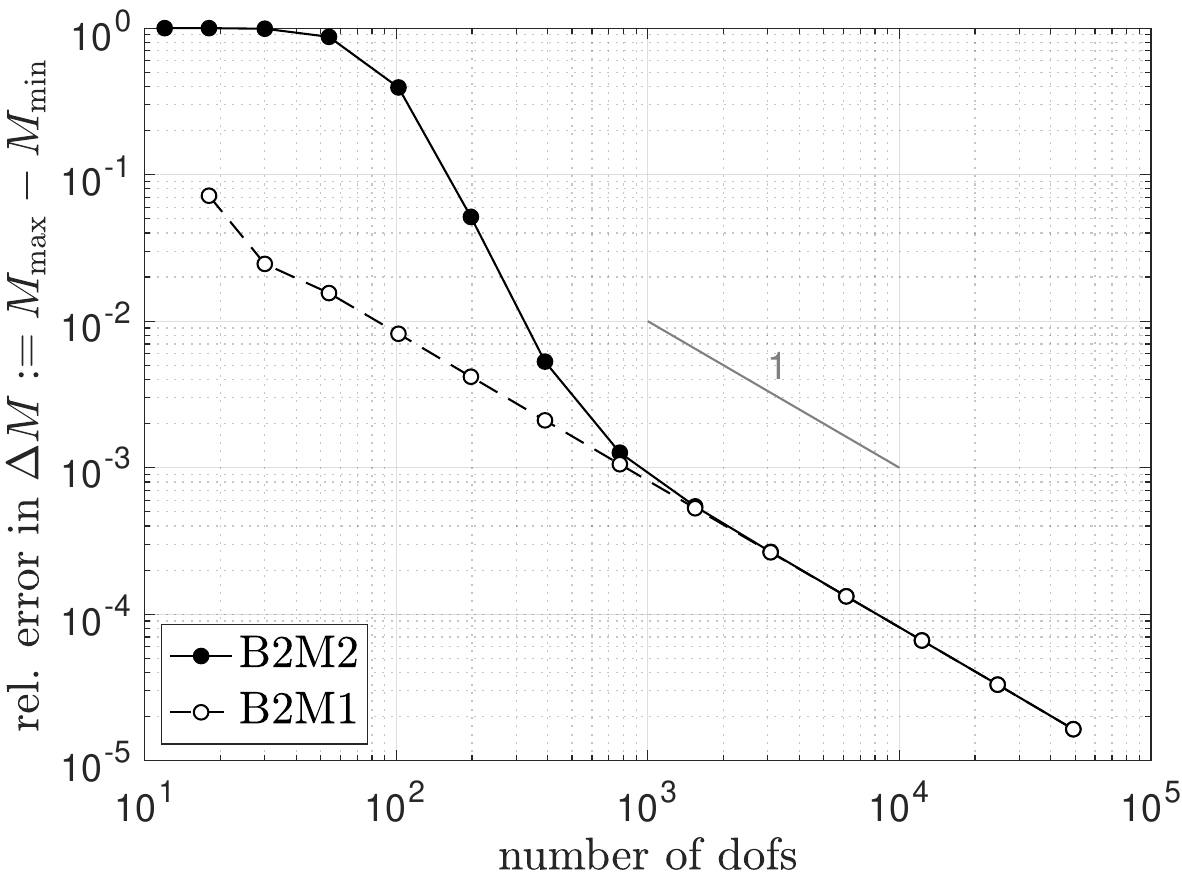}}
\put(0.2,-.1){\includegraphics[height=58mm]{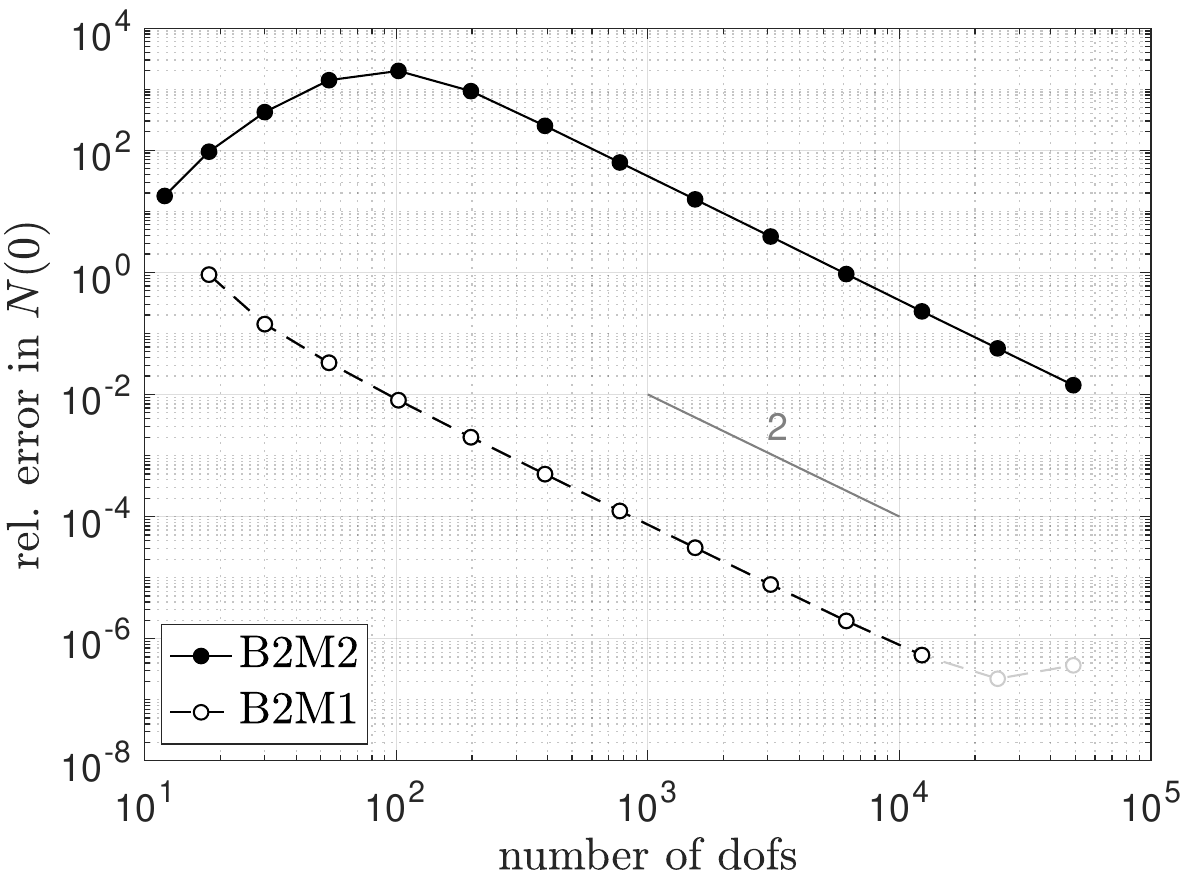}}
\put(-7.95,6.0){\footnotesize (a)}
\put(0.15,6.0){\footnotesize (b)}
\put(-7.95,0.0){\footnotesize (c)}
\put(0.15,0.0){\footnotesize (d)}
\end{picture}
\caption{Curved cantilever strip -- large deformation case: Maximum error convergence of (a) displacement $u_\mathrm{A}$, 
(b) raw effective membrane stress $\sigma$, (c) raw bending moment $M$ and (d) Cauchy membrane stress $N$ at $\theta=0$; 
all for $R/T=1000$ and with $m=2, 4, 8, 16, ..., 16384$ elements. 
The displacements and moments of B2M1 are more accurate than those of B2M2 up to $m = 512$ (i.e.~$n_\mathrm{dof} = 1542$).
The membrane stresses of B2M1 are several orders of magnitude better than those of B2M2 for all meshes.}
\label{f:CC_NL}
\end{center}
\end{figure}

Now the load is increased by a factor of 10 (i.e.~$qR^3/(ET^3)=-1$), and the problem is solved nonlinearly using the Newton-Raphson method and 10 load steps.
The resulting deformations and stresses are shown in Fig.~\ref{f:CC_usNL} for mesh $m=32$.
Here, $u_r$ is the displacement component with respect to the original radial direction (as defined by Eq.~\eqref{e:bu}).
Therefore the horizontal tip displacement is $u_\mathrm{A} = u_r(\pi/2)$.
Accurate reference results are obtained from the B5M5 discretization with very fine meshes. 
This gives 
$u_\mathrm{A}=-10.1288687743$, 
$\Delta\sig := \sig_\mathrm{max} - \sig_\mathrm{min} = 4.57049$, 
$\Delta M := M_\mathrm{max} - M_\mathrm{min} = 11.319864$ and
$\Delta N := N_\mathrm{max} - N_\mathrm{min} = 1.90083$
for $q=-1$ and $L=1$.
Further, equilibrium dictates that the Cauchy stress $N$ at the support ($\theta=0$) must be exactly equal to $q$, which can be used to asses the error free from any discretization inaccuracies stemming from the reference solution.
The convergence behavior of B2M1 vs.~B2M2 is shown in Fig.~\ref{f:CC_NL}.
Again, B2M1 is usually much more accurate than B2M2, especially for coarse discretizations.
In case of membrane stresses $\sig$ and $N$, B2M1 is also more accurate than B2M2 for fine discretizations.
For B2M2, stress errors can be very large even when displacement errors are small, which does not happen for B2M1.
The observed convergence rates are the same as those of the linear case, see Fig.~\ref{f:CC_max}.
Interestingly, the curves converge further than those of the linear case (see Fig.~\ref{f:CC_max}), such that a higher accuracy is obtained here than for the linear case.
This is attributed to the proper equilibrium enforcement by the Newton-Raphson iteration.

\subsection{Scordelis-Lo roof}\label{s:ScoLo}

\subsubsection{Problem setup and reference solution}

\begin{figure}[h]
\begin{center} \unitlength1cm
\begin{picture}(0,4.5)
\put(-7.9,-.2){\includegraphics[height=48mm]{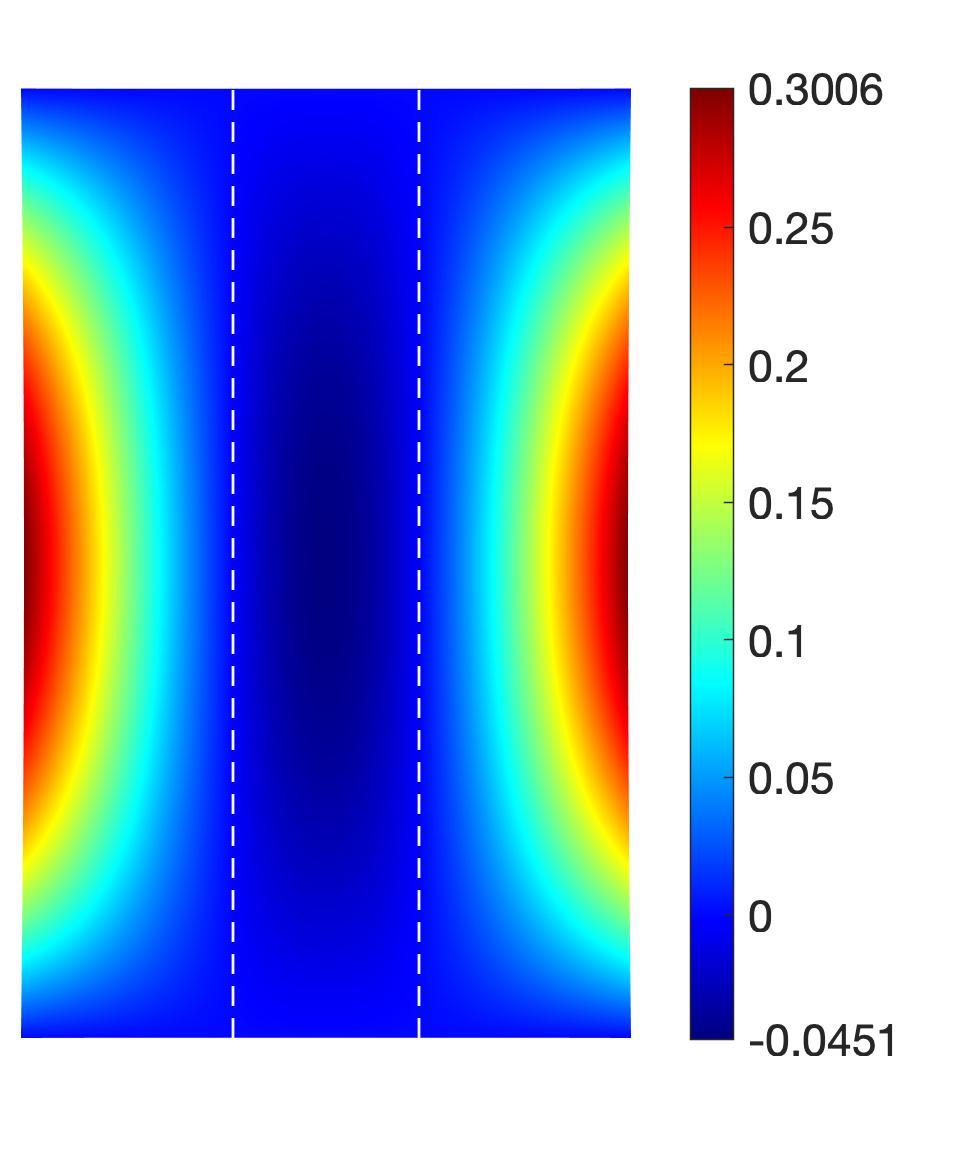}}
\put(-3.9,-.2){\includegraphics[height=48mm]{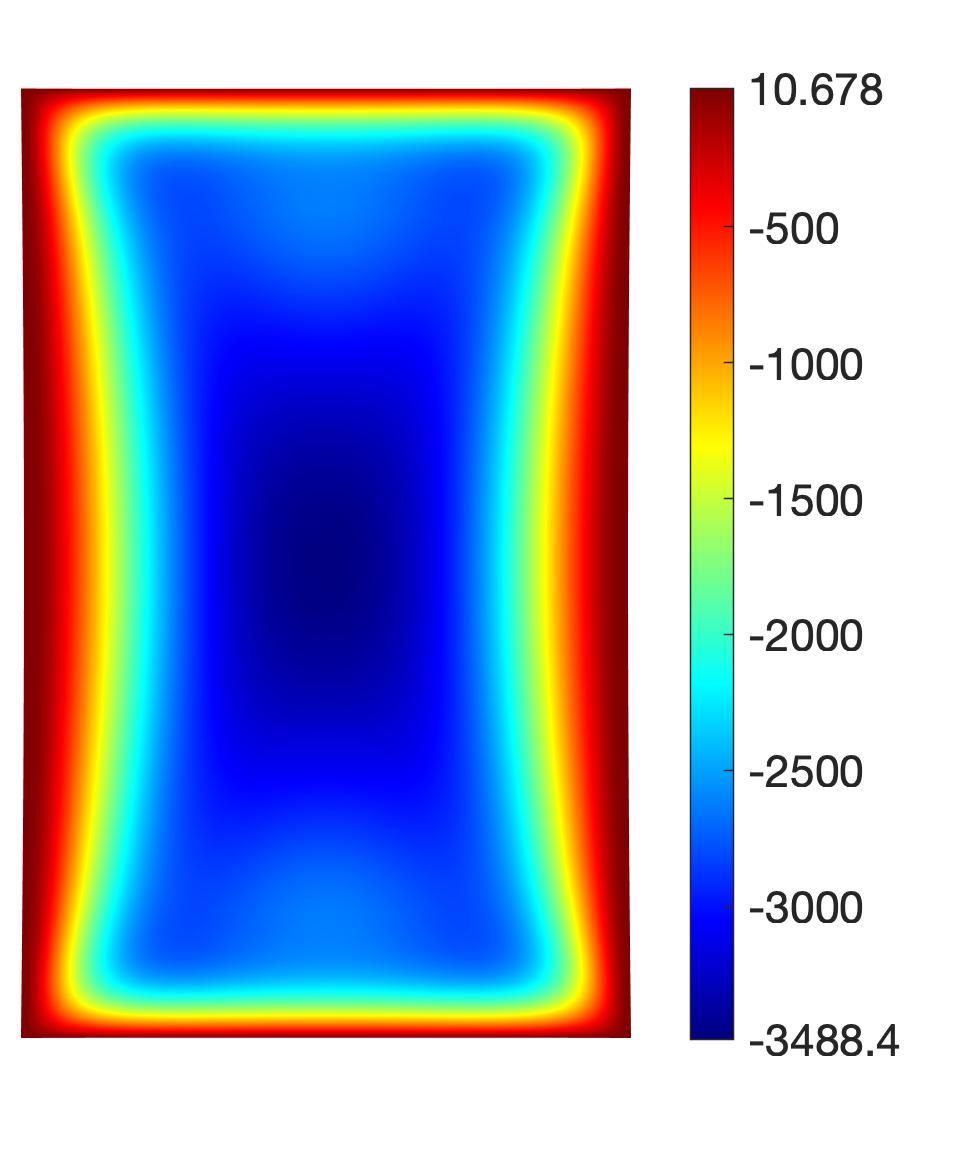}}
\put(0.1,-.2){\includegraphics[height=48mm]{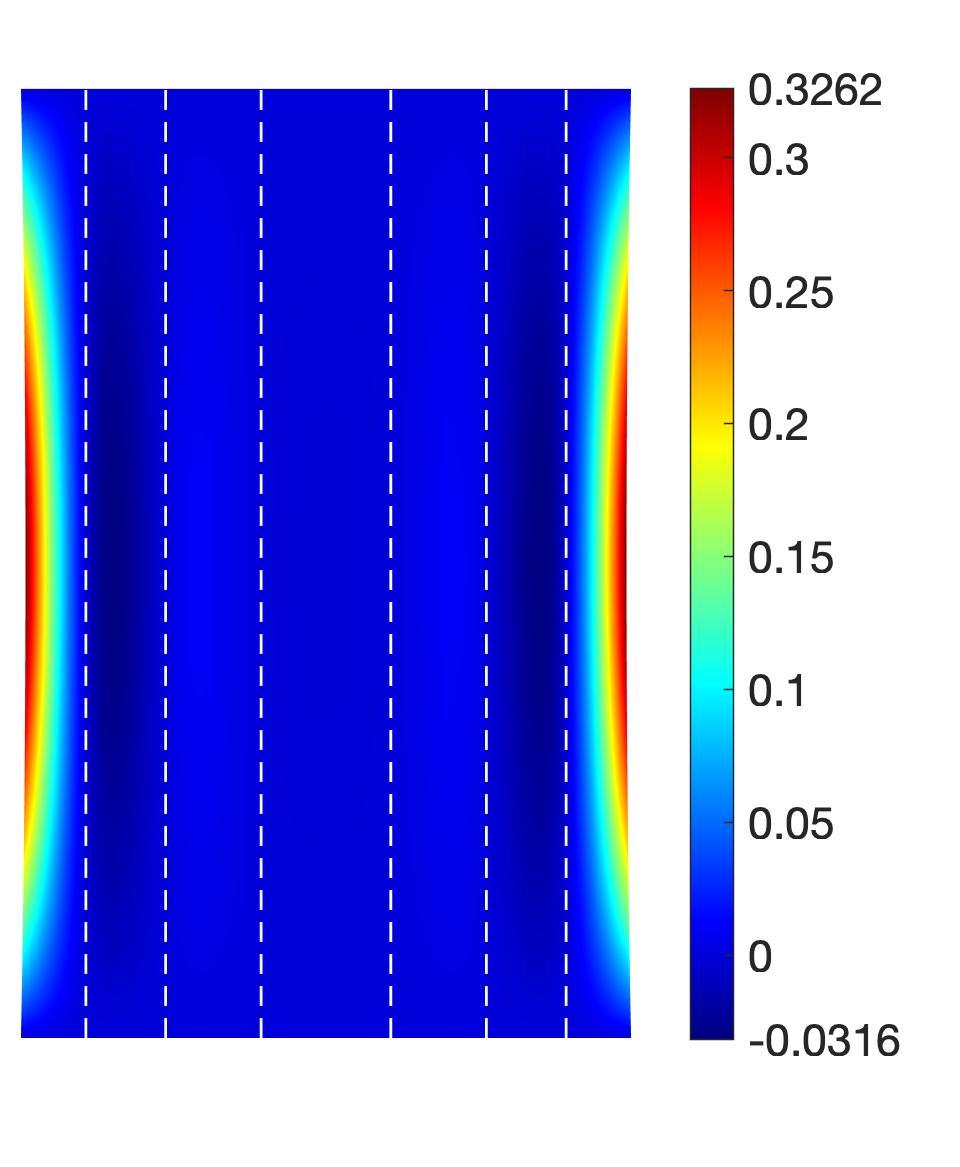}}
\put(4.1,-.2){\includegraphics[height=48mm]{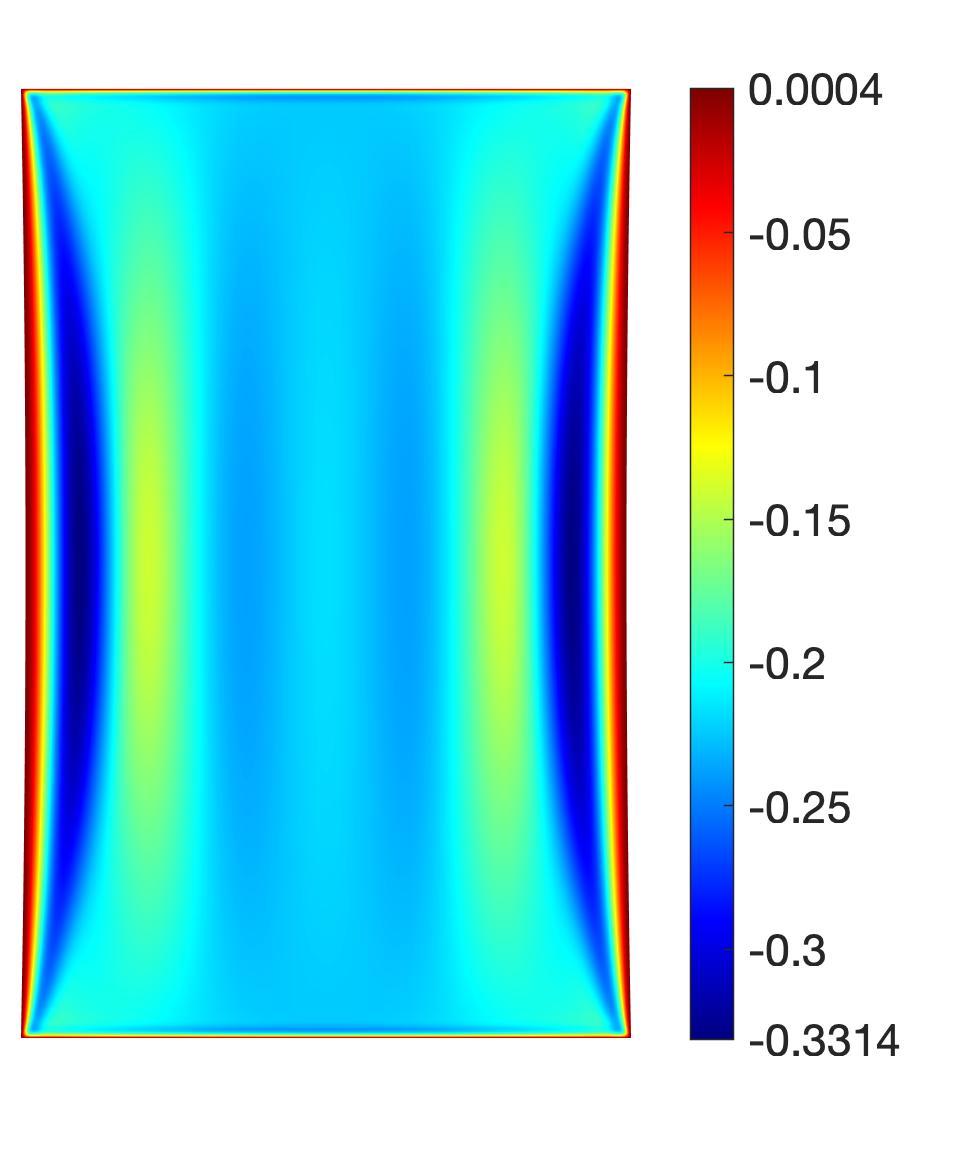}}
\put(-7.85,-.05){\footnotesize (a)}
\put(-3.85,-.05){\footnotesize (b)}
\put(0.15,-.05){\footnotesize (c)}
\put(4.15,-.05){\footnotesize (d)}
\end{picture}
\caption{Reference solution for the Scordelis-Lo roof: 
$u_3$ and $\sig_{11}$ for $R/T=100$ (a \& b, respectively), and $u_3$ and $\sig_{11}$ for $R/T=10^4$ (c \& d, respectively). All obtained with discretization $m=128$ and $p=5$.
The dashed white lines in (a) and (c) mark locations of zero displacement.
They show that the displacement field becomes more and more oscillatory as $R/T$ increases.\\[-9mm]}
\label{f:SLref}
\end{center}
\end{figure}
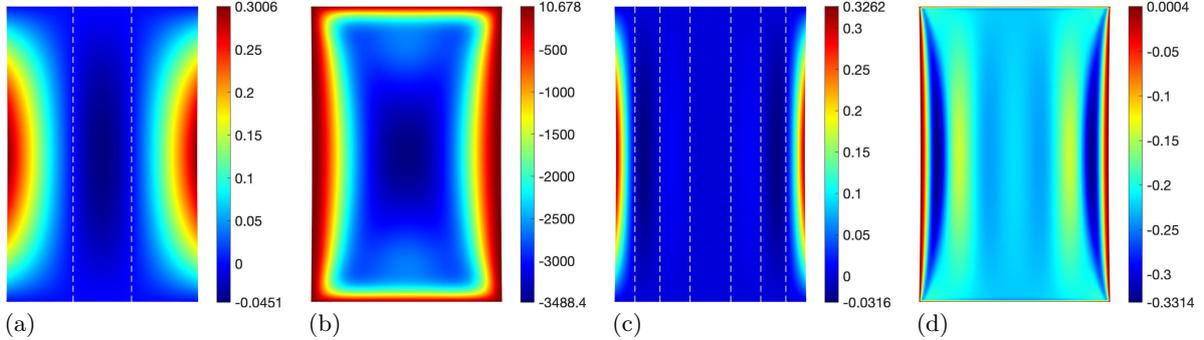

The Scordelis-Lo roof is a classical benchmark example initially proposed by \citet{scordelis64} and subsequently studied by many authors.
It consists of an $80^\circ$ cylinder surface segment with length $L=50L_0$, radius $R=25L_0$ and thickness $T$, see Fig.~\ref{f:B2M1s}.
The roof is discretized into $m\times1.5m$ biquadratic NURBS and $(m+1)\times(1.5m+1)$ bilinear Lagrange elements, with $m=4,\,8,\,16,\,32,\,...,\,512$.
See Fig.~\ref{f:B2M1s} for $m=4$.
The roof is made of linear elastic isotropic material with Young's modulus $E=4.32E_0$ and Poisson's ratio $\nu = 0$.
It is supported along the curved edges at $Y=0$ and $Y=L$ keeping the $X$- and $Z$- (but not $Y$-) direction fixed.
It is loaded by the uniform surface load $q = -1.44\cdot10^{-5}\,E_0\,(T/L_0)^2$ along the $Z$-direction.
Two slendernesses are considered: $R/T=100$ and $R/T=10^4$, i.e.~$T=L_0/4$ and $T=L_0/400$.
Fig.~\ref{f:SLref} shows the reference solution for the downward displacement $u_3$ and the effective membrane stress\footnote{Considering an orthonormal surface basis, such that $\sig_{\alpha\beta}=\sig^\alpha_\beta=\sig^{\alpha\beta}$.} $\sigma_{11}$ obtained from quintic IGA with $L_0=E_0=1$.

\subsubsection{Displacement accuracy}

\begin{figure}[h]
\begin{center} \unitlength1cm
\begin{picture}(0,11.7)
\put(-8,5.9){\includegraphics[height=58mm]{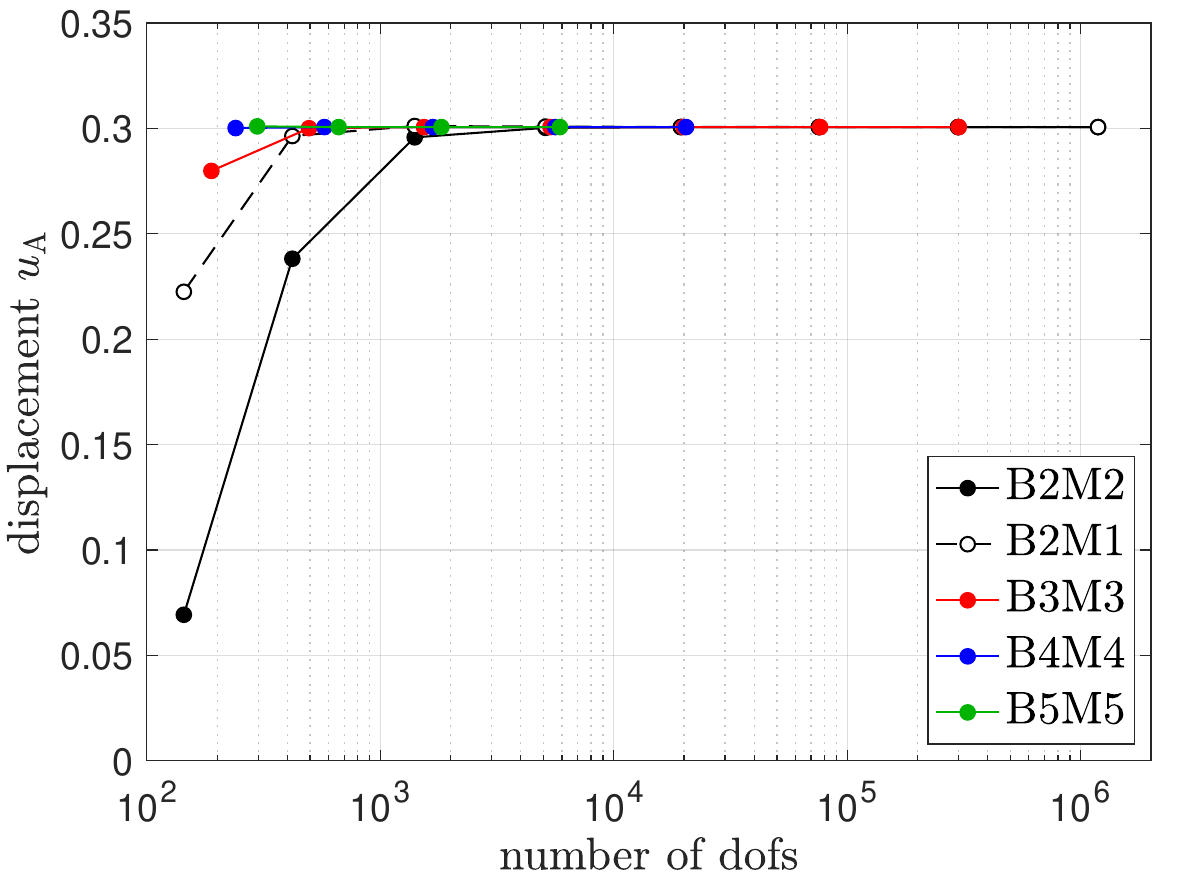}}
\put(-8,-.1){\includegraphics[height=58mm]{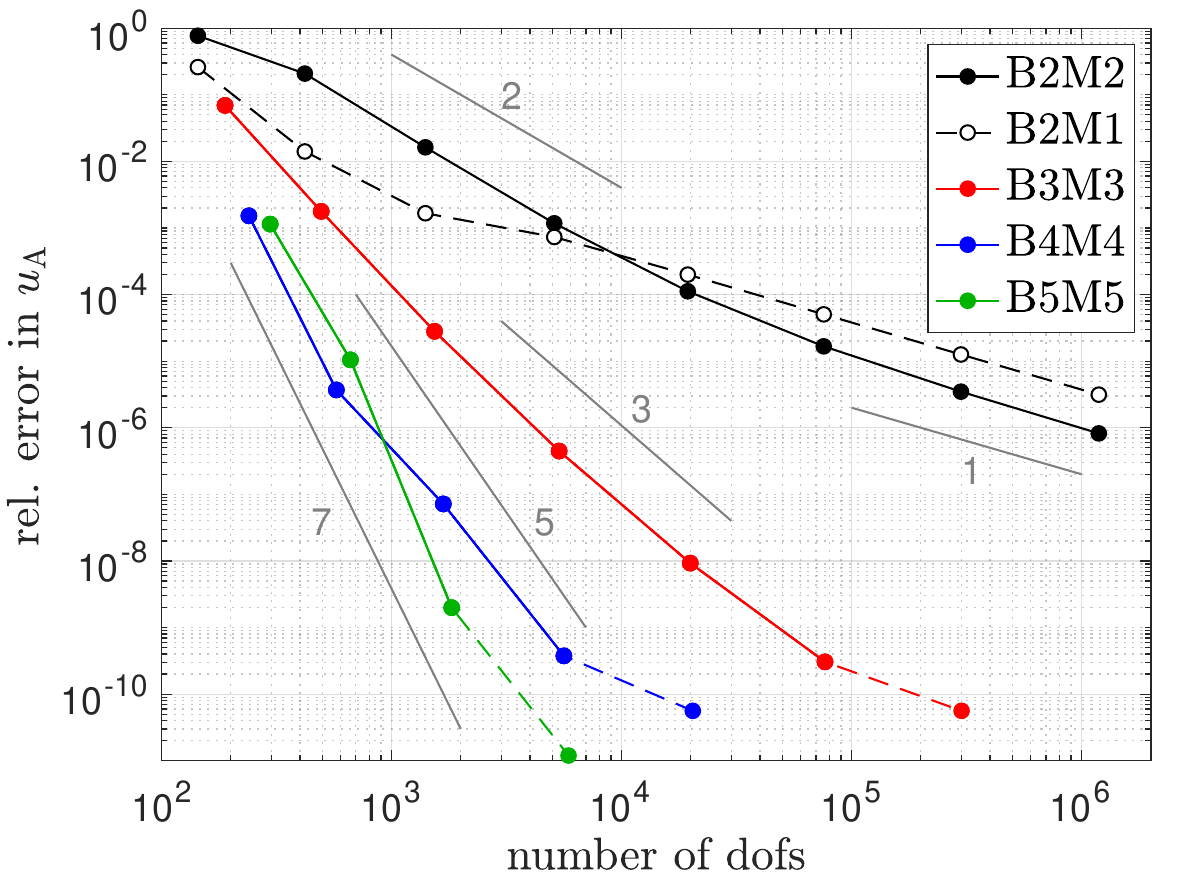}}
\put(0.2,5.9){\includegraphics[height=58mm]{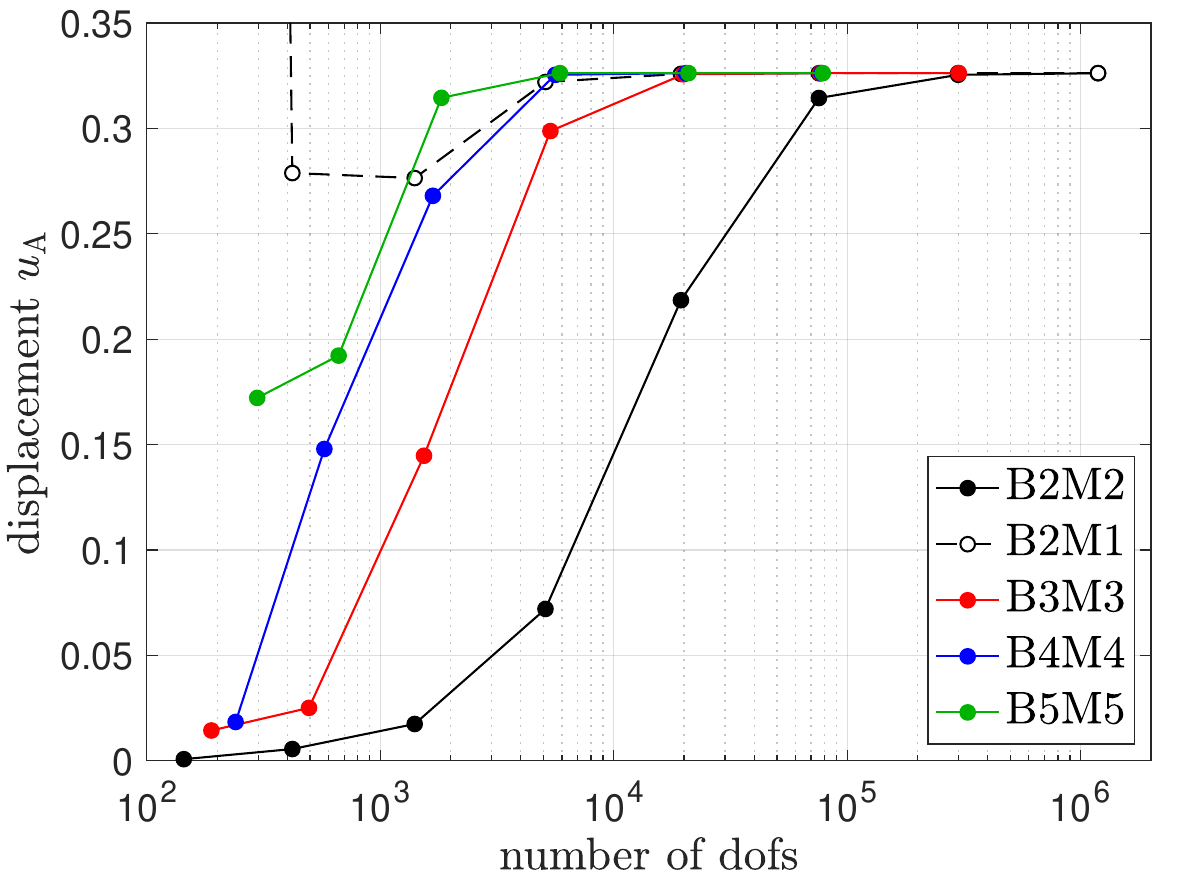}}
\put(0.2,-.1){\includegraphics[height=58mm]{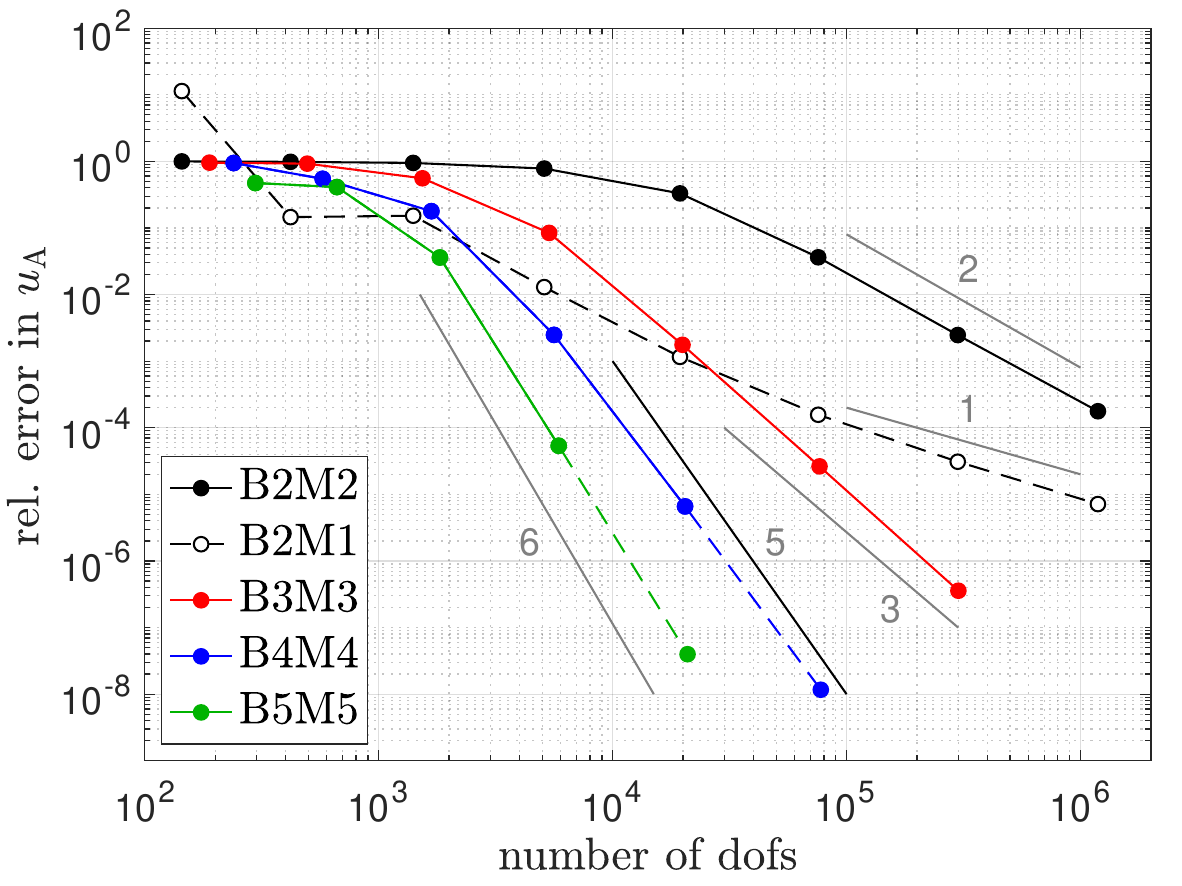}}
\put(-7.95,6.0){\footnotesize (a)}
\put(0.15,6.0){\footnotesize (b)}
\put(-7.95,0.0){\footnotesize (c)}
\put(0.15,0.0){\footnotesize (d)}
\end{picture}
\caption{Scordelis-Lo roof: Convergence of displacement $u_\mathrm{A}$ for $R/T=100$ (left side) and $R/T=10^4$ (right side) using $m=4, 8, 16, ..., 512$.  
The top row shows the absolute values of $u_\mathrm{A}$;
the bottom row its relative error. 
For coarse meshes and/or high $R/T$, B2M1 is much more accurate than B2M2.}
\label{f:SLue}
\end{center}
\end{figure}
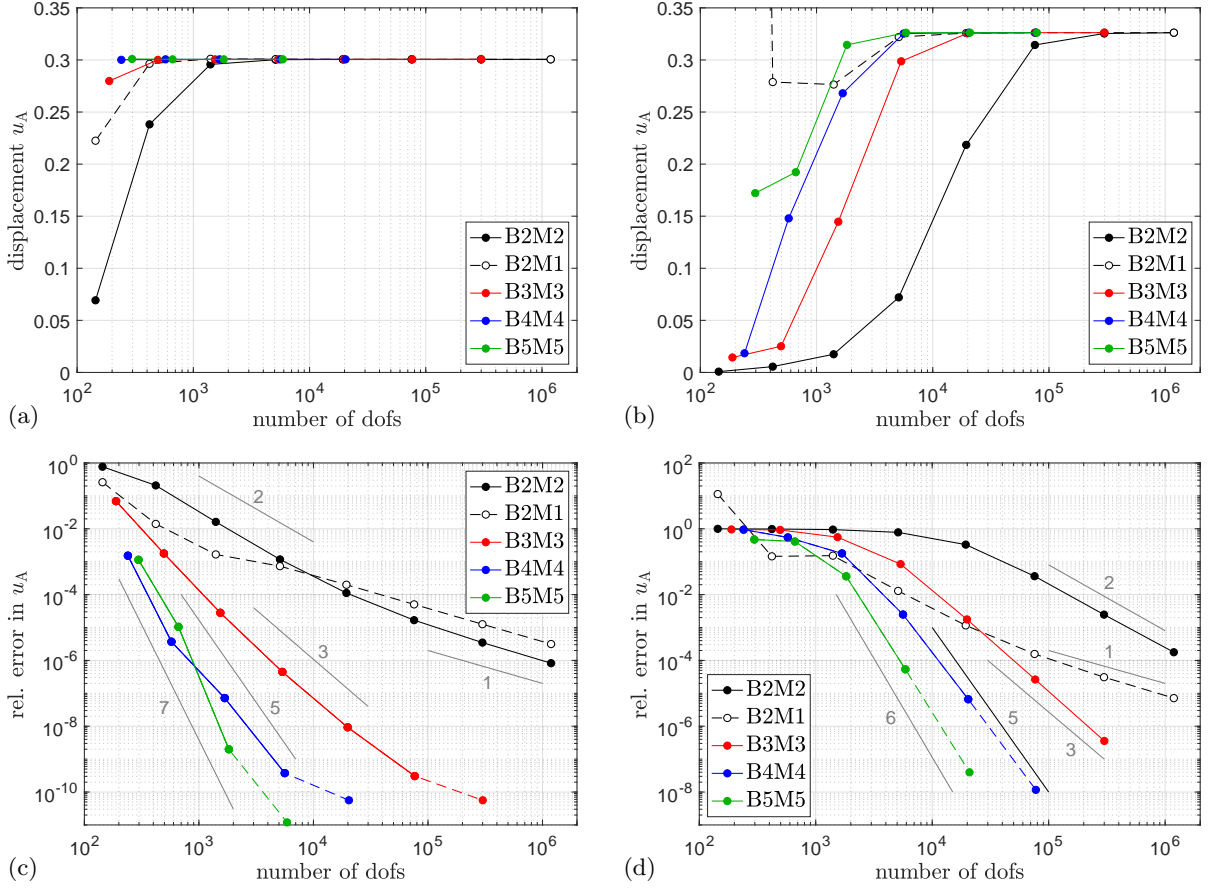

According to the reference solution from above, the reference values for the downward displacement at point A (see Fig.~\ref{f:B2M1s}) are 
$u_\mathrm{A} = 0.3005924566$ for $R/T=100$ and $u_\mathrm{A} = 0.32620099$ for $R/T=10^4$.
The first is close to $u_\mathrm{A} = 0.300592497$, the value given by \citet{mi21}.
The convergence of classical IGA (i.e.~the B$p$M$p$ discretization) and the B2M1 discretization to these reference values is shown in Fig.~\ref{f:SLue}.
The behavior is consistent to the one of the curved cantilever strip, see Fig.~\ref{f:CC_max}a:
B2M1 is usually more accurate than B2M2, especially for coarse meshes and high $R/T$.
For some meshes B2M1 even beats B3M3, B4M4 and B5M5.
Interestingly, for the coarsest mesh, the B2M1 response for $R/T=10^4$ is overly soft instead of overly stiff
-- see first data point at $n_\mathrm{dof} = 144$ in Fig.~\ref{f:SLue}b \& d.
This indicates that for very coarse meshes and large slenderness ratios, the B2M1 discretization may become unstable.
A minimum number of elements is needed to avoid this.
For $R/T=10^4$, $8\times12$ B2 elements are already sufficient, which is still very coarse in view of the highly oscillatory displacement field of this case (see Fig.~\ref{f:SLref}c). 

As in the curved cantilever strip (see Fig.~\ref{f:CC_max}a), locking expresses itself as the initial regime where the convergence behavior stagnates -- i.e.~the initial flat plateau, most prominently seen in Fig.~\ref{f:SLue}d.
The observed convergence rates are now half as large as in the curved cantilever strip, since the number of dofs now increase with $m^2$, as opposed to $m$, as is the case in Fig.~\ref{f:CC_max}a.

\subsubsection{Stress accuracy}

\begin{figure}[H]
\begin{center} \unitlength1cm
\begin{picture}(0,5.8)
\put(-8,-.1){\includegraphics[height=58mm]{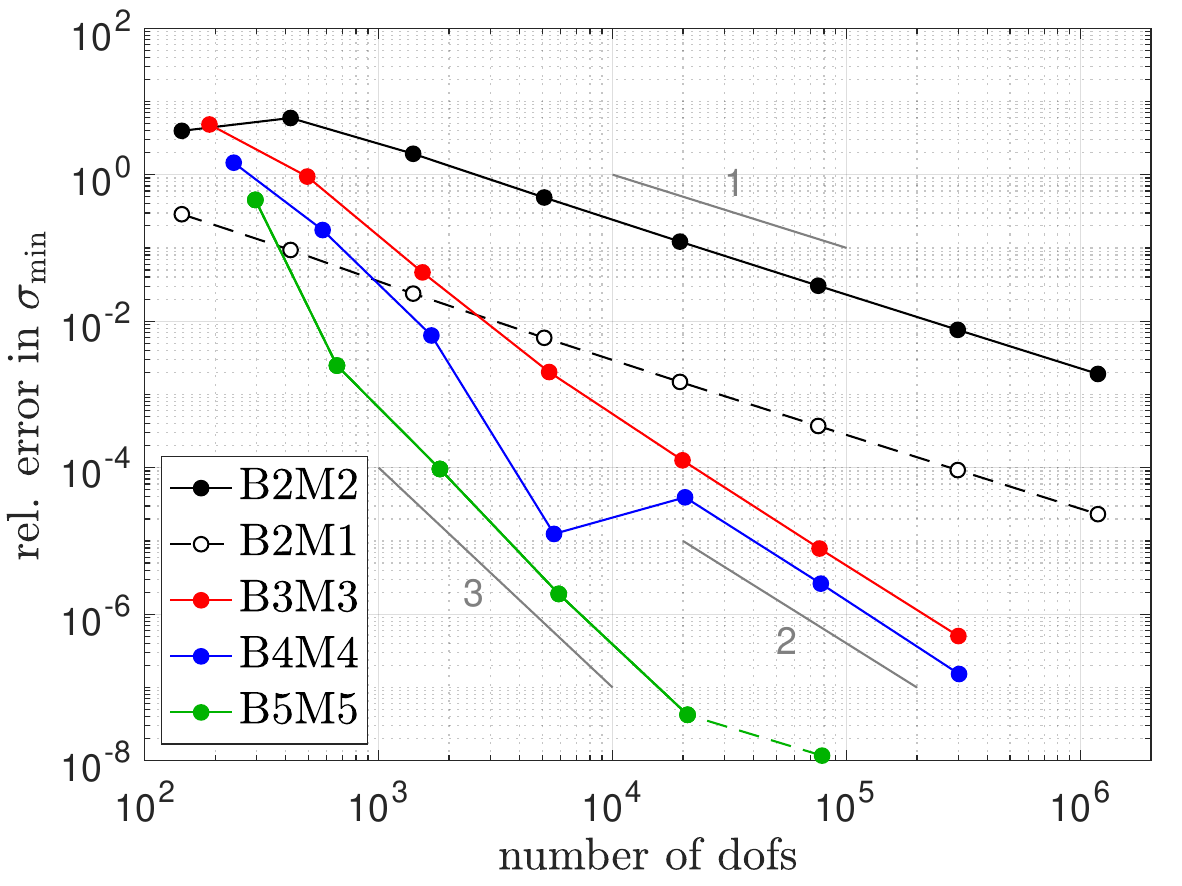}}
\put(0.2,-.1){\includegraphics[height=58mm]{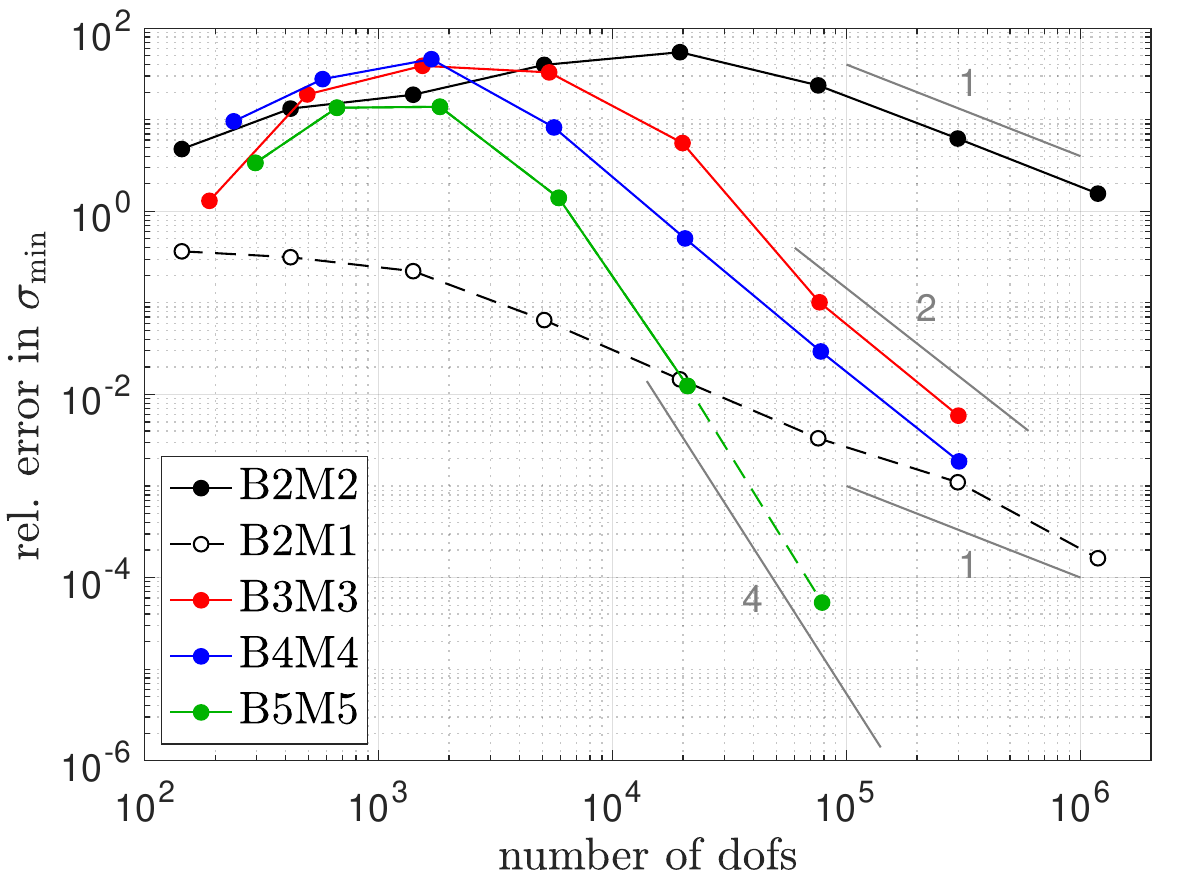}}
\put(-7.95,0.0){\footnotesize (a)}
\put(0.15,0.0){\footnotesize (b)}
\end{picture}
\caption{Scordelis-Lo roof: 
Convergence of the stress $\sig_\mathrm{min}$ for (a) $R/T=100$ and (b) $R/T=10^4$, $m=4, 8, ..., 512$.
For coarse meshes and/or high $R/T$, B2M1 is much more accurate than all other cases.}
\label{f:SLse}
\end{center}
\end{figure}

\begin{figure}[H]
\begin{center} \unitlength1cm
\begin{picture}(0,8.6)
\put(-7.9,4){\includegraphics[height=48mm]{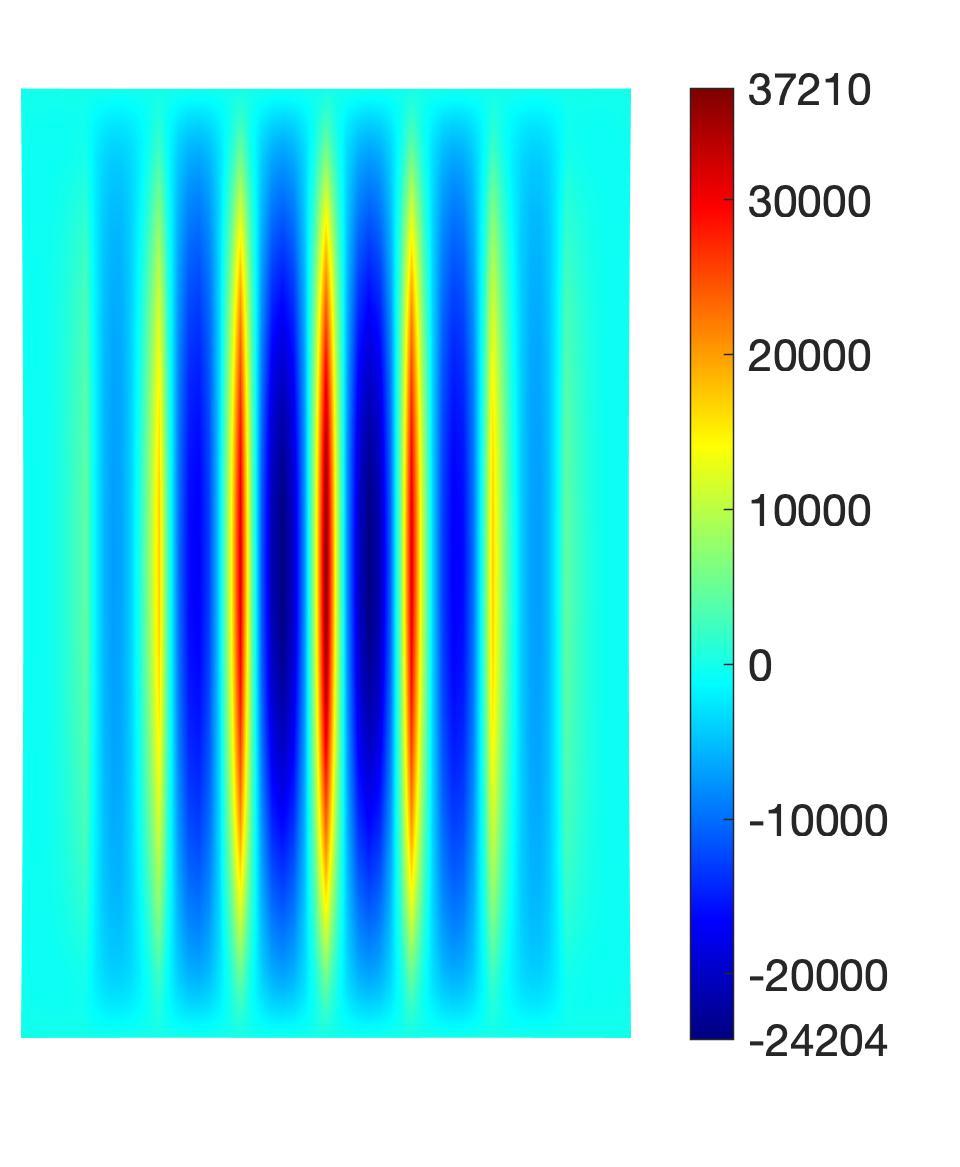}}
\put(-3.9,4){\includegraphics[height=48mm]{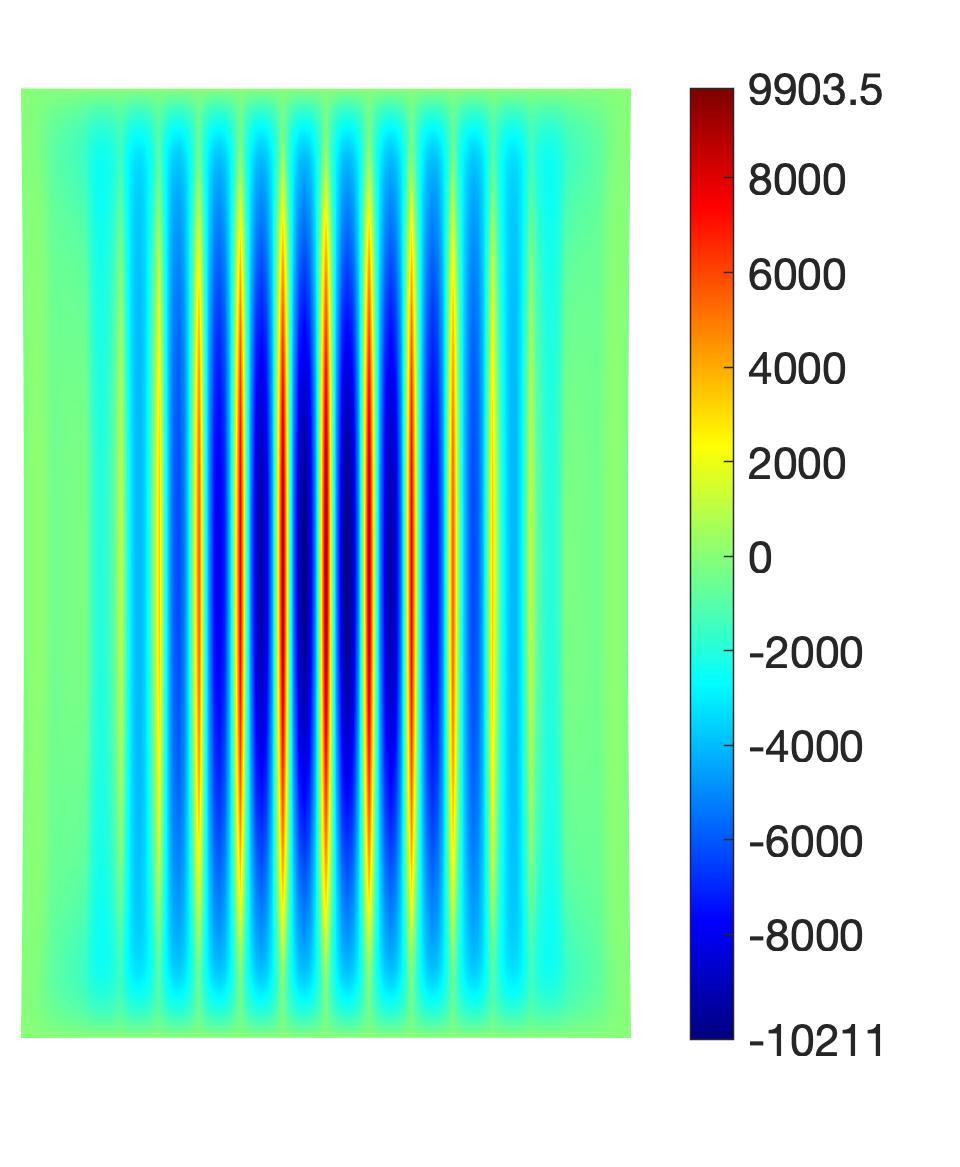}}
\put(0.1,4){\includegraphics[height=48mm]{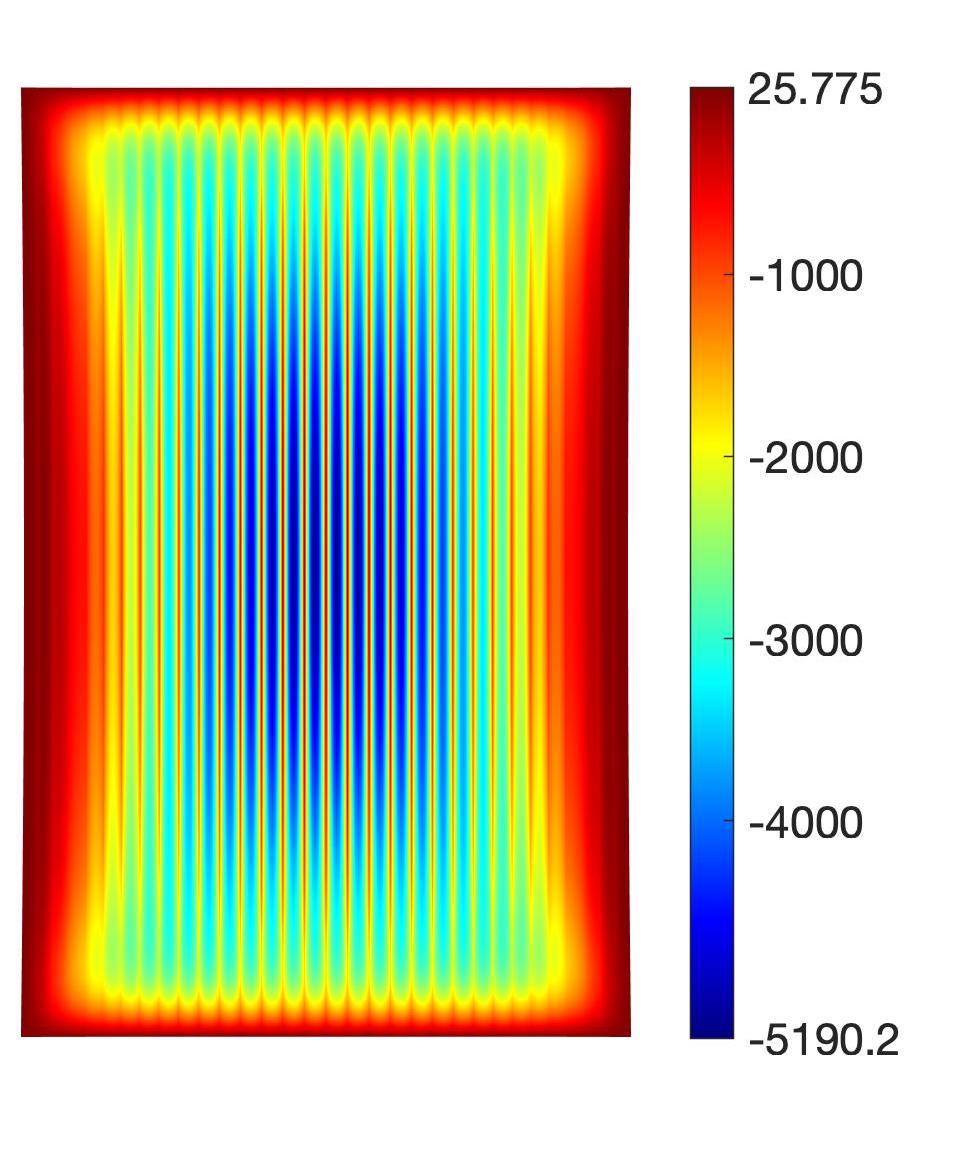}}
\put(4.1,4){\includegraphics[height=48mm]{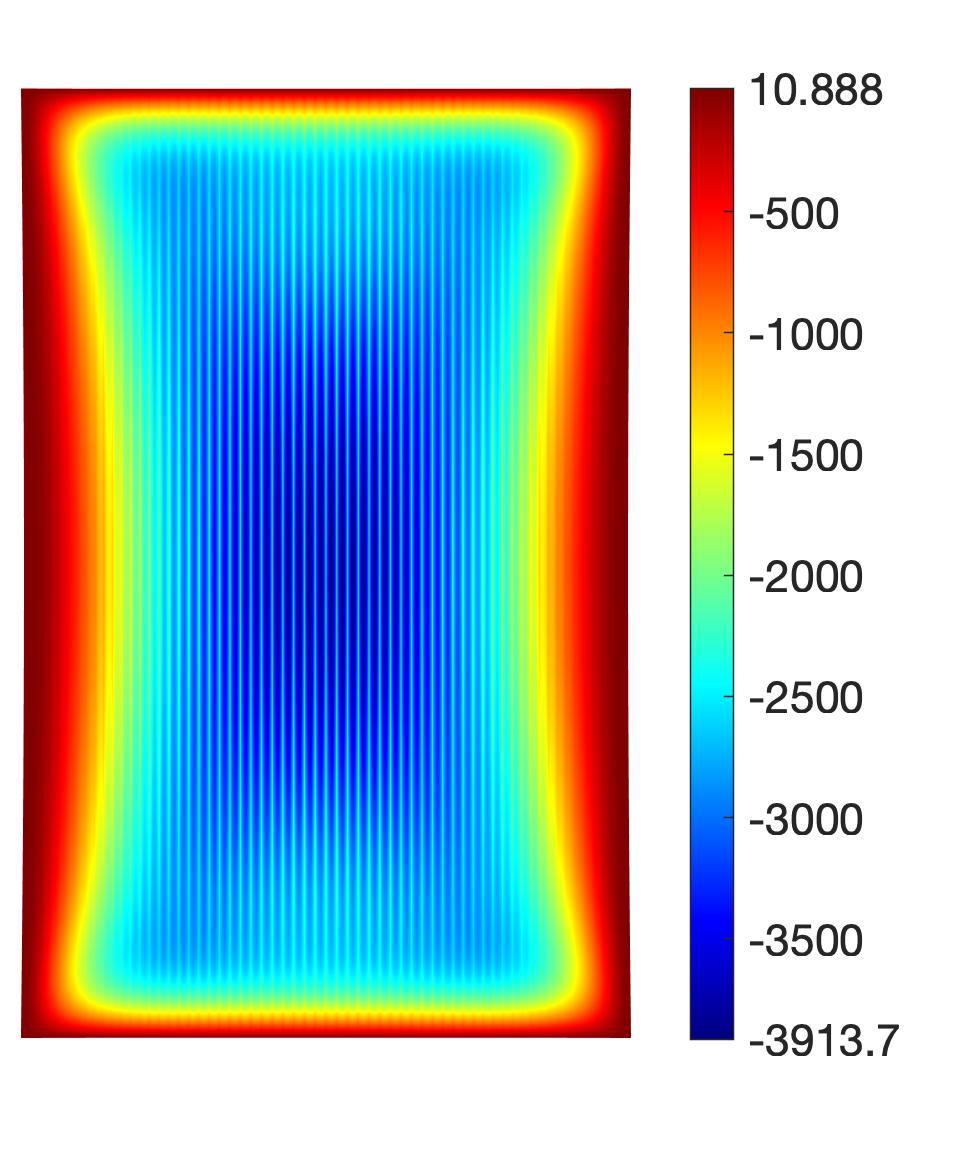}}
\put(-7.9,-.5){\includegraphics[height=48mm]{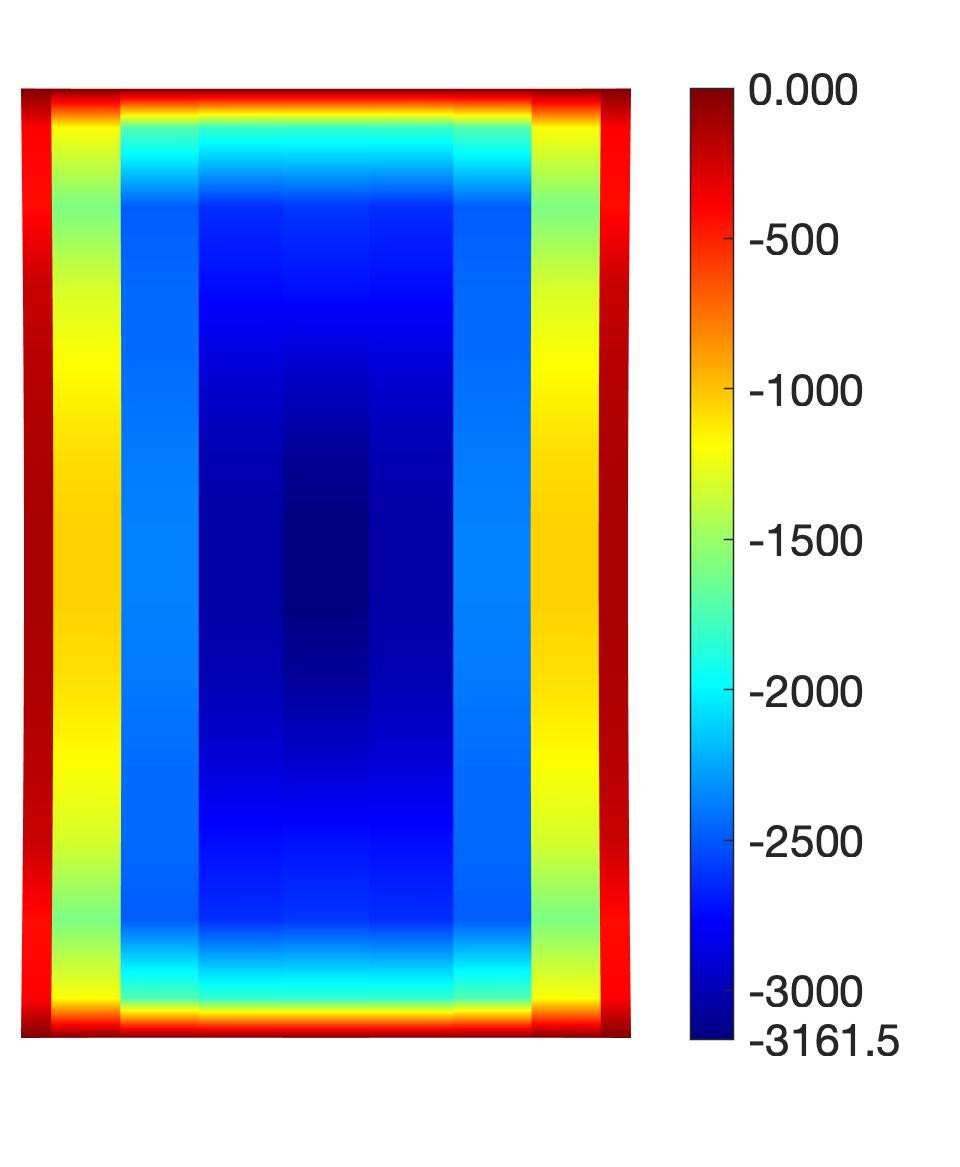}}
\put(-3.9,-.5){\includegraphics[height=48mm]{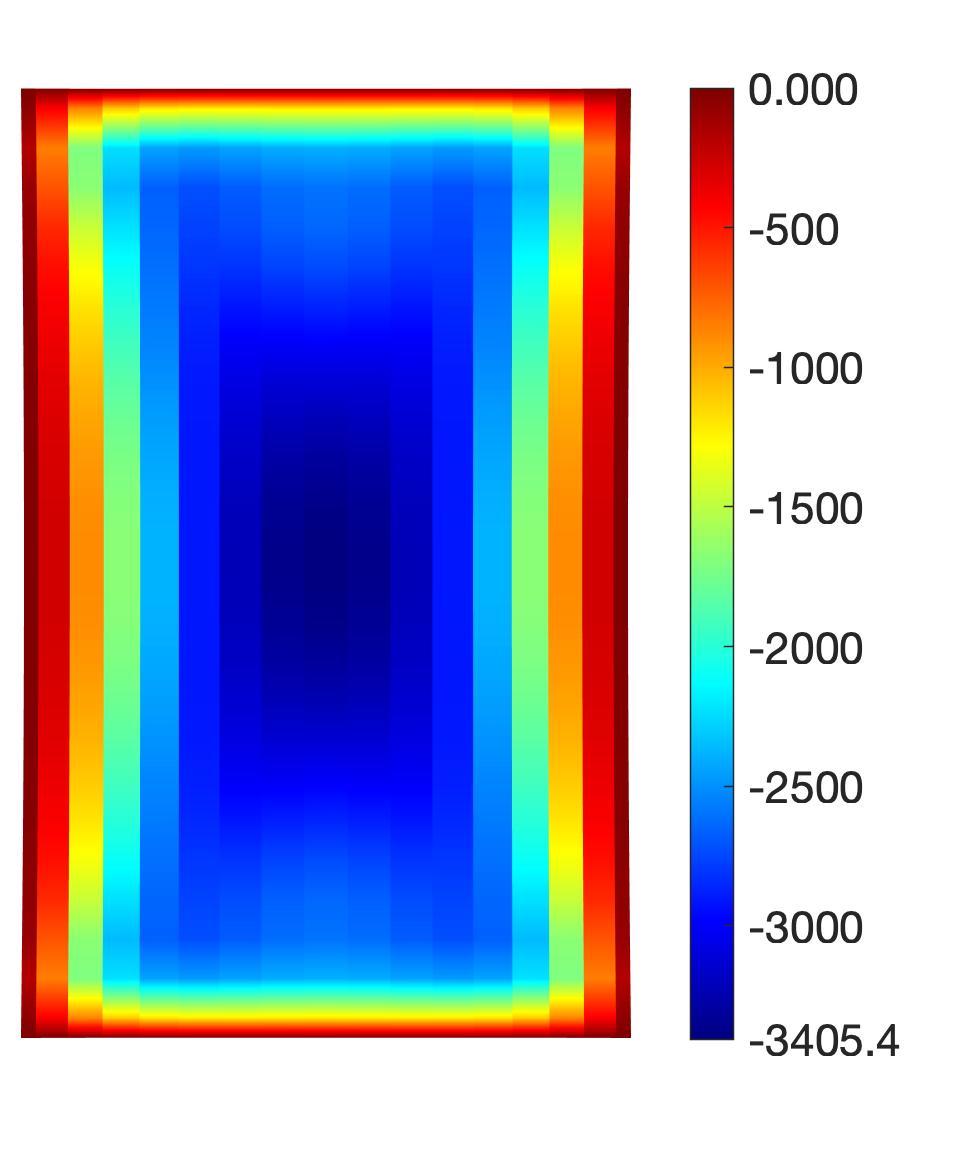}}
\put(0.1,-.5){\includegraphics[height=48mm]{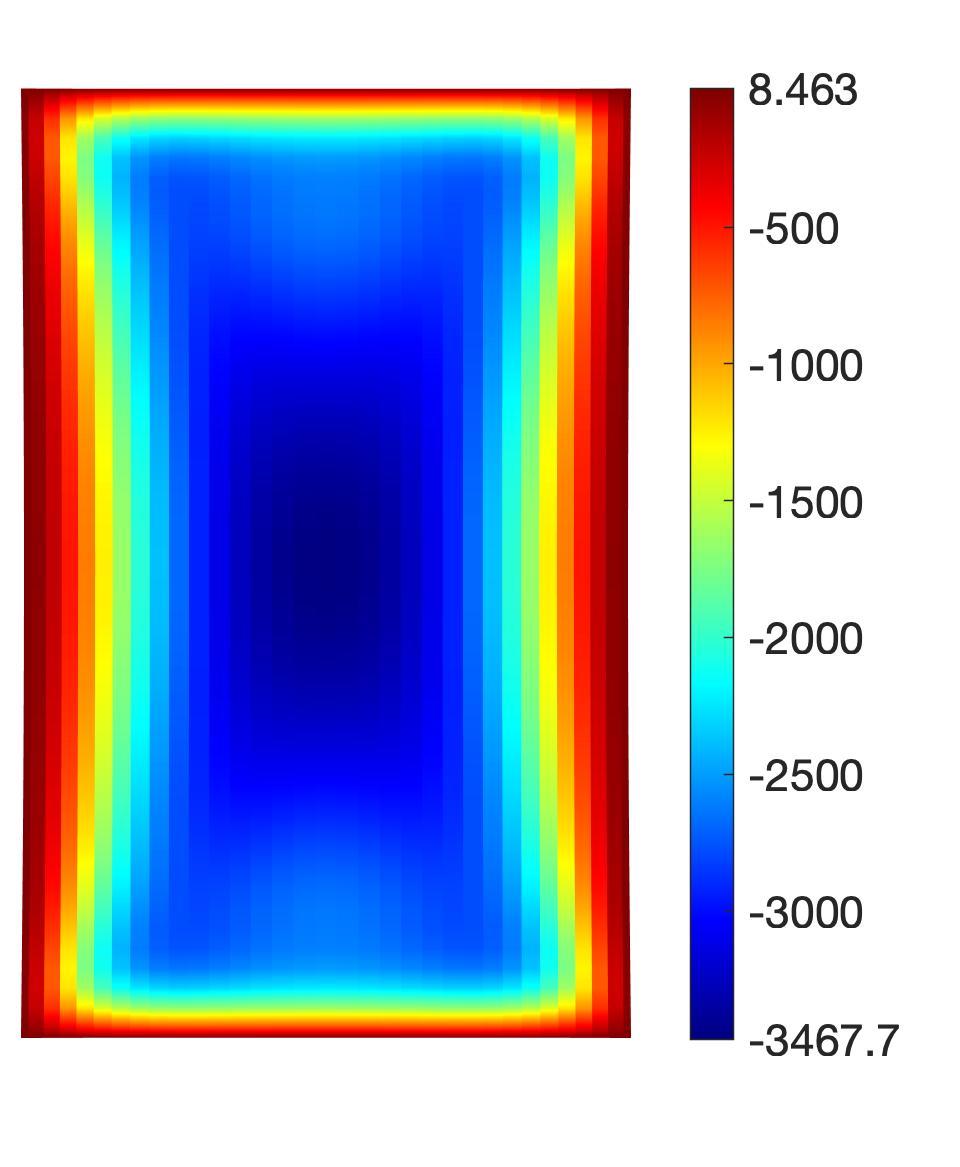}}
\put(4.1,-.5){\includegraphics[height=48mm]{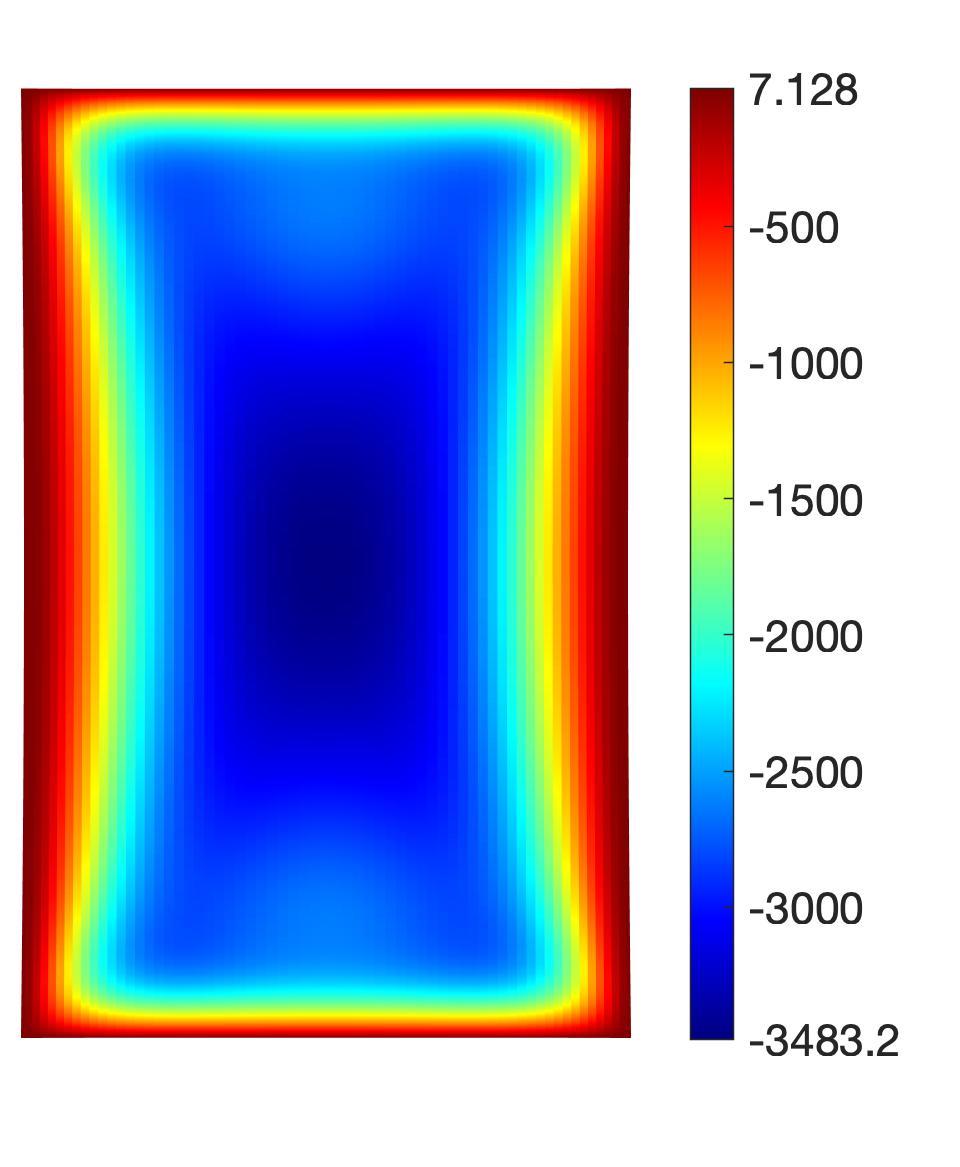}}
\end{picture}
\caption{Scordelis-Lo roof with $R/T=100$: Raw effective membrane stress~$\sig_{11}$ for the B2M2 discretization (top row) and B2M1 discretization (bottom row) for meshes $m = 8,\,16,\,32,\,64$ (left to right).
In case of B2M2, the stresses oscillate severely, necessitating very fine meshes.
In case of B2M1, no stress oscillation appear, allowing for good coarse mesh accuracy.}
\label{f:SL2sig}
\end{center}
\end{figure}

\begin{figure}[H]
\begin{center} \unitlength1cm
\begin{picture}(0,8.6)
\put(-7.9,4){\includegraphics[height=48mm]{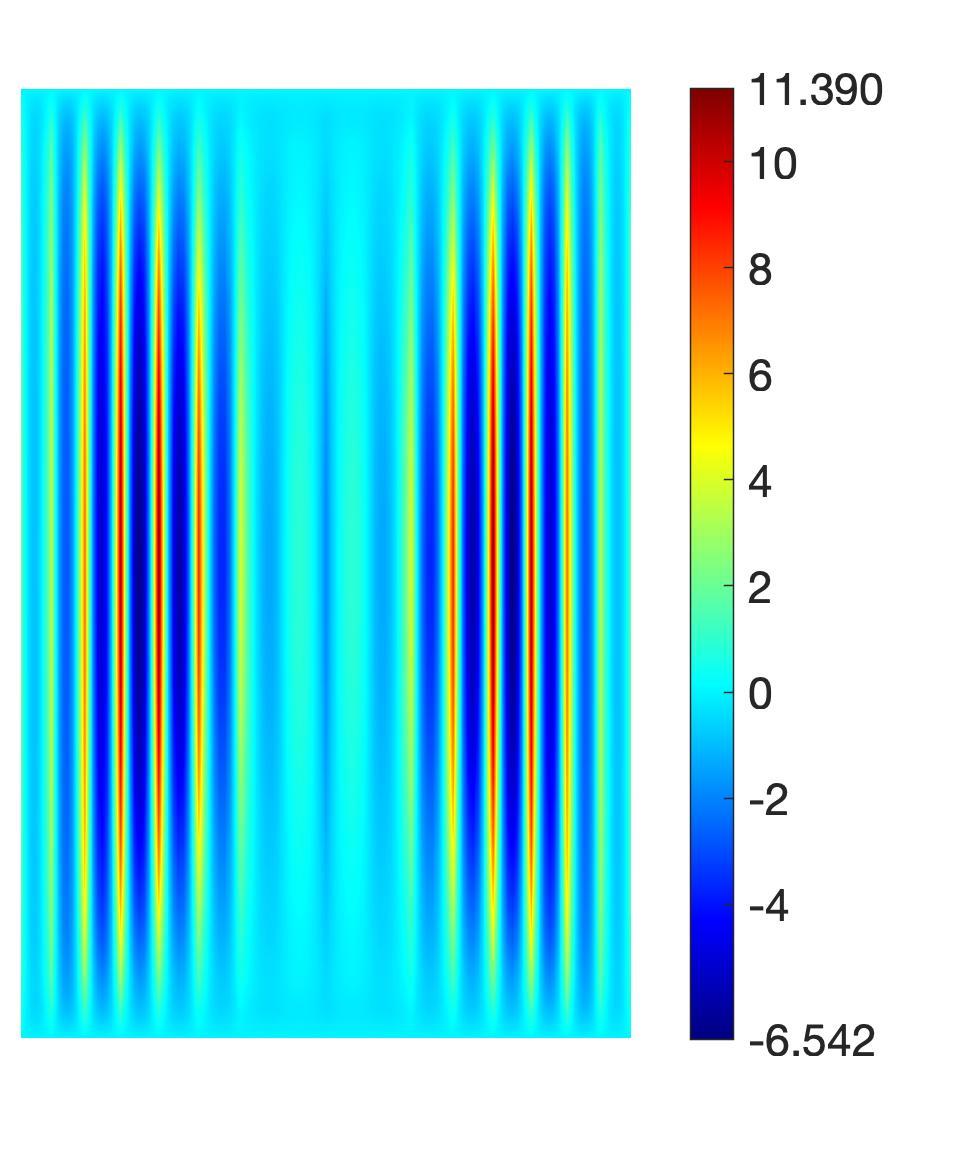}}
\put(-3.9,4){\includegraphics[height=48mm]{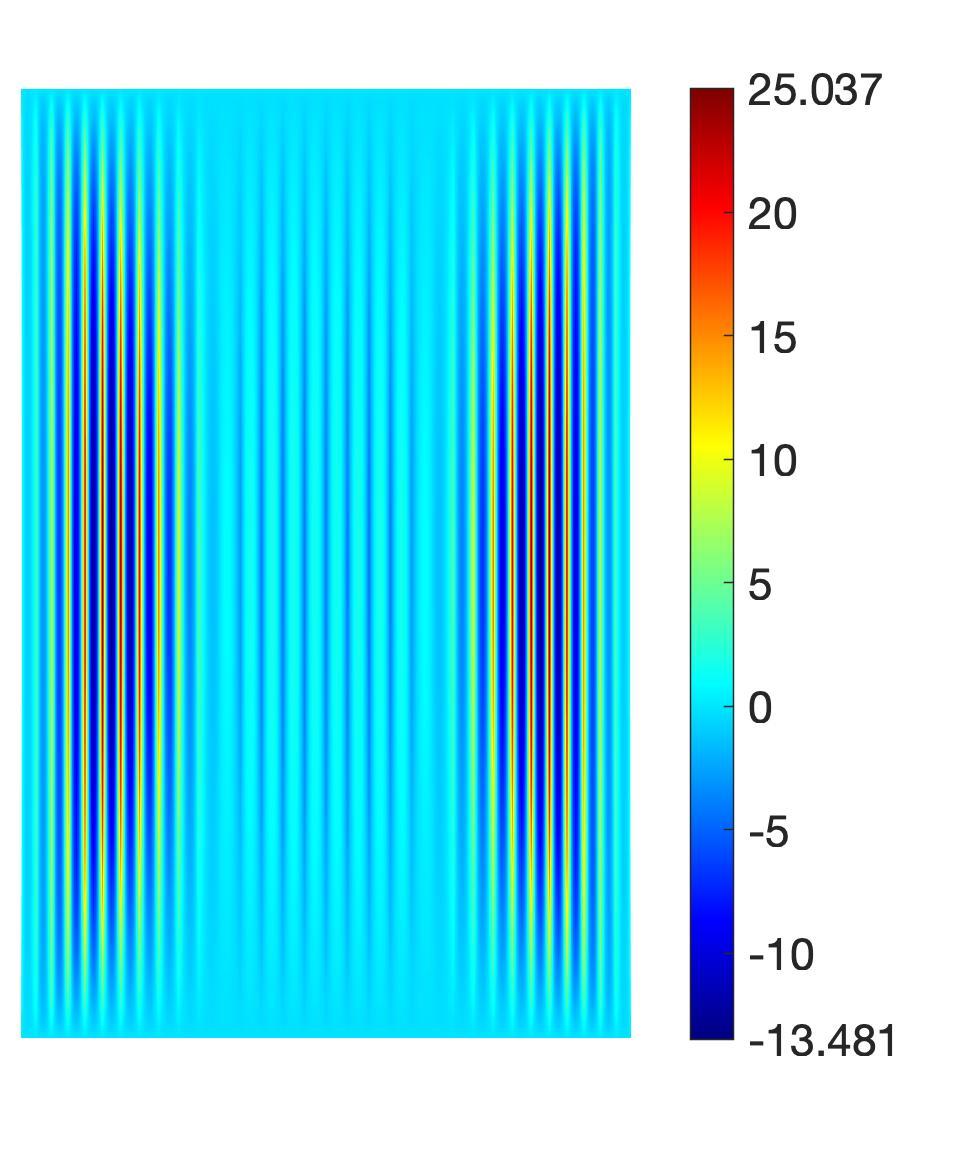}}
\put(0.1,4){\includegraphics[height=48mm]{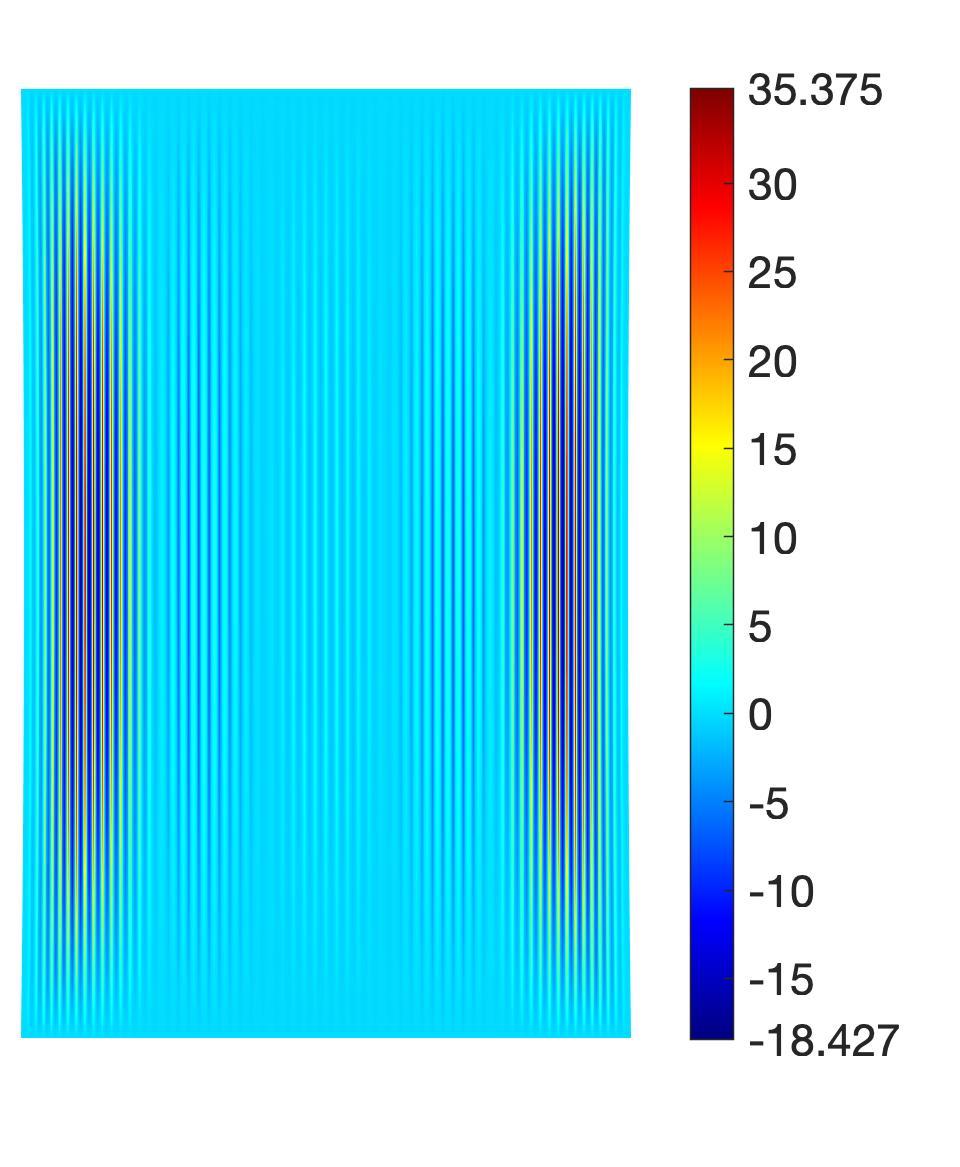}}
\put(4.1,4){\includegraphics[height=48mm]{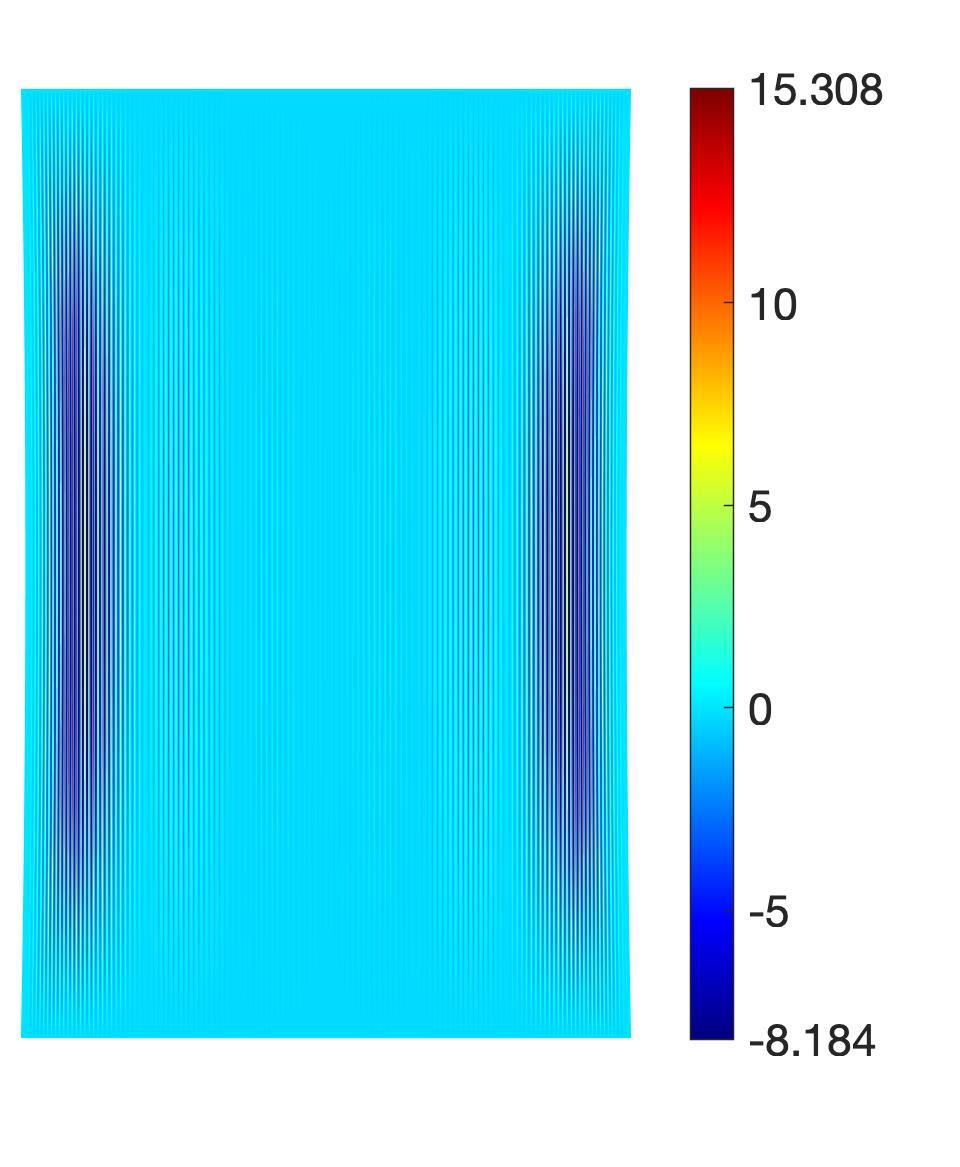}}
\put(-7.9,-.5){\includegraphics[height=48mm]{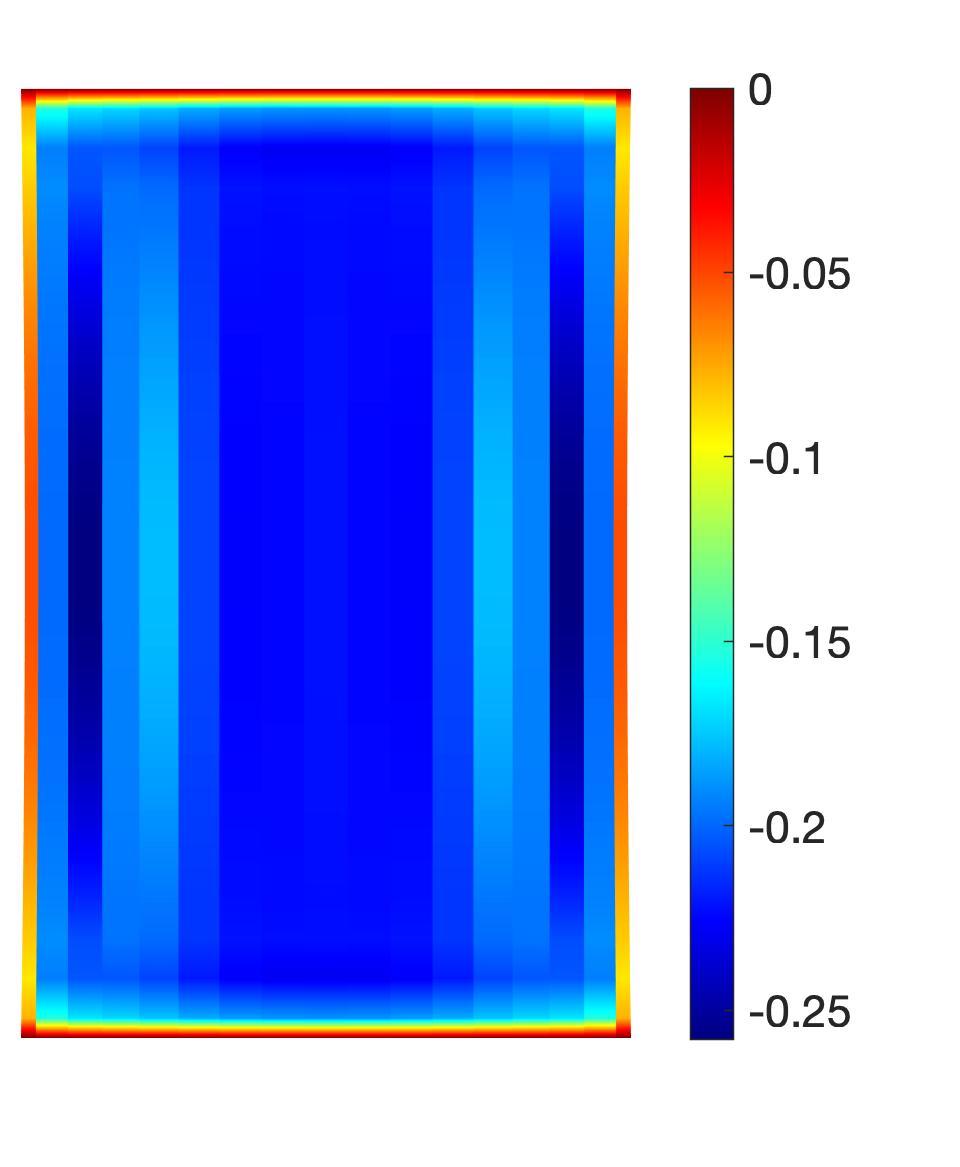}}
\put(-3.9,-.5){\includegraphics[height=48mm]{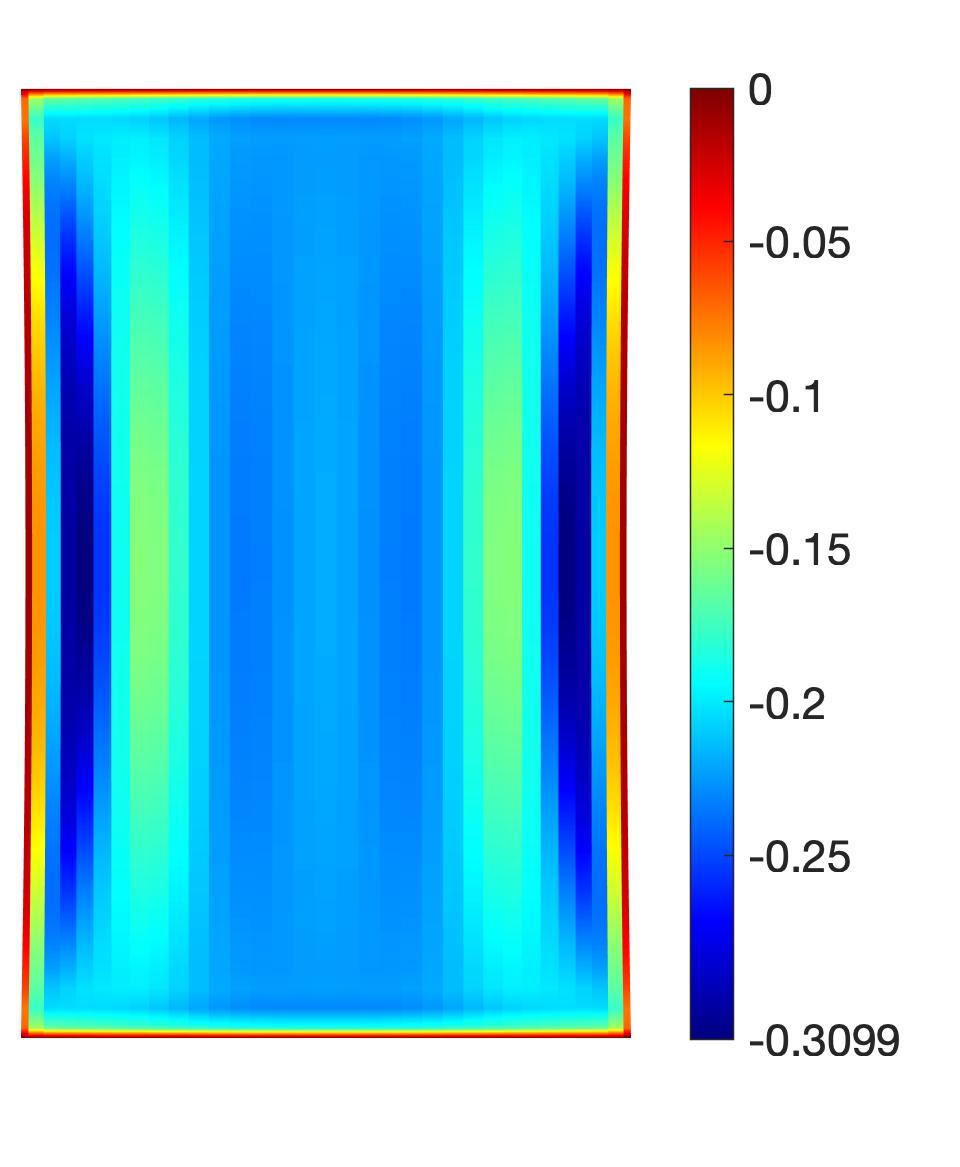}}
\put(0.1,-.5){\includegraphics[height=48mm]{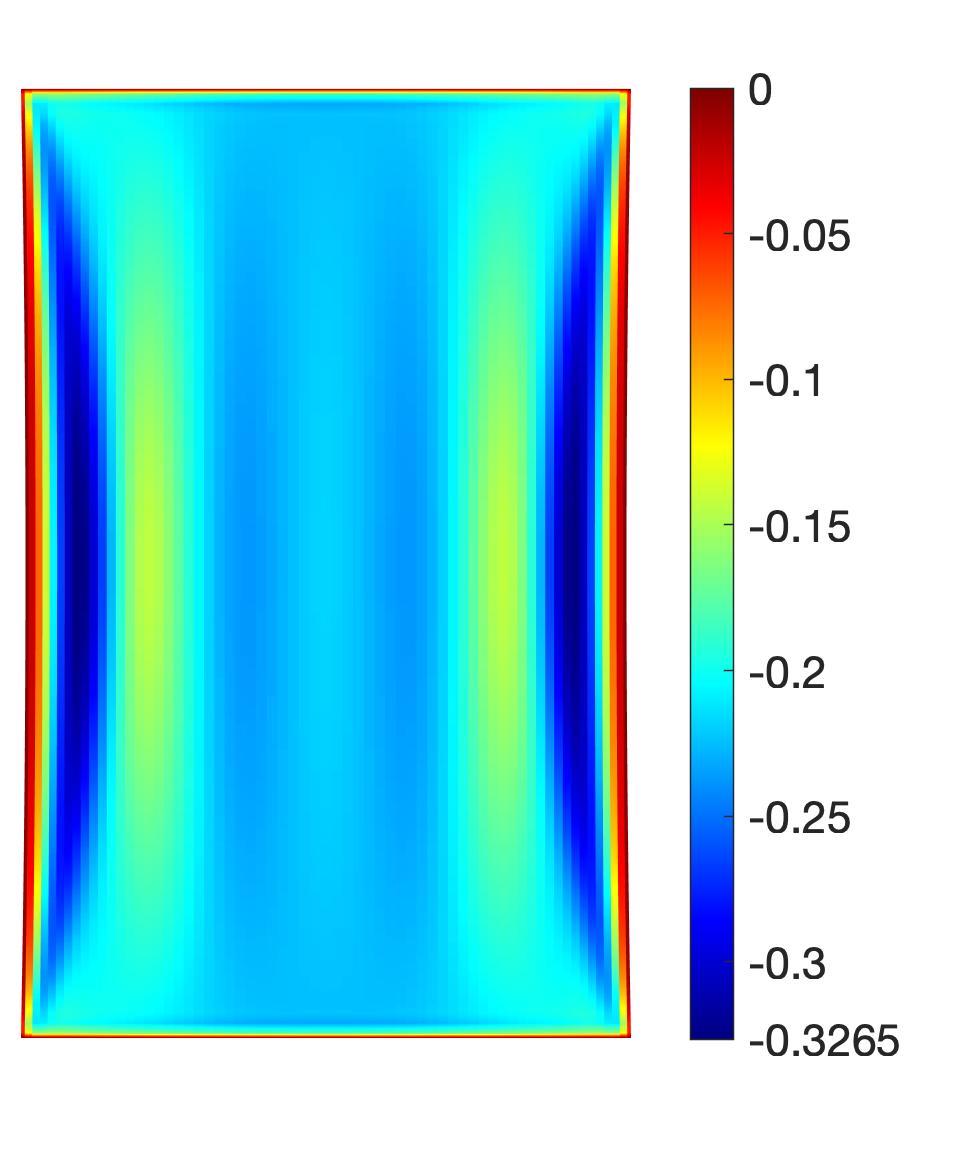}}
\put(4.1,-.5){\includegraphics[height=48mm]{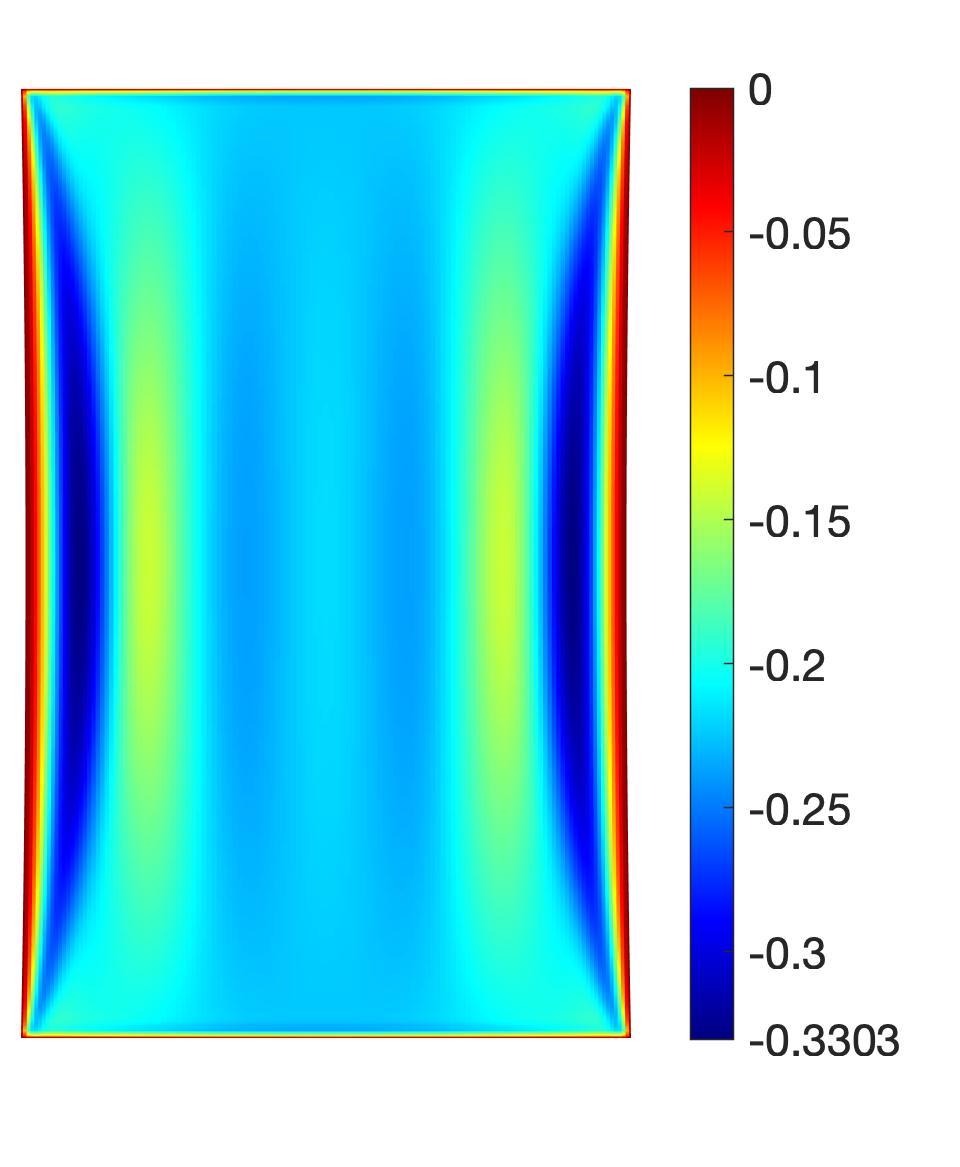}}
\end{picture}
\caption{Scordelis-Lo roof with $R/T=10^4$: Raw effective membrane stress~$\sig_{11}$ for the B2M2 discretization (top row) and B2M1 discretization (bottom row) for meshes $m = 16,\,32,\,64,\,128$ (left to right).
For B2M1, mesh $m=32$ already yields good results, while for B2M2, even mesh $m=128$ is still very bad.}
\label{f:SL4sig}
\end{center}
\end{figure}

The reference solution from above has the membrane stress minima $\sig_{11}^\mathrm{min} = -3488.3750$ for $R/T=100$ and $\sig_{11}^\mathrm{min} = -0.331359$ for $R/T=10^4$.
They are located at the center and close to the edge, respectively, as Fig.~\ref{f:SLue} shows.
The stress convergence of classical IGA (i.e.~the B$p$M$p$ discretization) and the B2M1 discretization to these reference values is shown in Fig.~\ref{f:SLse}.
Again, the behavior is consistent to the one seen in Fig.~\ref{f:CC_max}b:
B2M1 is much more accurate than B2M2, especially for coarse meshes and high $R/T$.
It converges from the beginning, while B$p$M$q$ first increases and thus exhibits very large stress error even when displacement errors are small.
For $R/T = 10^4$ B2M1 beats even quintic IGA up to $m=64$ ($n_\mathrm{dof}=2\cdot10^4$).
The initial increase of the stress in classical IGA is caused by severe membrane stress oscillations that only vanish for very fine meshes.
Those are shown in the top row of Figs.~\ref{f:SL2sig} and \ref{f:SL4sig} for $R/T=100$ and $R/T=10^4$, respectively.
In contrast, the B2M1 discretization does not show any stress oscillations and provides very good coarse mesh accuracy.
This is seen in the bottom row of Figs.~\ref{f:SL2sig} and \ref{f:SL4sig}.

\subsubsection{Influence of mesh distortion}

\begin{figure}[h]
\begin{center} \unitlength1cm
\begin{picture}(0,4.1)
\put(-7.6,-.25){\includegraphics[height=48mm]{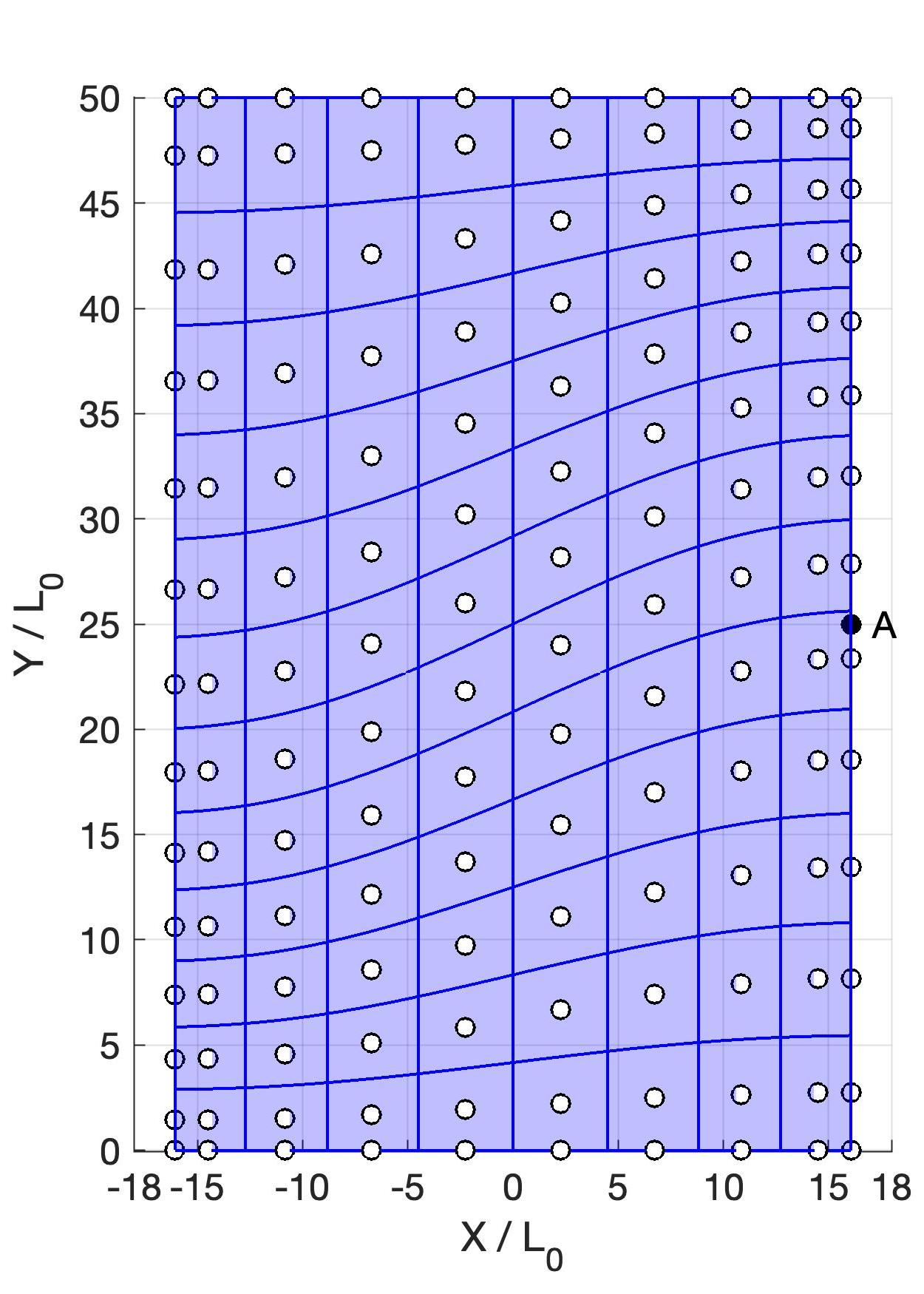}}
\put(-3.85,-.25){\includegraphics[height=48mm]{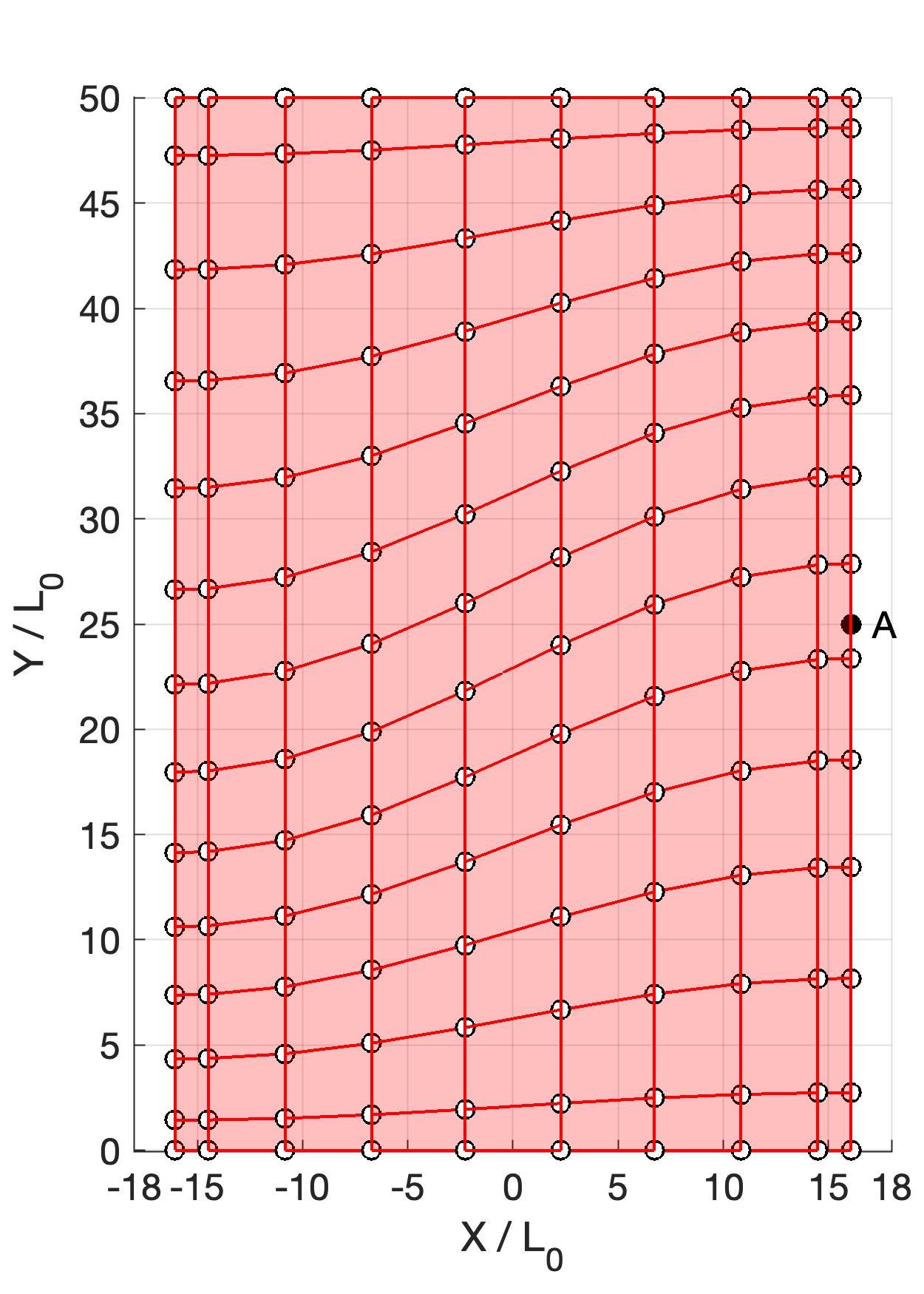}}
\put(0.65,-.25){\includegraphics[height=48mm]{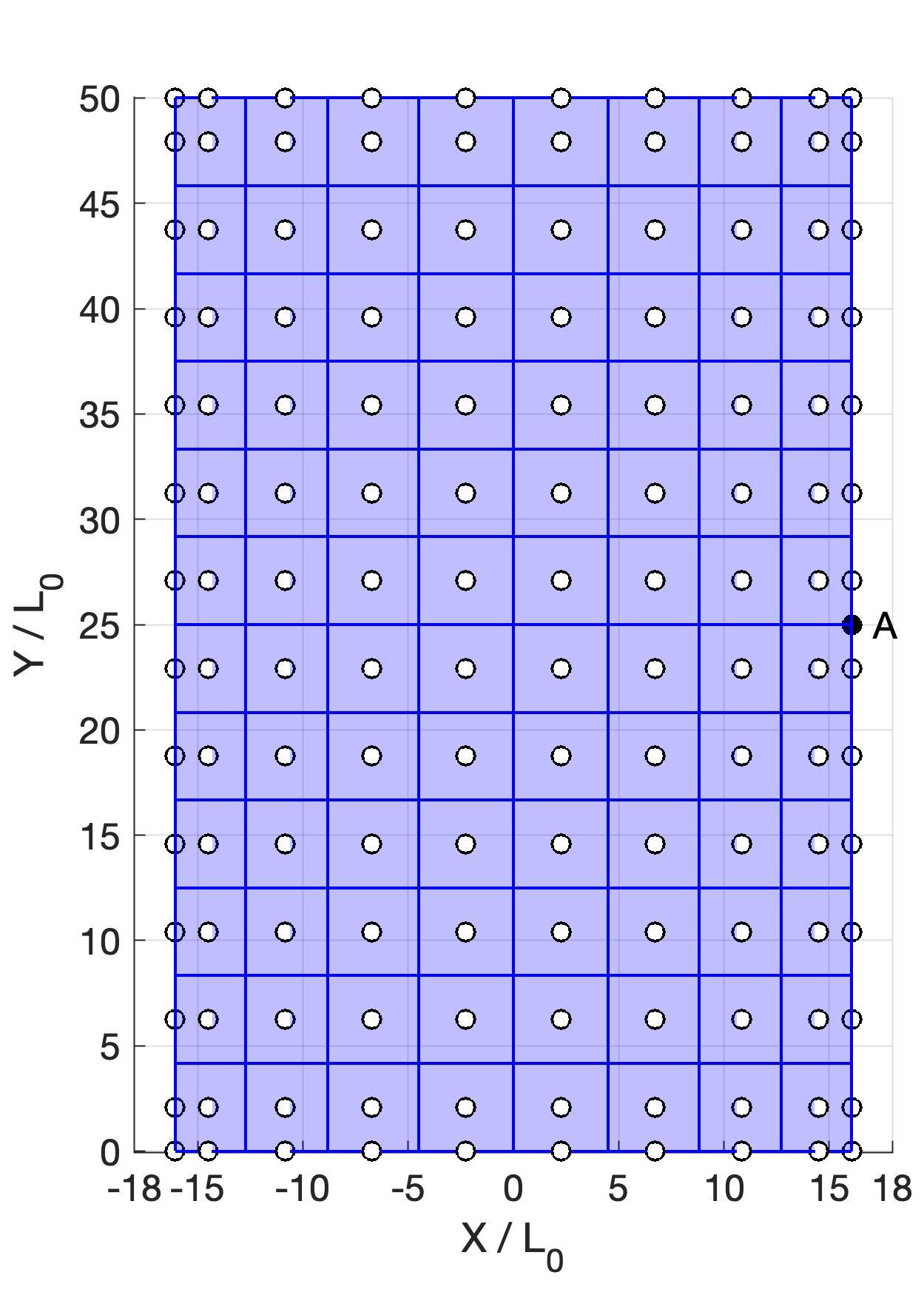}}
\put(4.4,-.25){\includegraphics[height=48mm]{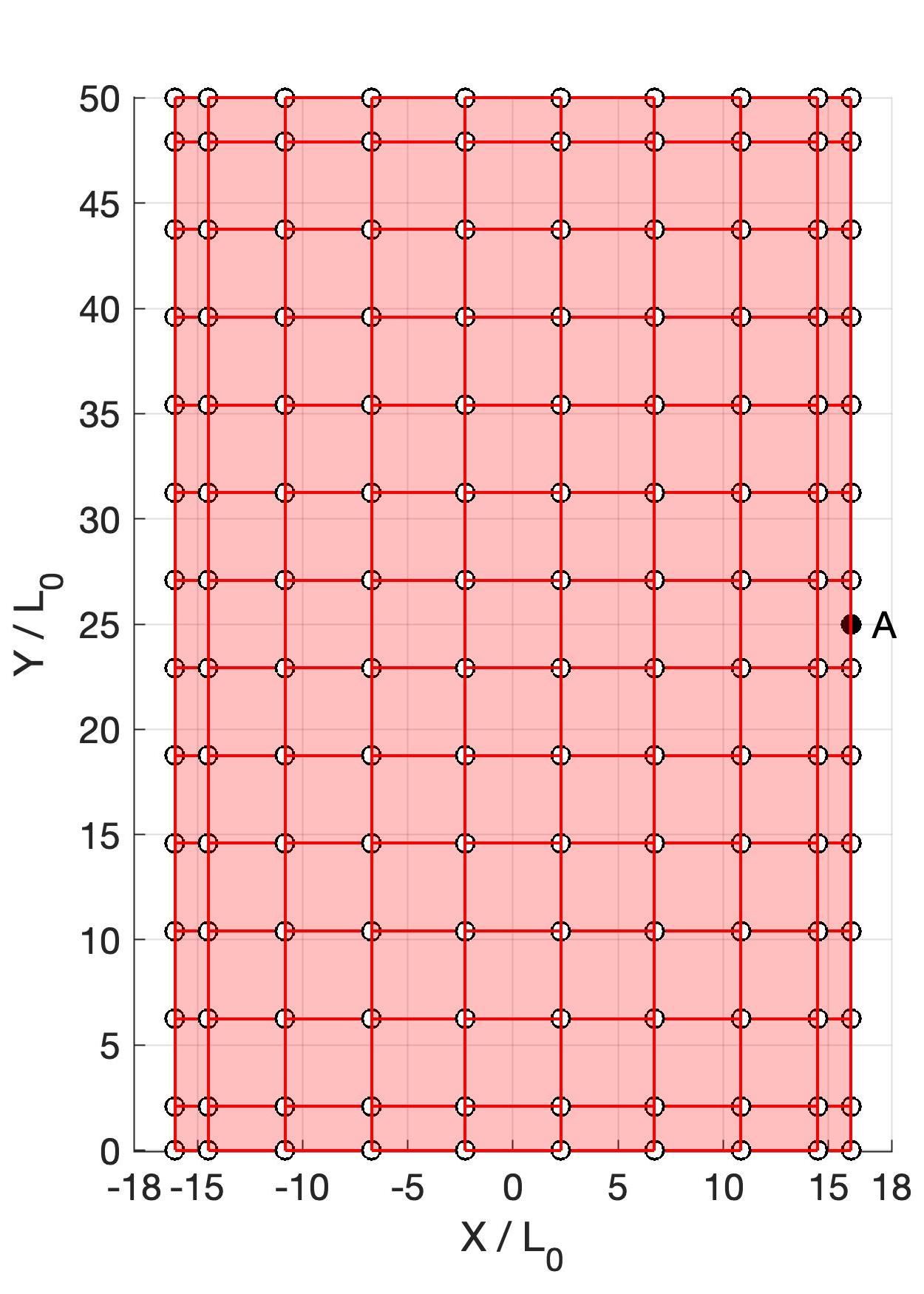}}
\put(-7.95,0.0){\footnotesize (a)}
\put(0.3,0.0){\footnotesize (b)}
\end{picture}
\caption{Scordelis-Lo roof with skew mesh: 
(a) Distorted and (b) original B2 and M1 meshes ($m=8$).
\vspace{-7.5mm}}
\label{f:SLskewmesh}
\end{center}
\end{figure}

Next, the influence of mesh distortion is examined.
Therefore, a skew mesh is created by modifying the $Y$-position of the control points (see Fig.~\ref{f:SLskewmesh}a) according to 
\eqb{l}
Y^\mathrm{skew}_I = \ds Y_I
+ 5 \sin\bigg(\frac{\pi X_I}{2X_{n_\mathrm{no}}}\bigg) \sin\bigg(\frac{\pi Y_I}{L}\bigg)\,,\quad
X_{n_\mathrm{no}} = R \sin 40^\circ\,.
\label{e:skew}\eqe
\begin{figure}[h]
\begin{center} \unitlength1cm
\begin{picture}(0,11.7)
\put(-8,5.9){\includegraphics[height=58mm]{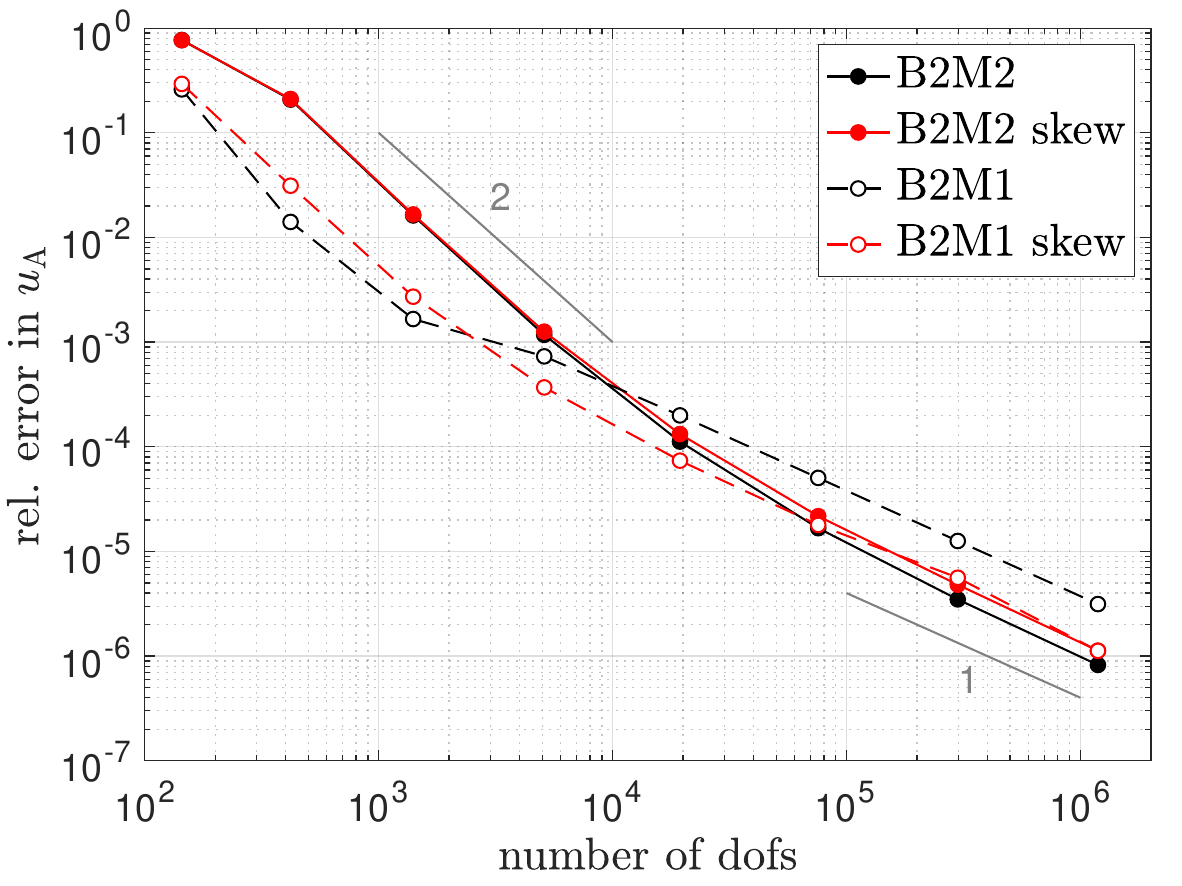}}
\put(0.2,5.9){\includegraphics[height=58mm]{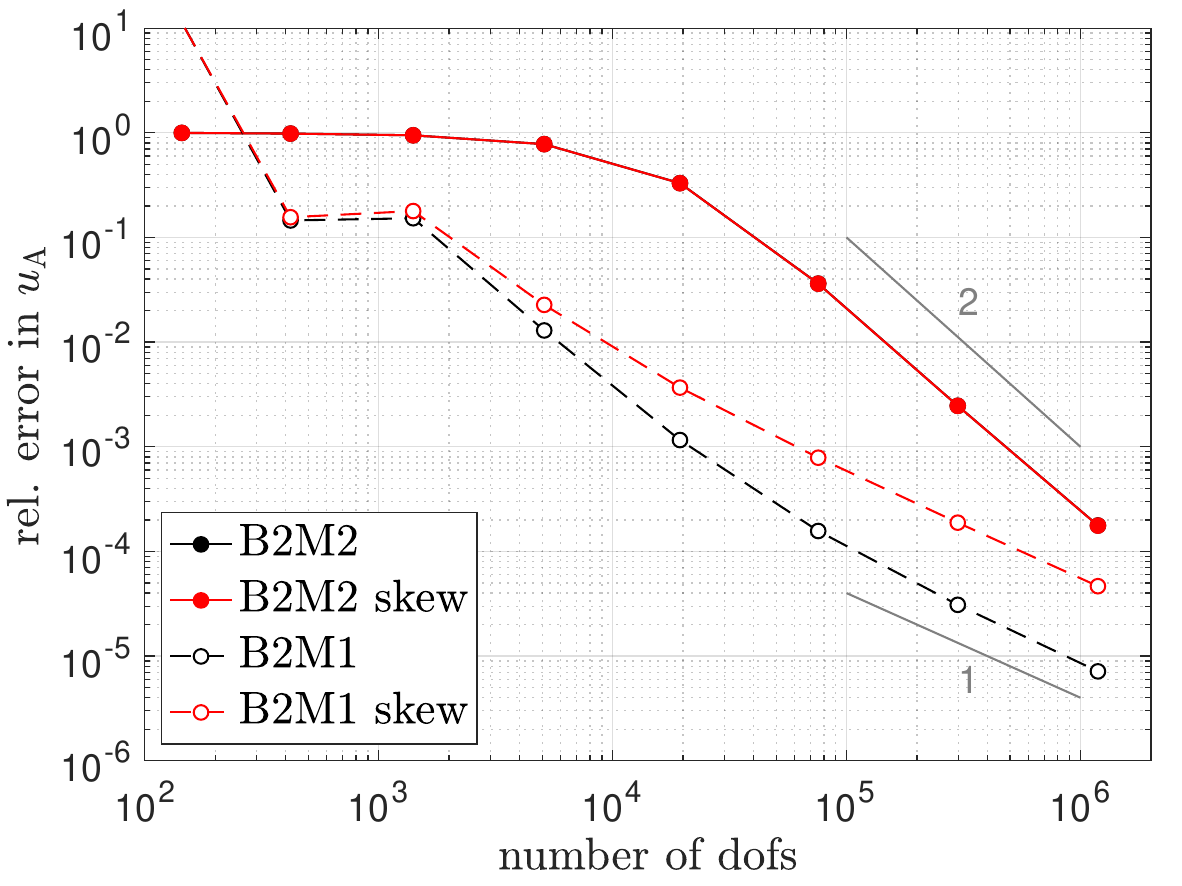}}
\put(-8,-.1){\includegraphics[height=58mm]{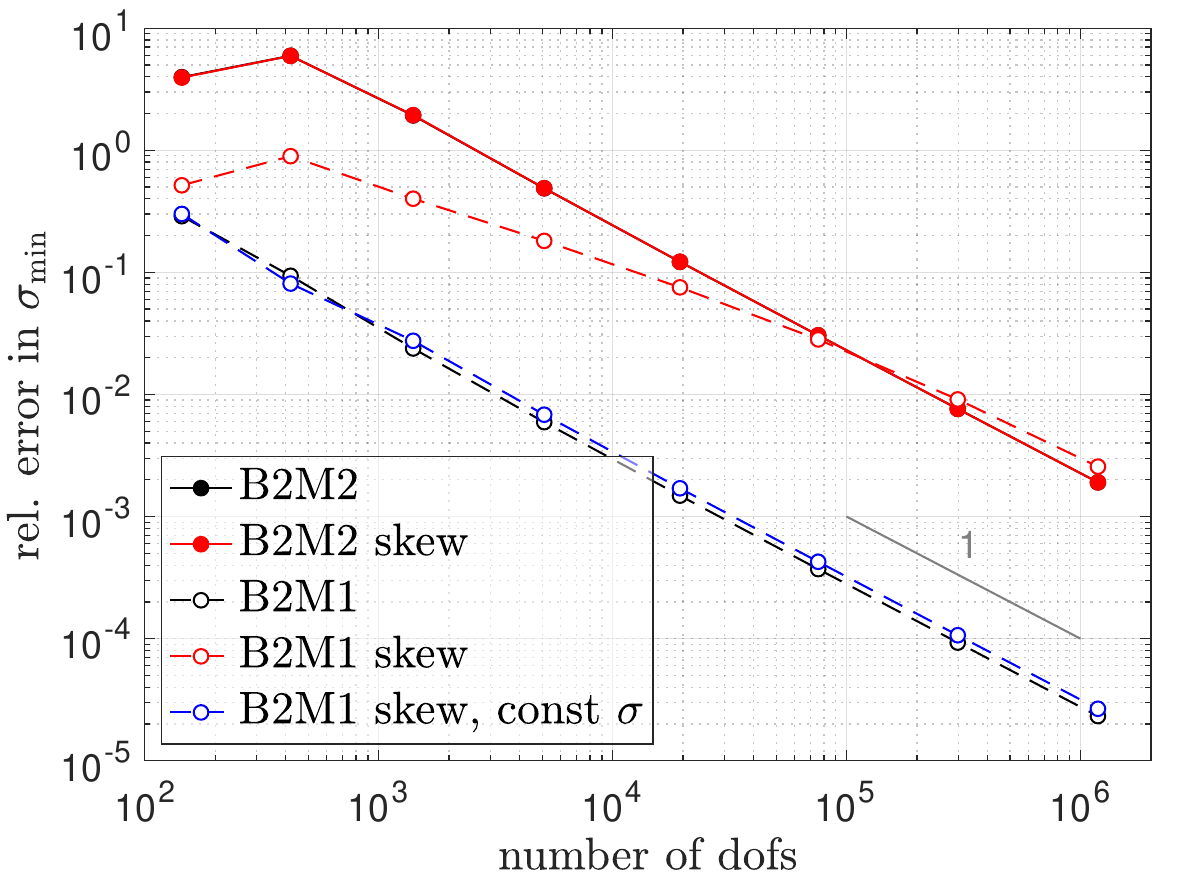}}
\put(0.2,-.1){\includegraphics[height=58mm]{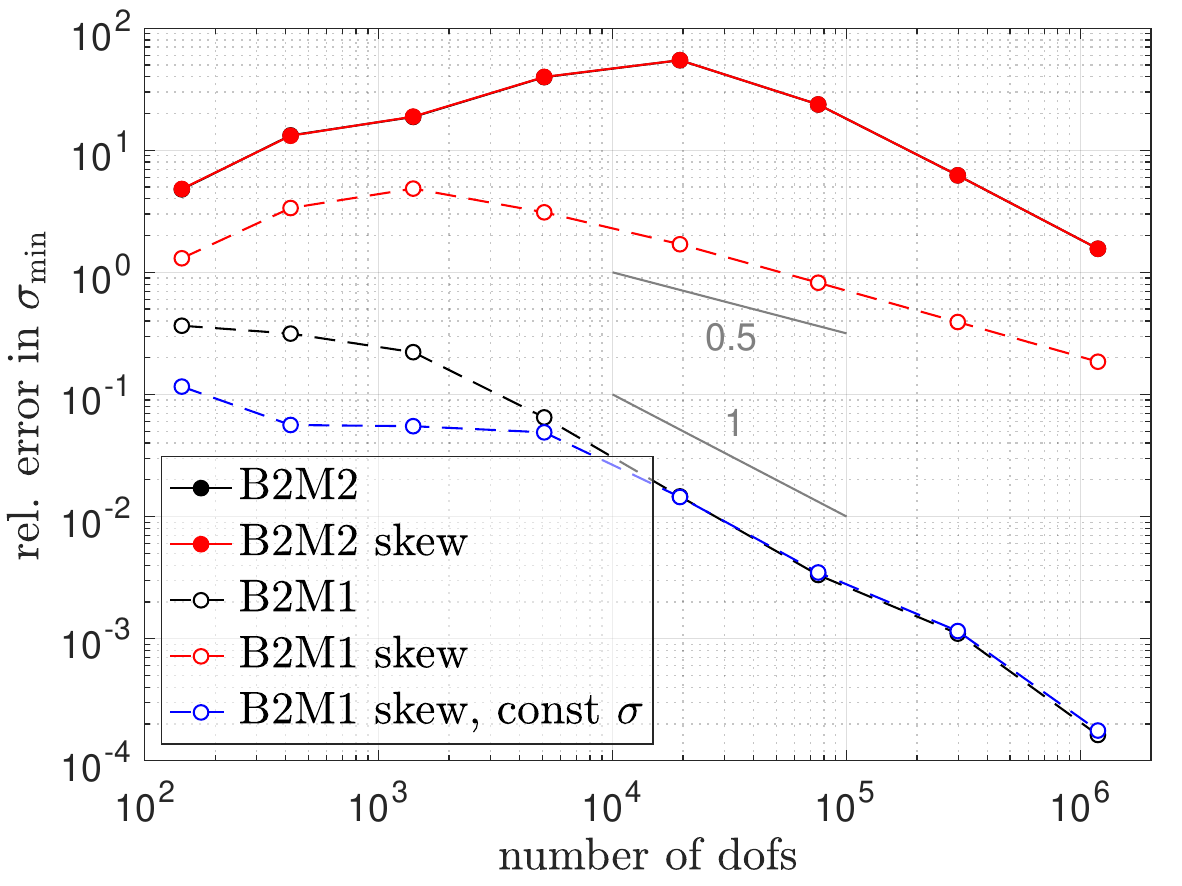}}
\put(-7.95,6.0){\footnotesize (a)}
\put(0.15,6.0){\footnotesize (b)}
\put(-7.95,0.0){\footnotesize (c)}
\put(0.15,0.0){\footnotesize (d)}
\end{picture}
\caption{Scordelis-Lo roof with skew mesh: Convergence behavior of $u_\mathrm{A}$ (top row) and $\sigma_\mathrm{min}$ (bottom row) for $R/T=10^2$ (left side) and $R/T=10^4$ (right side), using $m=4, 8, 16, ..., 512$. 
Even for this mesh distortion, $u_\mathrm{A}$ is more accurate with B2M1 than with B2M2 -- even more accurate than regular B2M1 in some cases. 
The stress accuracy of skew B2M1 is not as good as regular B2M1, unless the stresses are taken to be constant within each element (dashed blue line).
\vspace{-3mm}}
\label{f:SLskewe}
\end{center}
\end{figure}

Fig.~\ref{f:SLskewe} shows the influence of this mesh distortion on $u_\mathrm{A}$ and $\sig_\mathrm{min}$ considering $R/T=10^2$ and $R/T=10^4$.
The figure shows that skew meshes strongly affect the accuracy and convergence rates for B2M1 but not B2M2.
While the displacements for skew B2M1 are still more accurate than those of B2M2 for coarse meshes and/or large $R/T$, the accuracy gain is less than for regular meshes:
For $R/T = 10^4$ and $m=128$ ($n_\mathrm{dof}=75660$) for instance, regular B2M1 is 231 times more accurate than B2M2, while skew B2M1 is still 46 times more accurate than B2M2.
The stresses of skew B2M1 are more oscillatory than those of regular B2M1, leading to larger stress errors. 
A remedy for these stress errors in B2M1, is to only look at the average stress (the constant part obtained at the element center ($\xi^\alpha=0$)) during post processing.
This stress converges similarly well as the stresses for regular B2M1, as the dashed blue and dashed black lines in Figs.~\ref{f:SLskewe}c and d show.
This is also seen in Fig.~\ref{f:SLskew2}:
\begin{figure}[h]
\begin{center} \unitlength1cm
\begin{picture}(0,8.6)
\put(-7.9,4){\includegraphics[height=48mm]{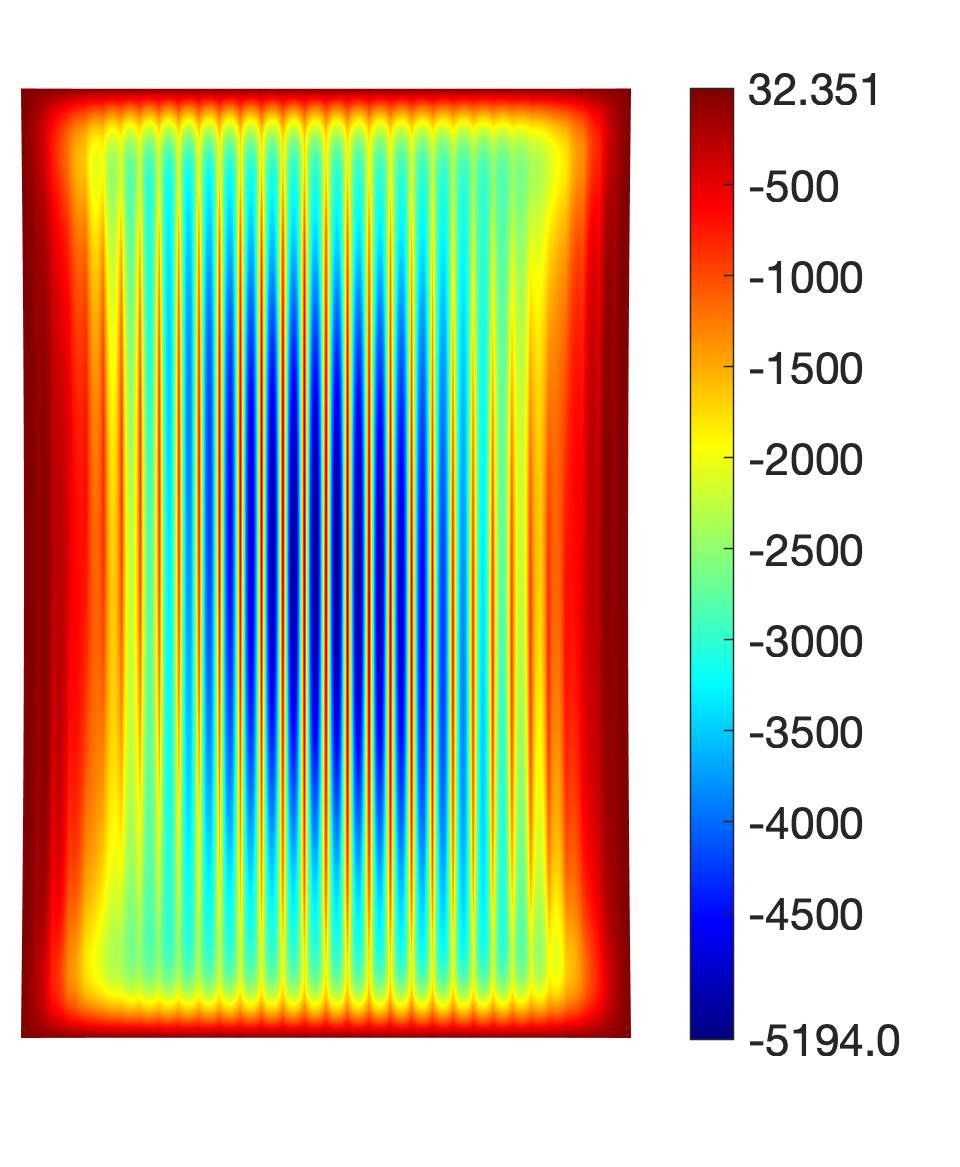}}
\put(-3.9,4){\includegraphics[height=48mm]{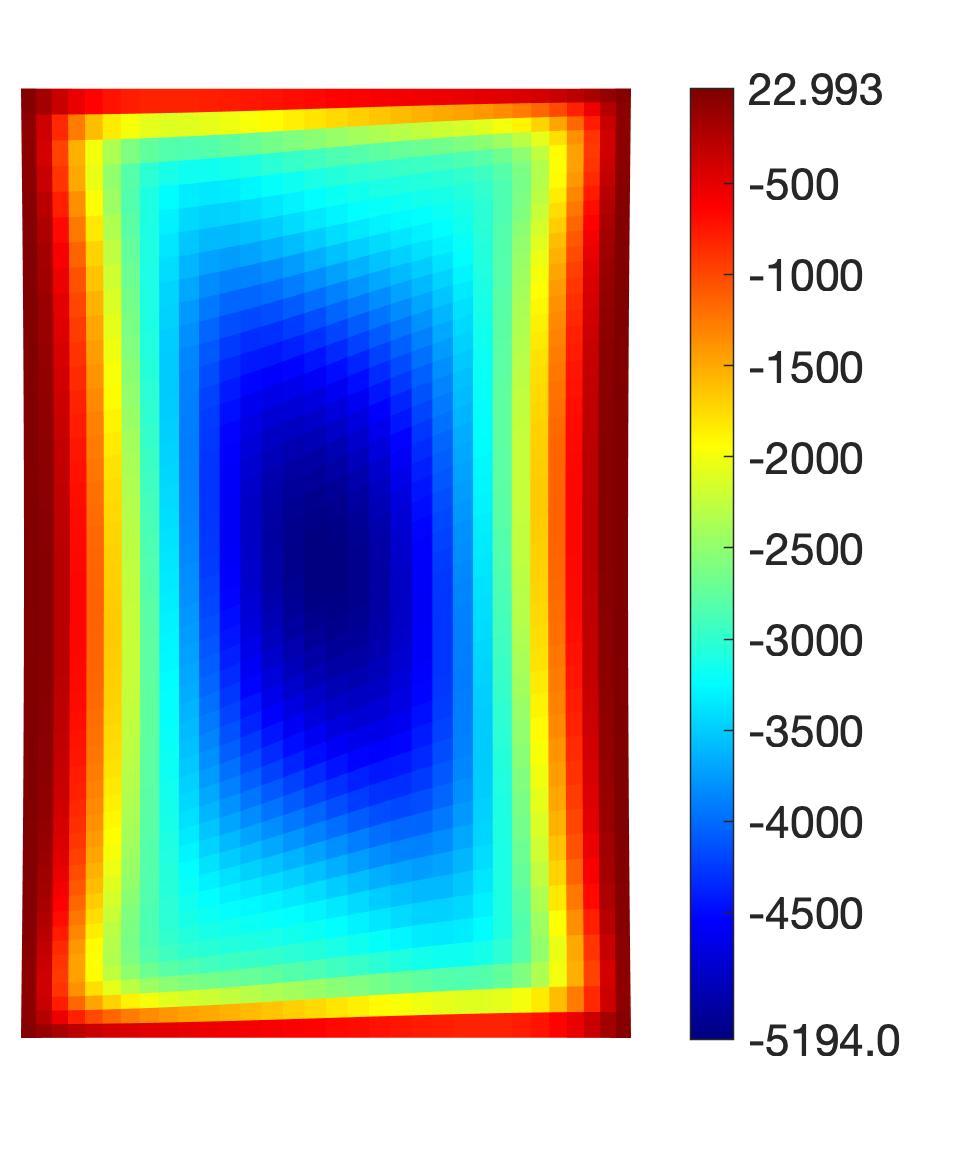}}
\put(0.1,4){\includegraphics[height=48mm]{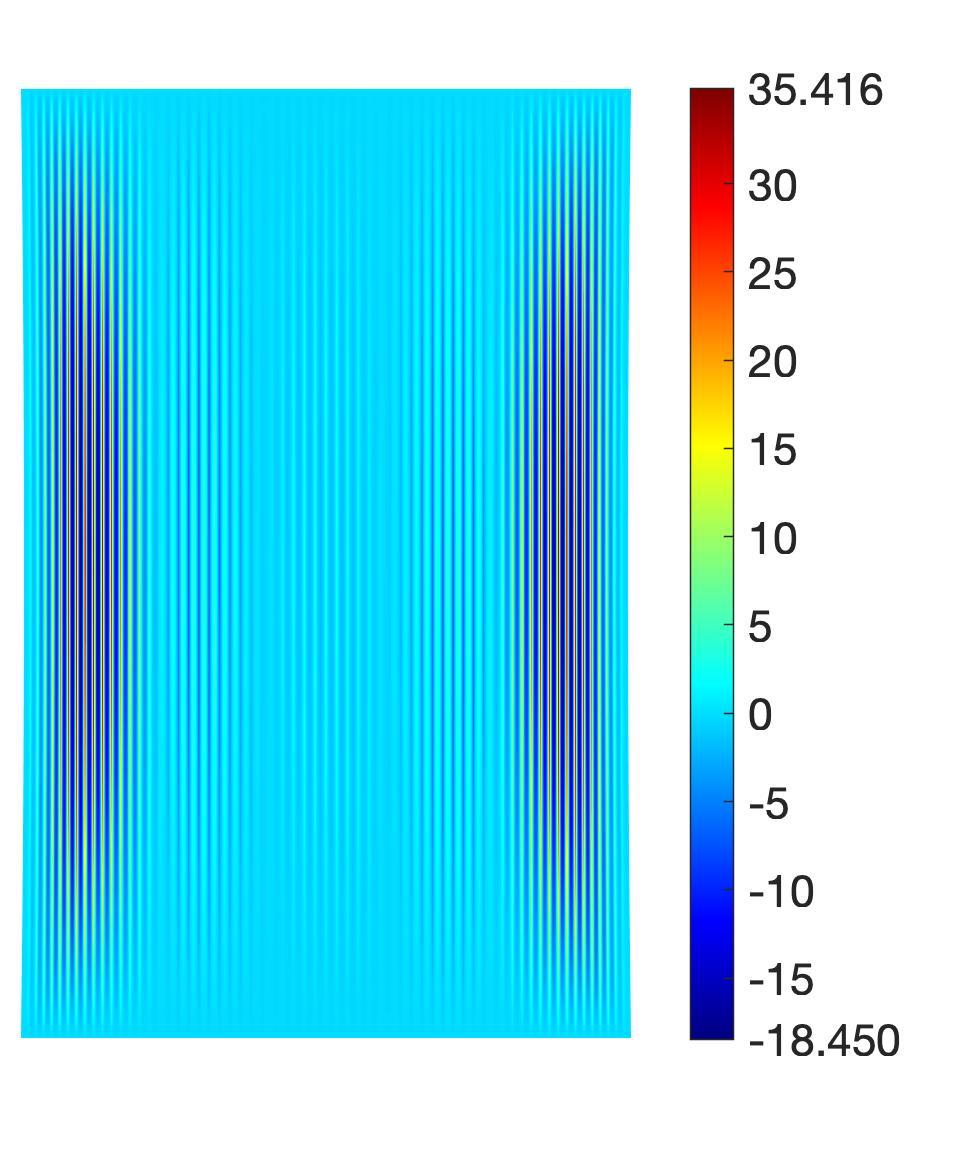}}
\put(4.1,4){\includegraphics[height=48mm]{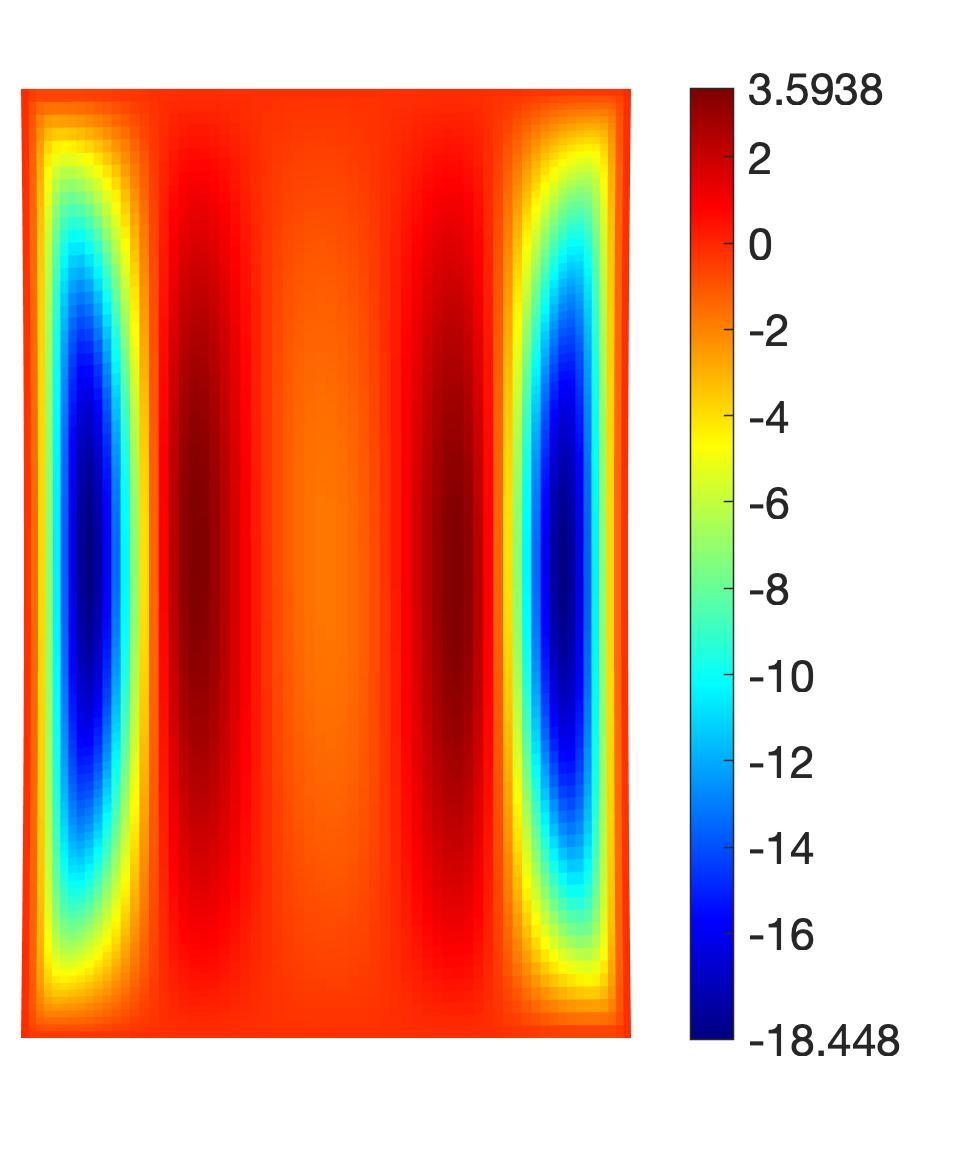}}
\put(-7.9,-.5){\includegraphics[height=48mm]{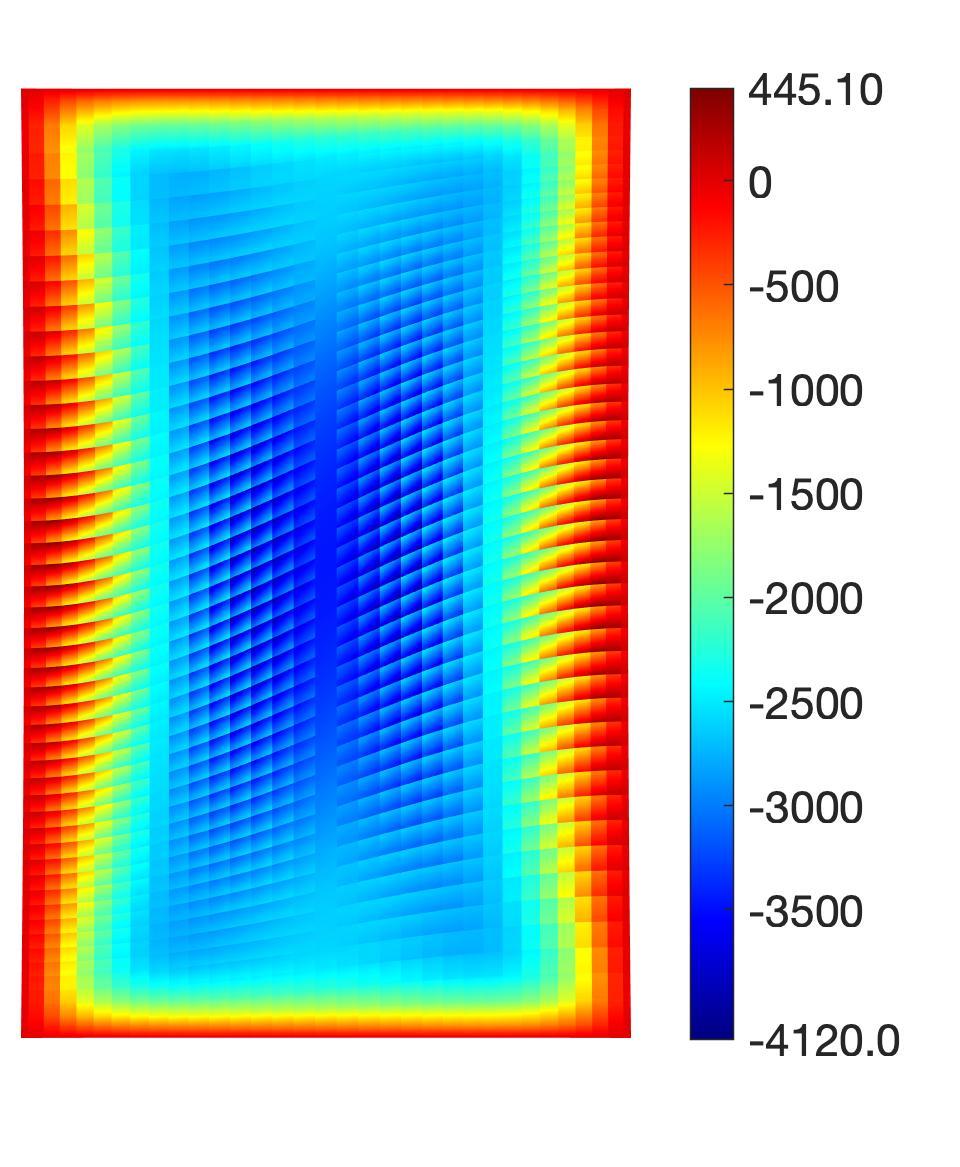}}
\put(-3.9,-.5){\includegraphics[height=48mm]{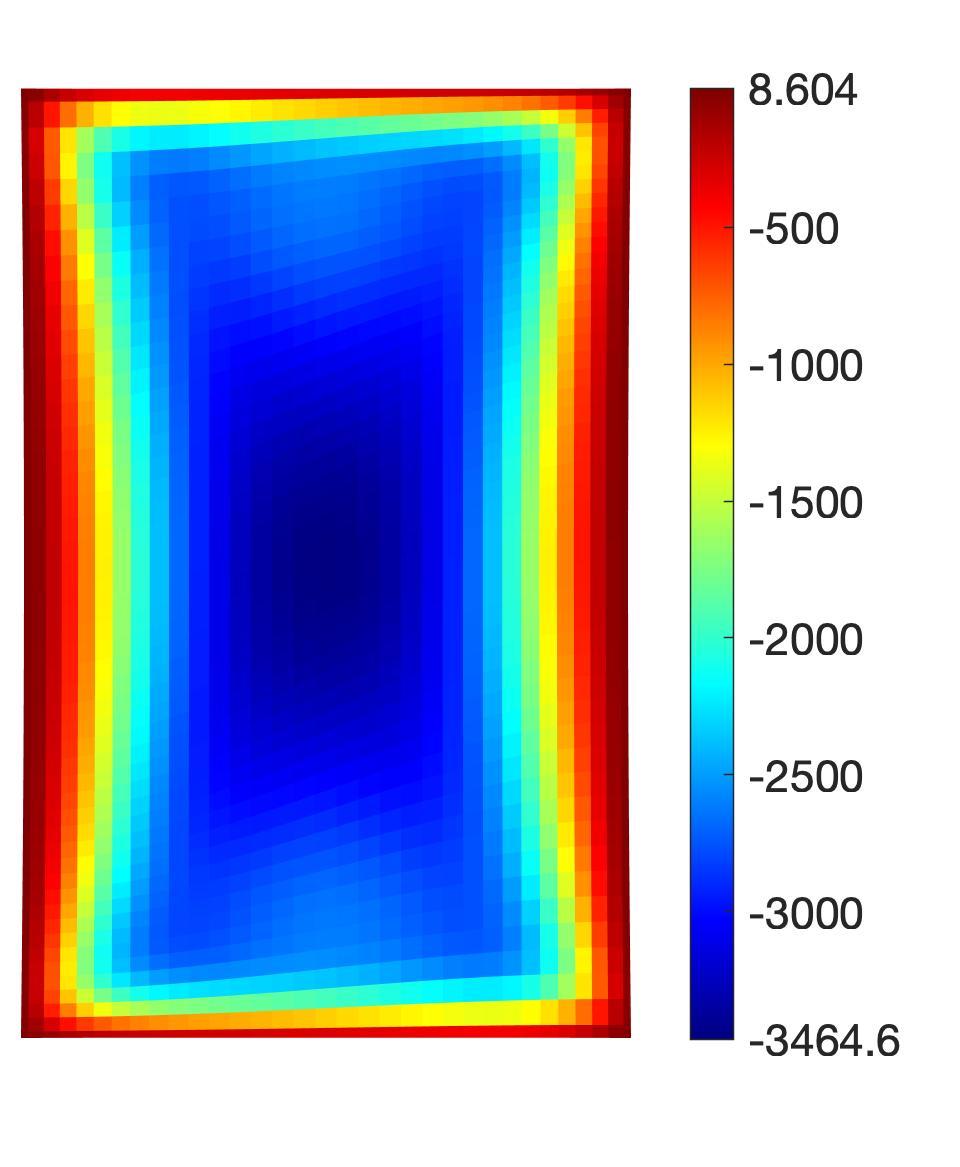}}
\put(0.1,-.5){\includegraphics[height=48mm]{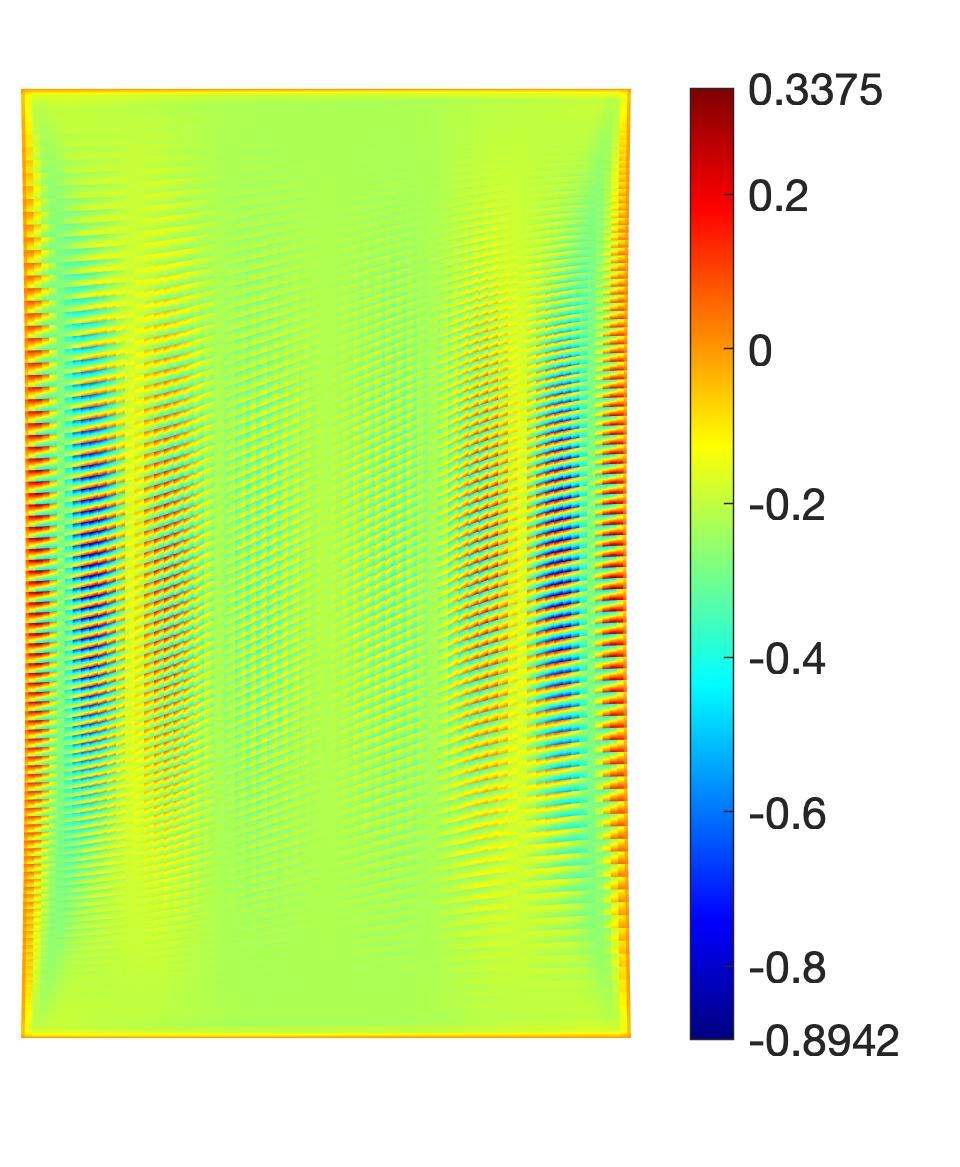}}
\put(4.1,-.5){\includegraphics[height=48mm]{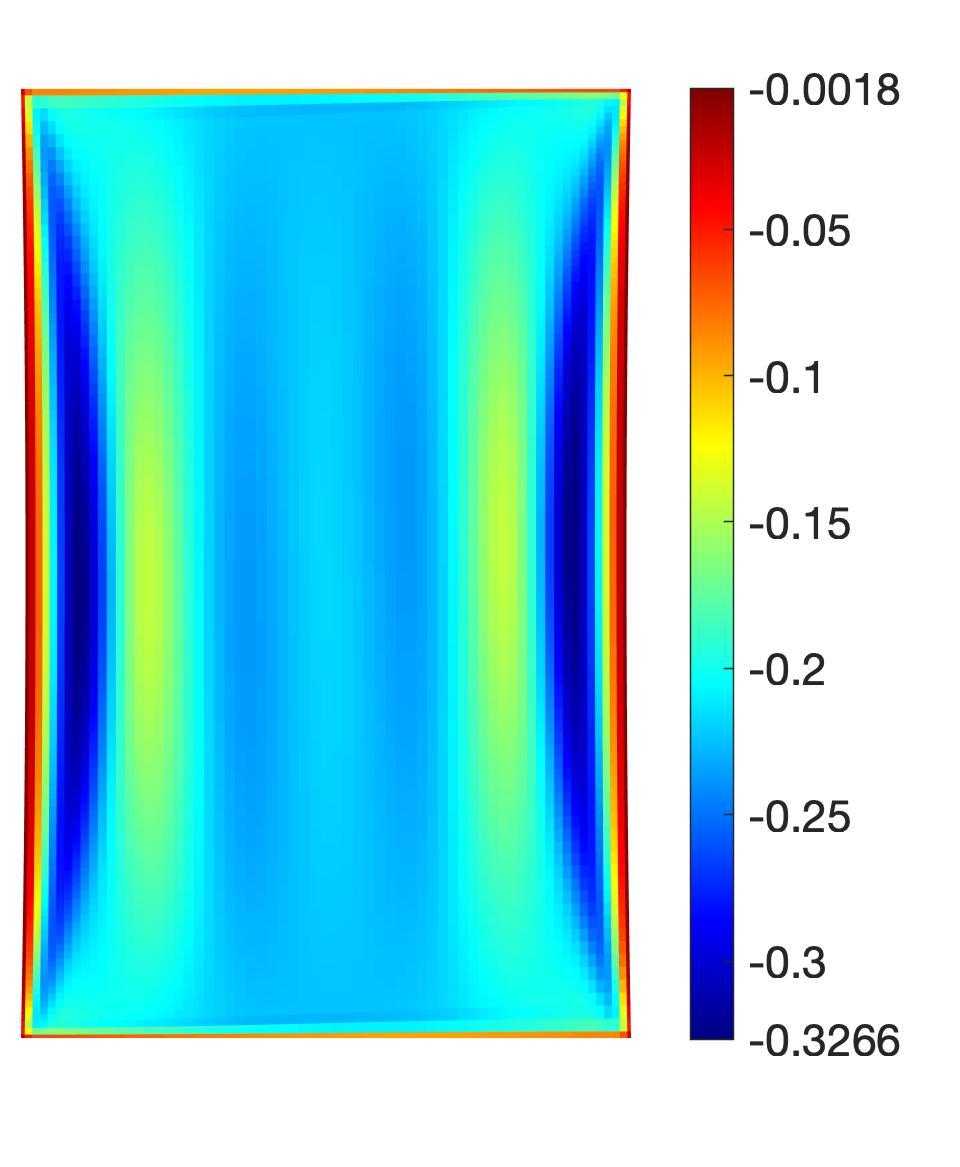}}
\end{picture}
\caption{Scordelis-Lo roof with skew mesh: Raw effective membrane stress~$\sig_{11}$ for the B2M2 discretization (top row) and B2M1 discretization (bottom row) for skew meshes. Left to right: 1.~raw stress $\sigma_{11}$ for $R/T = 100$ \& $m=32$, 2.~raw constant stress $\sigma^0_{11}$ for $R/T = 100$ \& $m=32$, 3.~raw stress $\sigma_{11}$ for $R/T = 10^4$ \& $m=64$, 4.~raw constant stress $\sigma^0_{11}$ for $R/T = 10^4$ \& $m=64$. 
Only the constant stresses $\sigma^0_{11}$ for the B2M1 discretization are accurate, while all other stresses, in particular all B2M2 ones, are highly inaccurate. \\[-2mm]}
\label{f:SLskew2}
\end{center}
\end{figure}
Using constant stress post processing leads to large accuracy gains in B2M1 but not in B2M2. 
Thus one can obtain accurate stresses with B2M1 even for distorted meshes and large slenderness ratios.

In all Scordelis-Lo roof results no force redistribution has been applied, since redistribution only affects the domain boundaries according to App.~\ref{s:W}. 
However, this is for regular meshes.
Force redistribution for skew mesh can be expected to be different, which should be investigated in future work.
Force redistribution may thus allow for further accuracy gains in distorted B2M1 meshes.

\subsubsection{Efficiency gains of B2M1}\label{s:SLa}

\begin{figure}[h]
\begin{center} \unitlength1cm
\begin{picture}(0,5.7)
\put(-8,-.1){\includegraphics[height=58mm]{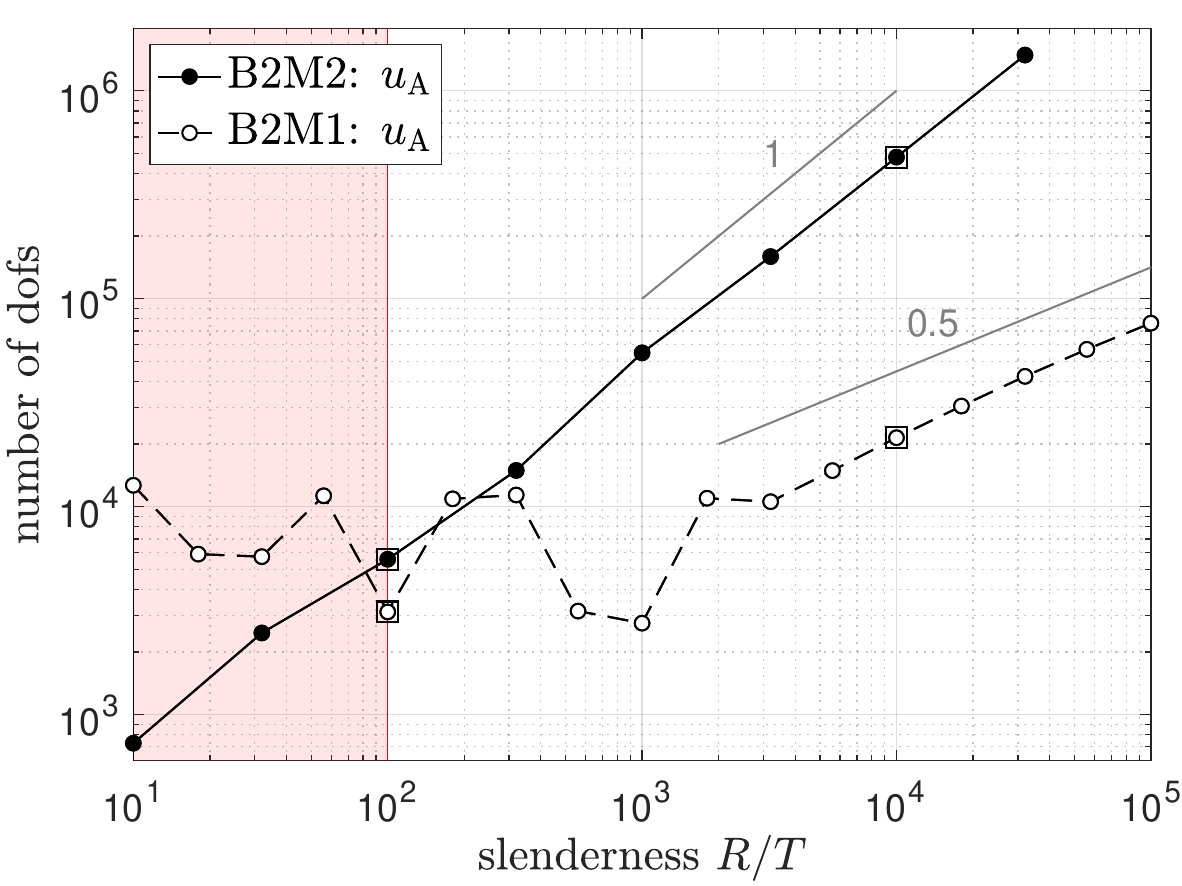}}
\put(0.2,-.1){\includegraphics[height=58mm]{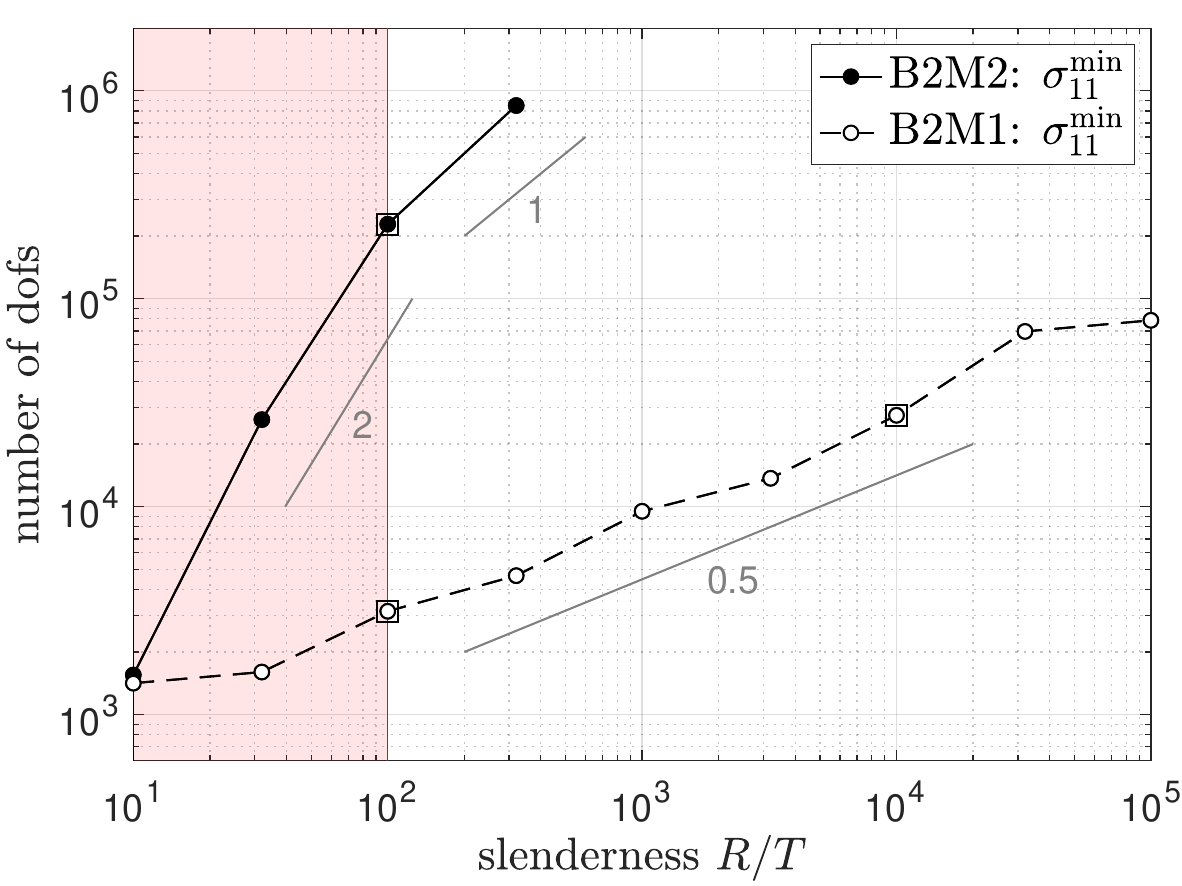}}
\put(-7.95,0.0){\footnotesize (a)}
\put(0.15,0.0){\footnotesize (b)}
\end{picture}
\caption{Efficiency gain of B2M1 vs.~B2M2 for the Scordelis-Lo roof. 
The diagrams show the minimum number of dofs, as a function of slenderness $R/T$, required to achieve a given accuracy in (a) $u_\mathrm{A}$ and (b) $\sig^\mathrm{min}_{11}$ -- here for the relative error falling below (a) $10^{-3}$ and (b) $10^{-2}$.
For B2M2 the dofs increase proportionally to $R/T$, while for B2M1 the dofs increase at most with $\sqrt{R/T}$.
The squares show the cases discussed in Sec.~\ref{s:ScoLo}. 
For $R/T = 10^4$ and $u_\mathrm{A}$ (a), for instance, B2M2 requires about 22 times more dofs than B2M1.
The convergence rates for B2M1 can be non-const and non-monotonic, especially for coarse meshes, leading to the jumps seen in (a).\\[-2mm]}
\label{f:SLac}
\end{center}
\end{figure}

In order to assess the performance of B2M1 vs.~B2M2, their efficiency is examined following the protocol used for the curved cantilever strip in Sec.~\ref{s:CCeff}.
The corresponding accuracy diagrams for $u_\mathrm{A}$ and $\sig^\mathrm{min}_{11}$ are shown Fig.~\ref{f:SLac}.
In case of $u_\mathrm{A}$ (left figure), the required dofs increase linearly with $R/T$ for B2M2, while for B2M1 the required dofs remain at a constant level up to $R/T\approx 3000$, and then increase with $\sqrt{R/T}$.
B2M2 is only more efficient in the regime where the Kirchhoff-Love assumptions break down ($R/T<100$, marked in pink).
In case of $\sig^\mathrm{min}_{11}$ (right figure), the required dofs increase at least linearly and become very large for B2M2, while for B2M1 the required dofs remain low and only increase with $\sqrt{R/T}$.  
The figure demonstrate that membrane locking is greatly alleviated by the B2M1 discretization.
In the regime where the Kirchhoff-Love assumptions hold ($R/T>100$, beyond the pink region), B2M1 is always more efficient than B2M2.

\subsubsection{Influence of large deformations}

\begin{figure}[h]
\begin{center} \unitlength1cm
\begin{picture}(0,4.1)
\put(-7.9,-.5){\includegraphics[height=48mm]{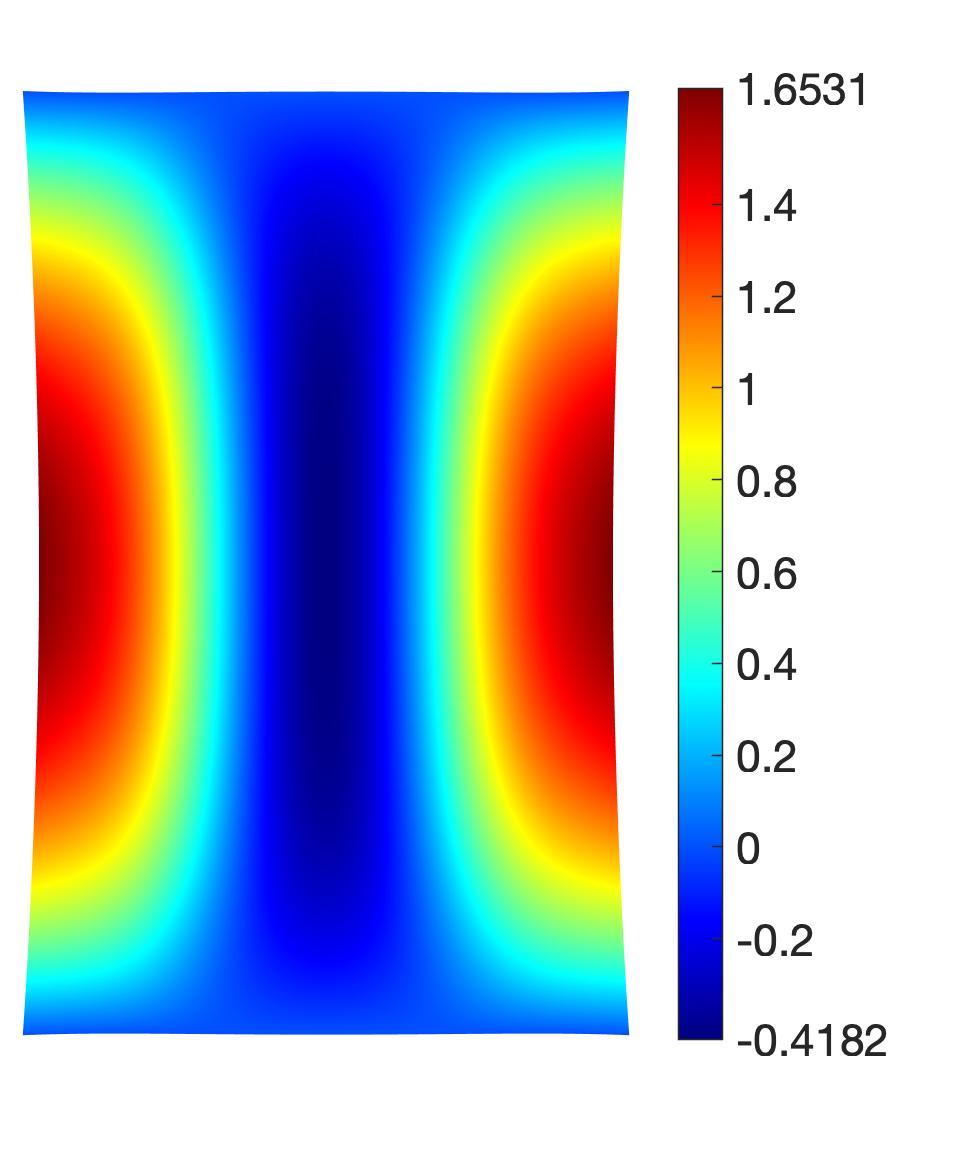}}
\put(-3.9,-.5){\includegraphics[height=48mm]{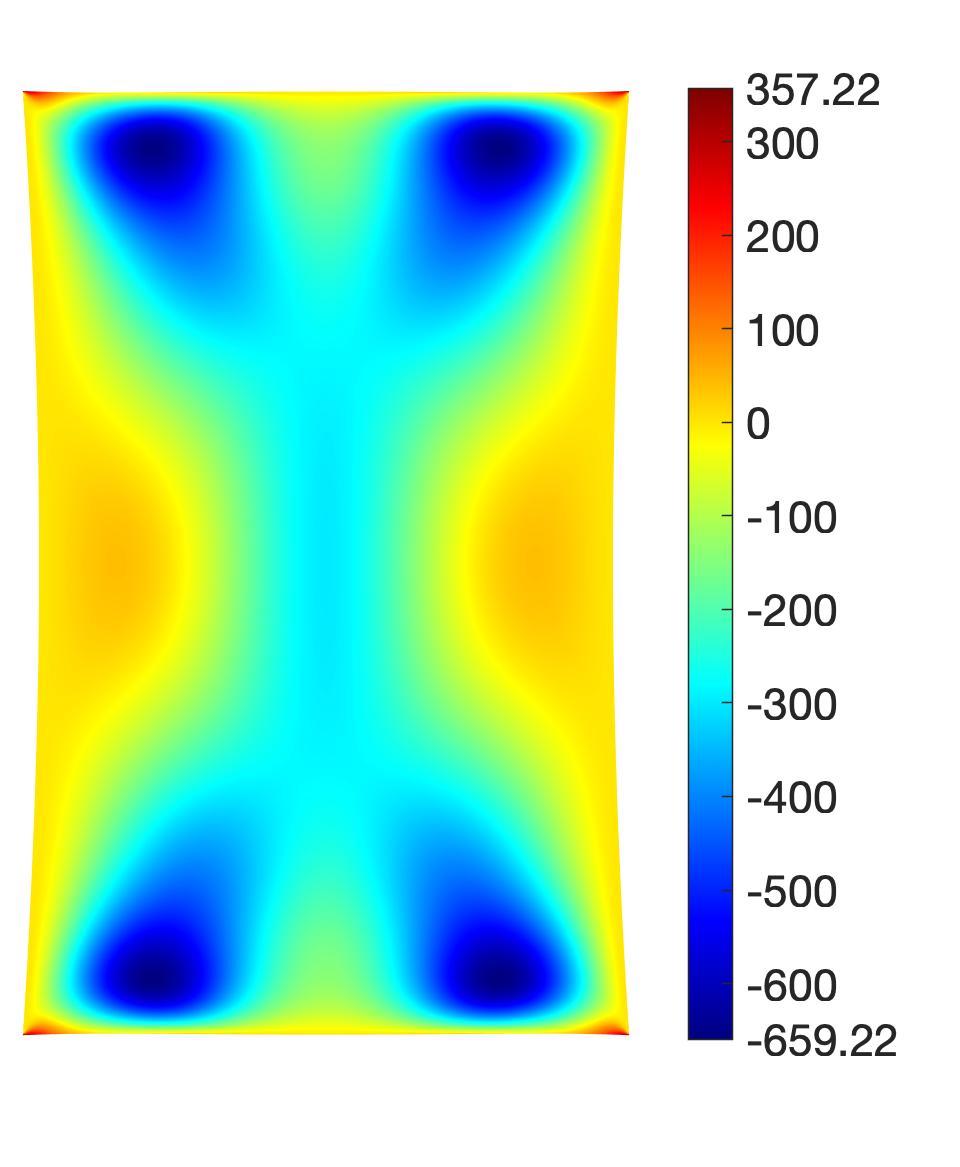}}
\put(0.1,-.5){\includegraphics[height=48mm]{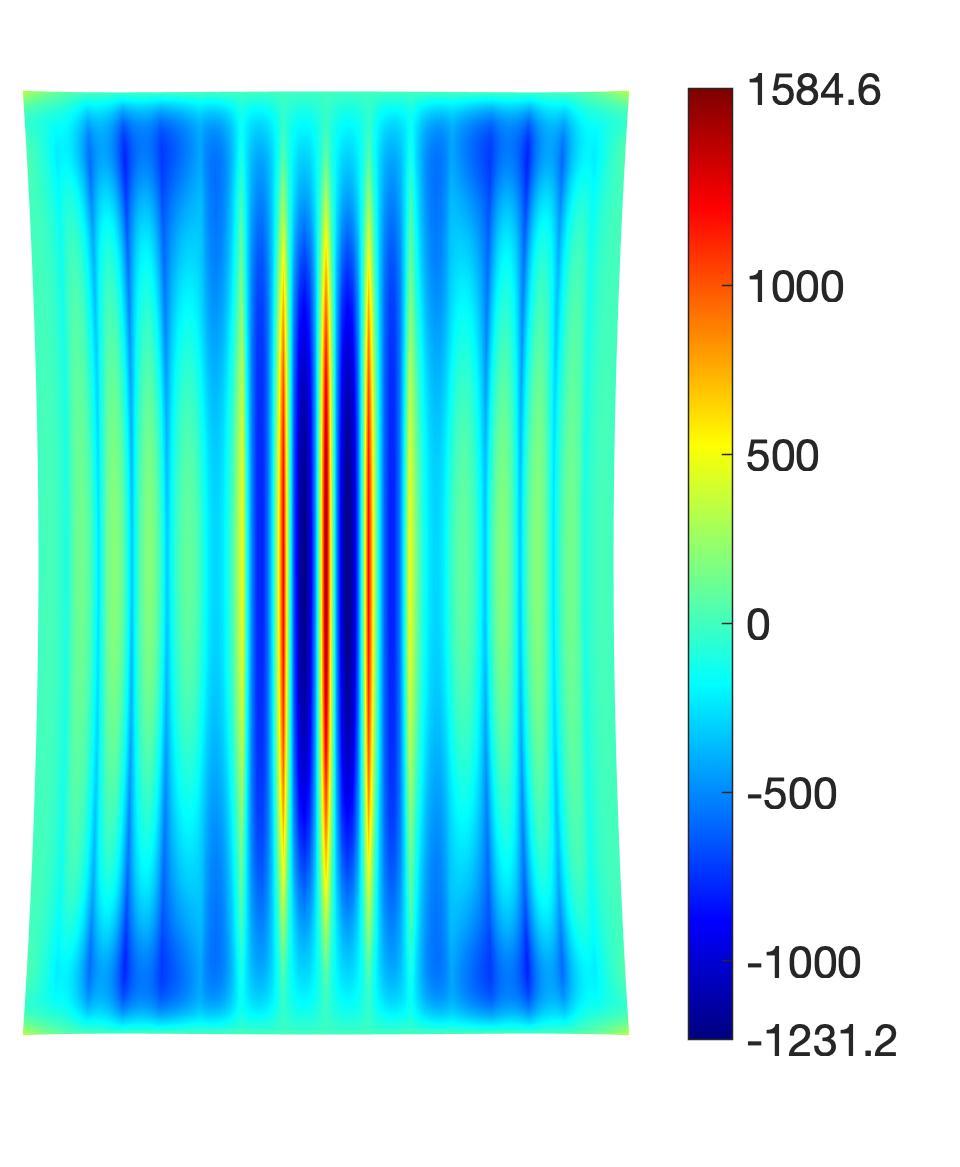}}
\put(4.1,-.5){\includegraphics[height=48mm]{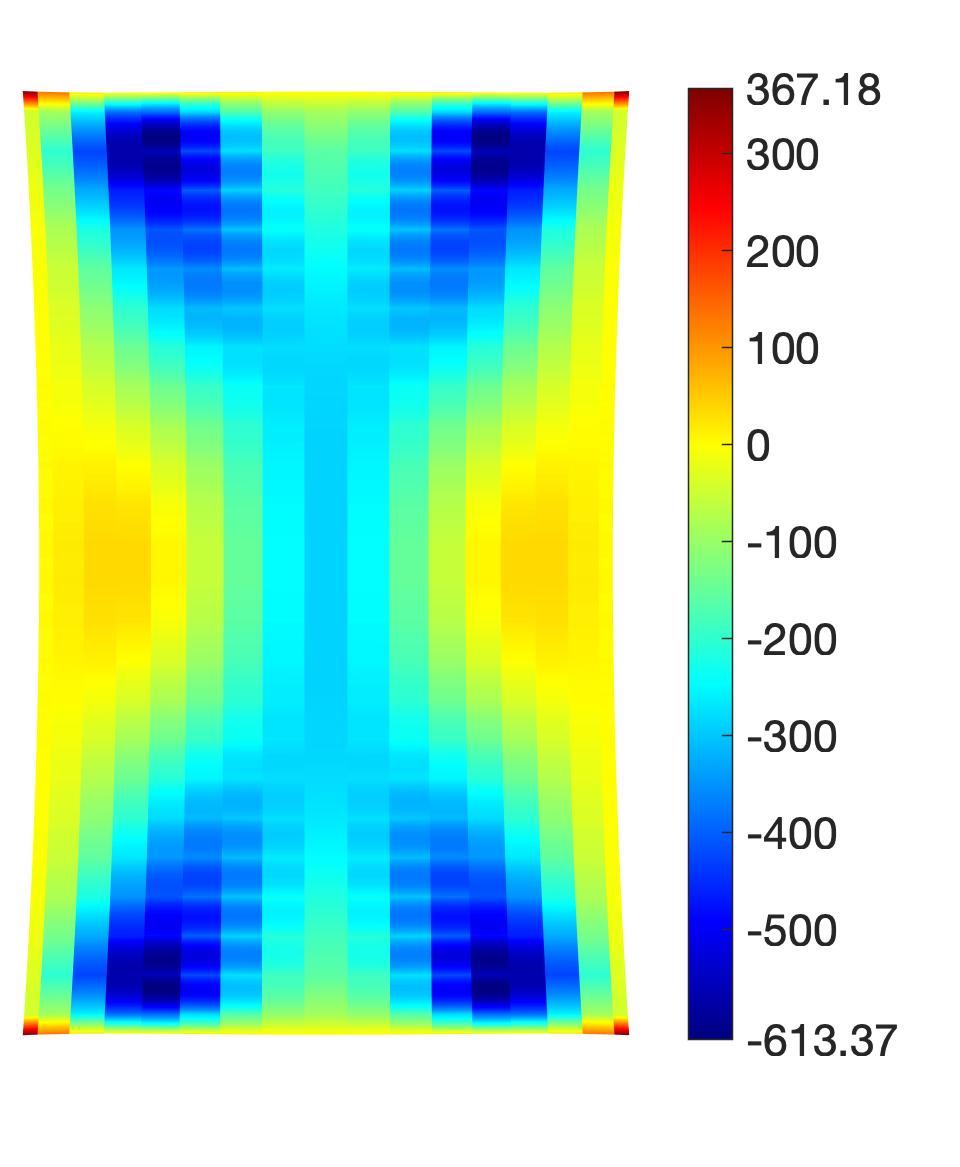}}
\end{picture}
\caption{Scordelis-Lo roof with large deformations for $R/T = 100$:
Reference result for displacement $u_3$ and raw stress $\sigma^1_1$ (left two figures; B5M5, $m=64$), $\sigma^1_1$ for B2M2 (third figure, $m=16$) and $\sigma^1_1$ for B2M1 (fourth figure, $m=16$). 
As for small deformations, B2M2's stresses are oscillatory and inaccurate, while the B2M1 discretization gives much more accurate results.\\[-2mm]}
\label{f:SLlarge}
\end{center}
\end{figure}

\begin{figure}[h]
\begin{center} \unitlength1cm
\begin{picture}(0,5.5)
\put(-8,-.1){\includegraphics[height=58mm]{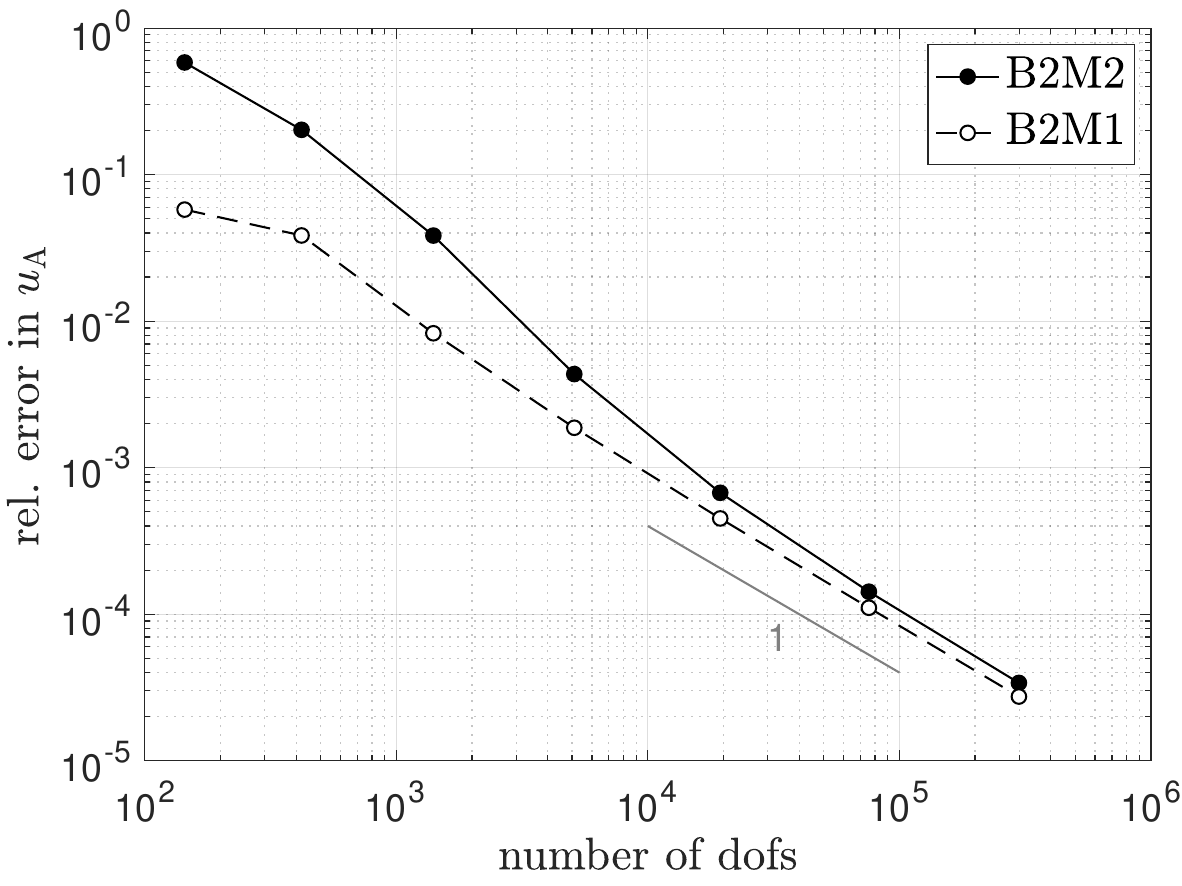}}
\put(0.2,-.1){\includegraphics[height=58mm]{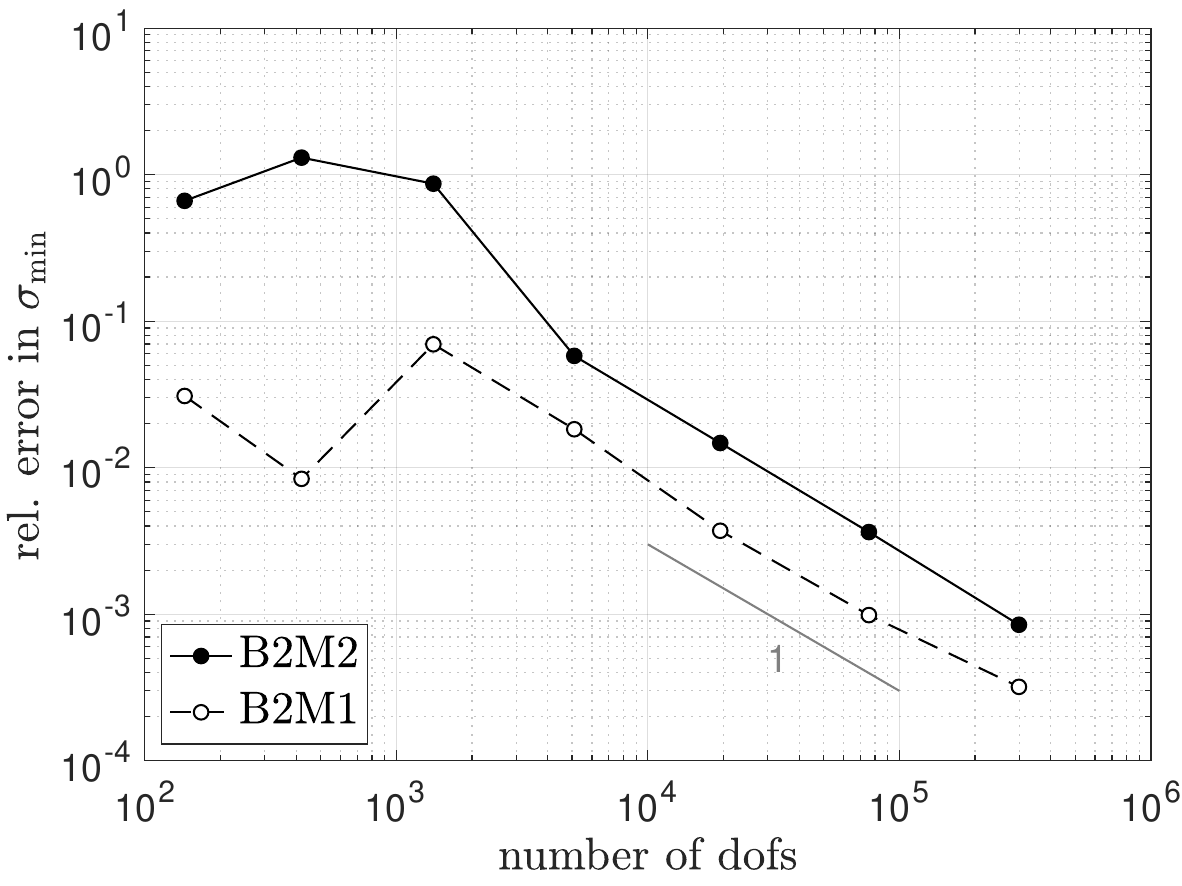}}
\put(-7.95,0.0){\footnotesize (a)}
\put(0.15,0.0){\footnotesize (b)}
\end{picture}
\caption{Scordelis-Lo roof with large deformations ($R/T = 100$): 
Convergence of (a) displacement $u_\mathrm{A}$ and (b) stress $\sig^1_{1\,\mathrm{min}}$, using $m=4, 8, 16, ..., 512$.
Again B2M1 is more accurate than B2M2, especially for coarse meshes.\\[-1mm]}
\label{f:SLlargeue}
\end{center}
\end{figure}

Now Young's modulus is decreased by a factor of 15 and the problem is solved nonlinearly.
For slenderness $R/T=100$, the reference values $u_\mathrm{A} = 1.65314024$ and $\sig^1_{1\,\mathrm{min}} = -659.22021$ are obtained with discretization $p=5$ and $m=128$.\footnote{The case $R/T=10^4$ does not run quasi-statically anymore for the decreased stiffness, due to the appearance of multiple solutions (i.e.~buckling patterns).}
Fig.~\ref{f:SLlarge} shows the stress field solution for the B2M2 and B2M1 discretizations in comparison to the reference solution.
Again, B2M1 shows good accuracy, while B2M2 is inaccurate due to strong oscillations.
Even the sharp stress peaks in the corner are captured quite well by B2M1. 
The accuracy gain of B2M1 over B2M2 is also seen in the convergence plots of Fig.~\ref{f:SLlargeue}.
The observed convergence rates are equal to those of the infinitesimal case, see Figs.~\ref{f:SLue} and \ref{f:SLse}, with B2M1 being consistently more accurate than B2M2 now.

\subsection{Hemisphere with hole}

\subsubsection{Problem setup and reference solution}\label{s:HHsetup}

\begin{figure}[h]
\begin{center} \unitlength1cm
\begin{picture}(0,3.9)
\put(2.1,-.5){\includegraphics[height=44mm]{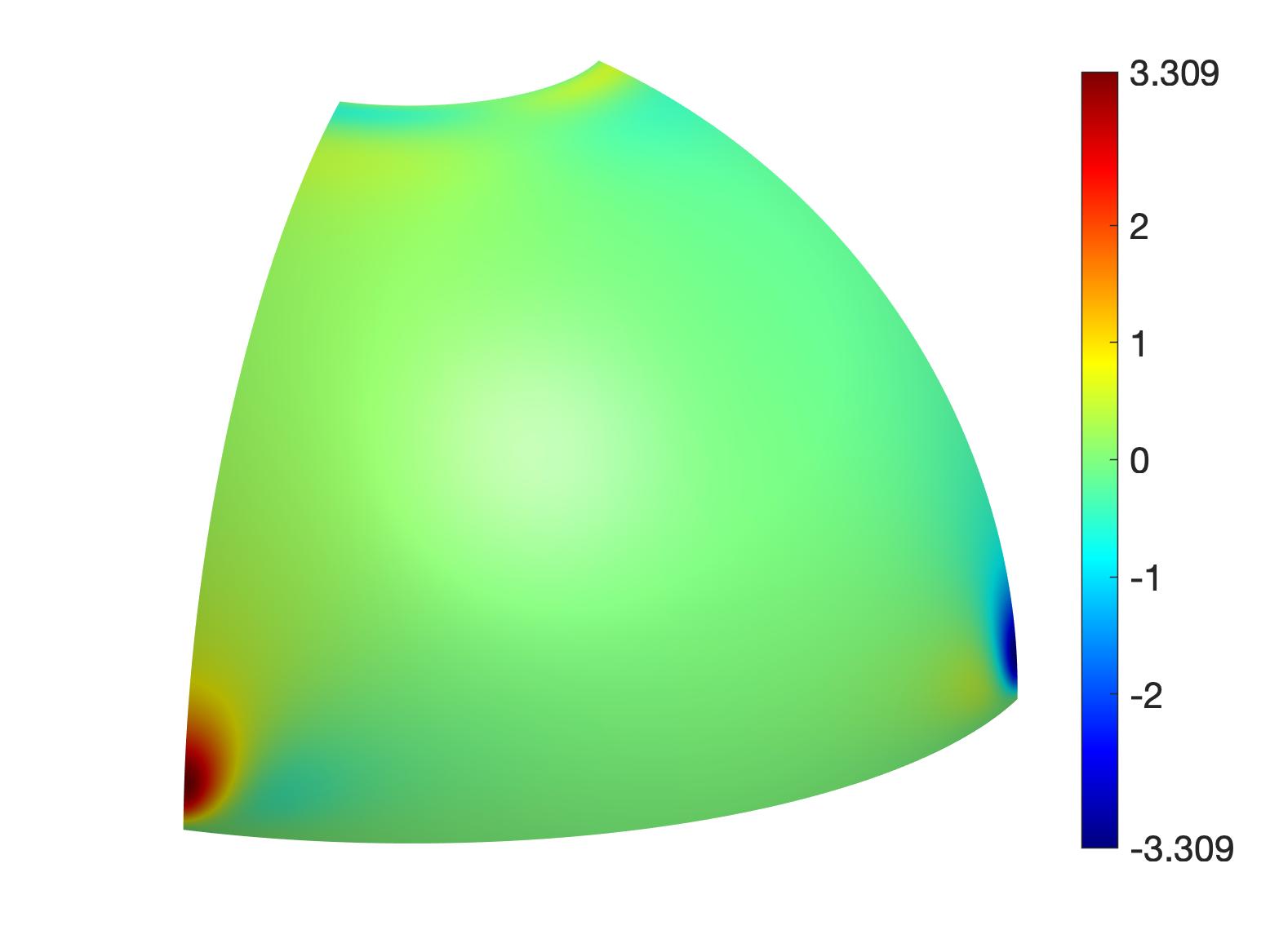}}
\put(-3.25,-.5){\includegraphics[height=44mm]{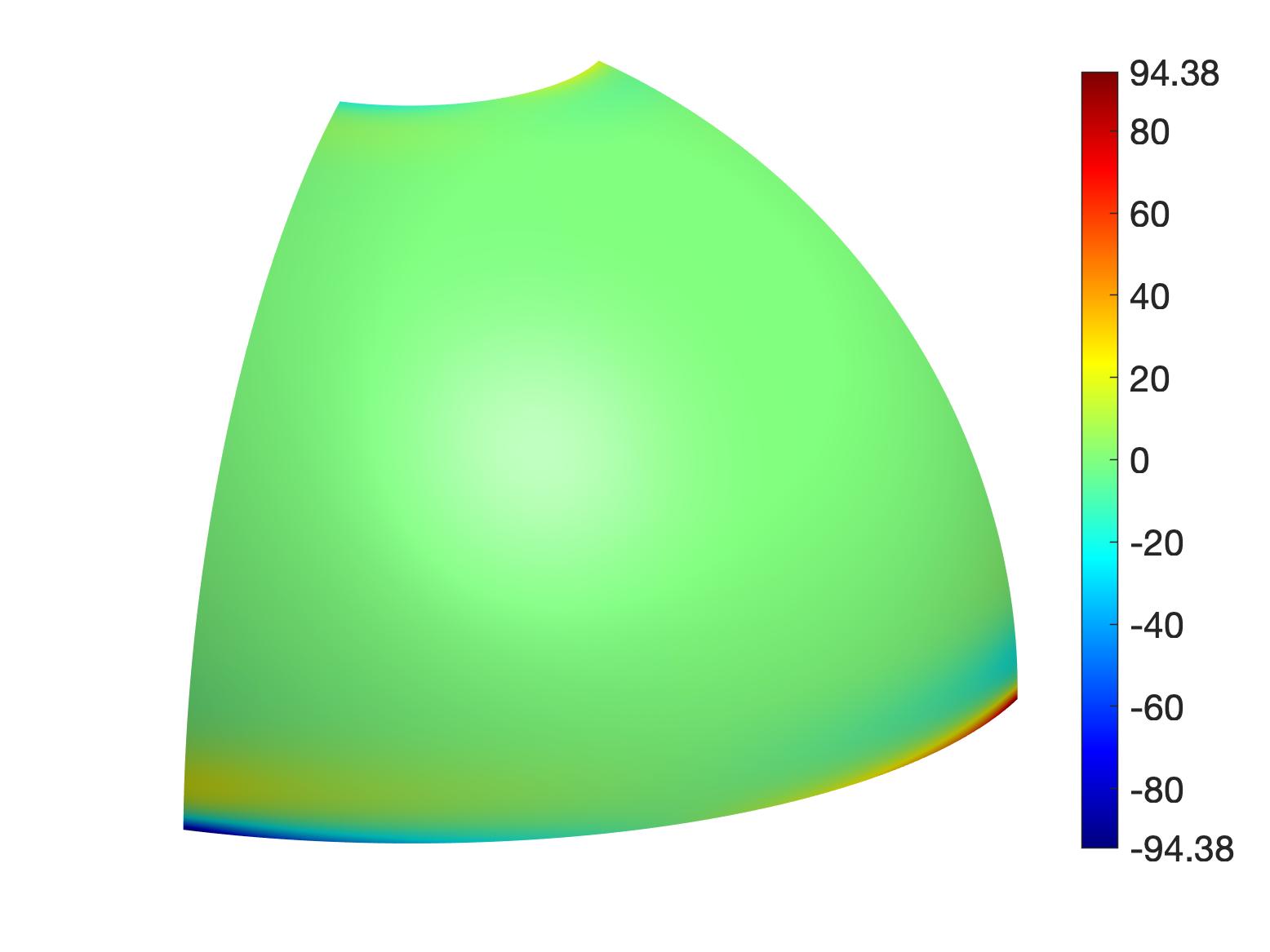}}
\put(-8.7,-.5){\includegraphics[height=44mm]{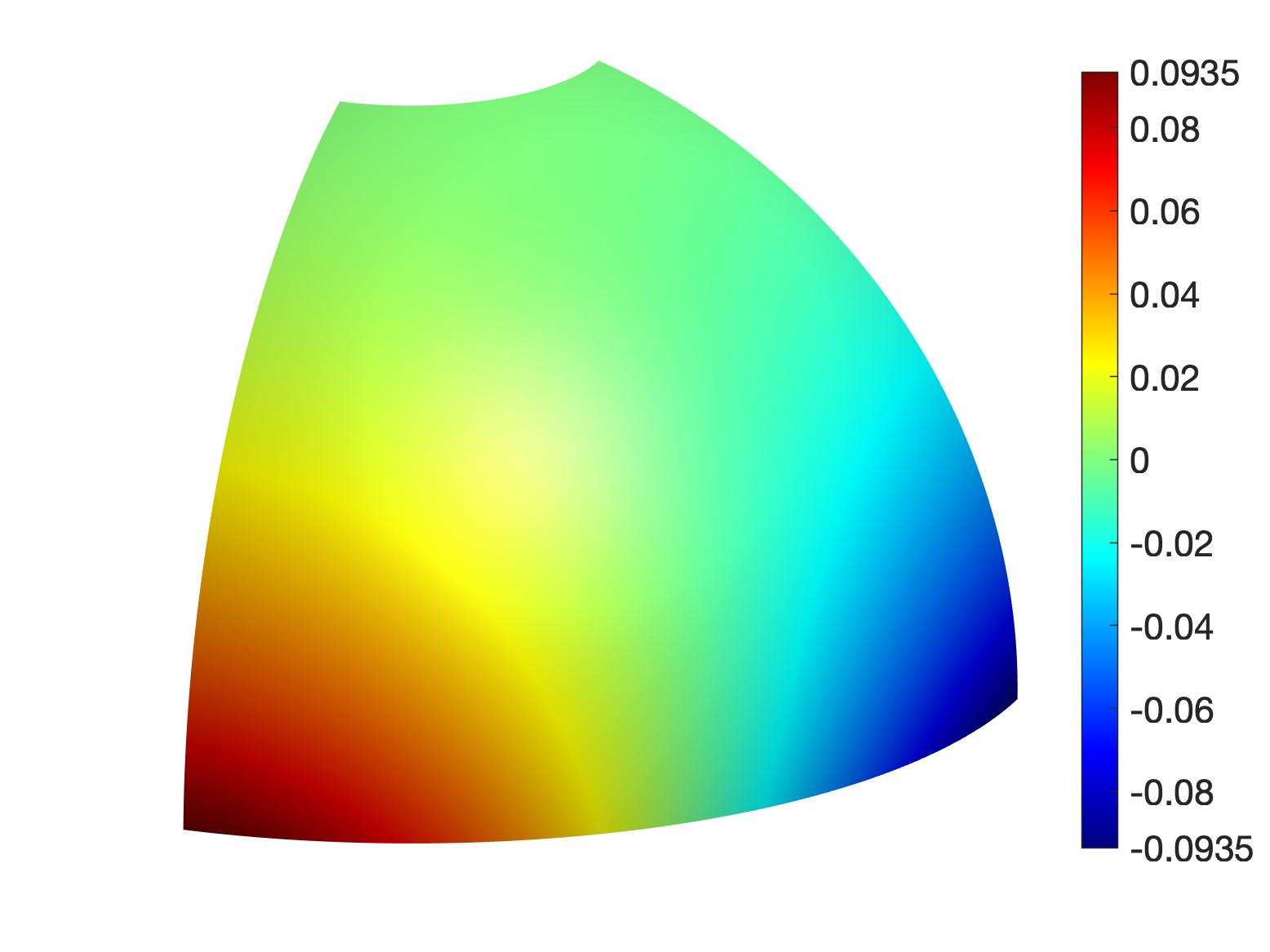}} 
\end{picture}
\caption{Hemisphere with hole: 
Reference result for displacement $u_r$ (left), raw hoop stress $\sigma_{11}$ (center) and raw azimuthal stress $\sigma_{22}$ (right). 
Left \& center for $m=64$, $p=5$, right for $m=128$, $p=5$.
}
\label{f:HHref}
\end{center}
\end{figure}

The hemisphere with hole problem proposed by \citet{macneal85} is another classical problem for investigating thin shells. 
It consists of a hemisphere with a $18^\circ$ hole at the top that is loaded by four alternating radial forces: 
Tensile at point $A$ and compressive at point $B$, see Fig.~\ref{f:B2M1h}.
The classical setup considers radius $R = 10L$, Young's modulus $E=6.825E_0$, Poisson's ratio $\nu = 0.3$ and radial force 
$P = 2\cdot 10^{-7} E_0L^2 (25T/L)^3$.
Exploiting symmetry, only a quarter is meshed using $m \times m$ quadratic NURBS and $(m+1)\times(m+1)$ linear Lagrange elements, see Fig.~\ref{f:B2M1h}.
This results in $3(m+p)^2$ dofs.
First the slenderness $R/T=250$ is used, as in the literature. 
A variation of $R/T$ is then considered later, see Sec.~\ref{s:HHa}.
Reference results for the displacement and stress field are shown in Fig.~\ref{f:HHref}.
The reference values for the extremal displacement and stresses are $u_\mathrm{A} = -u_\mathrm{B} = 0.09352155 $, $\sig^\mathrm{max}_{11} = 94.4$ and $\sig^\mathrm{max}_{22} = 3.3086$, respectively.
The relative accuracy of these values is about $10^{-7}$, $10^{-3}$ and $10^{-5}$, respectively.
The accuracy in $\sig^\mathrm{max}_{11}$ is much lower, due to an earlier onset of ill-conditioning in those stresses.

\subsubsection{Displacement accuracy}\label{s:HHu}

\begin{figure}[h]
\begin{center} \unitlength1cm
\begin{picture}(0,5.7)
\put(-8,-.1){\includegraphics[height=58mm]{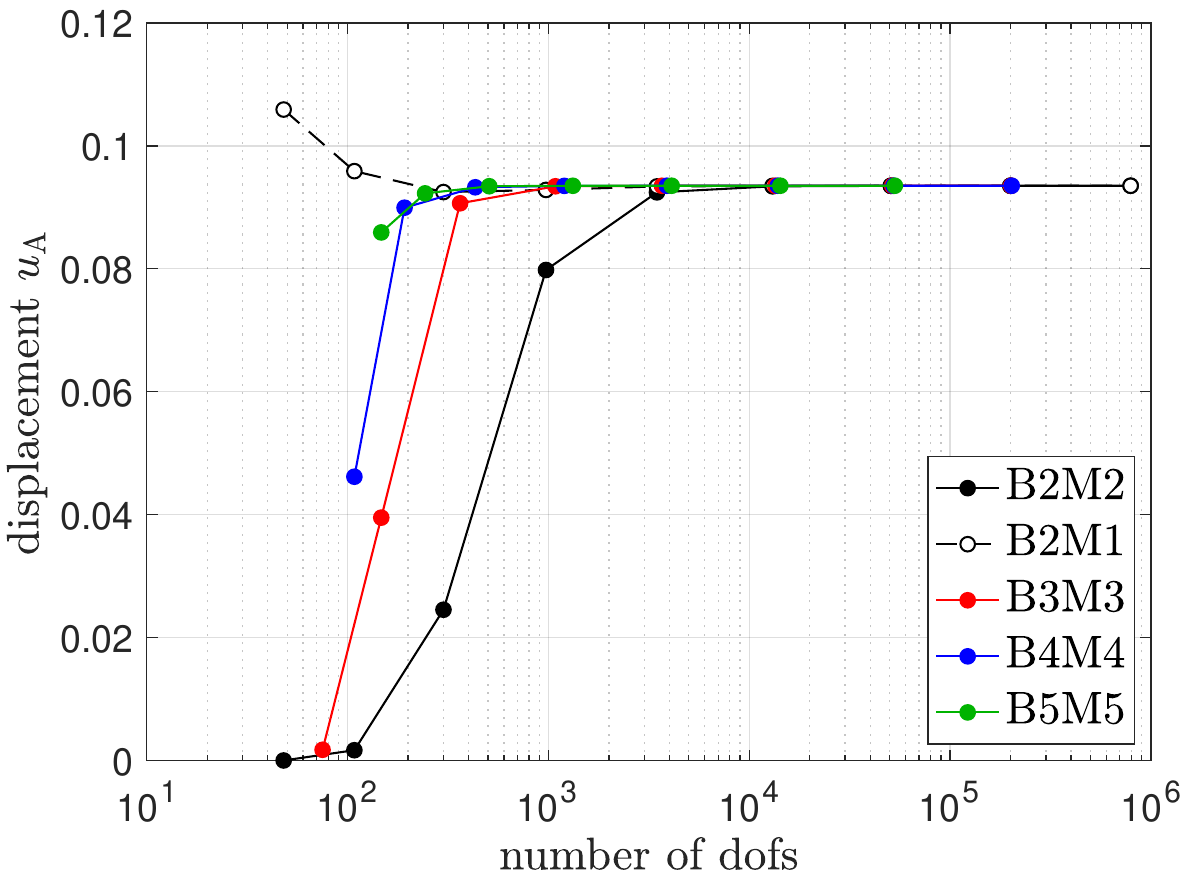}}
\put(0.2,-.1){\includegraphics[height=58mm]{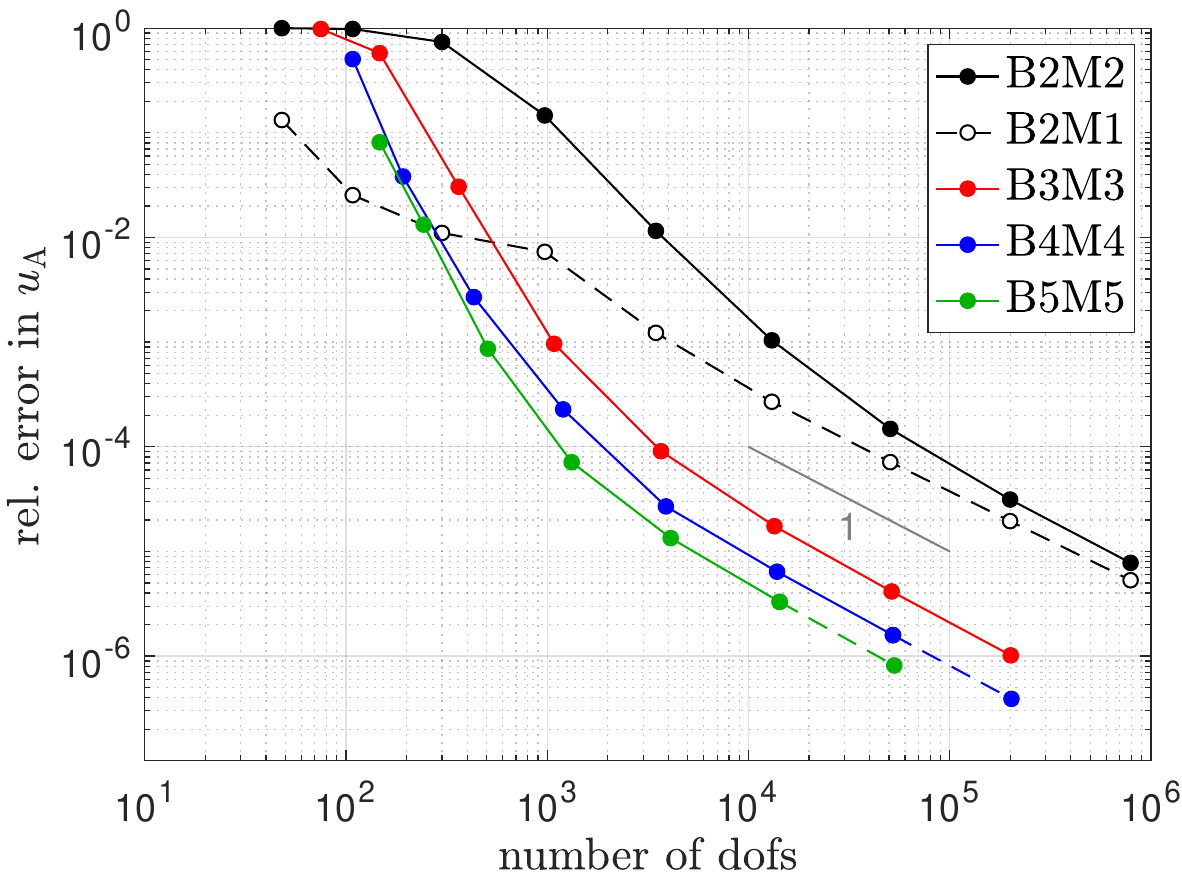}}
\put(-7.95,0.0){\footnotesize (a)}
\put(0.15,0.0){\footnotesize (b)}
\end{picture}
\caption{Hemisphere with hole: 
Convergence of the displacement $u_\mathrm{A}=-u_\mathrm{B}$;
(a) absolute values; (b)~relative error; $m=2, 4, 8, ..., 512$.
Again B2M1 is more accurate than B2M2, especially for coarse meshes.}
\label{f:HHue}
\end{center}
\end{figure}

Fig.~\ref{f:HHue} shows the convergence of displacement $u_A$ with mesh refinement.
The observed convergence rates are linear in all cases, due to the point loads.\footnote{If the loads are distributed over a finite area, the proper rates are obtained. But this changes the problem.}
The B2M1 results are consistently better than those of B2M2, and even surpass all other cases (B3M3, B4M4 \& B5M5) for coarse meshes, which confirms the findings in the preceding two examples.

\subsubsection{Stress accuracy}\label{s:HHs}

\begin{figure}[H]
\begin{center} \unitlength1cm
\begin{picture}(0,8)
\put(2.1,3.6){\includegraphics[height=44mm]{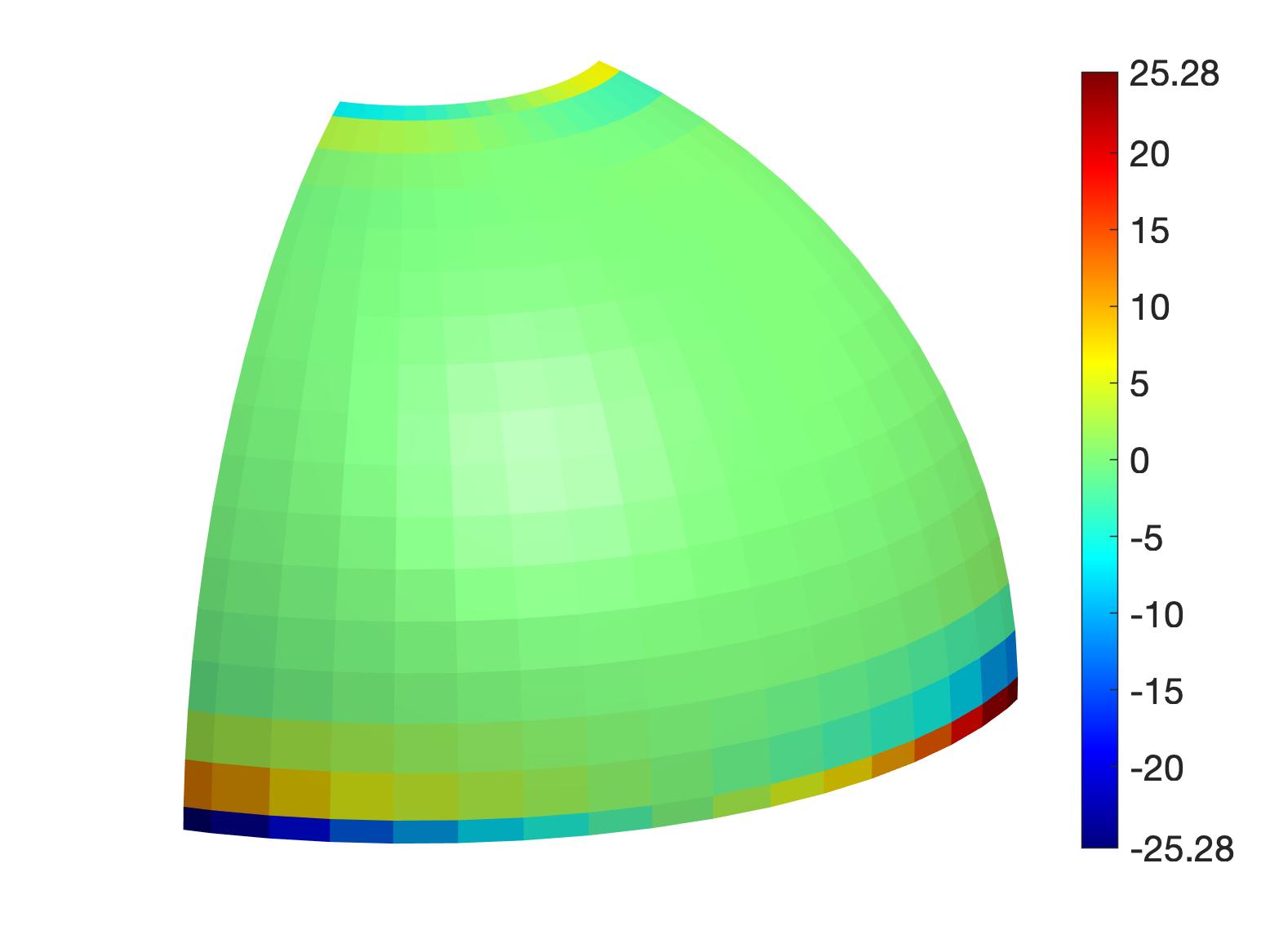}}
\put(-3.25,3.6){\includegraphics[height=44mm]{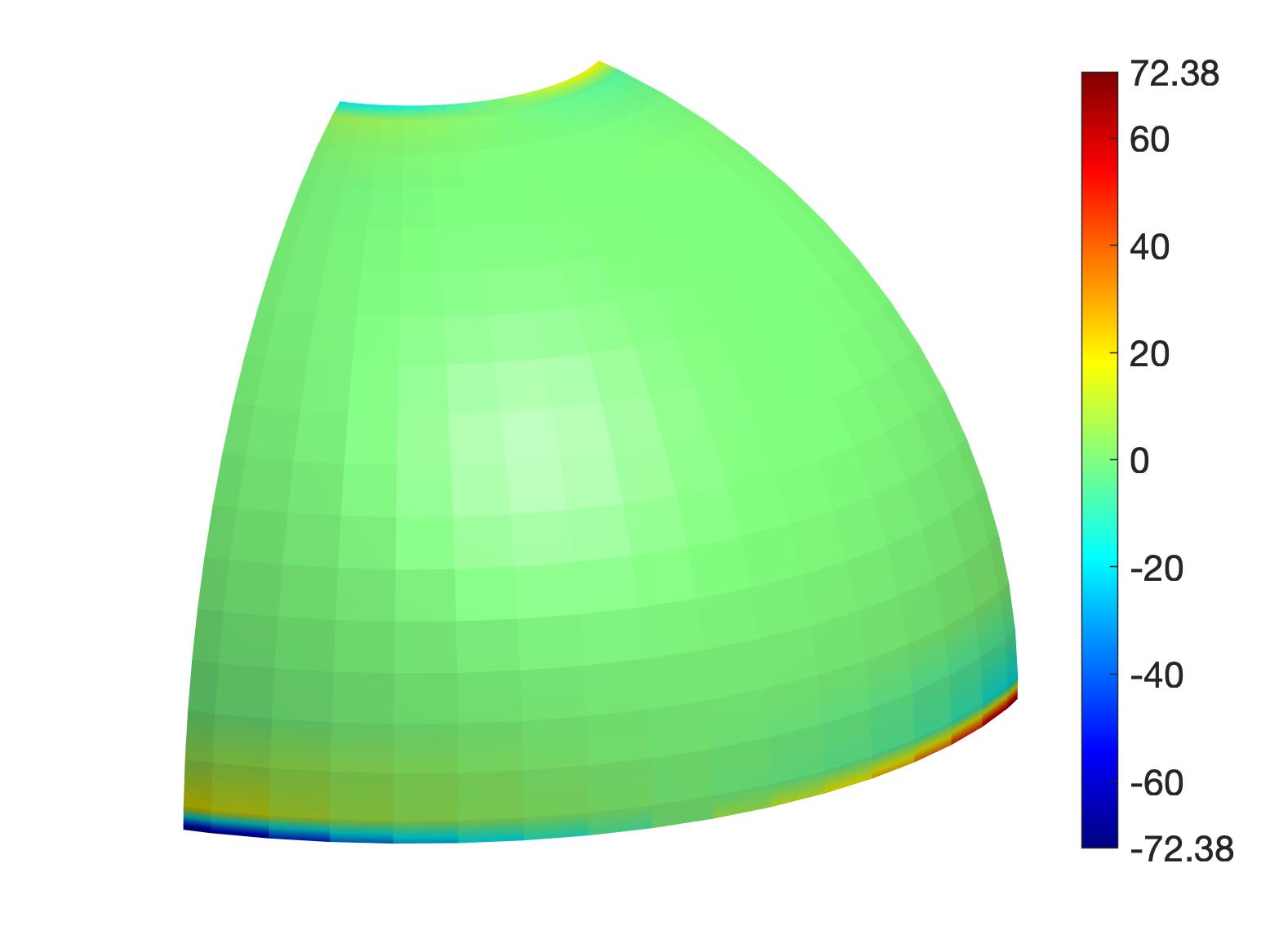}}
\put(-8.7,3.6){\includegraphics[height=44mm]{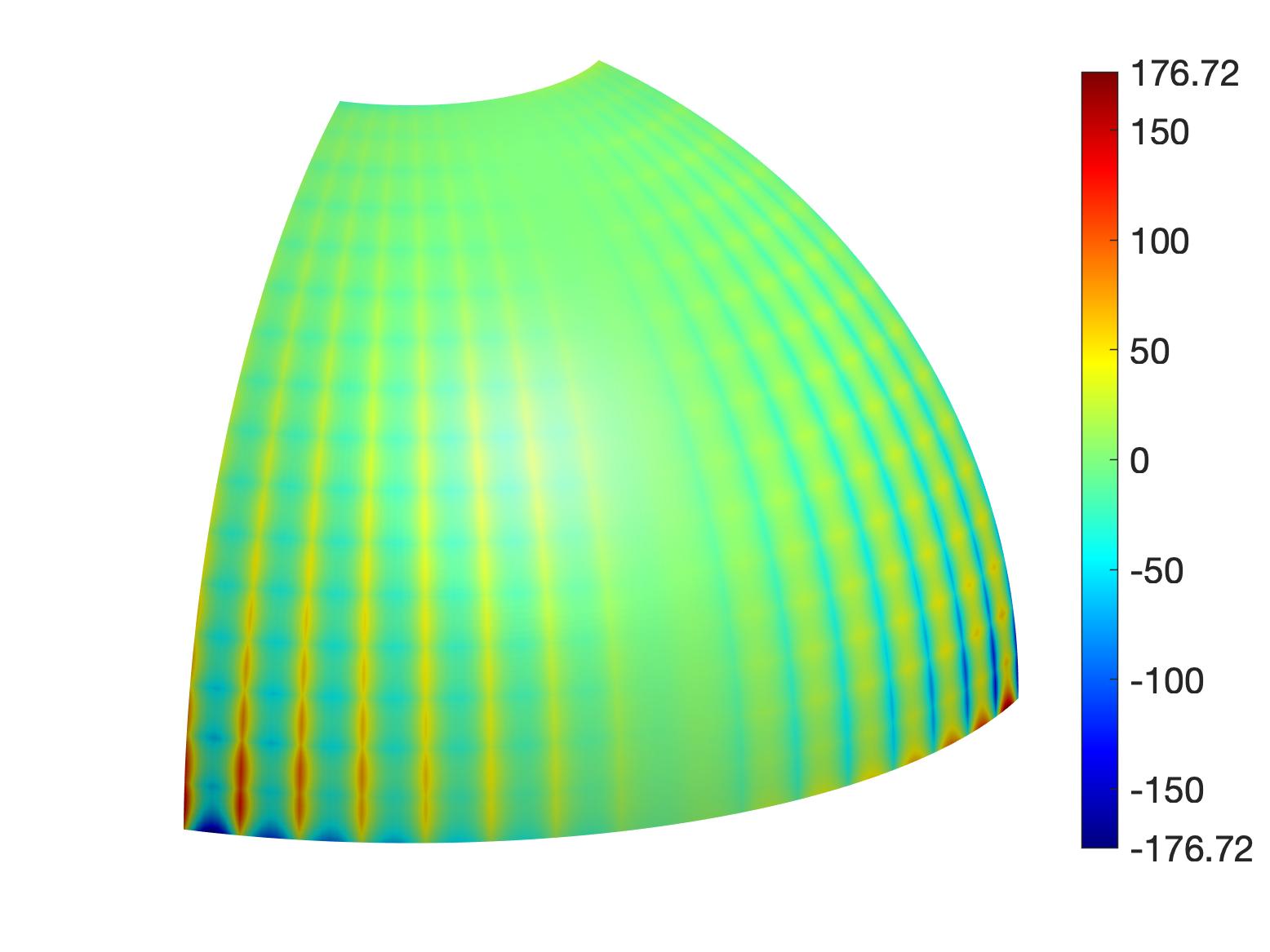}}
\put(2.1,-.5){\includegraphics[height=44mm]{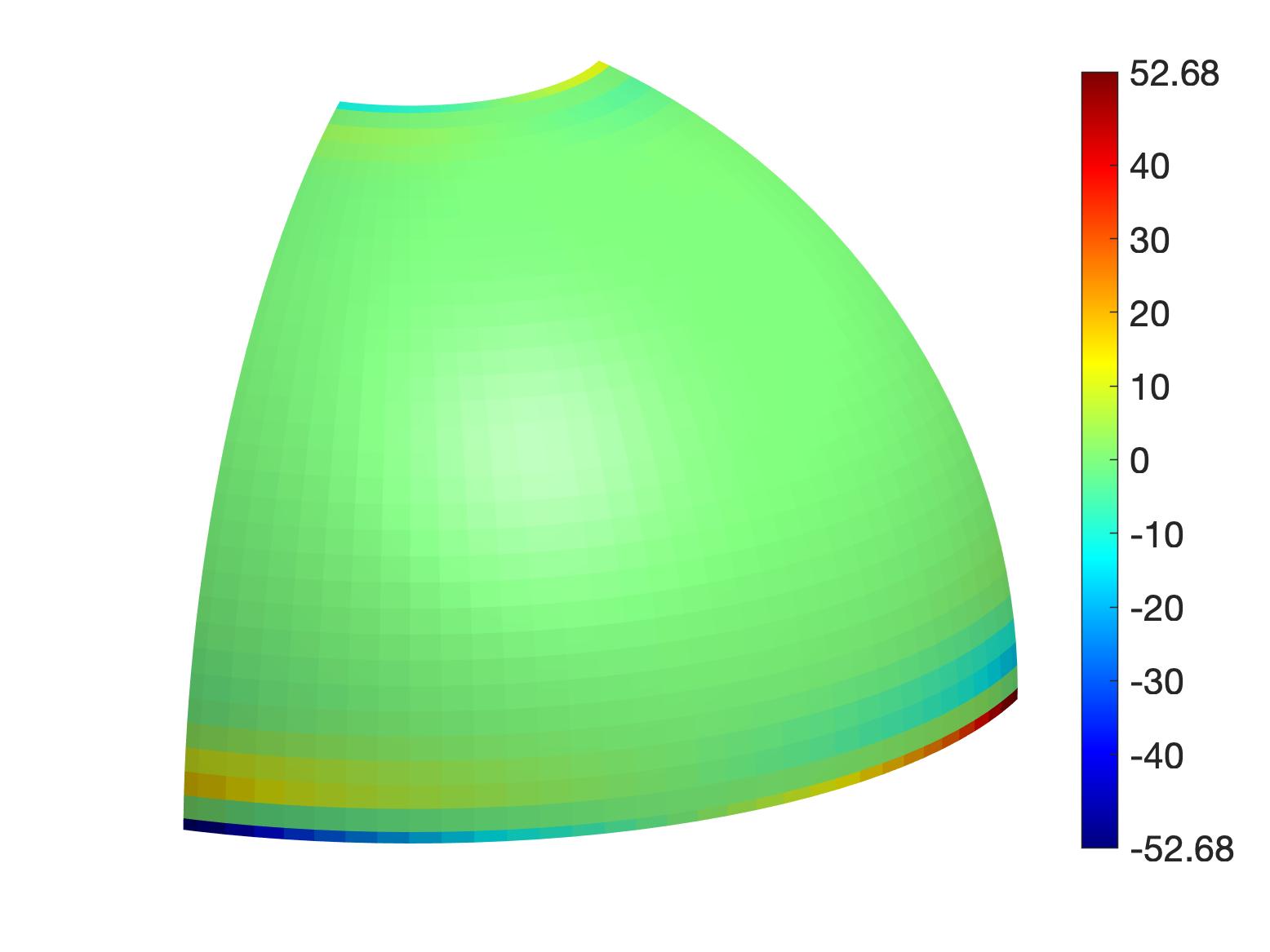}}
\put(-3.25,-.5){\includegraphics[height=44mm]{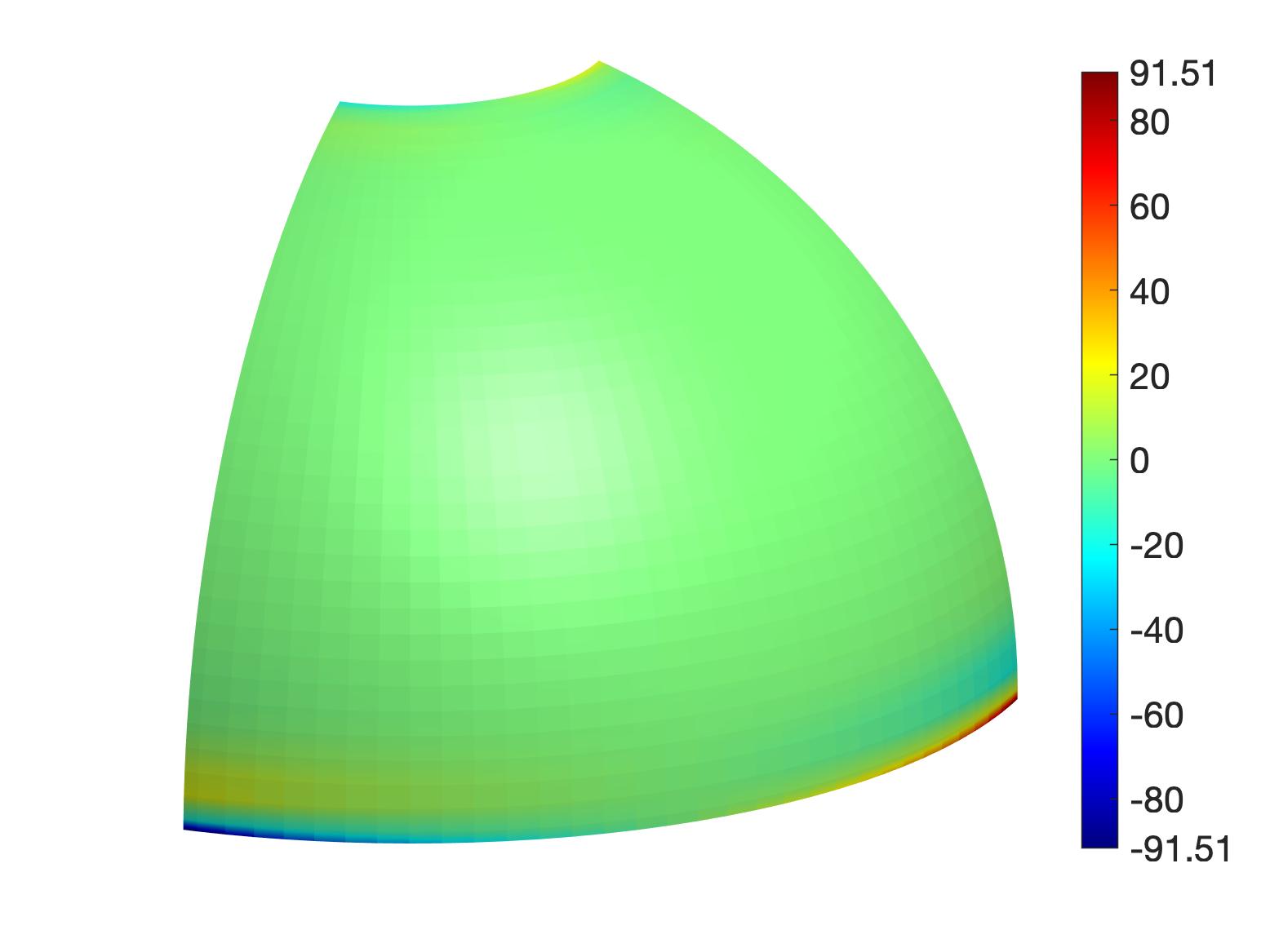}}
\put(-8.7,-.5){\includegraphics[height=44mm]{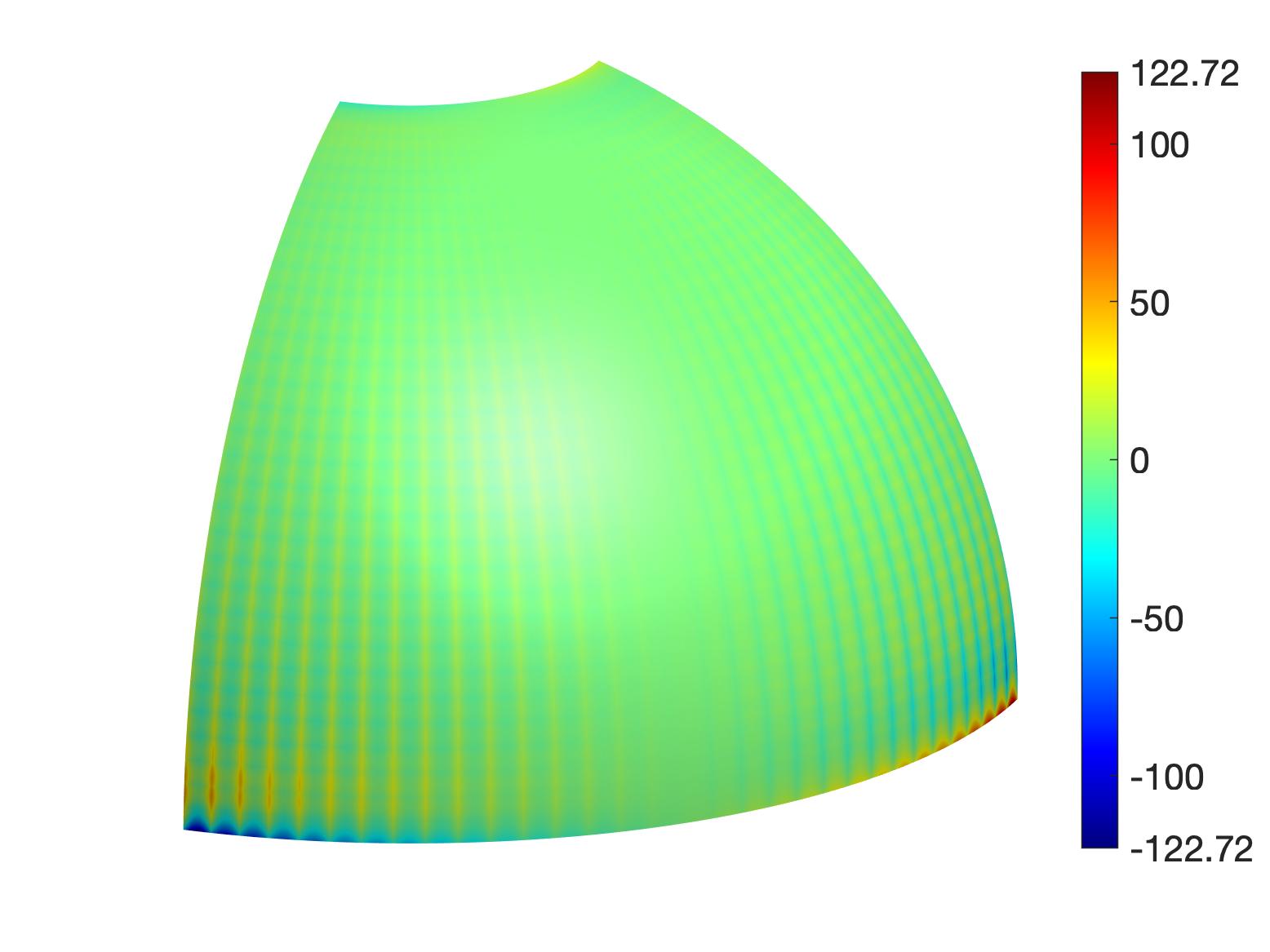}}
\end{picture}
\caption{Hemisphere with hole: Raw effective membrane stress $\sigma_{11}$ for B2M2 (left column), B2M1 (center column) and B2M1 with constant stress post processing (right column) using $m=16$ (top row) and $m=32$ (bottom row).
B2M2's stresses are highly oscillatory, while B2M1's stresses are much more accurate.
Taking only the constant stresses for B2M1 does not improve stress $\sigma_{11}$.}
\label{f:HHsig1}
\end{center}
\end{figure}

\begin{figure}[H]
\begin{center} \unitlength1cm
\begin{picture}(0,12.1)
\put(2.1,7.7){\includegraphics[height=44mm]{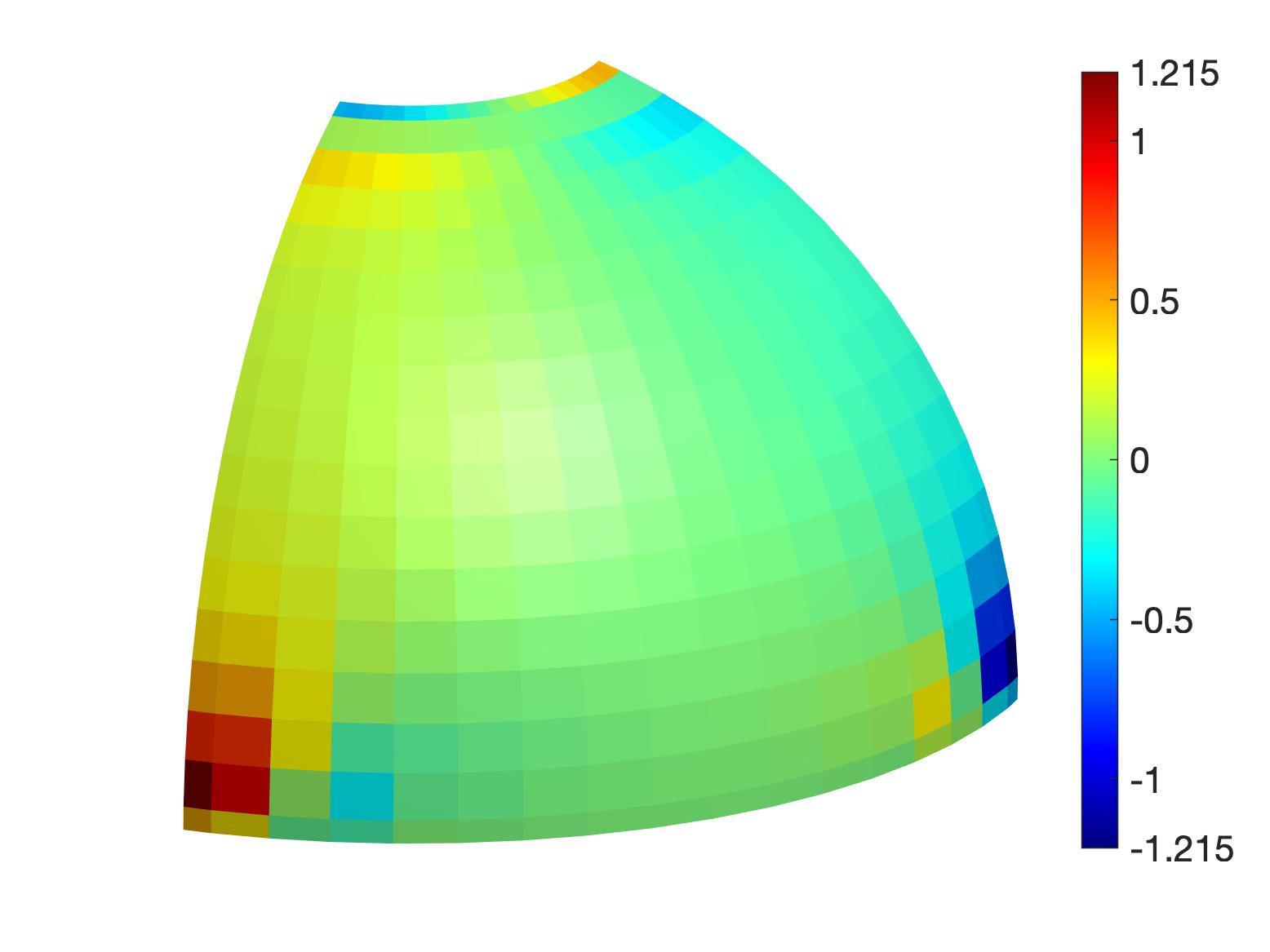}}
\put(-3.25,7.7){\includegraphics[height=44mm]{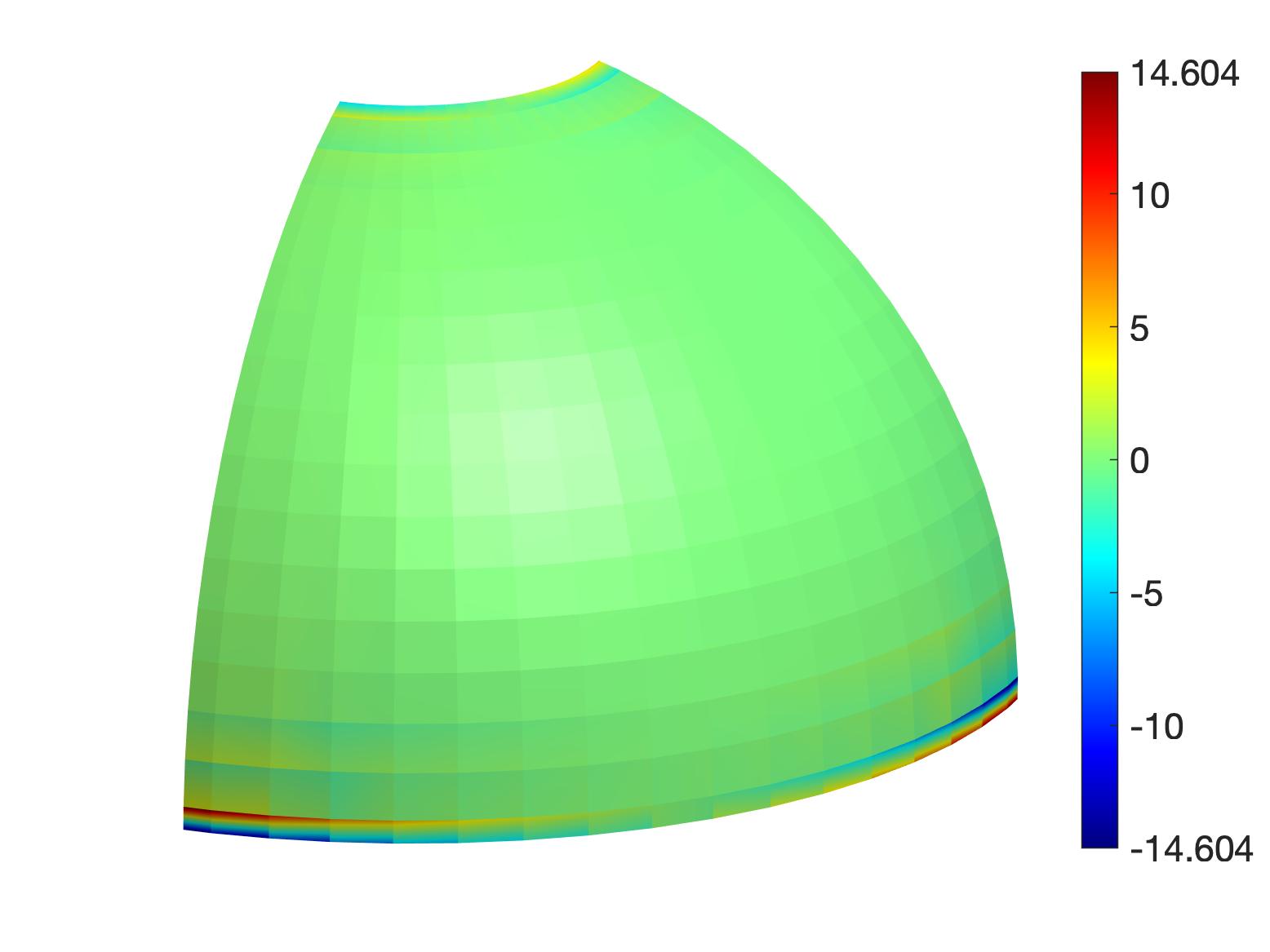}}
\put(-8.7,7.7){\includegraphics[height=44mm]{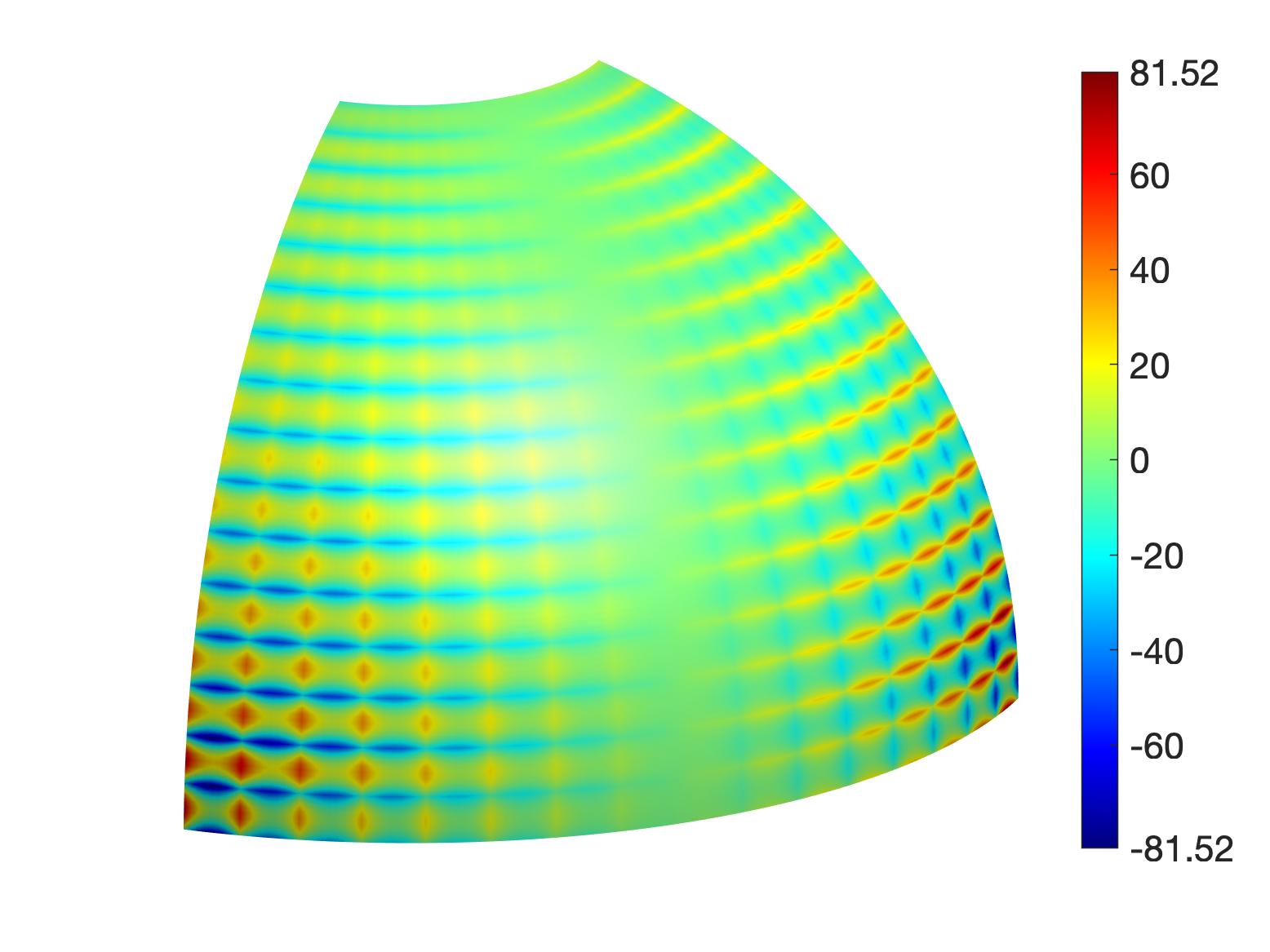}}
\put(2.1,3.6){\includegraphics[height=44mm]{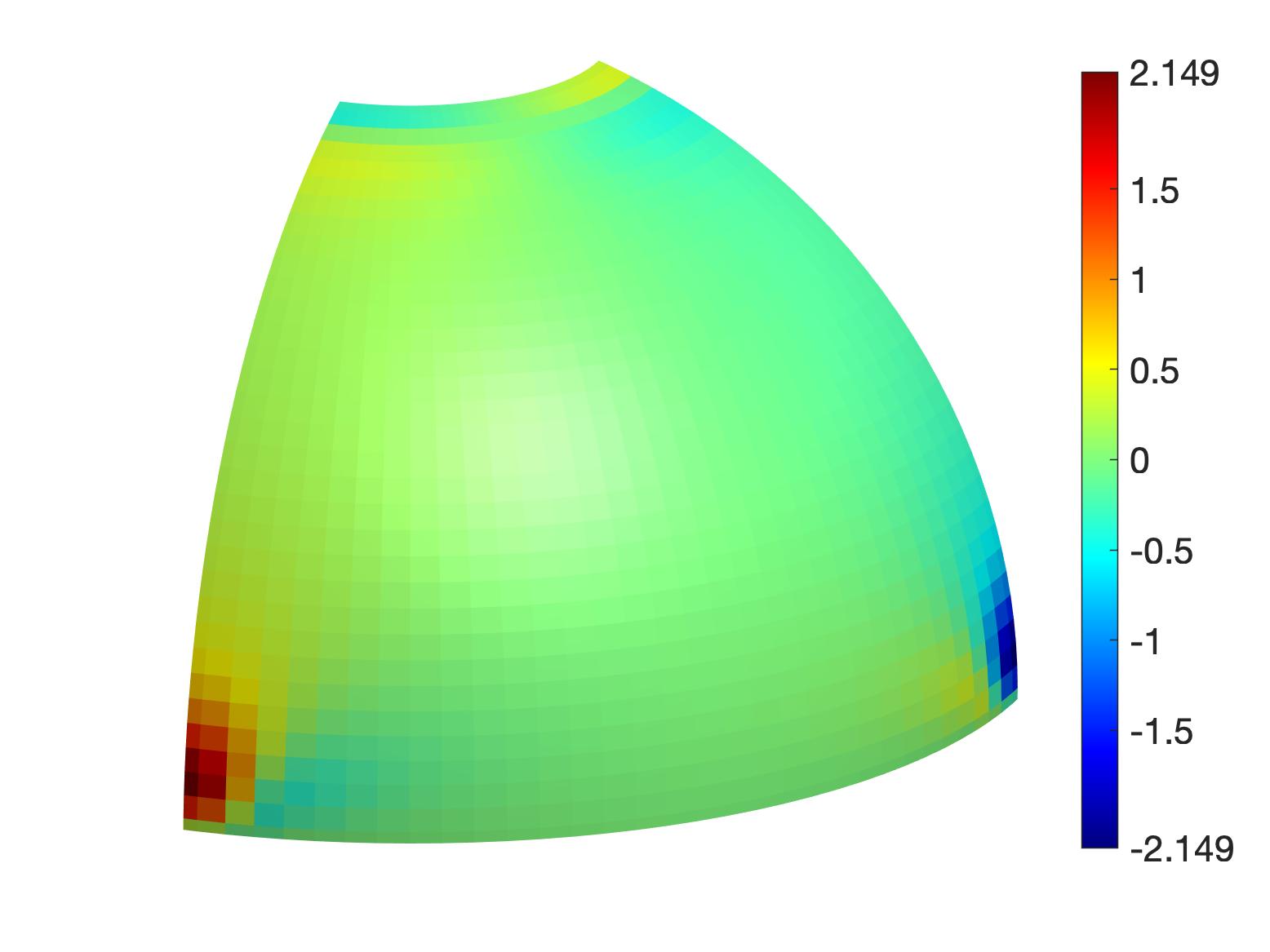}}
\put(-3.25,3.6){\includegraphics[height=44mm]{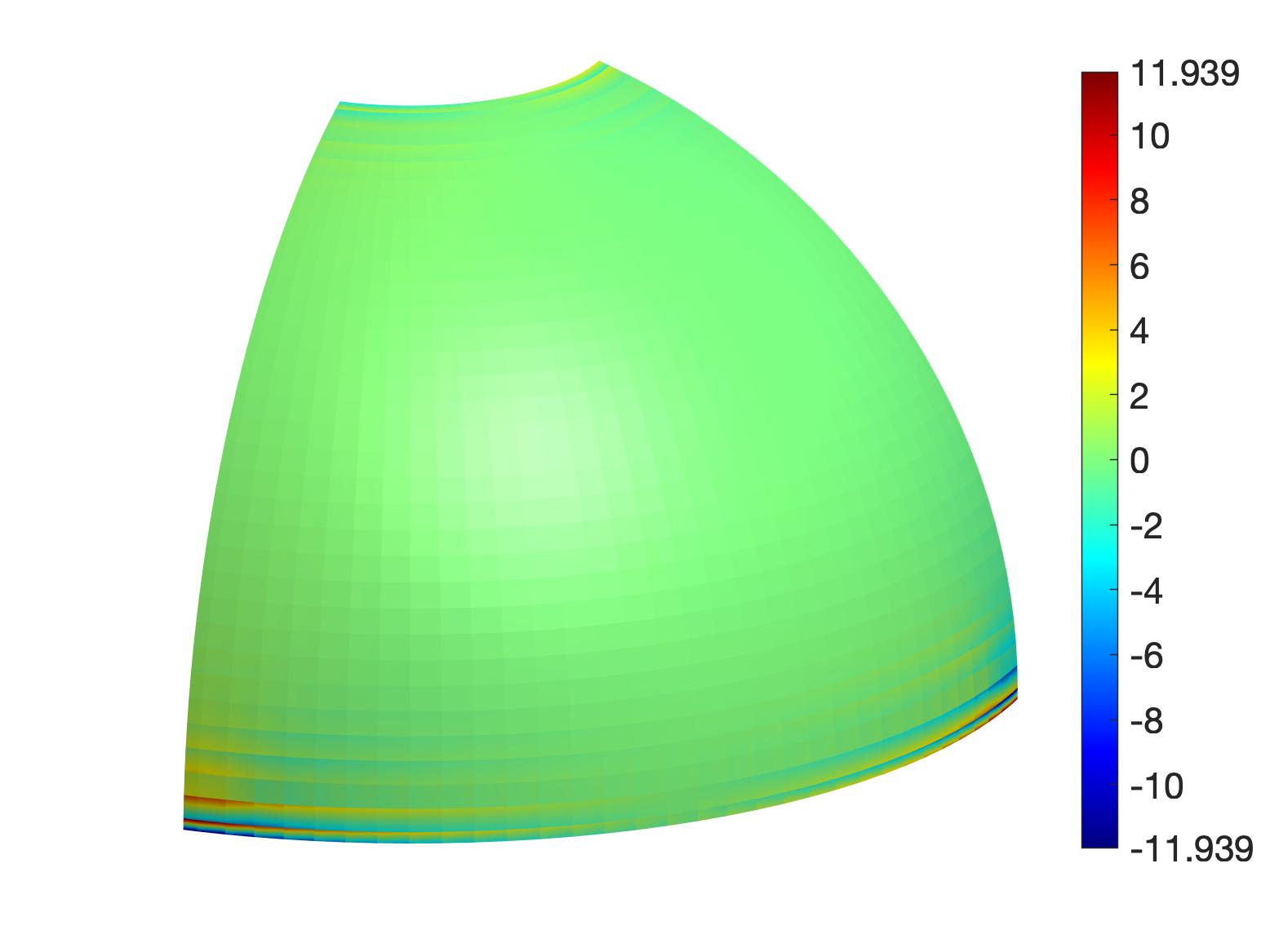}}
\put(-8.7,3.6){\includegraphics[height=44mm]{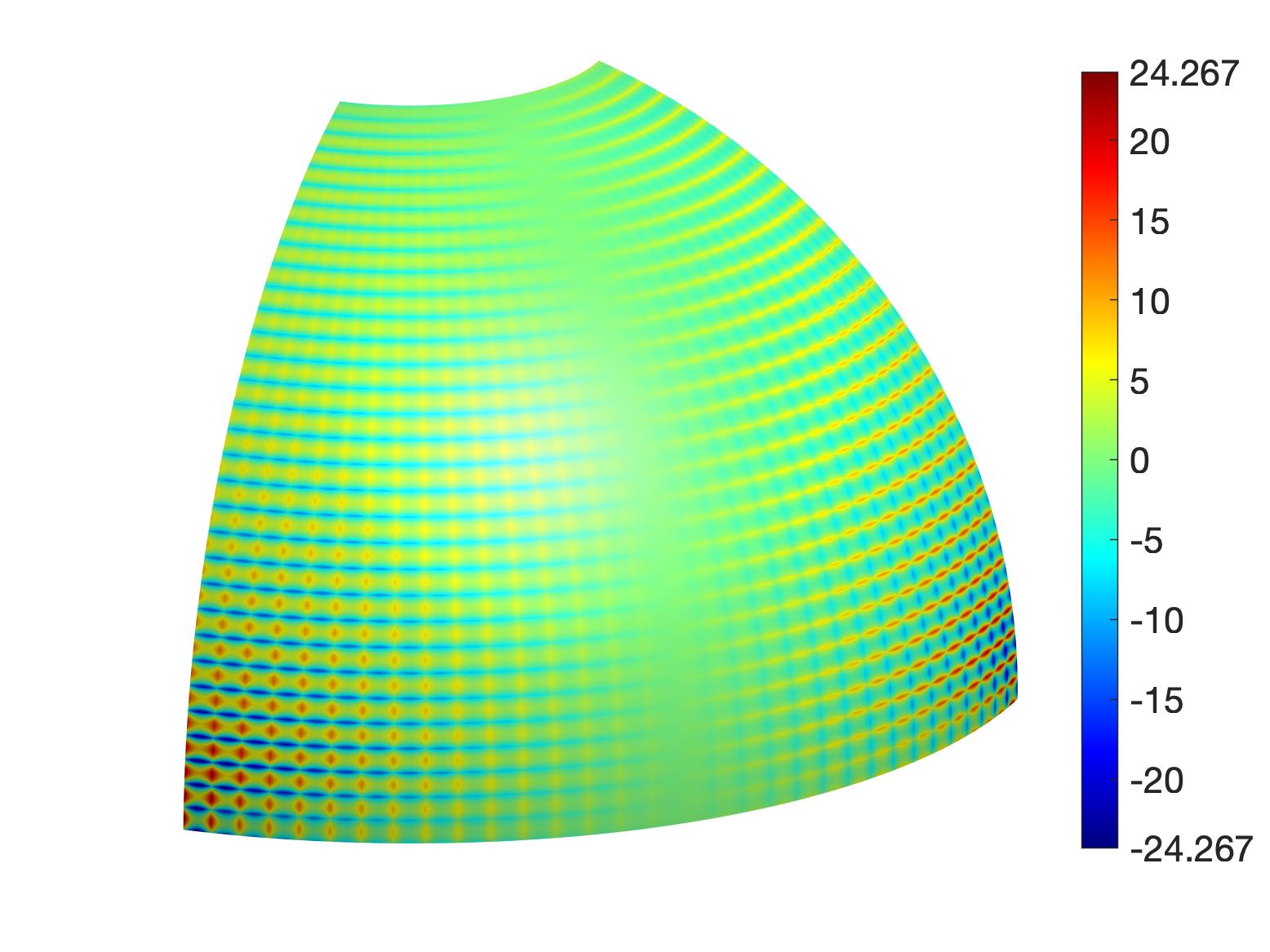}}
\put(2.1,-.5){\includegraphics[height=44mm]{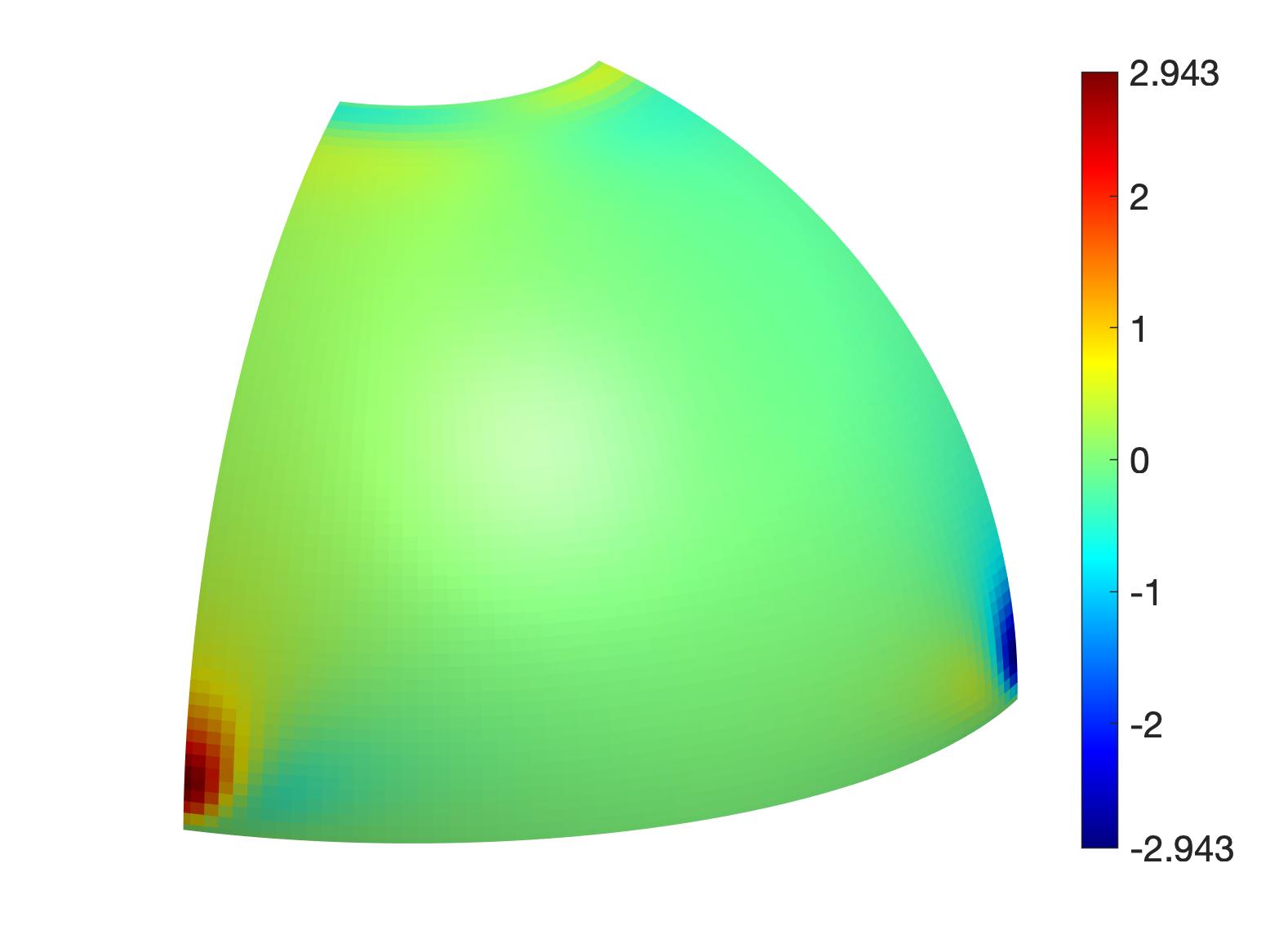}}
\put(-3.25,-.5){\includegraphics[height=44mm]{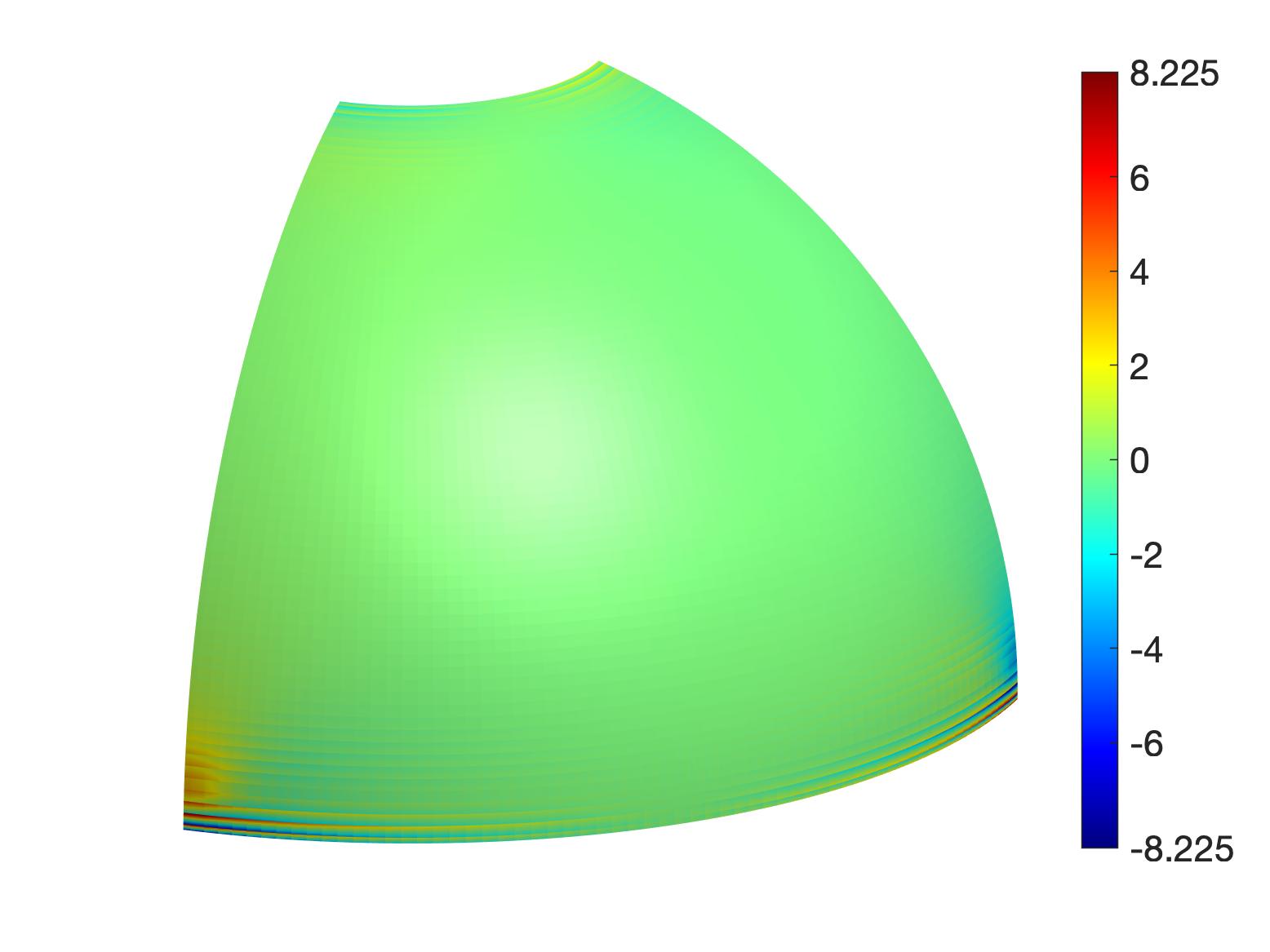}}
\put(-8.7,-.5){\includegraphics[height=44mm]{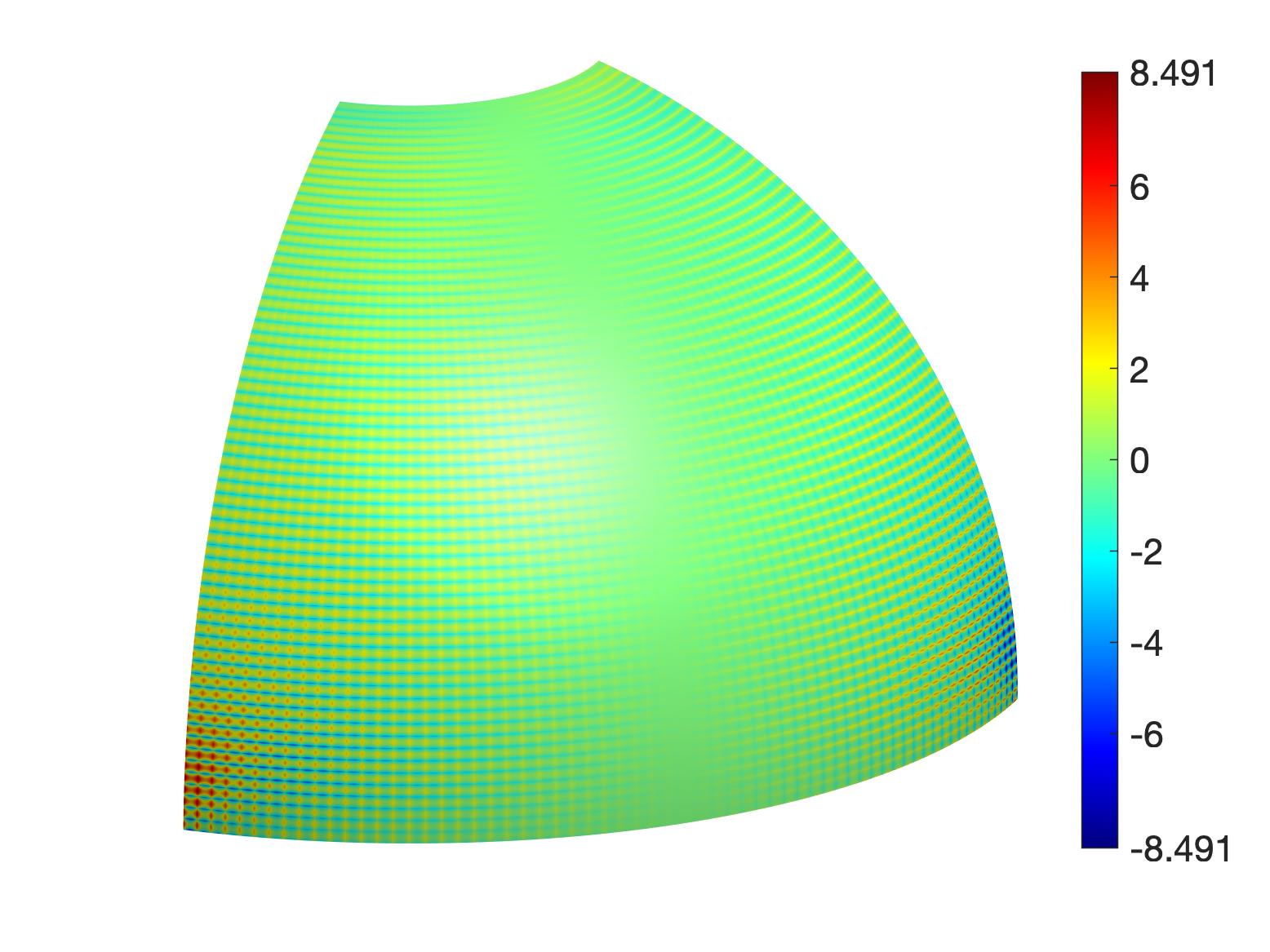}}
\end{picture}
\caption{Hemisphere with hole: Raw effective membrane stress $\sigma_{22}$ for B2M2 (left column), B2M1 (center column) and B2M1 with constant stress post processing (right column) using $m=16$ (top row), $m=32$ (center row) and $m=64$ (bottom row).
B2M2's stresses are highly oscillatory. 
Also B2M1's stresses do not look much better, unless only the constant part is examined.
That is quite accurate even for coarse meshes.
This becomes more clear, when plotted at equal color scale, see Fig.~\ref{f:HHsig2a}.}
\label{f:HHsig2}
\end{center}
\end{figure}

\begin{figure}[H]
\begin{center} \unitlength1cm
\begin{picture}(0,3.9)
\put(2.1,-.5){\includegraphics[height=44mm]{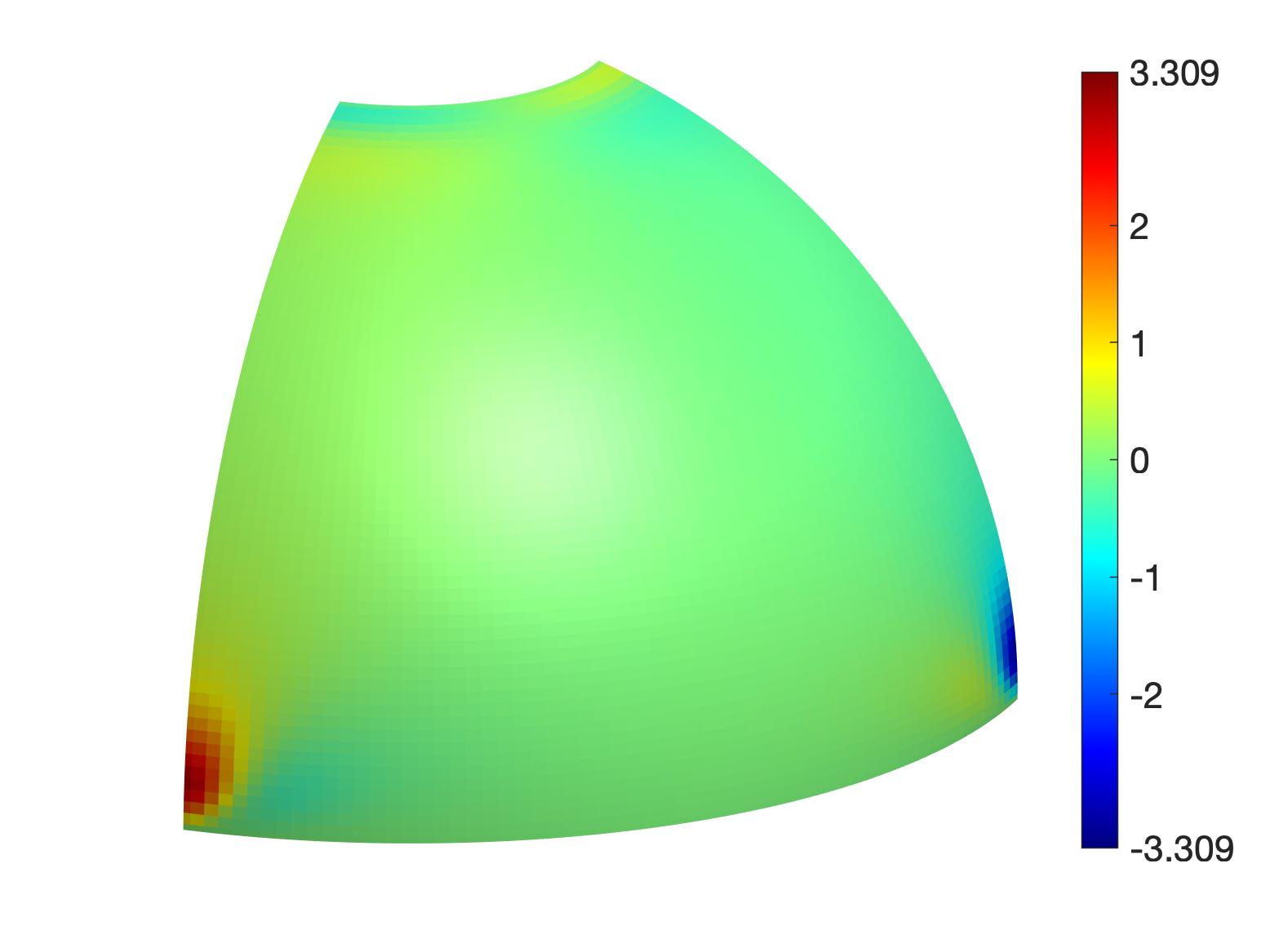}}
\put(-3.25,-.5){\includegraphics[height=44mm]{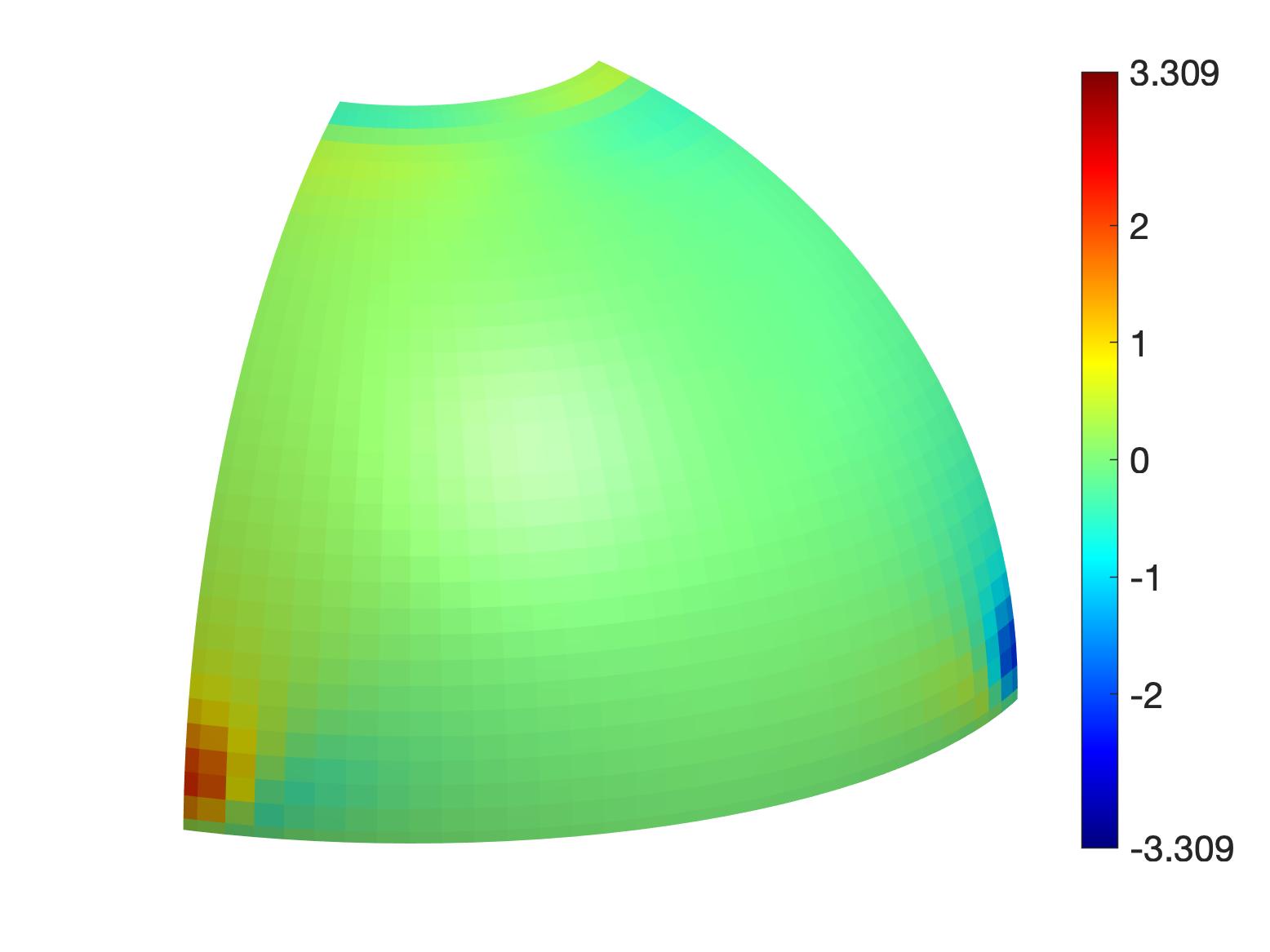}}
\put(-8.7,-.5){\includegraphics[height=44mm]{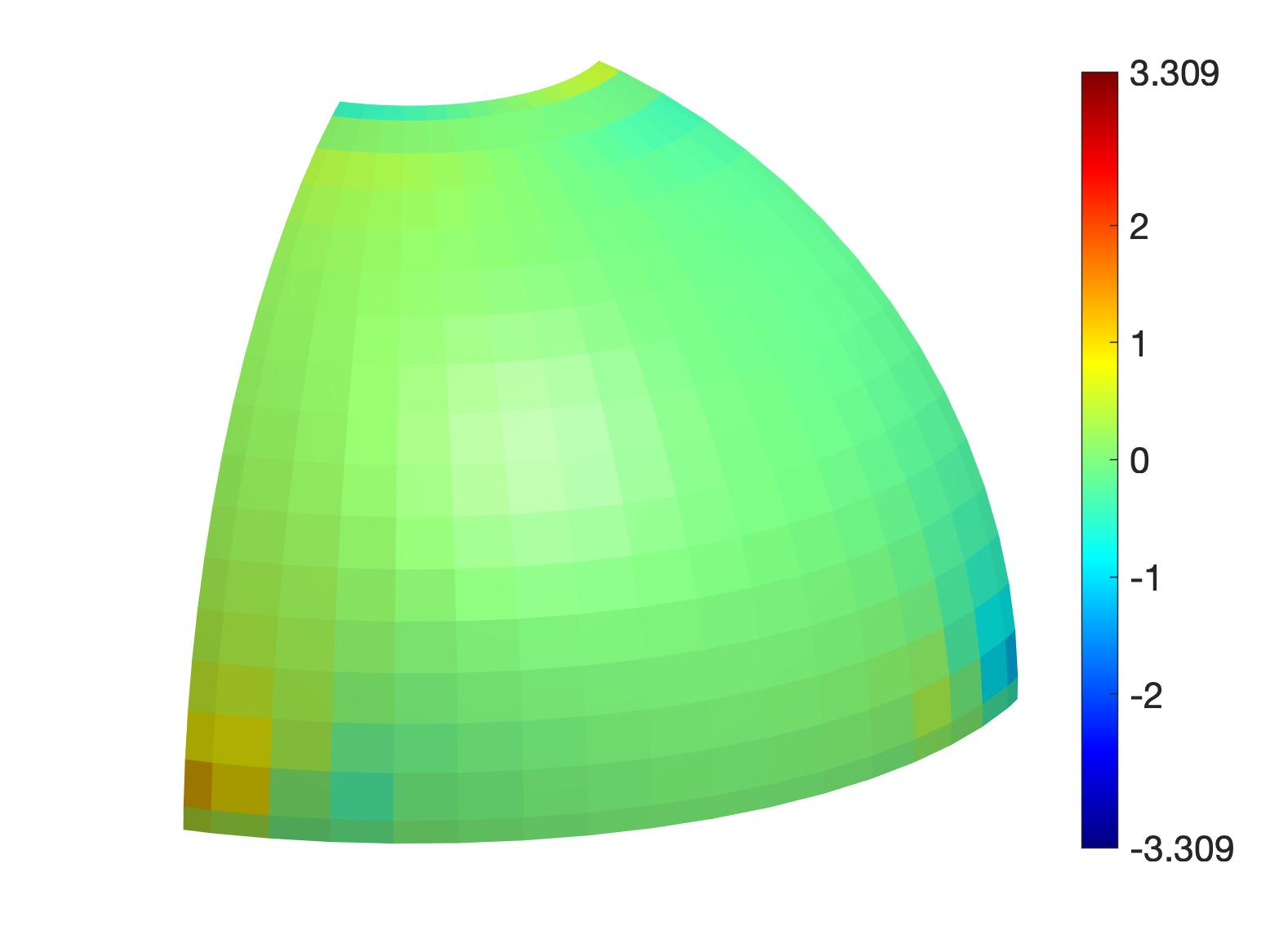}}
\end{picture}
\caption{Hemisphere with hole: Raw effective membrane stress $\sigma_{22}$ for B2M1 with constant stress post processing, plotted at equal color scale using $m=16$ (left), $m=32$ (center) and $m=64$ (right).}
\label{f:HHsig2a}
\end{center}
\end{figure}

\begin{figure}[h]
\begin{center} \unitlength1cm
\begin{picture}(0,5.7)
\put(-8,-.1){\includegraphics[height=58mm]{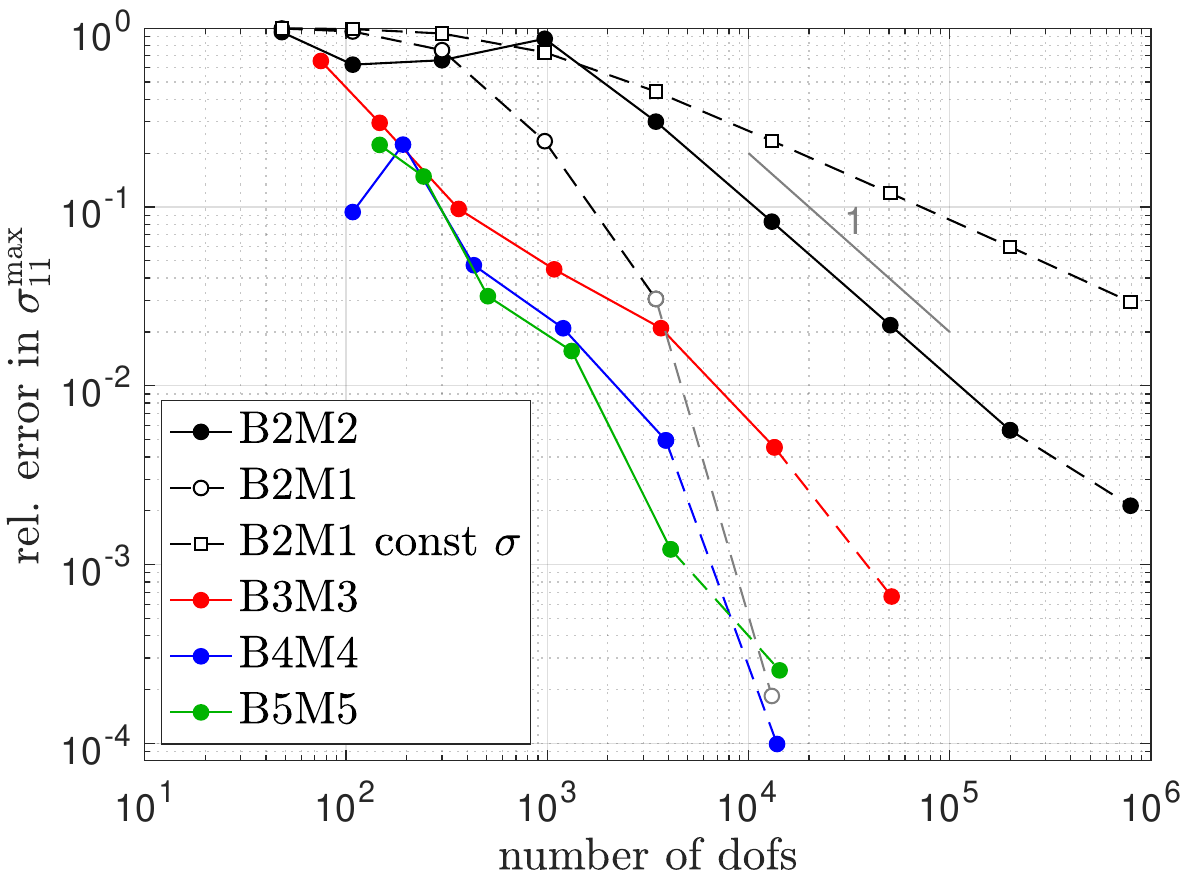}}
\put(0.2,-.1){\includegraphics[height=58mm]{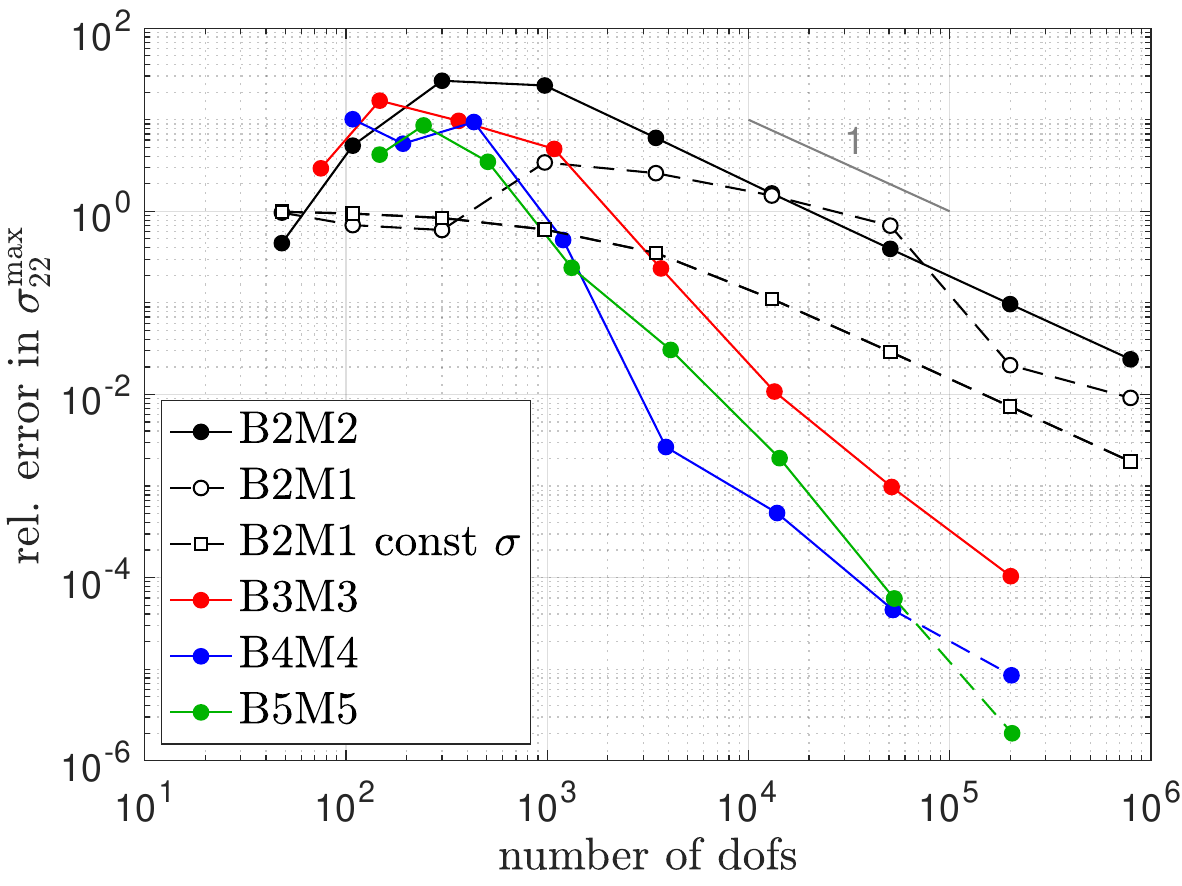}}
\put(-7.95,0.0){\footnotesize (a)}
\put(0.15,0.0){\footnotesize (b)}
\end{picture}
\caption{Hemisphere with hole: Convergence of the stress maxima $\sig_{11}^\mathrm{max}$ (a) and $\sig_{22}^\mathrm{max}$ (b) for $m=2, 4, 8, ..., 512$.}
\label{f:HHse}
\end{center}
\end{figure}

Figs.~\ref{f:HHsig1}-\ref{f:HHsig2a} show the stress plots of B2M2 and B2M1 for various mesh densities.
As in the prece-
ding examples, B2M2 exhibits large stress oscillations, even for fine meshes, while B2M1 yields much more accurate results. 
In case of $\sigma_{11}$ (see Fig.~\ref{f:HHsig1}), the regular raw stresses of B2M1 yield the best results, while the constant raw stresses (evaluated at the element center during post processing) yield the best results for $\sigma_{22}$ (see Fig.~\ref{f:HHsig2}). 
This can also be seen in the convergence plots of Fig.~\ref{f:HHse}.
As was already noted above, the relative accuracy of $\sig^\mathrm{max}_{11}$ is much lower than that of  $\sig^\mathrm{max}_{22}$.
This may be a result of their locations.
The figures show that the former occurs inside the domain, while the latter occurs at the boundary, right beneath the point load.

\subsubsection{Efficiency gains of B2M1}\label{s:HHa}

\begin{figure}[h]
\begin{center} \unitlength1cm
\begin{picture}(0,5.7)
\put(-8,-.1){\includegraphics[height=58mm]{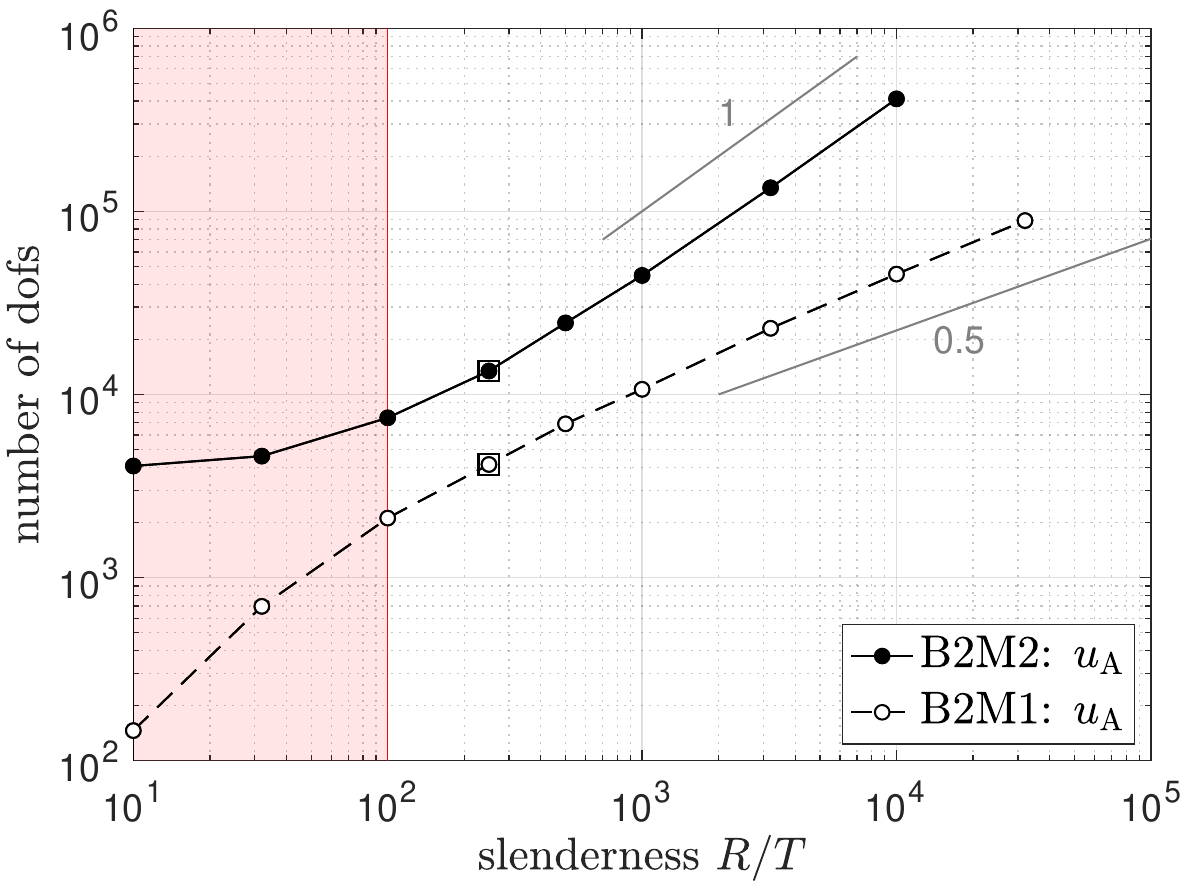}}
\put(0.2,-.1){\includegraphics[height=58mm]{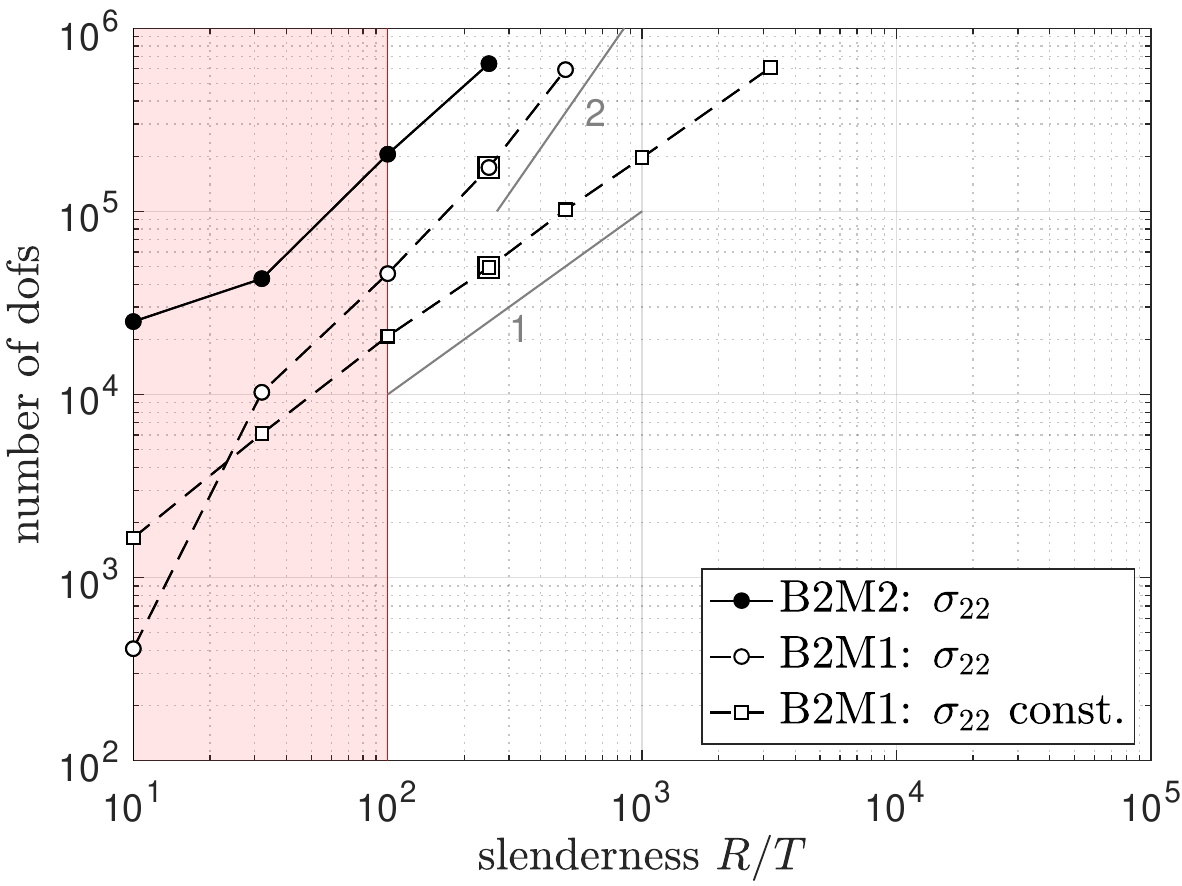}}
\put(-7.95,0.0){\footnotesize (a)}
\put(0.15,0.0){\footnotesize (b)}
\end{picture}
\caption{Efficiency gain of B2M1 vs.~B2M2 for the hemisphere with hole: 
The diagrams show the minimum number of dofs, as a function of slenderness $R/T$, required to achieve a given accuracy in (a)~$u_\mathrm{A}$ and (b) $\sig_{22}^\mathrm{max}$ -- here for the relative error falling below (a) $10^{-3}$ and (b) $3\cdot10^{-2}$.
For B2M2 the required dofs in (a) increase proportionally to $R/T$, while for B2M1 the required dofs increase with $\sqrt{R/T}$.
In (b) the rate of increase is similarly large for B2M2 and B2M1, but the latter is at a much lower level, especially for constant stress post processing.
The squares show the basic case discussed in Sec.~\ref{s:HHsetup}-\ref{s:HHs}. 
For $R/T = 10^4$ and $u_\mathrm{A}$ (a), for instance, B2M2 requires about 9 times more dofs than B2M1.}
\label{f:HHac}
\end{center}
\end{figure}

Fig.~\ref{f:HHac} shows the accuracy diagrams of $u_\mathrm{A}$ and $\sig^\mathrm{max}_{22}$ following the protocol used in Secs.~\ref{s:CCeff} and \ref{s:SLa}.
A lower accuracy level is taken for the stress now, as stress accuracy is harder to obtain for this example.
The accuracy in $\sig_{11}$ is too low altogether for this diagram, especially for low $R/T$.
The picture for $u_A$ is similar to the corresponding one of the Scordelis-Lo roof, see Fig.~\ref{f:SLac}a: 
For B2M2 the required dofs increase linearly with $R/T$, while for B2M1 it is only $\sqrt{R/T}$.
In case of $\sig_{22}$, the increase for B2M2 and B2M1 with constant stress post processing is linear, while regular B2M1 increases faster.
Thus the B2M1 stresses, contrary to the B2M1 displacements, are not as good for the hemisphere as for the Scodelis-Lo roof.
The reason is expected to lie in the double curvature, which affects the M1 mesh in a similar way than mesh distortion:
The elements are tapered towards the pole.
Accordingly, the azimuthal stress $\sig_{22}$ is affected more strongly than the hoop stress $\sig_{11}$.
This should be investigated further in future work.

\subsubsection{Influence of large deformations}\label{s:HHnl}

\begin{figure}[h]
\begin{center} \unitlength1cm
\begin{picture}(0,4.5)
\put(-7.9,-.2){\includegraphics[height=48mm]{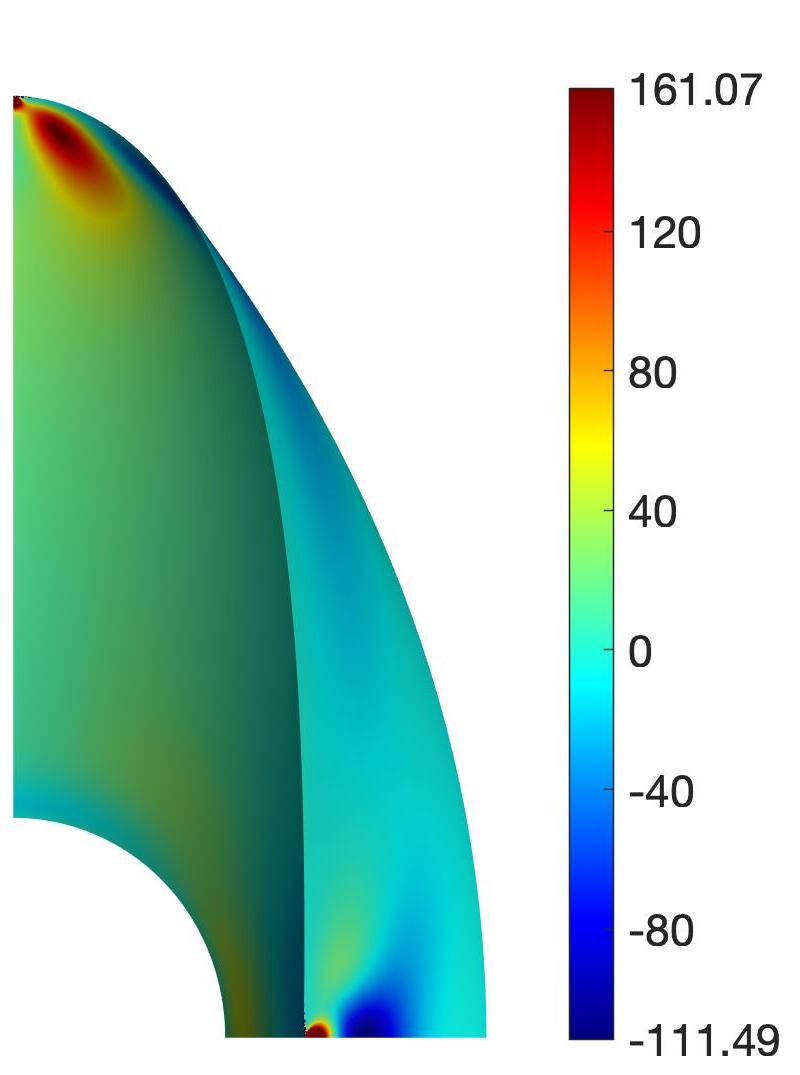}}
\put(-3.8,-.2){\includegraphics[height=48mm]{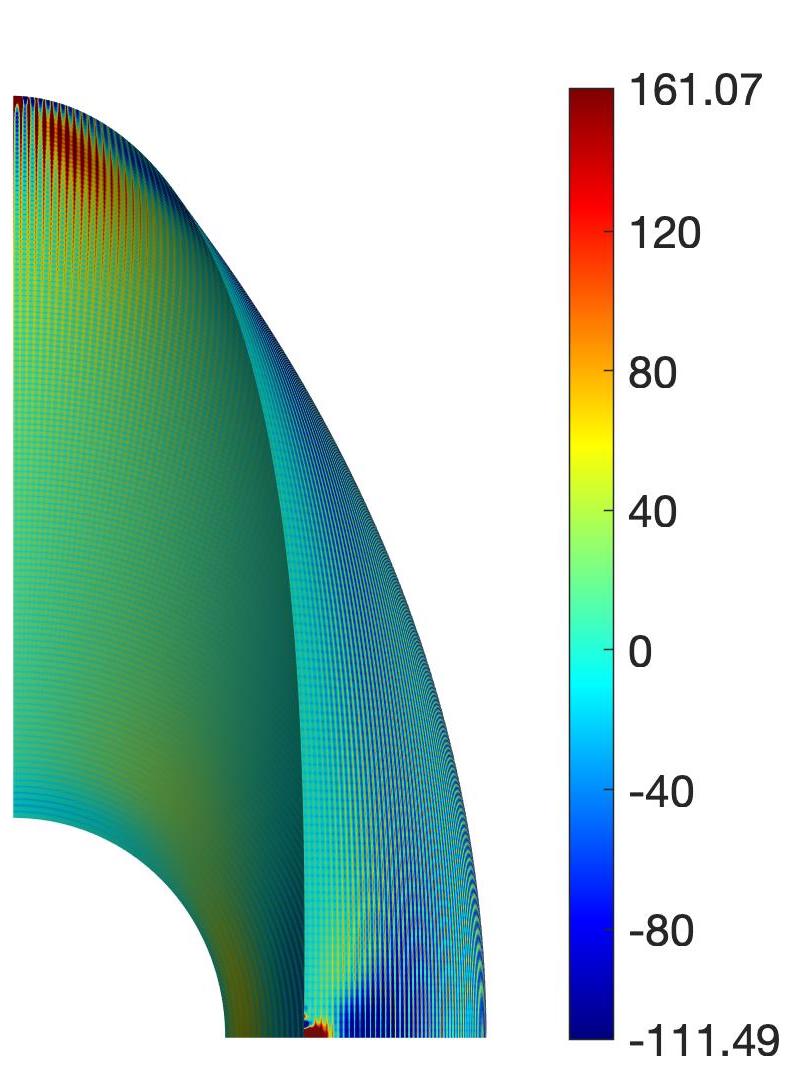}}
\put(0.3,-.2){\includegraphics[height=48mm]{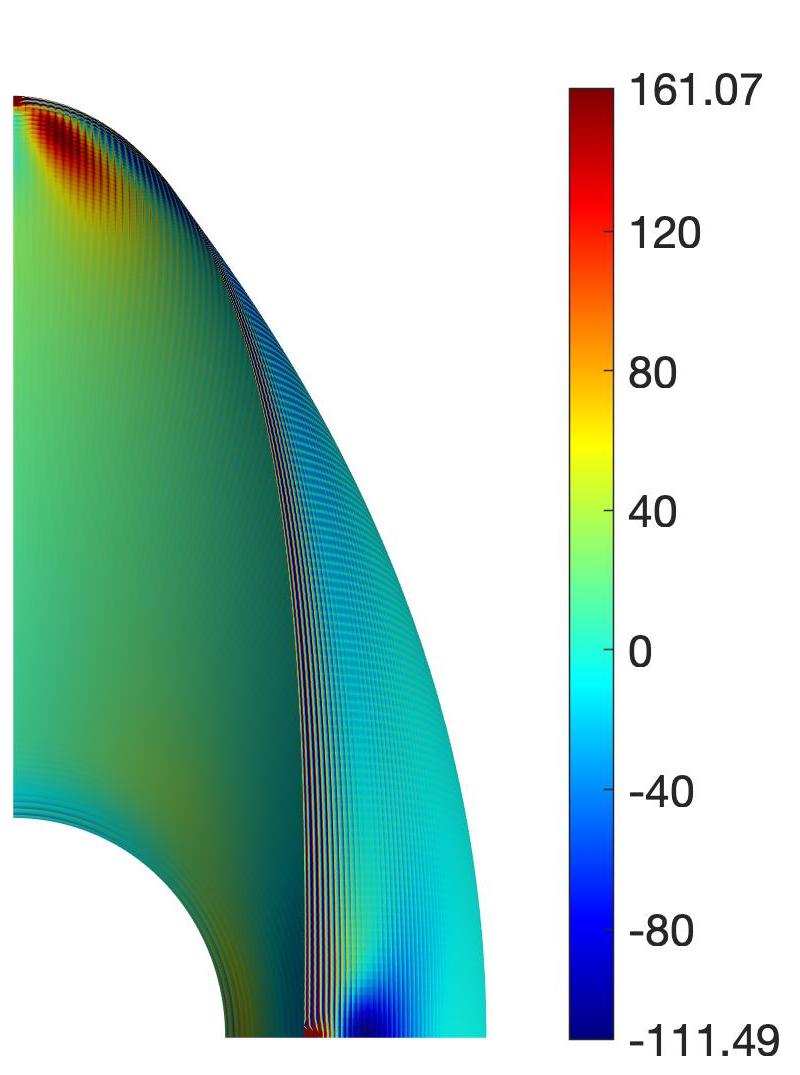}}
\put(4.4,-.2){\includegraphics[height=48mm]{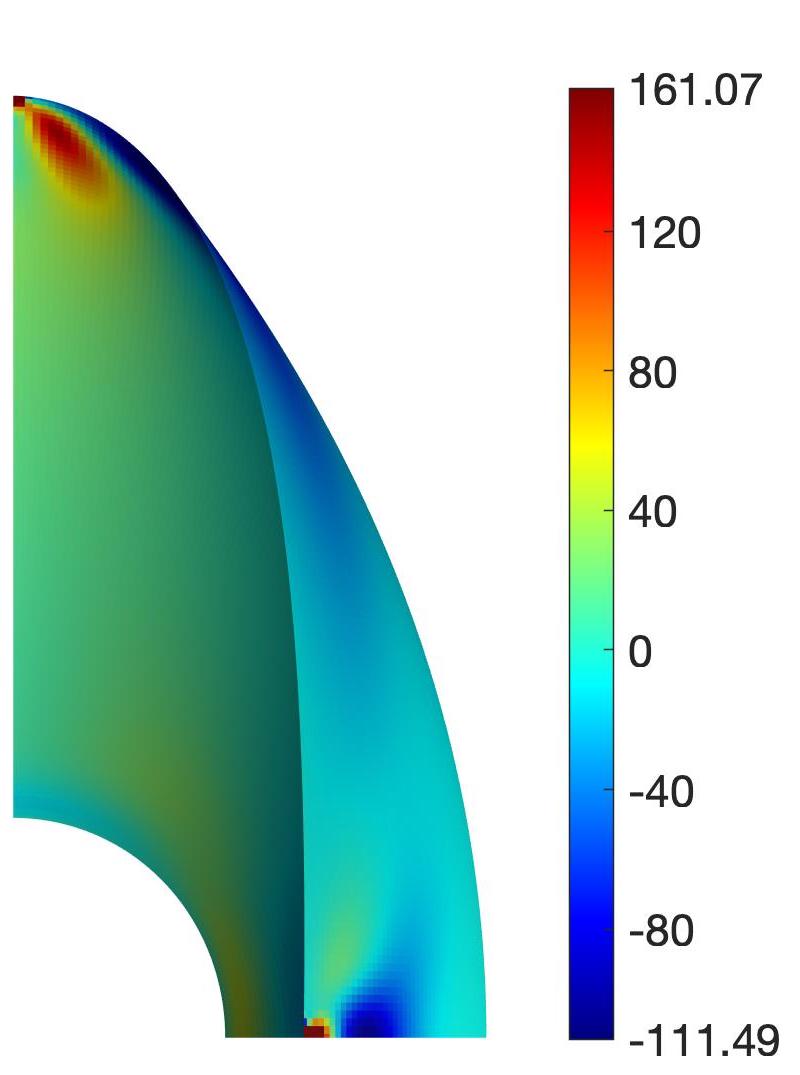}}
\end{picture}
\caption{Hemisphere with hole for large deformations ($R/T = 250$):  
Bottom view, from left to right: 
Raw membrane stress $\sig^2_2$ for B5M5 ($m=256$, reference result), B2M2 ($m=128$), regular B2M1 ($m=128$) and B2M1 with constant stress post processing ($m=128$). 
B2M2 shows major stress oscillations. 
These decrease in B2M1, especially for constant stress post processing.
The same colorscale is used in all results, truncating stress oscillations and singularities.}
\label{f:HHlarge}
\end{center}
\end{figure}

\begin{figure}[h]
\begin{center} \unitlength1cm
\begin{picture}(0,5.8)
\put(-8,-.1){\includegraphics[height=58mm]{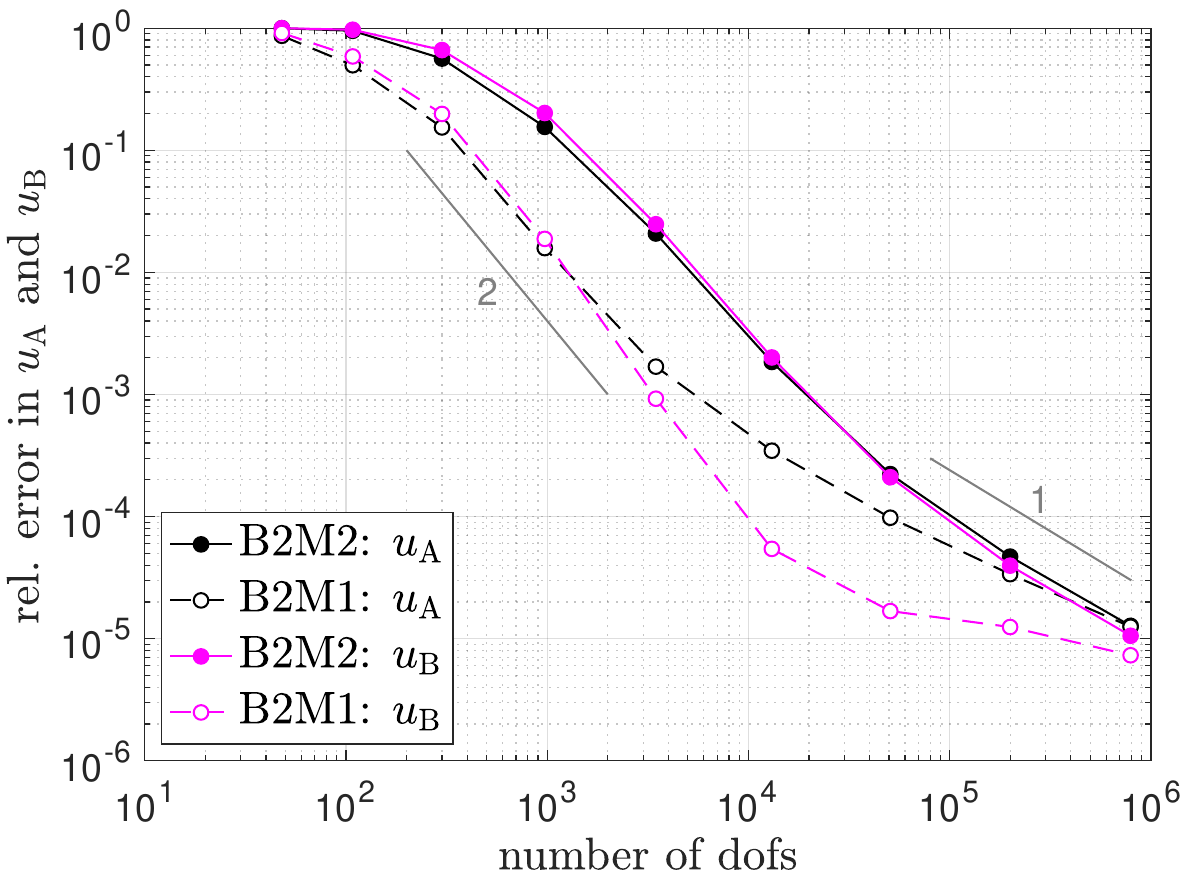}}
\put(0.2,-.1){\includegraphics[height=58mm]{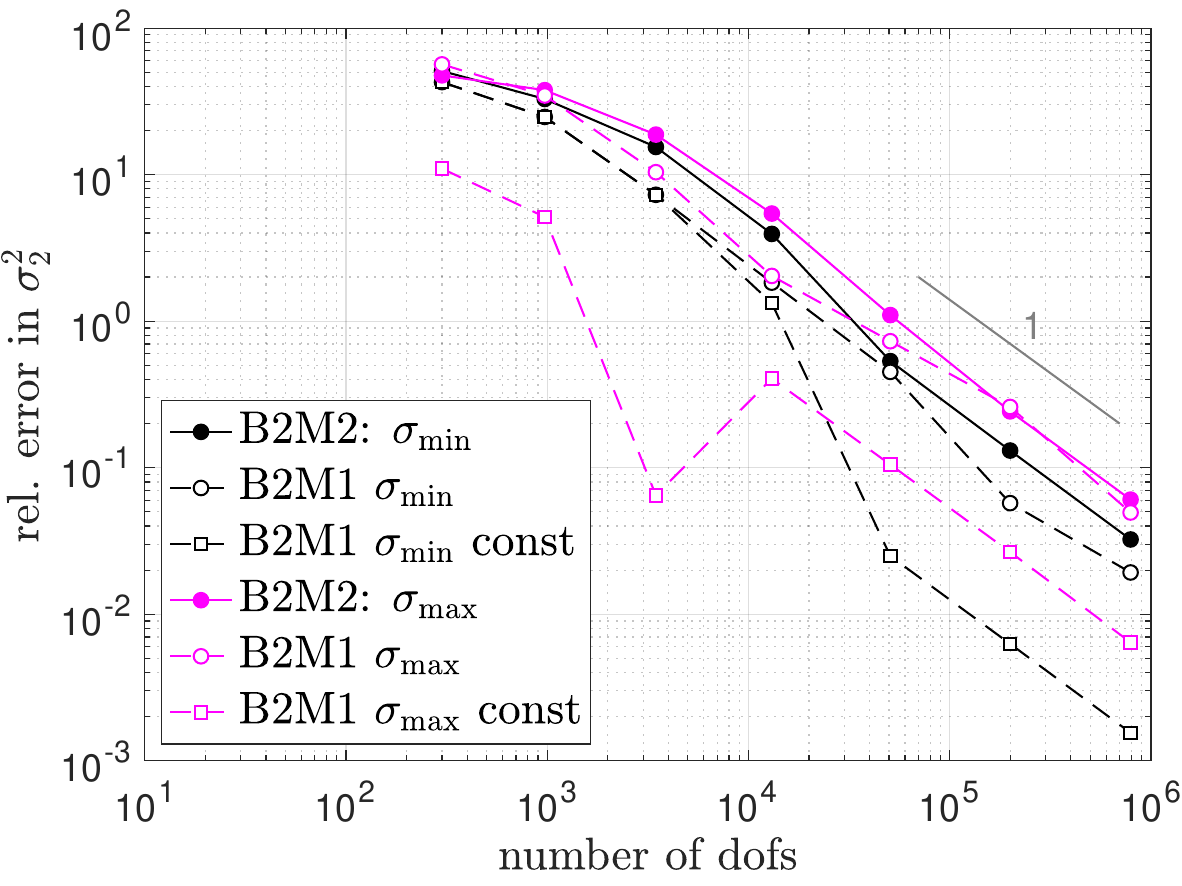}}
\put(-7.95,0.0){\footnotesize (a)}
\put(0.15,0.0){\footnotesize (b)}
\end{picture}
\caption{Hemisphere with hole for large deformations ($R/T = 250$): 
Convergence of (a) displacements $u_\mathrm{A}$ \& $u_\mathrm{B}$ and (b) stress $\sig^2_{2}$ for $m=2, 4, 8, ..., 512$.
Again B2M1 is more accurate than B2M2, especially for coarse meshes and constant stress post processing.}
\label{f:HHnl}
\end{center}
\end{figure}

Finally, the large deformation case is examined.
Therefore the load is increased by a factor of 100 and the problem is solved nonlinearly.
For slenderness $R/T=250$, the reference values 
$u_\mathrm{A} = 3.407360$ and $u_\mathrm{B} = -5.863051$ are obtained with discretization $p=5$ and $m=256$.
The membrane stresses are now singular, due to the point loads at points A and B, which now have in-plane components.
Apart from these singularities, the azimuthal stress\footnote{$\sigma^2_2$ is the stress along $\ba_2$ on the surface with normal $\ba^2$ (or vice versa). These directions do not coincide with $\bA_2$ and $\bA^2$ anymore, but they are parallel to the boundary normal $\bnu$ along the free boundaries.}  
field $\sigma^2_2$ has local maxima and minima close to points A and B, as seen in Fig.~\ref{f:HHlarge}.
The first image shows the reference results from B5M5.
The second and third image show the result from B2M2 and regular B2M1, respectively, which both exhibit oscillations.
The fourth image shows the result from B2M1 with constant stress post processing, where oscillations are absent.
This is consistent to the small deformation picture seen in Fig.~\ref{f:HHsig2}.
The convergence behavior for various displacement and stress components 
is shown in Fig.~\ref{f:HHnl}.
The displacement convergence follows the behavior already seen before (e.g.~in Figs.~\ref{f:CC_NL}a, \ref{f:SLlargeue}a \& \ref{f:HHue}b):
First stagnation, then a steep decrease with slope 2, followed by a slower decrease with slope 1, and finally ill-conditioning.
B2M1 outperforms B2M2 until the mesh becomes very dense, and it thus alleviates locking.
At $m=32$ ($n_\mathrm{dof} = 3468$), for instance, B2M1 is at least 10 times more accurate than B2M2 in the displacements.
Also the stress convergence follows that seen earlier (e.g.~in Figs.~\ref{f:SLlargeue}b \& \ref{f:HHse}b).
At $m=128$ ($n_\mathrm{dof} = 50700$), for instance, B2M1 with const.~$\sigma$ is at least 10 times more accurate than B2M2 in $\sig^2_2$.
These accuracy gains in the stresses are smaller than for the Scordelis-Lo roof and curved cantilever strip, in particular, which is attributed to the influence of tapered elements noted above.
Further gains are anticipated with force redistribution applied there.

\section{Conclusion}\label{s:concl}

The proposed new hybrid discretization approach B2M1 has been tested for small and large deformations on three classical benchmark examples that are characterized by increasing complexity. 
All examples give a consistent picture.
B2M1 generally outperforms the classical quadratic NURBS discretization B2M2, especially for coarse meshes and high slenderness ratios.
As Table~\ref{t:gains} shows, the accuracy gains in the displacements can exceed two orders of magnitude, 
while those in the stresses reach up to six orders of magnitude.
\begin{table}[h]
\centering
\vspace{2mm}
\begin{tabular}{|l|rl|rl|}
   \hline
   Problem & \multicolumn{2}{|c|}{Accuracy gain in $u$} & \multicolumn{2}{|c|}{Accuracy gain in $\sigma$} \\[0mm] \hline 
   & & & & \\[-4mm]   
   Curved cantilever strip -- linear &  ~~~~216 & (Fig.~\ref{f:CC_L2}a) & $5.5 \cdot 10^4$ & (Fig.~\ref{f:CC_L2}b)  \\ [0mm] 
   Curved cantilever strip -- nonlin. &  146 & (Fig.~\ref{f:CC_NL}a) & $1.1 \cdot 10^6$ & (Fig.~\ref{f:CC_NL}b) \\ [0mm] 
   Scordelis-Lo roof -- linear & 285 & (Fig.~\ref{f:SLue}d) & 9588 & (Fig.~\ref{f:SLse}b) \\ [0mm] 
   Scordelis-Lo roof -- skew mesh & 90 & (Fig.~\ref{f:SLskewe}b) & 8824 & (Fig.~\ref{f:SLskewe}d) \\ [0mm] 
   Scordelis-Lo roof -- nonlinear & 10 & (Fig.~\ref{f:SLlargeue}a) & 12 & (Fig.~\ref{f:SLlargeue}b) \\ [0mm] 
   Hemisphere -- linear & 67 & (Fig.~\ref{f:HHue}b) & 37 & (Fig.~\ref{f:HHse}b)  \\[0mm] 
   Hemisphere -- nonlinear & 37 & (Fig.~\ref{f:HHnl}a) & 21 & (Fig.~\ref{f:HHnl}b) \\[0mm] 
   \hline
\end{tabular}
\caption{Maximum accuracy gains of the B2M1 discretization over classical quadratic NURBS for the examples of Sec.~\ref{s:Nex}. 
The values denote the factors by which the errors decrease, and they follow from the mentioned figures, disregarding outliers.\\[-1mm]}
\label{t:gains}
\end{table}
Even large deformations and distorted meshes yield accuracy gains of more than one order of magnitude.
The accuracy gains of the B2M1 discretization allow for a more efficient solution, especially for large slendernesses, as Figs.~\ref{f:CC_acc}, \ref{f:SLac} and \ref{f:HHac} have demonstrated.
For particularly large slendernesses only B2M1 is actually able to solve the problem.
The accuracy gains are highest in the curved cantilever strip and lowest in the hemisphere with hole problem.
So they tend to decrease with problem complexity.
Further gains are anticipated with force redistribution, especially on distorted meshes.

There are several extensions of the proposed discretization approach that can be studied in future work.
One is the refinement of the present formulation:
It can be used with other membrane and bending material models -- also anisotropic ones,
it can be combined with multi-patch discretizations and corresponding continuity constraints \citep{patchshell},
it can be applied to large-scale examples,
and one can try extending it to higher orders.

Another extension is the application to other constraints and corresponding locking phenomena.
An obvious candidate is near-incompressibility, both in 2D and 3D.
Further candidates are shear and thickness locking in Reissner-Mindlin and Cosserat shells.
Also gradient elasticity could be investigated with the proposed new discretization.
The idea would be to use quadratic NURBS elements for the gradient parts together with linear Lagrange elements for the remaining parts. 
These applications can be expected to further benefit from higher-order versions of the proposed discretization approach.

Future work should also include further analysis of the proposed approach, especially for nonlinear problems and in comparison to existing mixed and assumed natural strain methods.

\bigskip

{\Large{\bf Acknowledgements}}

RAS acknowledges the support of EUROPIUM project funding from Gda\'{n}sk University of Technology.
The authors further thank Eshwar Savitha for his help at the onset of this research.

\appendix

\section{Nodal force redistribution coefficients and matrices}\label{s:W}

\begin{figure}[h!]
\begin{center} \unitlength1cm
\begin{picture}(0,18.3)
\put(-5.8,15.2){\includegraphics[height=30mm]{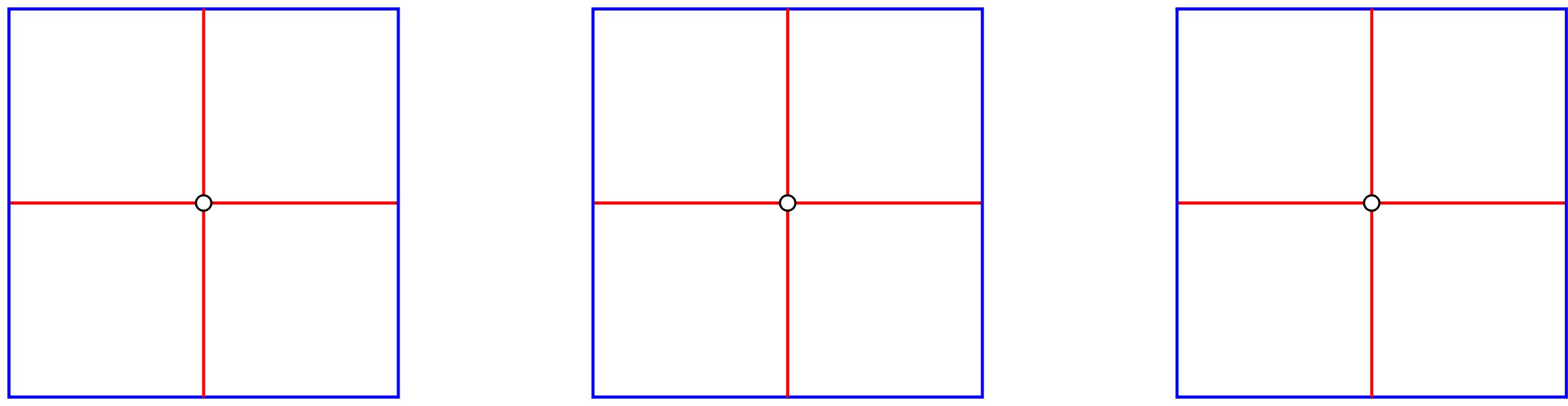}}
\put(-5.8,11.4){\includegraphics[height=30mm]{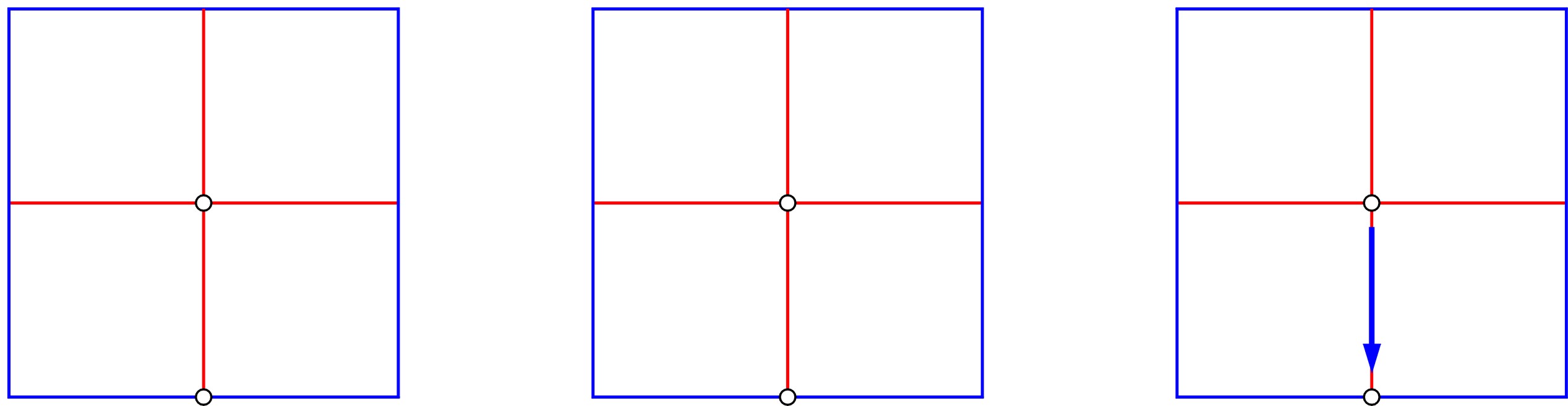}}
\put(-5.8,7.6){\includegraphics[height=30mm]{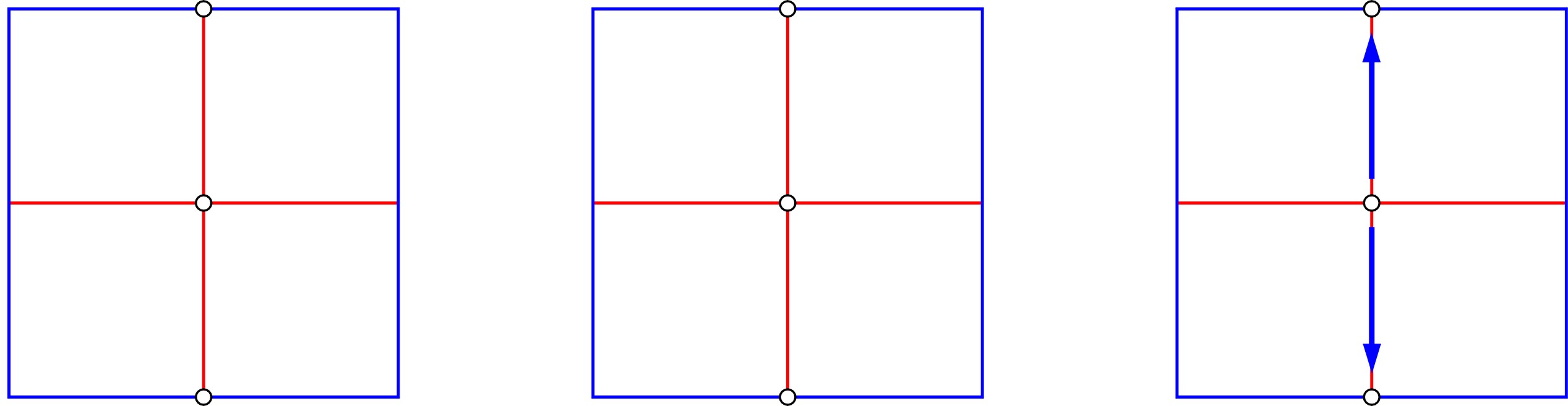}}
\put(-5.8,3.8){\includegraphics[height=30mm]{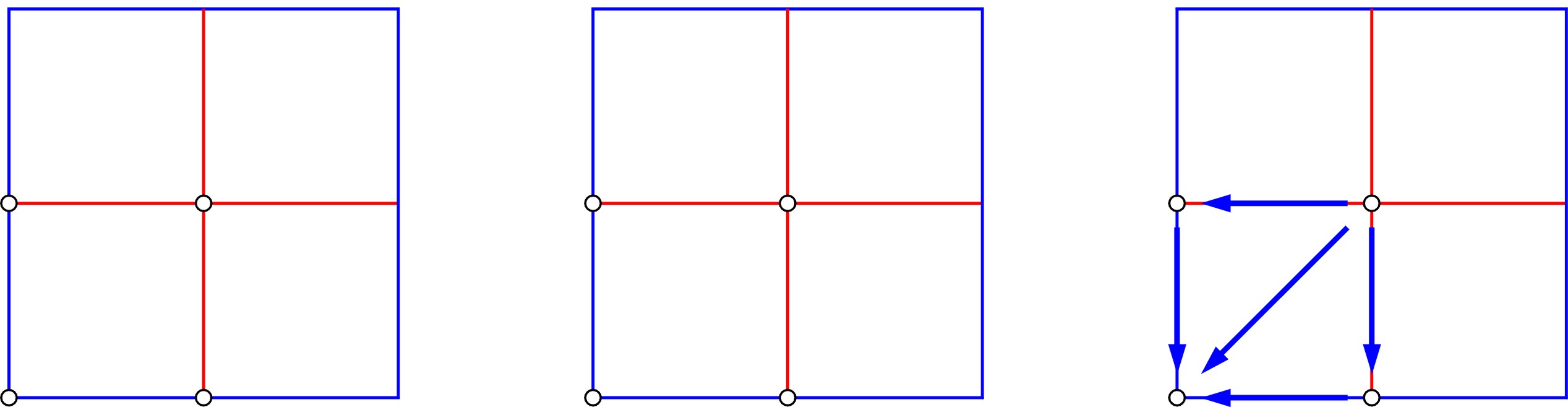}}
\put(-5.8,0){\includegraphics[height=30mm]{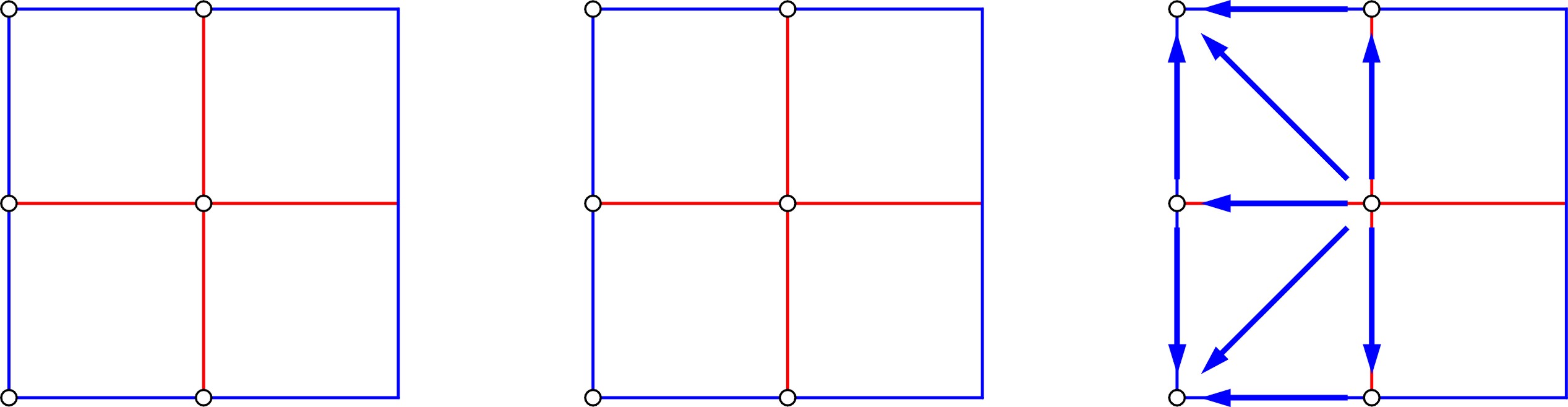}}
\put(-4.53,16.4){\scriptsize 1}
\put(-4.22,16.84){\footnotesize\textcolor{blue}{$1$}}
\put(0.09,16.84){\footnotesize\textcolor{red}{$1$}}
\put(4.42,16.84){\footnotesize{$\times 1$}}
\put(-4.53,11.57){\scriptsize 1}
\put(-4.55,13.0){\scriptsize 2}
\put(-4.22,13.11){\footnotesize\textcolor{blue}{$\frac{2}{3}$}}	
\put(-4.22,11.68){\footnotesize\textcolor{blue}{$\frac{1}{3}$}}
\put(0.09,13.11){\footnotesize\textcolor{red}{$\frac{3}{4}$}}		
\put(0.09,11.68){\footnotesize\textcolor{red}{$\frac{1}{4}$}}
\put(4.42,13.11){\footnotesize $\times\frac{8}{9}$}  	
\put(4.42,11.6){\footnotesize $\times 1$}
\put(4.42,12.25){\footnotesize\textcolor{blue}{$+\frac{1}{9}$}}
\put(-4.53,7.77){\scriptsize 1}
\put(-4.55,9.2){\scriptsize 2}
\put(-4.55,10.63){\scriptsize 3}
\put(-4.22,10.74){\footnotesize\textcolor{blue}{$\frac{1}{3}$}}	
\put(-4.22,9.31){\footnotesize\textcolor{blue}{$\frac{1}{3}$}}	
\put(-4.22,7.88){\footnotesize\textcolor{blue}{$\frac{1}{3}$}}
\put(0.09,10.74){\footnotesize\textcolor{red}{$\frac{1}{4}$}}	
\put(0.09,9.31){\footnotesize\textcolor{red}{$\frac{1}{2}$}}		
\put(0.09,7.88){\footnotesize\textcolor{red}{$\frac{1}{4}$}}
\put(4.42,10.66){\footnotesize $\times 1$}
\put(4.42,9.31){\footnotesize $\times\frac{2}{3}$}  	
\put(4.42,7.8){\footnotesize $\times 1$}
\put(4.42,10){\footnotesize\textcolor{blue}{$+\frac{1}{6}$}}
\put(4.42,8.45){\footnotesize\textcolor{blue}{$+\frac{1}{6}$}}
\put(-5.97,3.97){\scriptsize 1}
\put(-5.99,5.4){\scriptsize 3}
\put(-4.55,3.97){\scriptsize 2}
\put(-4.55,5.4){\scriptsize 4}
\put(-5.67,5.51){\footnotesize\textcolor{blue}{$\frac{2}{9}$}}	
\put(-5.67,4.08){\footnotesize\textcolor{blue}{$\frac{1}{9}$}}
\put(-4.22,5.51){\footnotesize\textcolor{blue}{$\frac{4}{9}$}}	
\put(-4.22,4.08){\footnotesize\textcolor{blue}{$\frac{2}{9}$}}
\put(-1.37,5.51){\footnotesize\textcolor{red}{$\frac{3}{16}$}}	
\put(-1.37,4.08){\footnotesize\textcolor{red}{$\frac{1}{16}$}}
\put(0.07,5.51){\footnotesize\textcolor{red}{$\frac{9}{16}$}}	
\put(0.07,4.08){\footnotesize\textcolor{red}{$\frac{3}{16}$}}
\put(2.3,5.23){\footnotesize $\times\frac{8}{9}$}  	
\put(2.35,3.8){\footnotesize $\times 1$}
\put(4.4,5.51){\footnotesize $\times\frac{64}{81}$}  	
\put(4.42,4.08){\footnotesize $\times\frac{8}{9}$}
\put(2.3,4.55){\footnotesize\textcolor{blue}{$+\frac{1}{9}$}}
\put(3.05,4.77){\footnotesize\textcolor{blue}{$+\frac{1}{81}$}}
\put(3.25,5.51){\footnotesize\textcolor{blue}{$+\frac{8}{81}$}}
\put(3.55,4.08){\footnotesize\textcolor{blue}{$+\frac{1}{9}$}}
\put(4.4,4.75){\footnotesize\textcolor{blue}{$+\frac{8}{81}$}}
%
%
\put(-5.97,.17){\scriptsize 1}
\put(-5.99,1.6){\scriptsize 3}
\put(-5.99,3.03){\scriptsize 5}
\put(-4.55,.17){\scriptsize 2}
\put(-4.55,1.6){\scriptsize 4}
\put(-4.55,3.03){\scriptsize 6}
\put(-5.67,3.14){\footnotesize\textcolor{blue}{$\frac{1}{9}$}}	
\put(-5.67,1.71){\footnotesize\textcolor{blue}{$\frac{1}{9}$}}	
\put(-5.67,.28){\footnotesize\textcolor{blue}{$\frac{1}{9}$}}
\put(-4.22,3.14){\footnotesize\textcolor{blue}{$\frac{2}{9}$}}	
\put(-4.22,1.71){\footnotesize\textcolor{blue}{$\frac{2}{9}$}}	
\put(-4.22,.28){\footnotesize\textcolor{blue}{$\frac{2}{9}$}}
\put(-1.37,3.14){\footnotesize\textcolor{red}{$\frac{1}{16}$}}	
\put(-1.35,1.71){\footnotesize\textcolor{red}{$\frac{1}{8}$}}		
\put(-1.37,.28){\footnotesize\textcolor{red}{$\frac{1}{16}$}}
\put(0.07,3.14){\footnotesize\textcolor{red}{$\frac{3}{16}$}}	
\put(0.09,1.71){\footnotesize\textcolor{red}{$\frac{3}{8}$}}		
\put(0.07,.28){\footnotesize\textcolor{red}{$\frac{3}{16}$}}
\put(2.35,2.86){\footnotesize $\times 1$} 			
\put(2.3,1.43){\footnotesize $\times\frac{2}{3}$}  	
\put(2.35,0){\footnotesize $\times 1$}
\put(4.4,3.14){\footnotesize $\times\frac{8}{9}$}		
\put(4.42,1.71){\footnotesize $\times\frac{16}{27}$}  	
\put(4.4,.28){\footnotesize $\times\frac{8}{9}$}
\put(2.3,0.75){\footnotesize\textcolor{blue}{$+\frac{1}{6}$}}
\put(2.3,2.15){\footnotesize\textcolor{blue}{$+\frac{1}{6}$}}
\put(3.35,3.14){\footnotesize\textcolor{blue}{$+\frac{1}{9}$}}
\put(3.43,2.4){\footnotesize\textcolor{blue}{$+\frac{1}{54}$}}
\put(3.05,1.71){\footnotesize\textcolor{blue}{$+\frac{2}{27}$}}
\put(3.05,0.97){\footnotesize\textcolor{blue}{$+\frac{1}{54}$}}
\put(3.55,0.28){\footnotesize\textcolor{blue}{$+\frac{1}{9}$}}
\put(4.42,2.4){\footnotesize\textcolor{blue}{$+\frac{4}{27}$}}
\put(4.42,0.95){\footnotesize\textcolor{blue}{$+\frac{4}{27}$}}
\end{picture}
\caption{Nodal force redistribution for center, edge, double edge, corner and double corner elements (from top to bottom): 
Left: Force distribution on B2 elements (in blue; with local node numbers in black). 
Center: Force distribution on M1 elements (in red).
Right: Force redistribution coefficients for the M1 elements (in black \& blue).}
\label{f:B2M1FR}
\end{center}
\end{figure}

For coarse meshes, edge and corner elements require the redistribution of the nodal forces of M1 elements to be consistent with the nodal forces of B2 elements.
For meshes with more than one element, the five elemental redistribution cases shown in Fig.~\ref{f:B2M1FR} exist.
%
The corresponding redistribution matrices are
\eqb{l}
\mw^e_1 = 1\,,
\eqe
\eqb{l}
\mw^e_2 = \ds\frac{1}{9}\begin{bmatrix}
9 & 1 \\
0 & 8
\end{bmatrix},
\eqe
\eqb{l}
\mw^e_3 = \ds\frac{1}{6}\begin{bmatrix}
6 & 1 & 0 \\
0 & 4 & 0 \\
0 & 1 & 6
\end{bmatrix},
\eqe
\eqb{l}
\mw^e_4 = \ds\frac{1}{81}\begin{bmatrix}
81 & 9 & 9 & 1 \\
0 & 72 & 0 & 8 \\
0 & 0 & 72 & 8 \\
0 & 0 & 0 & 64
\end{bmatrix}
\eqe
and
\eqb{l}
\mw^e_5 = \ds\frac{1}{54}\begin{bmatrix}
54 & 6 & 9 & 1 & 0 & 0 \\
0 & 48 & 0 & 8 & 0 & 0 \\
0 & 0 & 36 & 4 & 0 & 0 \\
0 & 0 & 0 & 32 & 0 & 0 \\
0 & 0 & 9 & 1 & 54 & 6 \\
0 & 0 & 0 & 8 & 0 & 48
\end{bmatrix},
\eqe
based on the nodal numbering shown in Fig.~\ref{f:B2M1FR}.
Since the nodal force is a vector in $\bbR^3$, each entry still needs to be multiplied by the identity $\bone$.
Note that the column sum is always 1.

\section{Analytical solution for the curved cantilever strip} \label{s:anasol}

Introducing the cylindrical basis (see Fig.~\ref{f:CC0})
\eqb{lll}
\be_r \is \sin\theta\,\be_1 + \cos\theta\,\be_3\,, \\[1mm]
\be_\theta \is \cos\theta\,\be_1 - \sin\theta\,\be_3\,,
\eqe
with $\be_{r,\theta} = \be_\theta$ and $\be_{\theta,\theta} = -\be_r$,
the initial configuration becomes
\eqb{l}
\bX = R\,\be_r\,,
\eqe
which is only a function of angle $\theta$, since $R =$ const.
This leads to the tangent vectors $\bA_1 = \bX_{,1} = \bX_{,\theta} = R\,\be_\theta$ and $\bA^1 =\be_\theta/R$, and the surface normal $\bN = \be_r$.
The displacement field in the polar basis is
\eqb{l}
\bu = u_\theta\,\be_\theta + u_r\,\be_r\,.
\label{e:bu}\eqe
The infinitesimal strain $\eps := \eps^1_1 = \bA^1\cdot\bu_{,1}$ and curvature $\kappa := \kappa^1_1 = A^{11} \bu_{,11} \cdot \bN $ thus become
\eqb{l}
\eps = \ds\frac{u_{\theta,\theta} + u_r}{R}\,,\quad
\kappa = \ds\frac{u_{r,\theta\theta} - 2u_{\theta,\theta} - u_r}{R^2}\,.
\eqe
The known stress and moment distributions \eqref{e:sigana} and (\ref{e:NMana}.2), together with the constitutive equations 
\eqb{l}
\sig = ET\,\eps\,,\quad
M = \ds\frac{ET^3}{12}\,\kappa\,,
\eqe
then lead to the two coupled ODEs
\eqb{l}
u_r + u_{\theta,\theta} = \ds\frac{2qR}{ET}\cos\theta \,,\quad
u_{r,\theta\theta} - 2u_{\theta,\theta} - u_r = \ds\frac{12qR^3}{ET^3}\cos\theta\,.
\label{e:ODE1}\eqe
Eliminating $u_{\theta,\theta}$ then gives
\eqb{l}
u_{r,\theta\theta} + u_r = 2u_0\cos\theta \,,\quad u_0:= \ds\frac{qR}{ET}\bigg(2 + 6\frac{R^2}{T^2}\bigg)\,,
\eqe
which is solved by
\eqb{l}
u_r(\theta) = u_0\,\theta\sin\theta
\eqe
for the present boundary condition $u_r(0) = 0$.
This can be rewritten as
\eqb{l}
u_r(\theta) = u_\mathrm{A}\ds\frac{2\theta}{\pi}\sin\theta\,,
\eqe
where
\eqb{l}
u_\mathrm{A} := u_\mathrm{Ab} \bigg(1 + \ds\frac{T^2}{3R^2} \bigg)\,,\quad 
u_\mathrm{Ab} := 3\pi\ds\frac{qR^3}{ET^3}\,,
\eqe
is the horizontal tip displacement along $\be_1$.
For $T\ll R$ it is dominated by its bending contribution $u_\mathrm{Ab}$.
A common choice in the literature is $qR^3/(ELT^3) = -0.1$, leading to $u_\mathrm{Ab} = -0.3\pi L$. 
\\
From Eq.~(\ref{e:ODE1}.1) now follows
\eqb{l}
u_{\theta,\theta} = -u_\mathrm{A}\ds\frac{2\theta}{\pi}\sin\theta + \ds\frac{2qR}{ET}\cos\theta \,.
\eqe
Integrating this gives
\eqb{l}
u_\theta(\theta) = -w_\mathrm{A}\sin\theta + u_\mathrm{A}\ds\frac{2\theta}{\pi}\cos\theta\,,
\eqe
for the present boundary condition $u_\theta(0) = 0$.
Here,
\eqb{l}
w_\mathrm{A} := \ds\frac{6qR^3}{ET^3}
\eqe
denotes the vertical tip displacement along $\be_3$.

\bibliographystyle{apalike}
\bibliography{bibliography}

\end{document}

%% file: neco.tex
\newcommand{\bitm}{\begin{itemize}}
\newcommand{\eitm}{\end{itemize}}
\newcommand{\bnumr}{\begin{enumerate}}
\newcommand{\enumr}{\end{enumerate}}

\newcommand {\eqb}[1]{\begin{equation}\begin{array}{#1}}
\newcommand {\eqe}{\end{array}\end{equation}}

\newcommand {\esb}[1]{\begin{equation*}\begin{array}{#1}}
\newcommand {\ese}{\end{array}\end{equation*}}
\newcommand {\ds}{\displaystyle}

\newcommand {\back}{\! \! \!}
\newcommand {\is}{\back &=& \back}
\newcommand {\dis}{\back &:=& \back}

\newcommand {\norm}[1]{\|#1\|}

\newcommand {\dif}{\mathrm{d}}

\newcommand {\II}{{I\kern-.3em I}}
\newcommand {\III}{{I\kern-.3em I\kern-.3em I}}

\newcommand {\mra}{\mathrm{a}}
\newcommand {\mrb}{\mathrm{b}}
\newcommand {\mrc}{\mathrm{c}}
\newcommand {\mrd}{\mathrm{d}}

\newcommand {\mrm}{\mathrm{m}}

\newcommand {\mrr}{\mathrm{r}}

\newcommand{\mrT}{\mathrm{T}}

\newcommand {\mf}{\mathbf{f}}

\newcommand {\mk}{\mathbf{k}}

\newcommand {\muu}{\mathbf{u}}

\newcommand {\mw}{\mathbf{w}}
\newcommand {\mx}{\mathbf{x}}

\newcommand {\ba}{\boldsymbol{a}}

\newcommand {\be}{\boldsymbol{e}}
\newcommand {\bff}{\boldsymbol{f}}

\newcommand {\bn}{\boldsymbol{n}}

\newcommand {\bu}{\boldsymbol{u}}

\newcommand {\bx}{\boldsymbol{x}}

\newcommand {\bnu}{\mbox{\boldmath$\nu$}}

\newcommand {\mK}{\mathbf{K}}

\newcommand {\mN}{\mathbf{N}}

\newcommand {\mW}{\mathbf{W}}
\newcommand {\mX}{\mathbf{X}}

\newcommand {\bA}{\boldsymbol{A}}

\newcommand {\bM}{\boldsymbol{M}}
\newcommand {\bN}{\boldsymbol{N}}

\newcommand {\bT}{\boldsymbol{T}}

\newcommand {\bX}{\boldsymbol{X}}

\newcommand {\eps}{\varepsilon}
\newcommand {\sig}{\sigma}

\newcommand {\bsig}{\mbox{\boldmath$\sigma$}}

\newcommand {\bone}{\mathbf{1}}

\newcommand {\bbR}{\mathbb{R}}

\newcommand {\IR}{{\rm\kern.24em
   \vrule width.02em height1.53ex depth-.05ex
   \kern-.3em R}}
\newcommand {\ic}{{\rm\kern.20em
   \vrule width.02em height1.0ex depth-.05ex
   \kern-.22em c}}
\newcommand {\ia}{{\rm\kern.20em
   \vrule width.02em height1.05ex depth-.0ex
   \kern-.25em a}}
\newcommand {\IC}{{\rm\kern.24em
   \vrule width.02em height1.4ex depth-.05ex
   \kern-.26em C}}
\newcommand {\ID}{{\rm\kern.34em
   \vrule width.02em height1.5ex depth-.05ex
   \kern-.36em D}}
\newcommand {\IS}{{\rm\kern.24em
   \vrule width.02em height1.6ex depth.05ex
   \kern-.26em S}}
\newcommand {\IT}{{\rm\kern.50em
   \vrule width.02em height1.55ex depth-.05ex
   \kern-.52em T}}

\newcommand {\IE}{{\rm\kern.24em
   \vrule width.02em height1.55ex depth-.05ex
   \kern-.33em E}}
\newcommand {\IEa}{{\rm\kern.24em
   \vrule width.02em height1.55ex depth-.05ex
   \kern-.33em E}^{1}_{ijkl}}
\newcommand {\IEb}{{\rm\kern.24em
   \vrule width.02em height1.55ex depth-.05ex
   \kern-.33em E}^{2}_{ijkl}}

\newcommand {\sC}{\mathcal{C}}

\newcommand {\sH}{\mathcal{H}}

\newcommand {\sS}{\mathcal{S}}

\newcommand {\sV}{\mathcal{V}}

\newcommand {\Ass}[2]{\kern 0.9ex \vrule width0.45em height0.2ex depth0ex \kern -2.1ex \bigwedge_{#1}^{#2}}
\newcommand {\ASS}[2]{\kern 1.45ex \vrule width0.5em height0.2ex depth0ex \kern -2.65ex \bigwedge_{#1}^{#2}}




%% file: B2M1.bbl
\begin{thebibliography}{}

\bibitem[Adam et~al., 2014]{adam14}
Adam, C., Bouabdallah, S., Zarroug, M., and Maitournam, H. (2014).
\newblock Improved numerical integration for locking treatment in isogeometric
  structural elements. {P}art {I}: {B}eams.
\newblock {\em Comput. Methods Appl. Mech. Eng.}, {\bf 279}:1--28.

\bibitem[Adam et~al., 2015a]{adam15}
Adam, C., Bouabdallah, S., Zarroug, M., and Maitournam, H. (2015a).
\newblock Improved numerical integration for locking treatment in isogeometric
  structural elements. {P}art {II}: {P}lates and shells.
\newblock {\em Comput. Methods Appl. Mech. Eng.}, {\bf 284}:106--137.

\bibitem[Adam et~al., 2015b]{adam15a}
Adam, C., Hughes, T. J.~R., Bouabdallah, S., Zarroug, M., and Maitournam, H.
  (2015b).
\newblock Selective and reduced numerical integrations for {NURBS}-based
  isogeometric analysis.
\newblock {\em Comput. Methods Appl. Mech. Eng.}, {\bf 284}:732--761.

\bibitem[Andelfinger and Ramm, 1993]{andelfinger93}
Andelfinger, U. and Ramm, E. (1993).
\newblock {EAS}-elements for two-dimensional, three-dimensional, plate and
  shell structures and their equivalence to {HR}-elements.
\newblock {\em Int. J. Numer. Meth. Eng.}, {\bf 36}:1311--1337.

\bibitem[Antolin et~al., 2020]{antolin20}
Antolin, P., Kiendl, J., Pingaro, M., and Reali, A. (2020).
\newblock A simple and effective method based on strain projections to
  alleviate locking in isogeometric solid shells.
\newblock {\em Comput. Mech.}, {\bf 65}:1621--1631.

\bibitem[Bathe and Dvorkin, 1986]{bathe86}
Bathe, K.-J. and Dvorkin, E.~N. (1986).
\newblock A formulation of general shell elements -- {T}he use of mixed
  interpolation of tensorial components.
\newblock {\em Int. J. Numer. Meth. Eng.}, {\bf 22}:697--722.

\bibitem[Belytschko et~al., 1985]{belytschko85}
Belytschko, T., Stolarski, H., Liu, W.~K., Carpenter, N., and Ong, J. S.-J.
  (1985).
\newblock Stress projection for membrane and shear locking in shell finite
  elements.
\newblock {\em Comput. Methods Appl. Mech. Eng.}, {\bf 51}:221--258.

\bibitem[Benson et~al., 2011]{benson10}
Benson, D.~J., Bazilevs, Y., Hsu, M.-C., and Hughes, T. J.~R. (2011).
\newblock Isogeometric shell analysis: {T}he {R}eissner-{M}indlin shell.
\newblock {\em Comput. Methods Appl. Mech. Eng.}, {\bf 199}(5-8):276--289.

\bibitem[Bieber et~al., 2022]{bieber22}
Bieber, S., Oesterle, B., Bischoff, M., and Ramm, E. (2022).
\newblock Strategy for preventing membrane locking through reparametrization.
\newblock In Aldakheel, F., Hudobivnik, B., Soleimani, M., Wessels, H., {Wei\ss
  enfels}, C., and Marino, M., editors, {\em Current Trends and Open Problems
  in Computational Mechanics}, pages 61--73. Springer.

\bibitem[Bieber et~al., 2018]{bieber18}
Bieber, S., Oesterle, B., Ramm, E., and Bischoff, M. (2018).
\newblock A variational method to avoid locking -- independent of the
  discretization scheme.
\newblock {\em Int. J. Numer. Meth. Eng.}, {\bf 114}:801--827.

\bibitem[Bombarde et~al., 2022]{bombarde22a}
Bombarde, D.~S., Agrawal, M., Gautama, S.~S., and Nandy, A. (2022).
\newblock {H}ellinger-{R}eissner principle based stress–displacement
  formulation for three-dimensional isogeometric analysis in linear elasticity.
\newblock {\em Comput. Methods Appl. Mech. Eng.}, {\bf 394}:114920.

\bibitem[Borden et~al., 2011]{borden11}
Borden, M.~J., Scott, M.~A., Evans, J.~A., and Hughes, T. J.~R. (2011).
\newblock Isogeometric finite element data structures based on {B}ezier
  extraction of {NURBS}.
\newblock {\em Int. J. Numer. Meth. Eng.}, {\bf 87}:15--47.

\bibitem[Bouclier et~al., 2012]{bouclier12}
Bouclier, R., Elguedj, T., and Combescure, A. (2012).
\newblock Locking free isogeometric formulations of curved thick beams.
\newblock {\em Comput. Methods Appl. Mech. Eng.}, {\bf 245}:144--162.

\bibitem[Bouclier et~al., 2013]{bouclier13}
Bouclier, R., Elguedj, T., and Combescure, A. (2013).
\newblock Efficient isogeometric {NURBS}-based solid-shell elements: {M}ixed
  formulation and {$\bar B$}-method.
\newblock {\em Comput. Methods Appl. Mech. Eng.}, {\bf 267}:86--110.

\bibitem[Bouclier et~al., 2015]{bouclier15}
Bouclier, R., Elguedj, T., and Combescure, A. (2015).
\newblock An isogeometric locking-free {NURBS}-based solid-shell element for
  geometrically nonlinear analysis.
\newblock {\em Int. J. Numer. Meth. Eng.}, {\bf 101}:774--808.

\bibitem[Caseiro et~al., 2014]{caseiro14}
Caseiro, J.~F., Valente, R. A.~F., Reali, A., Kiendl, J., Auricchio, F., and
  de~Sousa, R. J.~A. (2014).
\newblock On the {A}ssumed {N}atural {S}train method to alleviate locking in
  solid-shell {NURBS}-based finite elements.
\newblock {\em Comput. Mech.}, {\bf 53}:1341--1353.

\bibitem[Casquero and Golestanian, 2022]{casquero22}
Casquero, H. and Golestanian, M. (2022).
\newblock Removing membrane locking in quadratic {NURBS}-based discretizations
  of linear plane {K}irchhoff rods: {CAS} elements.
\newblock {\em Comput. Methods Appl. Mech. Eng.}, {\bf 399}:115354.

\bibitem[Casquero and Golestanian, 2024]{casquero24}
Casquero, H. and Golestanian, M. (2024).
\newblock Vanquishing volumetric locking in quadratic {NURBS}-based
  discretizations of nearly-incompressible linear elasticity: {CAS} elements.
\newblock {\em Comput. Mech.}, {\bf 73}(6):1241--1252.

\bibitem[Casquero and Mathews, 2023]{casquero23}
Casquero, H. and Mathews, K.~D. (2023).
\newblock Overcoming membrane locking in quadratic {NURBS}-based
  discretizations of linear {K}irchhoff-{L}ove shells: {CAS} elements.
\newblock {\em Comput. Methods Appl. Mech. Eng.}, {\bf 417}:116523.

\bibitem[Ciarlet, 2005]{ciarlet05}
Ciarlet, P.~G. (2005).
\newblock An introduction to differential geometry with applications to
  elasticity.
\newblock {\em J. Elast.}, {\bf 78}:1--215.

\bibitem[Duong et~al., 2022]{textshell2}
Duong, T.~X., Itskov, M., and Sauer, R.~A. (2022).
\newblock A general isogeometric finite element formulation for rotation-free
  shells with in-plane bending of embedded fibers.
\newblock {\em Int. J. Numer. Meth. Eng.}, {\bf 123}(14):3115--3147.

\bibitem[Duong et~al., 2017]{solidshell}
Duong, T.~X., Roohbakhshan, F., and Sauer, R.~A. (2017).
\newblock A new rotation-free isogeometric thin shell formulation and a
  corresponding continuity constraint for patch boundaries.
\newblock {\em Comput. Methods Appl. Mech. Eng.}, {\bf 316}:43--83.

\bibitem[Echter et~al., 2013]{echter13}
Echter, R., Oesterle, B., and Bischoff, M. (2013).
\newblock A hierarchic family of isogeometric shell finite elements.
\newblock {\em Comput. Methods Appl. Mech. Eng.}, {\bf 254}:170--180.

\bibitem[Golestanian and Casquero, 2023]{golestanian23}
Golestanian, M. and Casquero, H. (2023).
\newblock Extending {CAS} elements to remove shear and membrane locking from
  quadratic {NURBS}-based discretizations of linear plane {T}imoshenko rods.
\newblock {\em Int. J. Numer. Meth. Eng.}, {\bf 124}(18):3997--4021.

\bibitem[Greco and Cuomo, 2016]{greco16}
Greco, L. and Cuomo, M. (2016).
\newblock An isogeometric implicit {$G^1$} mixed finite element for {K}irchhoff
  space rods.
\newblock {\em Comput. Methods Appl. Mech. Eng.}, {\bf 298}:325--349.

\bibitem[Greco et~al., 2018]{greco18}
Greco, L., Cuomo, M., and Contrafatto, L. (2018).
\newblock A reconstructed local {$\bar B$} formulation for isogeometric
  {K}irchhoff-{L}ove shells.
\newblock {\em Comput. Methods Appl. Mech. Eng.}, {\bf 332}:462--487.

\bibitem[Greco et~al., 2017]{greco17}
Greco, L., Cuomo, M., Contrafatto, L., and Gazzo, S. (2017).
\newblock An efficient blended mixed {B}-spline formulation for removing
  membrane locking in plane curved {K}irchhoff rods.
\newblock {\em Comput. Methods Appl. Mech. Eng.}, {\bf 324}:476--511.

\bibitem[Hiemstra et~al., 2023]{hiemstra23}
Hiemstra, R.~R., Fuentes, F., and Schillinger, D. (2023).
\newblock Fourier analysis of membrane locking and unlocking.
\newblock {\em Comput. Methods. Appl. Mech. Eng.}, {\bf 417}:116353.

\bibitem[Hu et~al., 2016]{Hu16}
Hu, P., Hu, Q., and Xia, Y. (2016).
\newblock Order reduction method for locking free isogeometric analysis of
  {T}imoshenko beams.
\newblock {\em Comput. Methods. Appl. Mech. Eng.}, {\bf 308}:1--22.

\bibitem[Hu et~al., 2020]{hu20}
Hu, Q., Xia, Y., Natarajan, S., Zilian, A., Hu, P., and Bordas, S. P.~A.
  (2020).
\newblock Isogeometric analysis of thin {R}eissner-{M}indlin shells: locking
  phenomena and {B}-bar method.
\newblock {\em Comput. Mech.}, {\bf 65}:1323--1341.

\bibitem[Hughes et~al., 2005]{hughes05}
Hughes, T. J.~R., Cottrell, J.~A., and Bazilevs, Y. (2005).
\newblock Isogeometric analysis: {CAD}, finite elements, {NURBS}, exact
  geometry and mesh refinement.
\newblock {\em Comput. Methods Appl. Mech. Eng.}, {\bf 194}:4135--4195.

\bibitem[Kiendl et~al., 2009]{kiendl09}
Kiendl, J., Bletzinger, K.-U., Linhard, J., and W{\"u}chner, R. (2009).
\newblock Isogeometric shell analysis with {K}irchhoff-{L}ove elements.
\newblock {\em Comput. Methods Appl. Mech. Eng.}, {\bf 198}:3902--3914.

\bibitem[Kikis and Klinkel, 2022]{kikis22}
Kikis, G. and Klinkel, S. (2022).
\newblock Two-field formulations for isogeometric {R}eissner-{M}indlin plates
  and shells with global and local condensation.
\newblock {\em Comput. Mech.}, {\bf 69}:1--21.

\bibitem[Kim et~al., 2022]{Kim22}
Kim, M.-G., Lee, G.-H., Lee, H., and Koo, B. (2022).
\newblock Isogeometric analysis for geometrically exact shell elements using
  {B}\'ezier extraction of {NURBS} with assumed natural strain method.
\newblock {\em Thin-Walled Struc.}, {\bf 172}:108846.

\bibitem[Koiter, 1966]{koiter66}
Koiter, W.~T. (1966).
\newblock On the nonlinear theory of thin elastic shells.
\newblock {\em Proc. Kon. Ned. Akad. Wetensch.}, {\bf B69}:1--54.

\bibitem[Leonetti et~al., 2019]{leonetti19}
Leonetti, L., Magisano, D., Madeo, A., Garcea, G., Kiendl, J., and Reali, A.
  (2019).
\newblock A simplified {K}irchhoff-{L}ove large deformation model for elastic
  shells and its effective isogeometric formulation.
\newblock {\em Comput. Methods Appl. Mech. Eng.}, {\bf 354}:369--396.

\bibitem[MacNeal and Harder, 1985]{macneal85}
MacNeal, R. and Harder, R. (1985).
\newblock A proposed standard set of problems to test finite element accuracy.
\newblock {\em Finite Elem. Anal. Des.}, {\bf 1}:3--20.

\bibitem[Mi and Yu, 2021]{mi21}
Mi, Y. and Yu, X. (2021).
\newblock Isogeometric {MITC} shell.
\newblock {\em Comput. Methods Appl. Mech. Eng.}, {\bf 377}:113693.

\bibitem[Naghdi, 1972]{naghdi72}
Naghdi, P.~M. (1972).
\newblock Theory of plates and shells.
\newblock In Truesdell, C., editor, {\em Handbuch der Physik}, pages 425--640,
  Berlin. Springer.

\bibitem[Nguyen et~al., 2022]{nguyen22}
Nguyen, T.-H., Hiemstra, R.~R., and Schillinger, D. (2022).
\newblock Leveraging spectral analysis to elucidate membrane locking and
  unlocking in isogeometric finite element formulations of the curved
  {E}uler-{B}ernoulli beam.
\newblock {\em Comput. Methods Appl. Mech. Eng.}, {\bf 388}:114240.

\bibitem[Oesterle et~al., 2016]{oesterle16}
Oesterle, B., Ramm, E., and Bischoff, M. (2016).
\newblock A shear deformable, rotation-free isogeometric shell formulation.
\newblock {\em Comput. Methods. Appl. Mech. Eng.}, {\bf 307}:235--255.

\bibitem[Park and Stanley, 1986]{park86}
Park, K.~C. and Stanley, G.~M. (1986).
\newblock A curved {$C^0$} shell element based on assumed natural-coordinate
  strains.
\newblock {\em J. Appl. Mech.}, {\bf 53}:278--290.

\bibitem[Paul and Sauer, 2022]{viscshell}
Paul, K. and Sauer, R.~A. (2022).
\newblock An isogeometric finite element formulation for boundary and shell
  viscoelasticity based on a multiplicative surface deformation split.
\newblock {\em Int. J. Numer. Meth. Eng.}, {\bf 123}(22):5570--5617.

\bibitem[Paul et~al., 2020]{patchshell}
Paul, K., Zimmermann, C., Duong, T.~X., and Sauer, R.~A. (2020).
\newblock Isogeometric continuity constraints for multi-patch shells governed
  by fourth-order deformation and phase field models.
\newblock {\em Comput. Methods Appl. Mech. Eng.}, {\bf 370}:113219.

\bibitem[Roohbakhshan and Sauer, 2017]{bioshell}
Roohbakhshan, F. and Sauer, R.~A. (2017).
\newblock Efficient isogeometric thin shell formulations for soft biological
  materials.
\newblock {\em Biomech. Model. Mechanobiol.}, {\bf 16}(5):1569--1597.

\bibitem[Sauer and Duong, 2017]{shelltheo}
Sauer, R.~A. and Duong, T.~X. (2017).
\newblock On the theoretical foundations of solid and liquid shells.
\newblock {\em Math. Mech. Solids}, {\bf 22}(3):343--371.

\bibitem[Sauer et~al., 2017]{liquidshell}
Sauer, R.~A., Duong, T.~X., Mandadapu, K.~K., and Steigmann, D.~J. (2017).
\newblock A stabilized finite element formulation for liquid shells and its
  application to lipid bilayers.
\newblock {\em J. Comput. Phys.}, {\bf 330}:436--466.

\bibitem[Savitha and Sauer, 2023]{savitha22}
Savitha, E. and Sauer, R.~A. (2023).
\newblock A new anisotropic bending model for nonlinear shells: Comparison with
  existing models and isogeometric finite element implementation.
\newblock {\em Int. J. Solids Struc.}, {\bf 268}:112169.

\bibitem[Scordelis and Lo, 1964]{scordelis64}
Scordelis, A.~C. and Lo, K.~S. (1964).
\newblock Computer analysis of cylindrical shells.
\newblock {\em J. Am. Concrete Inst.}, {\bf 61}:539--561.

\bibitem[Simo and Fox, 1989]{simo89}
Simo, J.~C. and Fox, D.~D. (1989).
\newblock On a stress resultant geometrically exact shell model. {P}art {I}:
  {F}ormulation and optimal parameterization.
\newblock {\em Comput. Methods Appl. Mech. Eng.}, {\bf 72}:267--304.

\bibitem[Steigmann, 1999]{steigmann99b}
Steigmann, D.~J. (1999).
\newblock Fluid films with curvature elasticity.
\newblock {\em Arch. Rat. Mech. Anal.}, {\bf 150}:127--152.

\bibitem[Stolarski and Belytschko, 1982]{stolarski82}
Stolarski, H. and Belytschko, T. (1982).
\newblock Membrane locking and reduced integration for curved elements.
\newblock {\em J. Appl. Mech.}, {\bf 49}:172--176.

\bibitem[Zou et~al., 2022]{zou22}
Zou, Z., Hughes, T. J.~R., Scott, M.~A., Miao, D., and Sauer, R.~A. (2022).
\newblock Efficient and robust quadratures for isogeometric analysis: Reduced
  {G}auss and {G}auss-{G}reville quadratures.
\newblock {\em Comput. Methods Appl. Mech. Eng.}, {\bf 392}:114722.

\bibitem[Zou et~al., 2021]{zou21}
Zou, Z., Hughes, T. J.~R., Scott, M.~A., Sauer, R.~A., and Savitha, E.~J.
  (2021).
\newblock Galerkin formulations of isogeometric shell analysis: {A}lleviating
  locking with {G}reville quadratures and higher-order elements.
\newblock {\em Comput. Methods Appl. Mech. Eng.}, {\bf 380}:113757.

\bibitem[Zou et~al., 2020]{zou20}
Zou, Z., Scott, M., Miao, D., Bischoff, M., Oesterle, B., and Dornisch, W.
  (2020).
\newblock An isogeometric {R}eissner-{M}indlin shell element based on
  {B}\'ezier dual basis functions: {O}vercoming locking and improved coarse
  mesh accuracy.
\newblock {\em Comput. Methods Appl. Mech. Eng.}, {\bf 370}:113283.

\end{thebibliography}
